\documentclass[onecolumn]{elsart3p}
\usepackage{graphicx,amssymb}
\usepackage{subfigure}

\usepackage[%
  breaklinks=true,%
  colorlinks=true,%
  linkcolor=blue,anchorcolor=blue,%
  citecolor=blue,filecolor=blue,%
  menucolor=blue,pagecolor=blue,%
  urlcolor=blue]{hyperref}
\usepackage{%
pstricks,
pst-plot,%
pst-coil,%
pst-node,%
multido,%
amssymb}                                  
\usepackage{axodraw}

\newcount\Farbe
\ifnum\Farbe=1
        \def\IfFarbe#1#2{#1}
\else
        \def\IfFarbe#1#2{#2 }
\fi

\def\amu{a_\mu}
\def\amuh{a_\mu^{{\mathrm{had}}}}
\def\amuhlo{a_\mu^{(4)}(\mathrm{vap,\,had})}
\def\amuhho{a_\mu^{(6)}(\mathrm{vap,\,had})}

\def\dah0{\Delta\alpha^{(5)}_{\rm had}(-s_0)}
\def\dah{\Delta\alpha^{(5)}_{\rm had}}

\def\order#1{{\cal O}\left(#1\right)}

\def\sf{spectral function}
\def\sfs{spectral functions}

\newcommand{\al}{\alpha}
\newcommand{\aldpib}{\left(\frac{\alpha}{\pi}\right)}
\newcommand{\aldpi}{\frac{\alpha}{\pi}}

\newcommand{\amm}{anomalous magnetic moment }

\newcommand{\Amu}{A_\mu}
\newcommand{\anu}{\bar{\nu}}
\newcommand{\ba}{\begin{eqnarray}}
\newcommand{\bary}{\begin{array}}
\newcommand{\bb}{\begin{thebibliography}}
\newcommand{\bdm}{\begin{displaymath}}
\newcommand{\bea}{\begin{eqnarray*}}
\newcommand{\be}{\begin{equation}}
\newcommand{\beq}{\begin{equation}}
\newcommand{\bet}{\begin{center} \begin{tabular}}

\newcommand{\bit}{\begin{itemize}}
\newcommand{\bpi}{\beta_\pi}
\newcommand{\bra}[1]{\langle{#1}|}
\newcommand{\braket}[1]{\langle{#1}\rangle}

\newcommand{\co}{\; \; ,}

\newcommand{\crn}{\nonumber \\}
\newcommand{\cs}{ \;,\;\;}

\newcommand{\css}{ \:,\;}

\newcommand{\der}{\Delta r}

\newcommand{\downrightarrow}{\hspace{-3.3ex} \raisebox{0.2ex}{$\stackrel{ \!\! \mid }{\raisebox{-0.4ex}[0ex][0ex]{$ \;\;\;\;\;\; \longrightarrow \;\; $}}$}}

\newcommand{\ea}{\end{eqnarray}}
\newcommand{\eary}{\end{array}}
\newcommand{\eb}{\end{thebibliography}}

\newcommand{\ecc}{\;\;,\;\;\;}

\newcommand{\edm}{\end{displaymath}}
\newcommand{\eea}{\end{eqnarray*}}
\newcommand{\ee}{\end{equation}}
\newcommand{\eeq}{\end{equation}}
\newcommand{\eit}{\end{itemize}}

\newcommand{\ent}{\end{tabular} \end{center}}

\newcommand{\epm}{e^+e^-}

\newcommand{\epo}{ \;.}

\newcommand{\eqp}{\;\;.}

\newcommand{\Fem}{F_\pi}
\newcommand{\FF}{{\cal F}_{\pi^{0*}\gamma^*\gamma^*}}
\newcommand{\FFa}{{\cal F}_{\pi^0\gamma^*\gamma^*}}

\newcommand{\FFc}{{\cal F}_{\pi^{0*}\gamma^*\gamma}}

\newcommand{\FFac}{{\cal F}_{\pi^0\gamma^*\gamma}}

\newcommand{\FFabc}{{\cal F}_{\pi^0\gamma\gamma}}

\newcommand{\fH}{\mbox{{\grfett\symbol{'110}}}}

\newcommand{\gafi}{\gamma_5}
\newcommand{\gamuc}{\gamma^{\mu}}

\newcommand{\gapprox}{\raisebox{-.2ex}{$\stackrel{\textstyle>}{\raisebox{-.6ex}[0ex][0ex]{$\sim$}}$}}
\newcommand{\g}{\gamma}
\newcommand{\gimu}{\gamma^{\mu}}

\newcommand{\gi}{$g$-$i$}
\newcommand{\gis}{$g$-$i$ }

\newcommand{\gv}{\mbox{GeV}}

\newcommand{\VVP}{\langle V V \! P\rangle}

\newcommand{\lag}{{\cal L}}
\newcommand{\ha}{\frac{1}{2}}

\newcommand{\intzi}{\int\limits_{0}^{\infty}\:}
\newcommand{\intzo}{\int\limits_{0}^{1}\:}
\newcommand{\ie}{\I \veps}
\newcommand{\Impa}{\mbox{Im} \:}

\newcommand{\ket}[1]{|{#1}\rangle}

\newcommand{\lapprox}{\raisebox{-.2ex}{$\stackrel{\textstyle<}{\raisebox{-.6ex}[0ex][0ex]{$\sim$}}$}}

\newcommand{\Liz}[1]{\mathrm{Li}_2(#1)}

\newcommand{\lsim}{\mbox{\raisebox{-0.3ex}{\footnotesize $\:\stackrel{<}{\sim}\:$}} }

\newcommand{\MSb}{$\overline{\mathrm{MS}}$ }

\newcommand{\mv}{\mbox{MeV}}

\newcommand{\mz}{M^2_Z}
\newcommand{\nn}{\nonumber}
\newcommand{\noi}{\noindent}

\newcommand{\ofx}{(x)}

\newcommand{\pamu}{\partial_\mu}

\newcommand{\power}[1]{\times 10^{#1}}

\newcommand{\ppm}{\pi^+ \pi^-}
\newcommand{\ppx}[2]{\frac{\partial {#1}}{\partial {#2}}}

\newcommand{\ra}{\rightarrow}

\newcommand{\Repa}{\mbox{Re} \:}

\newcommand{\sighad}{\sigma_{\rm tot} (\epm \ra \gamma^* \ra {\rm hadrons})}
\newcommand{\sigqua}{\sigma_{\rm tot} (\epm \ra \gamma^* \ra \sum_q q\bar{q},\,q\bar{q}g,\cdots)}

\newcommand{\sigha}{\sigma_{\rm had}}

\newcommand{\sigmmbp}{\sigma_{\mu\mu,\,0 }}

\newcommand{\sla}[1]{#1 \!\!\!/}
\newcommand{\SL}{{\cal L}}

\newcommand{\Sp}{\mathrm{Li}_2 }

\newcommand{\ttcb}[1]{\multicolumn{2}{c|}{#1}}
\newcommand{\ttc}[1]{\multicolumn{2}{c}{#1}}

\newcommand{\Tr}{\mbox{Tr\,}}
\newcommand{\tv}{\mbox{TeV}}

\newcommand{\veps}{\varepsilon}
\newcommand{\vphi}{\varphi}
\newcommand{\WTis}{Ward-Takahashi identities }

\newcommand{\wz}{\sqrt{2}}

\newcommand{\rhon}{\rho^{\,0}}
\newcommand{\coal}{\cos \alpha}

\newcommand{\cobet}{\cos \beta}

\newcommand{\sial}{\sin \alpha}

\newcommand{\sibet}{\sin \beta}

\newcommand{\amuSUSY}{a_\mu^{\rm SUSY}}
\newcommand{\amuSUOL}{a_\mu^{\rm SUSY,1L}}
\newcommand{\amuSUTLferm}{a_\mu^{\rm SUSY,ferm,2L}}
\newcommand{\amuSUTLbos}{a_\mu^{\rm SUSY,bos,2L}}
\newcommand{\tunit}{\times10^{-10}}
\newcommand{\MSUSY}{M_{\rm SUSY}}
\newcommand{\cA}{{\cal A}}

\newcommand{\cC}{{\cal C}}

\newcommand{\cK}{{\cal K}}
\newcommand{\cL}{{\cal L}}
\newcommand{\cM}{{\cal M}}

\newcommand{\cO}{{\cal O}}
\newcommand{\cP}{{\cal P}}
\newcommand{\cQ}{{\cal Q}}
\newcommand{\cR}{{\cal R}}

\newcommand{\NC}[1]{{Nuovo Cimento \textbf{#1}}}
\newcommand{\NCA}[1]{{Nuovo Cimento A \textbf{#1}}}

\newcommand{\np}[1]{{Nucl. Phys. B \textbf{#1}}}
\newcommand{\npb}[1]{{Nucl. Phys. B \textbf{#1}}}

\newcommand{\pl}[1]{{Phys. Lett. \textbf{#1}B}}        
\newcommand{\PLB}[1]{{Phys. Lett. B \textbf{#1}}}
\newcommand{\plb}[1]{{Phys. Lett. B \textbf{#1}}}       

\newcommand{\pr}[1]{{Phys. Rev. \textbf{#1}}}
\newcommand{\prd}[1]{{Phys. Rev. D \textbf{#1}}}

\newcommand{\prl}[1]{{Phys. Rev. Lett. \textbf{#1}}}

\newcommand{\ZPC}[1]{{Z. Phys. C \textbf{#1}}}

\newfont{\liste}{pzdr scaled 1100}
\newfont{\grfett}{cmmib10 scaled 1100}
\newcommand{\D}{\mathrm{d}}
\newcommand{\I}{\mathrm{i}}
\newcommand{\E}{\mathrm{e}}
\DeclareMathSymbol{\varPhi}{\mathalpha}{operators}{"08}
\DeclareMathSymbol{\varOmega}{\mathalpha}{operators}{"0A}

\makeatletter
\def\myfootnotesize{\@setsize\footnotesize{10pt}\ixpt\@ixpt} 

\@ifundefined{square}{}{\let\Box\square}
\makeatother

\begin{document}

\newcommand{\mysymb}[5]{
  \unitlength1mm 
  \begin{picture}(#4,0) 
    \put(0,0){ 
 }
  \end{picture} 
}
~
\begin{flushright}\begin{tabular}{c}
HU-EP-09/07, 
HRI-P-09-02-001,
RECAPP-HRI-2009-003
\end{tabular}
\end{flushright}
\begin{frontmatter}

\title{The Muon g-2}
\author[Berlin,Katowice]{Fred Jegerlehner\corauthref{cor}},
\corauth[cor]{Corresponding author.}
\ead{fjeger@physik.hu-berlin.de}
\ead[url]{www-com.physik.hu-berlin.de/\~{}fjeger/}
\author[Allahabad]{Andreas Nyf\/feler}
\ead{nyffeler@hri.res.in}
\ead[url]{www.hri.res.in/\~{}nyf\/feler}

\address[Berlin]{Humboldt-Universit\"at zu Berlin, Institut f\"ur Physik,
       Newtonstrasse 15, D-12489 Berlin, Germany}
\address[Katowice]{Institute of Physics, University of Silesia,
ul. Uniwersytecka 4, PL-40007 Katowice, Poland} 
\address[Allahabad]{Regional Centre for Accelerator-based Particle Physics,
Harish-Chandra Research Institute, \\
Chhatnag Road, Jhusi, Allahabad - 211~019, India}

\begin{abstract}
The muon anomalous magnetic moment is one of the most precisely measured
quantities in particle physics. In a recent experiment at Brook\-ha\-ven it
has been measured with a remarkable 14-fold improvement of the previous CERN
experiment reaching a precision of 0.54ppm. Since the first results were
published, a persisting ``discrepancy'' between theory and experiment of about
3 standard deviations is observed.  It is the largest ``established''
deviation from the Standard Model seen in a ``clean'' electroweak observable
and thus could be a hint for New Physics to be around the corner.  This
deviation triggered numerous speculations about the possible origin of the
``missing piece'' and the increased experimental precision animated a
multitude of new theoretical efforts which lead to a substantial improvement
of the prediction of the muon anomaly $a_\mu=(g_\mu-2)/2$. The dominating
uncertainty of the prediction, caused by strong interaction effects, could be
reduced substantially, due to new hadronic cross section measurements in
electron-positron annihilation at low energies. Also the recent electron $g-2$
measurement at Harvard contributes substantially to the progress in this
field, as it allows for a much more precise determination of the fine
structure constant $\alpha$ as well as a cross check of the status of our
theoretical understanding.

In this report we review the theory of the anomalous magnetic moments of the
electron and the muon.  After an introduction and a brief description of the
principle of the muon $g-2$ experiment, we present a review of the status of
the theoretical prediction and in particular discuss the role of the hadronic
vacuum polarization effects and the hadronic light--by--light scattering
correction, including a new evaluation of the dominant pion-exchange
contribution. In the end, we find a 3.2 standard deviation discrepancy between
experiment and Standard Model prediction. We also present a number of examples
of how extensions of the electroweak Standard Model would change the
theoretical prediction of the muon anomaly $a_\mu$. Perspectives for future
developments in experiment and theory are briefly discussed and critically
assessed.  The muon $g-2$ will remain one of the hot topics for further
investigations.
\end{abstract}

\end{frontmatter}

\newpage
\tableofcontents

\section{Introduction}
\label{sec:intro}
The electron's spin and magnetic moment were evidenced from the deflection of
atoms in an inhomogeneous magnetic field and the observation of fine structure
by optical spectroscopy~\cite{SternGerlach24,GUspin}.  Ever since, magnetic
moments and $g$--values of particles in general and the $g-2$ experiments with
the electron and the muon in particular, together with high precision atomic
spectroscopy, have played a central role in establishing the modern
theoretical framework for particle physics. That is relativistic quantum field
theory in general and quantum electrodynamics in particular, the prototype
theory which developed further into the so called ``Standard Model'' (SM) of
electromagnetic, weak and strong interactions based on a local gauge principle
and spontaneous symmetry breaking. The muon $g-2$ is one of the most precisely
measured and theoretically best investigated quantities in particle physics.
Our interest in very high precision measurements is motivated by our eagerness
to exploit the limits of our present understanding of nature and to find
effects which cannot be explained by the established theory. More than 30
years after its invention this is still the SM of elementary particle
interactions, a $SU(3)_c \otimes SU(2)_L \otimes U(1)_Y$ gauge theory broken
to $SU(3)_c \otimes U(1)_{\mathrm{em}}$ by the Higgs mechanism, which requires
a not yet discovered Higgs particle to exist.

Designed to be a local, causal and renormalizable quantum field theory, quarks
and leptons are allowed to come only in families in order to be anomaly free
and not to conflict with renormalizability.  So we have as the first family
the quark doublet $(u, d)$ of the up and down quarks accompanied by the lepton
doublet $(\nu_e, e^-)$ with the electron neutrino and the electron, with the
left-handed fields in the doublets and all their right-handed partners in
singlets. All normal matter is made up from these 1st family fermions.

Most surprisingly a second and even a third quark--lepton family exist in
nature, all with identical quantum numbers, as if nature would repeat
itself. The corresponding members in the different families only differ by
their mass where the mass scales span an incredible range, from
$m_{\nu_e}\lapprox 10^{-3}$ eV for the electron neutrino to $m_t\simeq 173$
GeV for the top quark. The existence of three families allows for an extremely
rich pattern of all kinds of phenomena which derive from the natural
possibility of mixing of the horizontal vectors in family space formed by the
members with identical quantum numbers. The most prominent new effects only
possible with three or more families is CP violation.

The first member of the second family that was discovered was the muon
($\mu$).  It was discovered in cosmic rays by Anderson \& Neddermeyer in
1936~\cite{Anderson36}, only a few years after Anderson~\cite{Anderson32} had
discovered (also in cosmic rays) in 1932 antimatter in form of the positron, a
``positively charged electron'' as predicted by Dirac in
1930~\cite{Diracanti}. The $\mu$ was another version of an electron, just a
heavier copy, and was extremely puzzling for physicists at that time. Its true
nature only became clear much later after the first precise $g-2$ experiments
had been performed.  In fact the muon turns out to be a very special object in
many respects as we will see and these particular properties make it to play
a crucial role in the development of elementary particle theory.

The charged leptons primarily interact electromagnetically with the
photon and weakly via the heavy gauge bosons $W$ and $Z$, as well as
very much weaker also with the Higgs. Puzzling enough, the three
leptons $e$, $\mu$ and $\tau$ have identical properties, except for
the masses which are given by $m_e=0.511~ \mv $, $m_\mu=105.658~ \mv $
and $m_\tau=1776.99~ \mv $, respectively. As masses differ by orders
of magnitude, the leptons show very different behavior, the most
striking being the very different lifetimes. Within the SM the
electron is stable on time scales of the age of the universe, while
the $\mu$ has a short lifetime of $\tau_\mu=2.197 \times 10^{-6}$
seconds and the $\tau$ is even more unstable with a lifetime
$\tau_\tau=2.906 \times 10^{-13}$ seconds only. Also, the decay
patterns are very different: the $\mu$ decays very close to 100\% into
electrons plus two neutrinos ($e
\bar{\nu}_e \nu_\mu$), however, the $\tau$ decays to about 65\%
into hadronic states $\pi^- \nu_\tau\cs \pi^- \pi^0\nu_\tau\cs,
\cdots$ while the main leptonic decay modes only account for 17.36\%
($\mu^-\bar{\nu}_\mu\nu_\tau$) and 17.85\% ($e^-\bar{\nu}_e\nu_\tau$), 
respectively. This has a dramatic impact on the possibility to study
these particles experimentally and to measure various properties
precisely. The most precisely studied lepton is the electron, but
the muon can also be explored with extreme precision. Since the muon turns out
to be much more 
sensitive to hypothetical physics beyond the SM than the electron
itself, the muon is much more suitable as a ``crystal ball'' which
could give us hints about not yet uncovered physics. The reason is that
some effects scale with powers of $m_\ell^2$, as we will see
below. Unfortunately, the $\tau$, where new physics effects would be even
better visible, is so short lived, that corresponding
experiments are not possible with present technology.

As important as charge, spin, mass and lifetime, 
are the magnetic and electric \textit{dipole moments} which are 
typical for spinning particles like the leptons. 
Both electrical and magnetic properties have their origin in the electrical
\textit{charges and their currents}. Magnetic monopoles are not
necessary to obtain magnetic moments.
On the classical level, an orbiting particle with electric charge $e$ and mass $m$
exhibits a magnetic dipole moment  
given by 
\be
\vec{\mu}_L= \frac{e}{2m}\:\vec{L}
\label{Orbitalmm}
\ee 
where $\vec{L}=m\:\vec{r} \times \vec{v}$ is the orbital angular
momentum ($\vec{r}$ position, $\vec{v}$ velocity). An electrical
dipole moment can exist due to relative displacements of the centers of
positive and negative electrical charge distributions.  
Magnetic and electric moments contribute to the electromagnetic 
interaction Hamiltonian with magnetic and electric fields
\be
{\cal H}=-\vec{\mu}_m \cdot \vec{B} - \vec{d}_e \cdot \vec{E}\cs
\label{diploeHamiltonian}
\ee 
where $\vec{B}$ and $\vec{E}$ are the magnetic and electric field
strengths and $\vec{\mu}_m$ and $\vec{d}_e$ the magnetic and electric
dipole moment operators. Usually, we measure magnetic moments in units
of the \textit{Bohr magneton} $\mu_B$ which is defined as follows 
\be
\mu_B=\frac{e}{2m_e}=5.788381804(39) \power{-11}~~\mv
\mathrm{T}^{-1} \epo
\label{mBhor}
\ee
Here T as a unit stands for 1 Tesla = $10^4$ Gauss.\footnote{We will
  use the SI system of units, where $\mathrm{T} = \mathrm{V s}
  \mathrm{m}^{-2}$ and the electric charge $e$ is measured in Coulomb. The
  Bohr magneton is then defined by $\mu_B=e\hbar / 2m_e$, but we will set
  $\hbar = c = \epsilon_0 = 1$ throughout this article.}  

For a particle with spin the magnetic moment is \textbf{intrinsic}
and obtained by replacing the the angular momentum operator $\vec{L}$
by the \textit{spin operator}
\be
\vec{S}=\frac{\vec{\sigma}}{2} \cs 
\label{spinop}
\ee 
where $\sigma_i$ ($i=1,2,3$) are the Pauli spin matrices. 
Thus, generalizing the classical form (\ref{Orbitalmm}) of the orbital
magnetic moment, one writes
\be \vec{\mu}_m = g\: Q\: \mu_0\:
\frac{\vec{\sigma}}{2} \ecc \vec{d}_e = \eta \: Q\: \mu_0\:
\frac{\vec{\sigma}}{2}\cs
\label{dipoleform}
\ee
where $\mu_0=e/2m$, $Q$ is the
electrical charge in units of $e$, $Q=-1$ for the leptons ($\ell=e,\mu,\tau$),
$Q=+1$ for the antileptons and $m$ is the mass. The equations define the
gyromagnetic ratio $g$ ($g$-factor) and its electric pendant
$\eta$, respectively, quantities exhibiting important dynamical
information about the leptons as we will see later. The
deviation from the Dirac value $g_\ell/2=1$, obtained at the classical level,
is 
\be
a_\ell \equiv \frac{g_\ell-2}{2} 
\label{amudef}
\ee 
the famous \textit{anomalous magnetic moment} and $\amu$ is the
quantity in the focus of this review.

The magnetic interaction term gives rise to the well known
\textit{Zeeman effect}: level splitting seen in atomic spectra. 
If spin is involved one calls it \textit{anomalous Zeeman
effect}. The latter obviously is suitable to study the magnetic moment 
of the electron by investigating atomic spectra in magnetic fields.

The most important condition for the \amm to be a useful monitor for
testing a theory is its unambiguous predictability within that theory. The
predictability crucially depends on the following properties
of the theory:
\bit
\item it must be a local relativistic quantum field theory and
\item it must be renormalizable.
\eit
This implies that $g-2$ vanishes at tree level and cannot be an
independently adjustable parameter in any renormalizable QFT. This in
turn implies that for a given theory [model] $g-2$ is an unambiguously
calculable quantity and the predicted value can be confronted with
experiments. Its model dependence makes $\amu$ a good monitor for the
detection of new physics contributions. The key point is that $g-2$ can
be both precisely predicted as well as experimentally measured with
very high accuracy.  By confronting precise theoretical predictions
with precisely measured experimental data it is possible to subject
the theory to very stringent tests and to find its possible
limitations.
 
The \amm of a lepton is a dimensionless quantity, a number, which in
QED may be computed order by order as an expansion in the fine
structure constant $\alpha$. Beyond QED, in the SM or extensions of it, 
weak and strong coupling contributions are calculable. As a matter of
fact, the interaction of the lepton with photons or other particles
induces an effective interaction term 
\be
\delta {\cal L}_{\mathrm{eff}}^{\mathrm{AMM}}=
-\frac{\delta g}{2}\,\frac{e}{4m}
\left\{ \bar{\psi}_L \ofx \:\sigma^{\mu \nu}F_{\mu
\nu} \ofx \:\psi_R \ofx +\bar{\psi}_R
\ofx \:\sigma^{\mu \nu}F_{\mu
\nu} \ofx\:\psi_L \ofx \right\}\: 
\label{Lammeff}
\ee
where $\psi_L$ and $\psi_R$ are Dirac fields of negative (left--handed
$L$) and positive (right--handed $R$) chirality and $F_{\mu
\nu}=\partial_\mu A_\nu-\partial_\nu A_\mu$ is the electromagnetic
field strength tensor.
It corresponds to a dimension 5
operator and since a renormalizable theory is constrained to exhibit
terms of dimension 4 or less only, such a term must be absent for any
fermion in any renormalizable theory at tree level.

The dipole moments are very interesting quantities for the study of
the discrete symmetries. A basic consequence of any relativistic local
QFT is \textit{charge conjugation} $C$, the particle--antiparticle
duality~\cite{Diracanti} or crossing property, which implies in the
first place that particles and antiparticles have identical masses and
spins. In fact, charge conjugation turned out not to be a universal
symmetry in nature. Since an antiparticle
may be considered as a particle propagating backwards in time,
charge conjugation has to be considered together with
\textit{time-reversal} $T$ (time-reflection), which in a relativistic
theory has to go together with \textit{parity} $P$ (space-reflection).
The $CPT$ theorem says: the product of the three discrete transformations,
$C$, $P$ and $T$, taken in any order, is a symmetry of any relativistic
local QFT. Actually, 
in contrast to the individual transformations $C$, $P$ and $T$, which
are symmetries of the electromagnetic-- and strong--interactions only,
$CPT$ is a universal symmetry and it is this symmetry which guarantees
that particles and antiparticles have identical masses
and lifetimes in theories like the SM, where $C$, $P$ and $T$ are
not conserved.

The properties of the dipole moments under $C$, $P$ and $T$
transformations may be obtained easily by inspecting the interaction
Hamiltonian in Eq.~(\ref{diploeHamiltonian}). Naively, one would expect that
electromagnetic (QED) and strong interactions (QCD) are giving the
dominant contributions to the dipole moments. However, both preserve
$P$ and $T$ and thus the corresponding contributions to
(\ref{diploeHamiltonian}) must conserve these symmetries as well.  On
the one hand, both the magnetic and the electric dipole moment
$\vec{\mu}_m$ and $\vec{d}_e$ are axial vectors as they are
proportional to the spin vector $\vec{\sigma}$. On the other hand, the
electromagnetic fields $\vec{E}$ and $\vec{B}$ transform as vector and
axial vector, respectively. An axial vector changes sign under $T$, but
not under $P$, while a vector changes sign under $P$, but not under
$T$.  Hence, in $P$ and/or $T$ conserving theories only the magnetic
term $-\vec{\mu}_m \cdot \vec{B}$ is allowed while an electric dipole
term $-\vec{d}_e \cdot \vec{E}$ is forbidden. Consequently, $\eta$ in
(\ref{dipoleform}) would have to vanish exactly. However, as the weak
interactions violate parity maximally, weak contributions to $\eta$
cannot be excluded by the parity argument. The actual constraint here
comes from $T$, which by the $CPT$--theorem is equivalent to $CP$. $CP$
is also violated by the weak interactions, but only via fermion family
mixing in the Yukawa sector of the SM.  Therefore, electron and muon
electric dipole moments are suppressed by approximate $T$ invariance
in the light fermion sector at the level of second order weak
interactions (for a theoretical review
see~\cite{Hoogeveen90,BeSuEDM90}). In fact experimental bounds tell us
that they are very tiny~\cite{eEDM}
\be
|d_e|<1.6 \times 10^{-27}\,e\cdot \mathrm{cm \ at \  90\,\%\,C.L.}
\label{eEDM}
\ee
This limit also plays an important role in the extraction of $\amu$ from
experimental data, as we will see later. A new dedicated experiment for
measuring the muon electric dipole moment (EDM) in a storage ring is under
discussion~\cite{NewEDMexp}.

Berestetskii's argument of a dramatically enhanced
sensitivity~\cite{Berestetskii56} for short distances and for heavy
new physics states attracted new attention for the muon anomalous magnetic
moment. One of 
the main features of the \amm of leptons is that it mediates helicity
flip transitions. For massless particles helicity would be conserved
by all gauge boson mediated interactions and helicity flips would be
forbidden.  For massive particles helicity flips are allowed and their
transition amplitude is proportional to the mass of the
particle. Since the transition probability goes with the modulus
square of the amplitude, for the lepton's \amm this implies that
quantum fluctuations due to heavier particles or contributions from
higher energy scales are proportional to
\be
\delta a_\ell \propto \frac{m_\ell^2}{M^2}
~~~~~~~~(M \gg m_\ell)\cs
\label{Berestetskiirelation}
\ee
where $M$ may be
\bit
\item the mass of a heavier SM particle, or
\item the mass of a hypothetical heavy state beyond the SM, or
\item an energy scale or an ultraviolet cut--off where the SM
ceases to be valid.
\eit
Since the sensitivity to ``new physics'' grows quadratically with the mass of
the lepton, the interesting effects are magnified in
$a_\mu$ relative to $a_e$ by a factor $(m_\mu/m_e)^2
\sim 4 \times 10^{4}$ at a given resolution (precision). Yet, the
heavier the state, the smaller the effect (it decouples quadratically
as $M \to \infty$).  Thus we have the best sensitivity for nearby new
physics, which has not yet been discovered by other experiments. This
is why $a_\mu$ is a predestinated ``monitor for new physics''. By far
the best sensitivity we would have for $a_\tau$, if we could measure
it with comparable precision. This, however, is beyond present
experimental possibilities, because of the very short lifetime of the
$\tau$.\footnote{No real measurement exists yet for $a_\tau$. Theory
predicts $a_\tau=117721(5)\times 10^{-8}$; the experimental
limit from the LEP experiments OPAL and L3 is $ -0.052 < a_\tau <
0.013$ at 95\% CL~\cite{tauammexp}.}

Until about 1975 searching for ``new physics'' via $a_\mu$ in fact
essentially meant looking for physics beyond QED. As we will see
later, also SM hadronic and weak interaction effect carry
the enhancement factor $(m_\mu/m_e)^2$, and this is good news and bad
news at the same time. Good news because of the enhanced sensitivity
to many details of SM physics like the weak gauge boson contributions,
bad news because of the enhanced sensitivity to the hadronic
contributions which are very difficult to control and in fact limit
our ability to make predictions at the desired precision. This is why 
the discussion of the hadronic contributions will cover a large fraction 
of this review.

The pattern of lepton \amm physics which emerges is the following:
$a_e$ is a quantity which is dominated by QED effects up to very high
precision, presently at the .24 parts per billion (ppb) level! The
sensitivity to hadronic and weak effects as well as the sensitivity to
physics beyond the SM is very small. This allows for a very solid and
model independent (essentially pure QED) high precision prediction of
$a_e$. The very precise experimental value and the very good control
of the theory part in fact allows us to determine the fine structure
constant $\alpha$ with the highest accuracy in comparison with other
methods. A very precise value for $\alpha$ of course is needed as an
input to be able to make precise predictions for other observables
like $a_\mu$, for example. While $a_e$, theory-wise, does not attract
too much attention, although it requires to push QED calculation to
high orders, $a_\mu$ is a much more interesting and theoretically
challenging object, sensitive to all kinds of effects and thus probing
the SM to much deeper level. Note that in spite of the fact that $a_e$
has been measured about 2250 times more precisely than $a_\mu$, the
sensitivity of the latter to ``new physics'' is still about 19 times
larger. The experimental accuracy achieved in the past few years at
BNL is at the level of 0.54 parts per million (ppm) and better than
the accuracy of the theoretical predictions which are still obscured
by hadronic uncertainties. A small discrepancy at the 2 to 3 $\sigma$
level persisted~\cite{BNL99}--\cite{BNL04} since the first new
measurement in 2000 up to the one in 2004 (four independent
measurements during this time), the last for the time being. 
The ``disagreement'' between theory and experiment, suggested by the
first BLN measurement, rejuvenated the interest in the subject and
entailed a reconsideration of the theory predictions. 
Soon afterwards, in Ref.~\cite{KnechtNyffeler01} a sign error was discovered
in previous calculations of the problematic hadronic light--by--light
scattering contribution. The change improved the agreement between theory and
experiment by about 1 $\sigma$. 
Problems with the hadronic $\epm$--annihilation data used to 
evaluate the hadronic vacuum polarization contribution led to a
similar shift in opposite direction, such that a small though
noticeable discrepancy persists. Once thought as a QED test, today the
precision measurement of the \amm of the muon is a test of most
aspects of the SM, including the electromagnetic, the strong and the weak
interaction effects. And more, if we could establish that
supersymmetry is responsible for the observed deviation, for example,
it would mean that we are testing a supersymmetric extension of the SM
and constraining its parameter space, already now. There are many
excellent and inspiring introductions and overviews on the
subject~\cite{BaileyPicasso70}--\cite{Eidelman08}
which were very useful in preparing this article.
The reader can find many more details in the book~\cite{Jegerlehner:2008zz}.

  \subsection{History}

In principle, the \amm is an \emph{observable} which can be relatively
easily studied experimentally from the precise analysis of the motion
of the lepton in an external magnetic field. For rather unstable
particles like the muon, not to talk about the $\tau$, obviously the
problems are more involved. In case of the electron the observation of
magnetic moments started with the Stern-Gerlach
experiment~\cite{SternGerlach24} in 1924 and with Goudsmit and
Uhlenbeck's~\cite{GUspin} postulate that an electron has an intrinsic
angular momentum $\ha$, and that associated with this spin
angular momentum there is a magnetic dipole moment equal to
$e/2m_e$. The quantum mechanical theory of the electron spin,
where $g$ remains a free parameter, was formulated by Pauli in
1927~\cite{Paulispin}.  Soon later, in 1928 Dirac presented his
relativistic theory of the electron~\cite{Diracmm}.

Unexpectedly but correctly, the Dirac theory predicted $g=2$ for a
free electron~\cite{Diracmm}, twice the value $g=1$ known to be
associated with orbital angular momentum. Already in 1934 Kinster and
Houston succeeded in confirming Dirac's prediction
$g_e=2$~\cite{KinsterHouston34}.  Their measurement strongly supported
the Dirac theory, although experimental errors were relatively large.
To establish that the electron's magnetic moment actually exceeds 2 by
about 0.12\%, required more than 20 years of experimental
efforts~\cite{Kusch}. Essentially as long as it took the theoreticians
to establish the first prediction of an ``anomalous'' contribution
Eq.~(\ref{amudef}) to the magnetic moment. Only after the breakthrough
in understanding and handling renormalization of QED (Tomonaga,
Schwinger, Feynman, and others around 1948~\cite{TSFD}) unambiguous
predictions of higher order effects became possible. In fact the
calculation of the leading (one--loop diagram) contribution to the
\amm by Schwinger in 1948~\cite{Schwinger48} was one of the very first
higher order QED predictions. The result
\be 
a^{\mathrm{QED}(2)}_\ell = \frac{\alpha}{2\pi} \cs
(\ell=e,\mu,\tau)
\label{Sch48}
\ee 
established in theory the effect from quantum fluctuations via virtual
electron photon interactions. In QED this value is universal for all
leptons.  Before theory solved that problem, in 1947 Nafe, Nelson and
Rabi~\cite{NNR47} reported an anomalous value of about 0.26 \% in the
hyperfine splitting of hydrogen and deuterium. The result was very
quickly confirmed by Nagle et al.~\cite{NJZ47}, and
Breit~\cite{Breit47} suggested that an anomaly $g\neq2$ of the
magnetic moment of the electron could explain the effect. Kusch and
Foley~\cite{Kusch48} presented the first precision determination of
the magnetic moment of the electron $g_e=2.00238(10)$ in 1948, just
before the theoretical result had been settled. They had studied the
hyperfine--structure of atomic spectra in a constant magnetic field.
Together with Schwinger's result $a_e^{(2)}=\alpha/(2\pi)\simeq
0.00116$ (which accounts for 99 \% of the anomaly) this provided one
of the first tests of the virtual quantum corrections, predicted by a
relativistic quantum field theory. At about the same time, the
discovery of the fine structure of the hydrogen spectrum (Lamb--shift)
by Lamb and Retherford~\cite{Lambshift_ex} in 1947 and the corresponding
calculations by Bethe, Kroll \& Lamb and Weisskopf
\& French~\cite{Lambshift_th} in 1949 provided the second triumph in testing 
QED by precision experiments beyond the tree level. These events had a
dramatic impact in establishing quantum field theory as a general
framework for the theory of elementary particles and for our
understanding of the fundamental interactions. It stimulated the
development of QED in particular and the concepts of quantum field
theory in general. The extension to non-Abelian gauge theories finally
lead us to the SM, at present our established basis for understanding
the world of elementary particles. All this structure today is
crucial for obtaining sufficiently precise predictions for the \amm of
the muon as we will see.

In 1956 Berestetskii et al.~\cite{Berestetskii56} pointed out that the
sensitivity of $a_\ell$ to short distance physics scales like
Eq.~(\ref{Berestetskiirelation}) where $M=\Lambda$ is an UV cut--off
characterizing the scale of new physics. At that time $a_e$ was
already well measured by Crane et al.~\cite{Crane}, but it was
clear that the \amm of the muon would be a much better probe for
possible deviations from QED. But how to measure $a_\mu$?

The breakthrough came in 1957 when Lee and Yang suggested parity
violation by weak interaction processes~\cite{Pviol}.  It immediately
became clear that muons produced in weak decays of the pion ($\pi^+
\to \mu^+ + $ neutrino) should be longitudinally polarized. In
addition, the decay positron of the muon ($\mu^+ \to e^++2 $
neutrinos) could indicate the muon spin direction. Garwin, Lederman
and Weinrich~\cite{Garwin57} and Friedman and
Telegdi~\cite{Friedman57}\footnote{The latter reference for the first
time points out that $P$ and $C$ are violated simultaneously, in fact
$P$ is maximally violated while $CP$ is to a very good approximation
conserved in this decay.} were able to confirm this pattern in a
convincing way. The first of the two papers for the first time
determined $g_\mu=2.00$ within 10\% by applying the muon spin
precession principle. Now the road was free to seriously think about
the experimental investigation of $a_\mu$.

The first measurement of the \amm of the muon was performed at
Columbia University in 1960~\cite{Columbia60}. The result $a_\mu=0.00122(8)$ at a
precision of about 5\% showed no difference with the electron. Shortly
after in 1961, the first precision determination was possible at the
CERN cyclotron (1958-1962)~\cite{CERN62,CERN65}. Surprisingly, nothing
special was observed within the 0.4\% level of accuracy of the
experiment. This provided the first real evidence that the muon was
just a heavy electron. It meant that the muon was a point--like double
of the electron and no extra short distance effects could be
seen. This latter point of course is a matter of accuracy and the
challenge to investigate the muon structure further was evident.

The idea of a \textit{muon storage ring} was put forward next.  A first one
was successfully realized at CERN
(1962-1968)~\cite{CERN66,CERN68,CERN72}.  It allowed to measure
$a_\mu$ for both $\mu^+$ and $\mu^-$ at the same machine. Results
agreed well within errors and provided a precise verification of the
CPT theorem for muons. An accuracy of 270 ppm was reached and an
insignificant 1.7 $\sigma$ deviation from theory was
found. Nevertheless the latter triggered a reconsideration of
theory. It turned out that in the estimate of the three--loop
$O(\alpha^3)$ QED contribution the leptonic light--by--light
scattering part (dominated by the electron loop) was missing. Aldins
et al.~\cite{LBLlep} then calculated this and after including it,
perfect agreement between theory and experiment was obtained.

The first successes of QED predictions and the growing precision of the $a_e$
experiments challenged many particle theorists to tackle the much more
difficult higher order calculations for $a_e$ as well as for $a_\mu$.  Many of
these calculations were strong motivations for inventing and developing
computer algebra codes as advanced tools to solve difficult problems by means
of computers. Also the dramatic increase of computer performance and the use
of more efficient computing algorithms have been crucial for the progress
achieved.

Already in 1959 a new formula for measuring $\amu$ was found by Bargmann,
Michel and Telegdi~\cite{BMT59}. At a particular energy, the magic
energy, which turned out to be at about 3.1 GeV, a number of systematic
difficulties of the existing experiment could be eliminated (see
the discussion in Sect.~\ref{ssec:theBNLexp}).  
This elegant method was realized with the
second muon storage ring at CERN (1969-1976)~\cite{CERN77}.  
The precision of 7 ppm reached
was an extraordinary achievement at that time. For the first time the
$m^2_\mu/m^2_e$--enhanced hadronic contribution came into play. Again
no deviations were found. With the achieved precision the muon $g-2$
remained a benchmark for beyond the SM theory builders ever
since. Only 20 years later the BNL experiment E821, again a muon
storage ring experiment run at the magic energy, was able to set new standards
in precision. This will be outlined in Sect.~\ref{ssec:theBNLexp}. 

Now, at the present level of accuracy, the complete SM is
needed in order to be able to make predictions at the appropriate
level of precision.  As already mentioned, at present further progress
is hampered to some extent by difficulties to include properly the
non--perturbative strong interaction part. At
a certain level of precision \textit{hadronic
effects} become important and we are
confronted with the question of how to evaluate them reliably. At low
energies QCD gets strongly interacting and a perturbative calculation
is not possible.  Fortunately, analyticity and unitarity allow us to
express the leading hadronic vacuum polarization contributions via a
dispersion relation (analyticity) in terms of experimental
data~\cite{CabibboGatto61}. The key relation here is the optical
theorem (unitarity) which determines the imaginary part of the vacuum
polarization amplitude through the total cross section for
electron--positron annihilation into hadrons. First estimations were
performed in~\cite{BM,Durand,KO67} after the discovery of the
$\rho$-- and the $\omega$--resonances, and
in~\cite{GdeR}, after first $\epm$ cross--section measurements were
performed at the $\epm$ colliding beam machines in
Novosibirsk~\cite{Budker67} and Orsay~\cite{Augustin69},
respectively. One drawback of this method is that now the precision of
the theoretical prediction of $\amu$ is limited by the accuracy of
experimental data. 
Much more accurate $\epm$--data from experiments
at the electron positron storage ring VEPP-2M at Novosibirsk allowed a
big step forward in the evaluation of the leading hadronic vacuum
polarization effects~\cite{KNO84,Barkov85,CLY85} (see
also~\cite{FJ86}). A more detailed analysis based on a complete
up--to--date collection of data followed about 10 years
later~\cite{EJ95}. Further improvements were possible thanks to new
hadronic cross section measurements by BES II~\cite{BES02} (BEPC
ring) at Beijing and by CMD-2~\cite{CMD203} at Novosibirsk. More
recently, cross section measurements via the radiative return
mechanism by KLOE~\cite{KLOE04} (DA$\varPhi$NE ring) at Frascati
and by BaBar at SLAC became available. This will be elaborated in much more detail in 
Sect.~\ref{sec:hadvap}.

Another important development was the discovery of reliable methods to control
strong interaction dynamics at low energies where perturbative QCD fails to
work. At very low energy, the well developed chiral perturbation theory
(CHPT)~\cite{CHPT} works. At higher energies, CHPT has been extended to a
resonance Lagrangian approach~\cite{EckerCPT}, which unifies to some extent
low energy effective hadronic models. These models play a role in the
evaluation of the hadronic light-by-light scattering contribution, which we
will discuss in Sect.~\ref{sec:lbl}.

Of course it was the hunting for deviations from theory and the
theorists speculations about ``new physics around the corner'' which
challenged new experiments again and again. The reader may find more
details about historical aspects and the experimental developments in
the interesting review: ``The 47 years of muon g-2'' by Farley
and Semertzidis~\cite{FaSe04}.

\subsection{Muon Properties}
Why the muon anomalous magnetic moment is so interesting and plays a
key role in elementary particle physics at its fundamental level is
due to the fact that it can be predicted by theory with very high
accuracy and at the same time can be measured as precisely in an 
unambiguous experimental setup. That the experimental conditions can
be controlled very precisely, with small systematic uncertainties, has
to do with the very interesting intrinsic properties of the muon,
which we briefly describe in the following. 

\subsubsection{Spin Transfer in Production and Decay of Muons}
The muon $g-2$ experiments observe the motion of the spin of the muons
on circular orbits in a homogeneous magnetic field. This requires the
muons to be polarized. After the discovery of the parity violation in
weak interaction it immediately became evident that weak decays of
charged pions are producing polarized muons. Thereby the maximal
parity violation of charged current processes provides the ideal
conditions. The point is that right--handed neutrinos $\nu_R$ are not
produced in the weak transitions mediated by the charged $W^\pm$ gauge
bosons. As a consequence the production rate of $\nu_R$'s in ordinary
weak reactions is practically zero which amounts to lepton number
conservation for all practical purposes in laboratory
experiments\footnote{Only in recent years phenomenon of neutrino
oscillations could be established unambiguously which proves that
lepton number in fact is not a perfectly conserved quantum
number. Neutrino oscillations are possible only if neutrinos have
masses which requires that right--handed neutrinos ($\nu_R$'s)
exist. In fact, the smallness of the neutrino masses explains the
strong suppression of lepton number violating effects.}.

Pions may be produced by shooting protons (accumulated in a proton
storage ring) on a target material where pions are the most
abundant secondary particles. The most effective pion production
mechanism proceeds via excitation and subsequent decay of baryon
resonances. For pions the dominating channel is the $\Delta_{33} \to N
\pi$ isobar.

All muon $g-2$ experiments are based on the decay chain 
\bea
\pi &\to& \mu + \nu_\mu \\
 && \downrightarrow e + \nu_e + \nu_\mu \cs
\eea
producing the polarized muons which decay into electrons which carry
along with their direction of propagation the muon's 
polarization (see e.g.~\cite{Scheck84}).\\

\noi {\bf 1) Pion decay:}\\
The $\pi^-$ is a pseudoscalar bound state $\pi^-=(\bar{u}\gafi d)$ of a $d$ 
quark and a $u$ antiquark $\bar{u}$. The main decay proceeds via
\begin{figure}[h]
\centering
\IfFarbe{%
\includegraphics{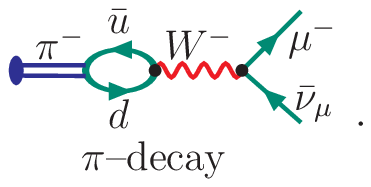}}{%
\includegraphics{pidecay.eps}}
\end{figure}

\begin{figure}[h]
\centering
\IfFarbe{%
\includegraphics[height=4cm]{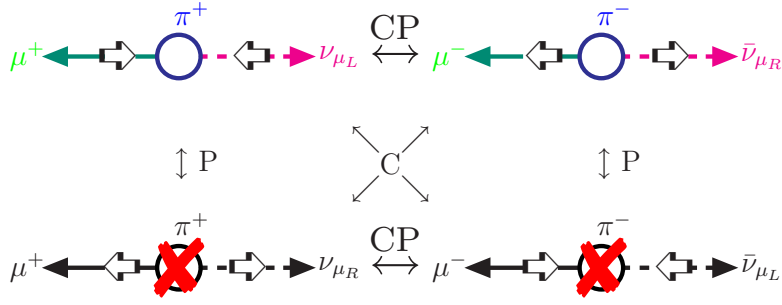}}{%
\includegraphics[height=4cm]{muonpro.eps}}
\caption{In the P violating weak pion decays leptons of definite handedness 
are produced depending on the given charge.  $\mu^-$ [$\mu^+$]
is produced with positive [negative] helicity
$h=\vec{S}\cdot\vec{p}/|\vec{p}|$. The physical $\mu^-$ and
$\mu^+$ decays are related by a CP transformation. The decays obtained by C or P
alone are inexistent.}
\label{fig:muonpro} 
\end{figure}

\noindent
Being a two--body decay, the lepton energy is fixed (monochromatic) and given
by $E_\ell=\sqrt{m_\ell^2+p_\ell^2}=\frac{m_\pi^2+m_\ell^2}{2m_\pi}\,,\;
p_\ell=\frac{m_\pi^2-m_\ell^2}{2m_\pi} \epo$ The part of
the Fermi type effective Lagrangian which describes this decay reads
\bea
\SL_{\rm eff, int}= -\frac{G_\mu}{\wz}\,V_{ud}\,
\left(\bar{\mu} \gamma^\alpha \:(1-\gafi)\: \nu_\mu \right)
\left(\bar{u} \gamma_\alpha \:(1-\gafi)\: d \right) + \mathrm{h.c.}
\eea 
where $G_\mu$ denotes the Fermi constant and $V_{ud}$ the first entry
in the CKM matrix. For our purpose $V_{ud} \sim 1$. 
The basic hadronic matrix element for pion decay is 
$\left<0|~{\bar{d} }\:{\gamma_\mu \gamma_5}\: {u}~|
{\pi(p) } \right> \doteq \I F_\pi p_\mu$
which defines the pion decay constant $F_\pi$.
The transition matrix--element for the process of our interest then reads
\bea
T &=&_{\rm out}\!\!<\mu^-,\anu _\mu | \pi^- >_{\rm in} 
=-\I \frac{G_\mu}{\wz}\,V_{ud}\, F_\pi 
\left(\bar{u}_\mu \gamma^\alpha \,(1-\gafi)\,v_{\nu_\mu} \right)\,p_{\alpha}\epo
\eea
Since the $\pi^+$ has spin 0 and the emitted neutrino is left--handed
($(1- \gafi)/2$ projector), by angular momentum conservation, the
$\mu^+$ must be left--handed as well. Only the axial part of the weak
charged $V-A$ current couples to the pion, as it is a pseudoscalar
state. In order to obtain the $\pi^-$ decay not only particles have to
be replaced by antiparticles (C) but also the helicities have to be
reversed (P), since a left--handed antineutrino (essentially) does not
exist. Note that the decay is possible only due to the non--zero muon
mass, which allows for the necessary helicity flip of the muon. How
the handedness is correlated with the charge is illustrated in
Fig.~\ref{fig:muonpro}.

\noi The pion decay rate is given by 
\ba
\Gamma_{\pi^- \ra \mu^- \bar{\nu}_\mu} = \frac{G_\mu^2}{8\pi} \; 
|V_{ud}|^2 F_\pi^2 \;
m_\pi \: m_\mu^2\:\left(1-\frac{m_\mu^2}{m_\pi^2} \right)^2 \times
\left(1+\delta_{\rm QED}\right)\cs
\ea
with $\delta_{\rm QED}$ the electromagnetic correction.\\
 
\noi {\bf 2) Muon decay:}\\
The muon is unstable and decays via the weak three body decay $\mu^- \to e^-
\bar{\nu}_e \nu_\mu$ 
\begin{figure}[h]
\centering
\IfFarbe{%
\includegraphics{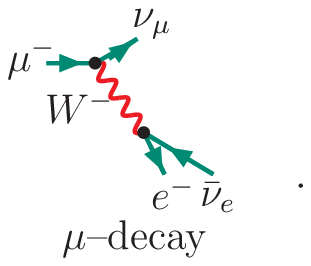}}{%
\includegraphics{mudecay.eps}}
\end{figure}

\begin{figure}[h]
\centering
\IfFarbe{%
\includegraphics[height=1.4cm]{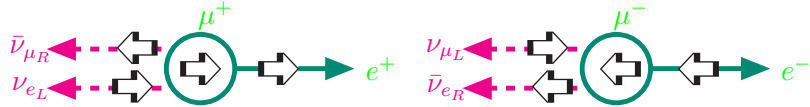}}{%
\includegraphics[height=1.4cm]{muondec.eps}}
\caption{In $\mu^-$ [$\mu^+$] decay the produced $e^-$ [$e^+$] has
negative [positive] helicity, respectively.}
\label{fig:muondec} 
\end{figure}

\noi The $\mu$--decay matrix element follows from the
relevant part of the effective Lagrangian which reads
\bea
\SL_{\rm eff, int}= -\frac{G_\mu}{\wz}
\left(\bar{e} \gamma^\alpha \:(1-\gafi)\: \nu_e \right)
\left(\anu _\mu \gamma_\alpha \:(1-\gafi)\: \mu \right) + \mathrm{h.c.}
\eea 
and is given by
\bea
T &=& _{\rm out}\!\!<e^-,\anu _e \nu_\mu | \mu^- >_{\rm in}
  = \frac{G_\mu}{\wz}
\left(\bar{u}_e \gamma^\alpha \:(1-\gafi)\: v_{\nu_e} \right)
\left(\bar{u}_{\nu_\mu} \gamma_\alpha \:(1-\gafi)\: u_\mu \right)\epo
\eea
This proves that the $\mu^-$ and the $e^-$ have both the same
left--handed helicity [the corresponding anti--particles are
right--handed] in the massless approximation. This implies the
decay scheme of Fig.~\ref{fig:muondec} for the muon.
\noi Again it is the P violation which prefers electrons emitted in the direction 
of the muon spin. Therefore, measuring the direction of the electron
momentum provides the direction of the muon spin.  After integrating
out the two unobservable neutrinos, the differential decay probability
to find an $e^\pm$ with reduced energy between $x_e$ and $x_e+dx_e$, emitted
at an angle between $\theta$ and $\theta + d \theta$, reads
\be
\frac{d^2\Gamma^\pm}{dx_e\:d\cos \theta} =
\frac{G_\mu^2m_\mu^5}{192\pi^3}\:x^2_e \:\left(3-2x_e\pm P_\mu\: \cos \theta \:
(2x_e-1) \right)
\label{mudecdist1}
\ee
and typically is strongly peaked at small angles.  The charge sign
dependent asymmetry in the production angle $\theta$ represents
the parity violation. The reduced $e^\pm$ energy is $x_e=E_e/W_{\mu e}$
with $W_{\mu e}=\mathrm{max~}E_e=(m_\mu^2+m_e^2)/2m_\mu$, the 
emission angle $\theta$ is the angle between the momentum
$\vec{p}_e$ of $e^\pm$ and the muon polarization vector $\vec{P}_\mu$. The
result above holds in the approximation $x_0=m_e/W_{e\mu}\sim 9.67 \times
10^{-3} \simeq 0$.

\subsection{Lepton Magnetic Moments}
Our particular interest is the motion of a lepton in an external field
under consideration of the full relativistic quantum behavior.  It is
controlled by the QED equations of motion with an
external field added
\ba
\begin{array}{rcl}
\left(\I  \gimu \pamu +Q_\ell e \gimu (\Amu \ofx +A^\mathrm{ext}_\mu
\ofx) -m_\ell \right)\: \psi_\ell \ofx & = & 0 \cs \\[2mm]
\left( \Box g ^{\mu \nu} - \left( 1 - \xi^{-1} \right) \partial ^\mu
\partial ^\nu \right) A_\nu (x) & = & - Q_\ell\, e \bar{\psi}_\ell (x) \g ^\mu
\psi_\ell (x) 
 \epo
\end{array}
\label{QEDfieldeq3}
\ea
What we are looking for is the solution of the Dirac equation with an
external field, specifically a constant magnetic field, as a
relativistic one--particle problem, neglecting the radiation field in
a first step. For slowly varying fields
$A^{\mu\:\mathrm{ext}}=(\Phi,\vec{A})$ the motion is essentially
determined by the generalized Pauli equation
(W.~Pauli 1927)
\be
 \I  \ppx{{\vphi}}{t}  = {\fH}\:{\vphi}=
\left(\frac{1}{2m} (\vec{p}-e \vec{A})^2 + e \:\Phi - \frac{e}{2m} \:
\vec{\sigma} \cdot \vec{B} \right)\: {\vphi}\cs
\ee
which up to the spin term is nothing but the non--relativistic
Schr\"odinger equation and which also serves as a basis for
understanding the role of the magnetic moment of a lepton on the
classical level. ${\vphi}$ is a non--relativistic two-component
Pauli--spinor. As we will see, in the absence of electrical fields
$\vec{E}$, the quantum correction miraculously may be subsumed in a
single number, the anomalous magnetic moment, which is the result of
relativistic quantum fluctuations.

To study radiative corrections we have to extend the discussion of the
preceding paragraph and consider the full QED interaction Lagrangian
\be
\SL^{\mathrm{QED}}_{\mathrm{int}}\ofx = - e \bar{\psi}(x)\gamma^\mu
\psi(x)\:A_\mu(x) 
\ee
for the case where the photon field is part of the dynamics but has an external
classical component $A^{\mathrm{ext}}_\mu(x)$:
$A_\mu(x) \to A_\mu(x) + A^{\mathrm{ext}}_\mu (x)\epo$
We are thus dealing with QED exhibiting an additional
external field insertion ``vertex'':

\unitlength1pt
\begin{picture}(30,30)(-40,-16)
\SetWidth{1}
\ArrowLine(00,00)(9.0,9.0)
\ArrowLine(9.0,-9.0)(00,00)
\Text(41,00)[]{$\otimes$~~~
$=\: - \I e\: \gamma^\mu \: \tilde{A}^{\mathrm{ext}}_\mu\;.$}
\end{picture}


Gauge invariance requires that a gauge transformation
of the external field
$A^{\mathrm{ext}}_\mu(x) \to A^{\mathrm{ext}}_\mu(x) -\partial_\mu
\alpha (x) ,$
for an arbitrary scalar classical field $\alpha (x)$, leaves physics
invariant.

The motion of the lepton in the external field is described by a
simultaneous expansion in the fine structure constant
$\alpha=e^2/4\pi$ and in the external field
$A^{\mathrm{ext}}_\mu(x)$ assuming the latter to be weak

\unitlength1pt
\SetWidth{1}
\begin{picture}(30,30)(-40,00)
\ArrowLine(-20,00)(00,00)
\ArrowLine(-00,00)(20,00)
\Text(-10,-10)[]{$p_1$}
\Text(10,-10)[]{$p_2$}
\Text(05,10)[]{$q$}
\Text(00,00)[]{$\otimes$}
\Text(27,00)[]{$~+$}
\end{picture}
\SetOffset(20,0)
\SetWidth{1}
\begin{picture}(30,30)(-40,00)
\ArrowLine(-20,00)(00,00)
\ArrowLine(-00,00)(20,00)
\Photon(00,00)(00,18){1}{3}
\COval(00,12)(5,3)(0){Black}{White}
\Text(00,18)[]{$\otimes$}
\Text(27,00)[]{$~+$}
\end{picture}
\SetOffset(40,0)
\SetWidth{1}
\begin{picture}(30,30)(-40,00)
\ArrowLine(-20,00)(-10,00)
\ArrowLine(10,00)(20,00)
\Line(-10,00)(10,00)
\PhotonArc(00,00)(10,180,360){1}{4}
\Text(00,00)[]{$\otimes$}
\Text(27,00)[]{$~+$}
\end{picture}
\SetOffset(60,0)
\SetWidth{1}
\begin{picture}(30,30)(-40,00)
\ArrowLine(-26,00)(-10,00)
\ArrowLine(10,00)(26,00)
\ArrowLine(-10,00)(10,00)
\Text(-10,00)[]{$\otimes$}
\Text(10,00)[]{$\otimes$}
\Text(32,00)[]{$~+$}
\Text(49,00)[]{$\cdots$}
\SetOffset(0,0)
\end{picture}

\vspace*{0.8cm}

\noi In the following we will use the more customary graphic representation

\unitlength1pt
\begin{picture}(30,30)(-40,-16)
\SetWidth{1}
\ArrowLine(00,00)(9.0,9.0)
\ArrowLine(9.0,-9.0)(00,00)
\Text(00,00)[]{$\otimes$}
\Text(20,00)[]{$\Rightarrow$}
\ArrowLine(50,00)(59.0,9.0)
\ArrowLine(59.0,-9.0)(50,00)
\Photon(36,00)(50,00){1}{3}
\end{picture}

\unitlength1mm

\noi of the external vertex, just as an amputated photon line at zero momentum.

The gyromagnetic ratio of the muon is defined by the ratio of the
magnetic moment which couples to the magnetic field in the Hamiltonian
and the spin operator in units of  $\mu_0=e/2m_\mu $
\be
\vec{\mu}={g_\mu}\:\frac{e}{2m_\mu }\:\vec{s}\;\; ;\;\;\;
{g_\mu=2\:(1+{a_\mu})}
\ee
and as indicated has a tree level part, the Dirac moment
$g^{(0)}_\mu=2$~\cite{Diracmm}, and a higher order part $a_\mu$ the
muon anomaly or anomalous magnetic moment.

In QED $\amu$ may be calculated in perturbation theory by considering the
matrix element
\bea
{\cal M}(x;p)=\bra{\mu^-(p_2,r_2)} j_\mathrm{em}^\mu(x) \ket{\mu^-(p_1,r_1)}
\eea
of the electromagnetic current for the scattering of an
incoming muon $\mu^-(p_1,r_1)$ of momentum $p_1$ and 3rd component
of spin $r_1$ to a muon $\mu^-(p_2,r_2)$ of
momentum $p_2$ and 3rd component
of spin $r_2$, in the classical limit of zero momentum transfer
$q^2=(p_2-p_1)^2 \to 0$.  In momentum space
we obtain
\bea
\tilde{{\cal M}}(q;p)&=&\int \D^4x \: \E ^{- \I qx} \bra{\mu^-(p_2,r_2)} j_\mathrm{em}^\mu(x) \ket{\mu^-(p_1,r_1)} \\
&=& (2\pi)^4\: \delta^{(4)}(q-p_2+p_1)\: \bra{\mu^-(p_2,r_2)} j_\mathrm{em}^\mu(0) \ket{\mu^-(p_1,r_1)}\,,
\eea
proportional to the $\delta$--function of four--momentum conservation.
The $T$--matrix element is then given by
\bea
\bra{\mu^-(p_2)} j_\mathrm{em}^\mu(0) \ket{\mu^-(p_1)}=
(-\I e)\:\bar{u}(p_2)\,\Gamma^\mu(P,q)\,u(p_1) \cs \ \ (P=p_1+p_2)\epo
\eea
In QED it has a relativistically covariant decomposition of  the form\\[4mm]

\unitlength1mm 

\ba
\mysymb{0}{2}{20}{20}{eesym20}~~~~~ =(-\I e)\:\bar{u}(p_2)\left[\gamma^\mu
F_\mathrm{E}(q^2)+\I \frac{\sigma^{\mu\nu}q_\nu}{2m_\mu}F_\mathrm{M}(q^2) \right]u(p_1)\;,
\label{QEDFF}
\ea
\begin{picture}(30,30)(-8.6,-25)
\put(04,18){\makebox(0,0)[t]{$\gamma(q)$}}
\put(23,20){\makebox(0,0)[t]{$\mu(p_2)$}}
\put(23,08){\makebox(0,0)[t]{$\mu(p_1)$}}
\end{picture}

\vspace*{-2.6cm}

\noi
where $q=p_2-p_1$ and $u(p)$ denote the Dirac
spinors. $F_\mathrm{E}(q^2)$ is the electric charge or Dirac form
factor and $F_\mathrm{M}(q^2)$ is the magnetic or Pauli form
factor. Note that the matrix $\sigma^{\mu
\nu}=\frac{\I}{2}[\gamma^\mu, \gamma^\nu]$ represents the spin $1/2$
angular momentum tensor. In the static (classical) limit we have
\be
F_\mathrm{E}(0)=1\;\;,\;\;\; F_\mathrm{M}(0)=a_\mu \cs
\ee
where the first relation is the \textit{charge renormalization
condition} (in units of the physical positron charge $e$, which by
definition is taken out as a factor), while the
second relation is the finite prediction for $\amu$, in terms of the
form factor $F_\mathrm{M}$ the calculation of which will be described
below. Instead of calculating the full vertex function $\Gamma_\mu(P,q)$ one can use
the projection technique described in~\cite{BR74} and expand the
vertex function to linear order in the external photon momentum $q$: 
\ba
\Gamma_\mu(P,q) \simeq \Gamma_\mu(P,0)+q^\nu
\frac{\partial}{\partial q^\nu}\left.\Gamma_\mu(P,q)\right|_{q=0}\equiv
V_\mu(p)+q^\nu\:T_{\nu \mu}(p) \cs
\ea
for fixed $P$. This allows us to simplify the calculation by working
directly in the limit $q \to 0$ afterwards. Since $\amu$ does not depend on the
direction of the muon momentum one can average over the direction of
$P$ which is orthogonal to $q$ ($P \cdot q=0$).  As a master formula one
finds
\ba
a_\mu&=&\frac{1}{8\:(d-2)(d-1)\:m_\mu} \:\Tr \left\{
(\sla{p}+m_\mu)\:[\gamma^\mu,\gamma^\nu]\: 
(\sla{p}+m_\mu)\:T_{\nu \mu}(p)\right\} \nonumber \\ 
& & +\frac{1}{4\:(d-1)\:m_\mu^2}
\:\Tr \left.
\left\{ \left[m_\mu^2\:\gamma^\mu -(d-1)\:m_\mu\: p^\mu -d\: \sla{p}\:p^\mu
  \:\right] 
V_{\mu}(p)\right\}\right|_{p^2=m_\mu^2}\cs 
\label{amumaster}
\ea
where $d=4-\veps$ is the space-time dimension. In case of UV divergences
the choice $\veps>0$ provides a dimensional regularization. The limit
$\veps \to 0$ is to be performed after renormalization.
The amplitudes $V_\mu(p)$ and $T_{\nu\mu}(p)$ depend on one on--shell
momentum $p=P/2$, only, and thus the problem reduces to the calculation of
on--shell self--energy type diagrams as the external photon momentum now can
be taken zero.

\begin{figure}[t]
\centering
\IfFarbe{%
\includegraphics[height=8.6cm]{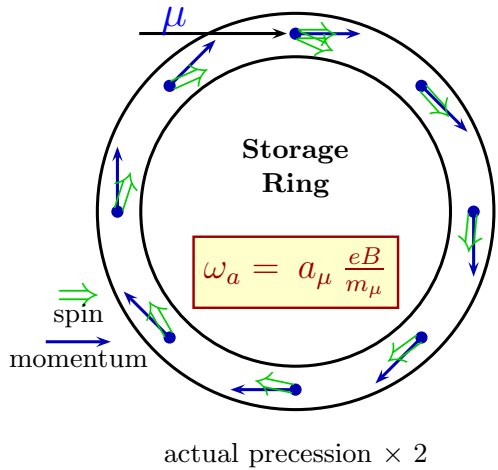}}{%
\includegraphics[height=8.6cm]{sprece.eps}}
\vspace*{-9mm}
\caption{Spin precession in the $g-2$ ring ($\sim 12^\circ$/circle).}
\label{fig:spinprecession}
\end{figure}

Note that in higher orders the form factors
in general aquire an imaginary part. One
may therefore write an effective dipole moment Lagrangian with
complex ``coupling''
\be
\cL^\mathrm{DM}_\mathrm{eff}=-\frac{1}{2} \left\{\bar{\psi}\,
\sigma^{\mu \nu}\,
\left[ D_\mu\, \frac{1+\gafi}{2}+ D^*_\mu\, \frac{1-\gafi}{2}\right]
\psi \right\}\, F_{\mu\nu}
\label{DMReIm}
\ee
with $\psi$ the muon field and
\be
\Repa D_\mu =a_\mu \,\frac{e}{2m_\mu}~~,~~~
\Impa D_\mu= d_\mu=\frac{\eta_\mu }{2}\frac{e}{2m_\mu}\epo
\ee
Thus the imaginary part of $F_\mathrm{M}(0)$ corresponds to an
electric dipole moment. The latter is non--vanishing only if we have
$T$ violation. The existence of a relatively large EDM would also
affect the extraction of $\amu$. This will be discussed towards the end of 
the next section.

\section{The Muon $g-2$ Experiments}

\subsection{The Brookhaven Muon $g-2$ Experiment}
\label{ssec:theBNLexp}
The measurement of $a_\mu$ in principle is simple. As illustrated in
Fig.~\ref{fig:spinprecession}, when polarized
muons travel on a circular orbit in a constant magnetic field, then
$a_\mu$ is responsible for the \textit{Larmor precession} of the
direction of the spin of the muon, characterized by the angular
frequency $\vec{\omega}_a$. Correspondingly, the principle of the BNL
muon $g-2$ experiment involves the study of the orbital and spin
motion of highly polarized muons in a magnetic storage ring. This
method has been applied in the last CERN experiment~\cite{CERNfinal}
already. The key improvements of the BLN experiment include the very
high intensity of the primary proton beam from the proton storage ring
AGS (Alternating Gradient Synchrotron), the injection of muons instead
of pions into the storage ring, and a super--ferric storage ring
magnet~\cite{BNLfinal} (see also the
reviews~\cite{FarleyPicasso90,Hughes01,FaSe04,HertzogMorse2004,Miller07}).

\begin{figure}[t]
\centering
\IfFarbe{%
\includegraphics[height=7.2cm]{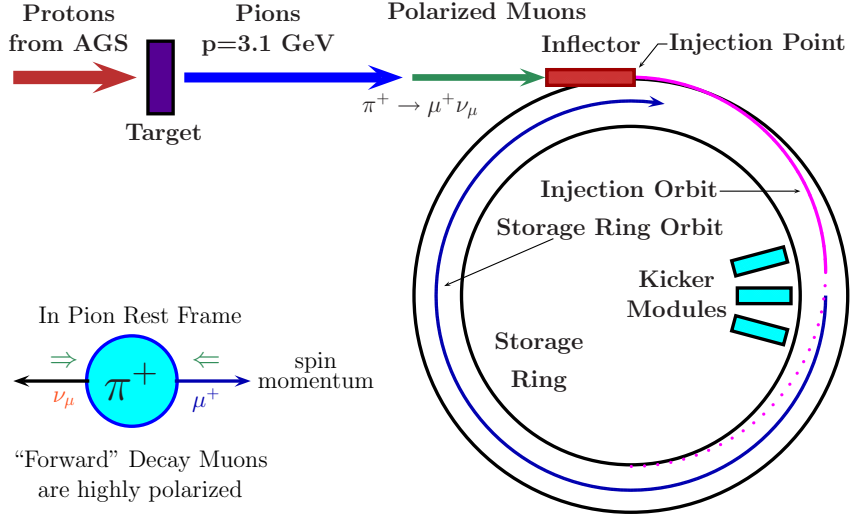}}{%
\includegraphics[height=7.2cm]{muonorbit.eps}}
\caption{The schematics of muon injection and storage in the $g-2$ ring.}
\label{fig:SRschema} 
\end{figure}

The muon $g-2$ experiment at Brookhaven works as illustrated in
Fig.~\ref{fig:SRschema}~\cite{ring,quads,kicker}. Protons of
energy 24 GeV from the AGS hit a target and produce pions. The pions
are unstable and decay into muons plus a neutrino where the muons
carry spin and thus a magnetic moment which is directed along the
direction of the flight axis. The longitudinally polarized muons from
pion decay are then injected into a uniform magnetic field $\vec{B}$
where they travel in a circle.  The ring\footnote{A picture of the BNL
muon storage ring may be found on the Muon $g-2$ Collaboration Web
Page {\tt http://www.g-2.bnl.gov/}} is a toroid--shaped structure with
a diameter of 14 meters, the aperture of the beam pipe is 90 mm, the
field is 1.45 Tesla and the momentum of the muon is $p_\mu =3.094$
GeV. In the horizontal plane of the
orbit the muons execute a relativistic cyclotron motion with angular
frequency $\omega_c$. By the motion of the muon magnetic moment in the
homogeneous magnetic field the spin axis is changed in a particular
way as described by the Larmor precession. After each circle the
muon's spin axis changes by 12' (arc seconds), while the muon is
traveling at the same momentum (see
Fig.~\ref{fig:spinprecession}). The muon spin is precessing with
angular frequency $\omega_s$, which is slightly bigger than $\omega_c$
by the difference angular frequency $\omega_a=\omega_s-\omega_c$.
\ba
&&\omega_c=\frac{e B}{m_\mu \, \gamma}\,,\;\;
\omega_s=\frac{e B}{m_\mu \, \gamma}+ a_\mu \, \frac{e B}{m_\mu}\,,\;\;
\omega_a=a_\mu \, \frac{e B}{m_\mu}\cs
\label{oscfreqs}
\ea
where $\gamma=1/\sqrt{1-v^2}$ is the relativistic Lorentz 
factor and $v$ the muon velocity. In the experiment
$\omega_a$ and $B$ are measured. The muon mass $m_\mu$ is obtained
from an independent experiment on muonium, which is a $(\mu^+e^-)$
bound system.  Note that if the muon would just have its Dirac
magnetic moment $g=2$ (tree level) the direction of the spin of the
muon would not change at all.

In order to retain the muons in the ring an electrostatic focusing
system is needed. Thus in addition to the magnetic field
$\vec{B}$ an electric quadrupole field $\vec{E}$ in the plane normal
to the particle orbit must be applied. This transversal electric field
changes the angular frequency according to
\be
\vec{\omega_a}=\frac{e}{m_\mu}\left(\amu \vec{B}-\left[\amu -
\frac{1}{\gamma^2-1}\right]\:\vec{v}\times \vec{E} \right)\epo
\label{omegaa}
\ee
This key formula for measuring $\amu$ was found by Bargmann, Michel
and Telegdi in 1959~\cite{BMT59,precession}.  Interestingly, one has the
possibility to choose $\gamma$ such that $\amu-1/(\gamma^2-1)=0$, in
which case $\omega_a$ becomes independent of $\vec{E}$. This is the
so--called \textit{magic} $\gamma$.  When running at the corresponding
magic energy, the muons are highly relativistic, the magic
$\gamma$-factor being $\gamma=\sqrt{1+1/a_\mu}=29.3$. The muons thus
travel almost at the speed of light with energies of about
$E_{\mathrm{magic}}=\gamma m_\mu \simeq 3.098$ GeV.  This rather
high energy, which is dictated by the requirement to minimize the
precession frequency shift caused by the electric quadrupole
superimposed upon the uniform magnetic field, also leads to a large
time dilatation. The lifetime of a muon at rest is 2.19711 $\mu$s,
while in the ring it is 64.435 $\mu$s (theory) [64.378 $\mu$s
(experiment)]). Thus, with their lifetime being much larger than at
rest, muons are circling in the ring many times before they decay into
a positron plus two neutrinos: $\mu^+ \to e^+ +\nu_e+\bar{\nu}_\mu$.
In this decay we have the necessary
strong correlation between the muon spin direction and the direction
of emission of the positrons. The differential decay rate for the muon
in the rest frame is given by Eq.~(\ref{mudecdist1}) which may be
written as
\be
d\Gamma=N(E_e)\:\left(1+\frac{1-2x_e}{3-2x_e}\:\cos\theta \right)\:d\Omega\epo
\label{mudecdist}
\ee 
Again, $E_e$ is the positron energy, $x_e$ is $E_e$ in units of the
maximum energy $m_\mu/2$, $N(E_e)$ is a normalization factor and
$\theta$ the angle between the positron momentum in the muon rest
frame and the muon spin direction. The $\mu^+$ decay spectrum is
peaked strongly for small $\theta$ due to the non--vanishing
coefficient of $\cos \theta$
\be
A(E_e) \doteq \frac{1-2x_e}{3-2x_e}\cs
\ee
the asymmetry factor which reflects the 
\textit{parity violation}.

\begin{figure}[t]
\centering
\IfFarbe{%
\includegraphics[height=8cm]{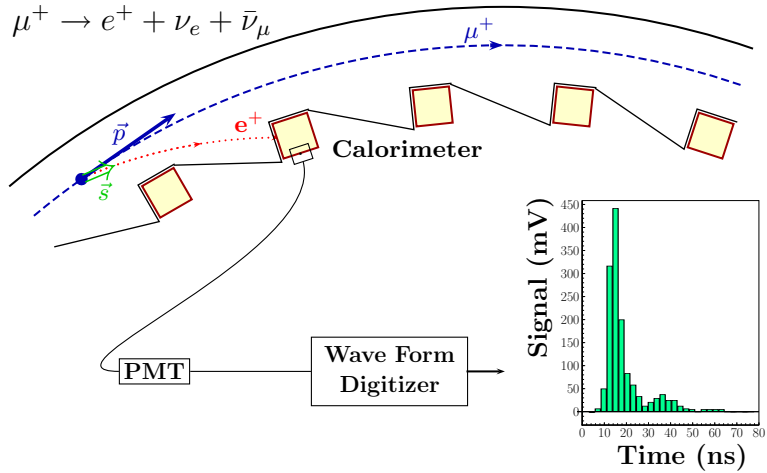}}{%
\includegraphics[height=8cm]{detection.eps}}
\caption{Decay of $\mu^+$ and detection of the 
emitted $e^+$ (PMT=Photomultiplier).}
\label{fig:detection} 
\end{figure}

\begin{figure}[t]
\centering
\IfFarbe{%
\includegraphics[height=8cm]{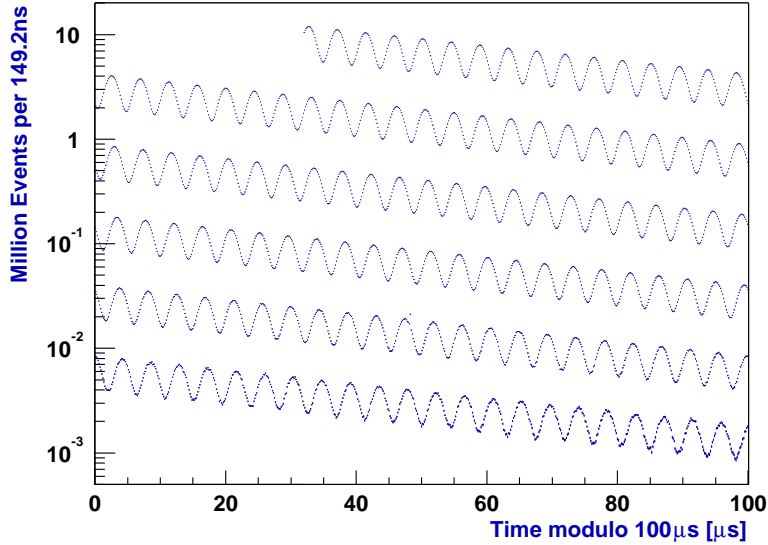}}{%
\includegraphics[height=8cm]{wiggle2004.eps}}
\caption{Distribution of counts versus time for the 3.6 billion decays
in the 2001 negative muon data--taking period [Courtesy of the E821
collaboration. Reprinted with permission from~\cite{BNLfinal}. Copyright (2007) by
the American Physical Society].}
\label{fig:wiggles2004} 
\end{figure}

The
positron is emitted with high probability along the spin axis of the muon as illustrated in 
Fig.~\ref{fig:detection}. The decay
positrons are detected by 24 
calorimeters evenly distributed inside the muon storage ring. These
counters measure the positron energy and allow to determine the
direction of the muon spin. A precession frequency dependent rate is
obtained actually only if positrons above a certain energy are selected 
(forward decay positrons).
The number of decay positrons with energy
greater than $E$ emitted at time $t$ after muons are injected into the
storage ring is given by
\be
N(t)=N_0(E)\:\exp \left(\frac{-t}{\gamma \tau_\mu}
\right)\:\left[1+A(E)\:\sin(\omega_a t+\phi(E)) \right]\cs
\label{timestructure}
\ee
where $N_0(E)$ is a normalization factor, $\tau_\mu$ the muon life
time (in the muon rest frame), and $A(E)$ is the asymmetry factor for
positrons of energy greater than $E$. Fig.~\ref{fig:wiggles2004} shows
a typical example for the time structure detected in the BNL experiment. 
As expected the exponential decay law for
the decaying muons is modulated by the $g-2$ angular frequency.  In
this way the angular frequency $\omega_a$ is neatly determined from
the time distribution of the decay positrons observed with the
electromagnetic calorimeters~\cite{BNL99}--\cite{BNL04}.

The second quantity which has to be measured very precisely  in the experiment is the magnetic
field. This is accomplished by \textit{Nuclear Magnetic Resonance}
(NMR) using a standard probe of H$_2$O~\cite{nmr}. This standard can
be related to the magnetic moment of a free proton
by
\be
B=\frac{\omega_p}{2 \mu_p} \cs
\label{omegaproton}
\ee
where $\omega_p$ is the Larmor spin precession angular velocity of a
proton in water.
Using $\omega_p$ and the frequency $\omega_a$ together with 
$\mu_\mu=(1+\amu)\:e/(2m_\mu)$, one obtains
\be
\amu=\frac{R}{\lambda-R}
\mathrm{ \ \ where \ \ }
R=\omega_a/\omega_p \mathrm{ \ \ and \ \ } \lambda=\mu_\mu/\mu_p \epo
\label{amuextraction}
\ee
The quantity $\lambda$ shows up because the value of the muon mass
$m_\mu$ is needed, and also because the $B$ field measurement involves
the proton mass $m_p$. Here the precision experiments on the microwave spectrum of
ground state muonium ($\mu^+e^-$)~\cite{Muonium99} performed at LAMPF at Los
Alamos provide the needed result. The measurements in combination with the theoretical 
prediction of the Muonium hyperfine splitting $\Delta \nu$~\cite{MuoniumKino98,MuoniumCEK02} 
(and references therein), allowed to extract the precise value 
\be
\lambda=\mu_\mu/\mu_p=3.183\,345\,39(10)~[30~\mathrm{ppb}]\cs
\label{lambdaHFS}
\ee
which is used by the E821 experiment to determine $\amu$
via Eq.~(\ref{amuextraction}).

Since the spin precession frequency can be measured very well, the
precision at which $g-2$ can be measured is essentially determined by
the possibility to manufacture a constant homogeneous magnetic field
$\vec{B}$ and to determine its value very precisely. Important but
easier to achieve is the tuning to the magic energy. Possible
deviations may be corrected by adjusting the effective magnetic field
appropriately.

Note that one of the reasons why the relativistic motion of the muons
is so well understood is the fact that the orbital motion of charged
particles in the storage ring may be investigated separately from the
spin motion. The forces associated with the anomalous magnetic moment
are very weak ($\amu \approx 1.16 \power{-3}$) in comparison to the
forces of the charge of the particle determining the orbital motion.
While the static magnetic field $\vec{B}(r,z)=(0,0,B_0)$ causes the
particles to move on a circle of radius $r_0=\gamma m /(e B_0)$ the
electric quadrupole field $\vec{E}=(E_r,E_\theta,E_z)=(\kappa x
,0,-\kappa z)$, (which produces a restoring force in the vertical
direction and a repulsive force in the radial direction)
leads to a superimposed oscillatory motion
\ba
&&x=A \, \cos (\sqrt{1-n}\,\omega_c\,t)\,,\;
z=B \, \cos (\sqrt{n}\,\omega_c\,t)\;,
\ea
of the muons about the central beam (assumed to move along the y-axis)
position. Here, $x=r-r_0$, $\kappa$ a positive constant and
$n=\frac{\kappa r_0}{\beta B_0}$ with $\beta=v$ is the field index.
This motion is called \textit{betatron
oscillation}. The amplitudes depend on the initial condition of the
particle trajectory. The betatron frequencies are $\omega_{y
\mathrm{BO}}=\sqrt{n}\,\omega_c$ and $\omega_{x
\mathrm{BO}}=\sqrt{1-n}\,\omega_c$ where $\omega_c=v/r_0$ is
the cyclotron frequency.  

The betatron motion also affects the anomalous magnetic precession 
Eq.~(\ref{omegaa}), which holds for transversal magnetic field  
$\vec{v}\cdot\vec{B}=0$. The latter, due to electrostatic focusing, is not
accurately satisfied such that the more general formula
\be
\vec{\omega}_a=-\frac{e}{m_\mu}\bigg\{
a_\mu \,\vec{B}-a_\mu \,\left(\frac{\gamma}{\gamma+1}\right)\,
(\vec{v}\cdot\vec{B}) \, \vec{v} +\left(a_\mu -\frac{1}{\gamma^2-1}
\right)\, 
\vec{E}\times \vec{v}\bigg\} \cs
\label{omegageneral}
\ee
has to be used as a starting point. Expanding about
$\vec{v}\cdot\vec{B}=0$ at the magic energy yields the \textit{Pitch
Correction} which for the BNL experiment amount to $C_P \simeq
0.3~\mathrm{ppm}$. Similarly, the deviation from the magic energy (beam
spread) requires a \textit{Radial Electric Field Correction}, for the BNL
experiment typically $C_E\simeq 0.5~ \mathrm{ppm}$. For more details on the
machine and the basics of the beam dynamics we refer
to~\cite{Miller07,Jegerlehner:2008zz}.

A possible correction of the magnetic precession could be due to an electric
dipole moment of the muon. If a large enough EDM 
\be
\vec{d}_e=\frac{\eta \,e}{2m_\mu } \vec{S}
\ee
would exist, where $\eta$ is the dimensionless constant equivalent 
of magnetic moment $g$-factors, the applied electric field
$\vec{E}$ (which is vanishing at the equilibrium beam position) and
the motional electric field induced in the muon rest frame
$\vec{E}^*=\gamma \,\vec{\beta} \times \vec{B}$ would add an extra
precession of the spin with a component along $\vec{E}$ and one about
an axis perpendicular to $\vec{B}$:
\ba
\vec{\omega}_{a}=\vec{\omega}_{a0} +\vec{\omega}_\mathrm{EDM}=
\vec{\omega}_{a0}-\frac{\eta\,e}{2m_\mu}\,\left(\vec{E}
+\vec{\beta} \times \vec{B}\right)
\ea
where $\vec{\omega}_{a0}$ denotes the would-be precession frequency
for $\eta=0$. The shift caused by a non-vanishing $\eta$ is
\bea
\Delta \vec{\omega}_a=-2d_\mu\,\left(\vec{\beta}\times \vec{B} \right)
-2d_\mu\,\vec{E} 
\eea
which, for $\beta \sim 1$ and $d_\mu\,\vec{E}\sim 0$, yields
\ba
\omega_a=B\,\sqrt{\left(\frac{e}{m_\mu}\,a_\mu\right)^2+\left(2d_\mu\right)^2}\epo
\ea
  The result is that the
plane of precession in no longer horizontal but tilted at an angle
\be
\delta\equiv \arctan \frac{\omega_{\rm EDM}}{\omega_{a0}}= \arctan
\frac{\eta\,\beta}{2a_\mu}\simeq 
\frac{\eta}{2a_\mu}  
\ee
and the precession frequency is increased by a factor 
\be \omega_a=\omega_{a0}\,\sqrt{1+\delta^2}\epo
\label{EDMomegashift}
\ee 
The tilt gives rise to an oscillating vertical component of the muon
polarization and may be detected by recording separately the
electrons which strike the counters above and below the mid--plane of
the ring. This measurement has been performed in the last CERN
experiment on $g-2$. The result 
$d_\mu=(3.7\pm 3.4) \power{-19}\,e\cdot\mathrm{cm}$
showed that it is negligibly small. The present experimental bound is 
$d_\mu < 2.7 \power{-19}\,e\cdot\mathrm{cm}$
while the SM estimate is $d_\mu  \sim 3.2 \power{-25}\,e\cdot\mathrm{cm}$.
One thus may safely assume $d_\mu$ to be too small to be able to affect
the extraction of $\amu$.

\subsection{Summary of Experimental Results}
\label{sec:results5}
Before the E821 experiment at Brookhaven presented their results in
the years from 2001 to 2004, the last of a series of measurements of
the anomalous $g$-factor at CERN was published about 30 years ago.  At
that time $a_\mu$ had been measured for muons of both charges in the
Muon Storage Ring at CERN. The two results,
\ba
a_{\mu^-} &=& 1165937(12)\power{-9}\cs \crn 
a_{\mu^+} &=& 1165911(11)\power{-9}\cs
\ea
are in good agreement with each other, and combine to give a mean
\be
a_\mu = 1165924.0(8.5) \power{-9}~~\mathrm{[7~ppm]}\cs
\ee
which was very close to the theoretical prediction $1165921.0(8.3)
\power{-9}$ at that time. The measurements thus confirmed the
remarkable QED calculation as well as a substantial hadronic photon
vacuum polarization contribution, and served as a precise verification
of the CPT theorem for muons.
\begin{table}[t]
\centering
\caption{Summary of CERN and E821 Results.}
\label{tab:allamudata}
\begin{tabular}{ccclcc}
&&&&&\\[-2mm]
\hline\noalign{\smallskip}\noalign{\smallskip}
Experiment&Year & Polarity & $a_\mu \times 10^{10}$ & ~~~Pre. [ppm]~~~ & Ref. \\
\noalign{\smallskip}\hline\noalign{\smallskip}
CERN I\,~&1961 & $\mu^+$ &   11\,450\,000(220000) & 4300 & \cite{CERNI61} \\
CERN II\,&1962-1968 & $\mu^+$ & 11\,661\,600(3100) & 270 & \cite{CERNII62_68} \\
CERN III &1974-1976 & $\mu^+$ & 11\,659\,100(110) & 10 & \cite{CERNfinal} \\
CERN III &1975-1976 & $\mu^-$ & 11\,659\,360(120) & 10 & \cite{CERNfinal} \\
BNL &1997 & $\mu^+$ &   11\,659\,251(150) & 13 & \cite{BNL99} \\
BNL &1998 & $\mu^+$ &   11\,659\,191(59) & 5 & \cite{BNL00} \\
BNL &1999 & $\mu^+$ &   11\,659\,202(15) & 1.3 & \cite{BNL01} \\
BNL &2000 & $\mu^+$ & 11\,659\,204(9) & 0.73 & \cite{BNL02} \\
BNL &2001 & $\mu^-$ & 11\,659\,214(9) & 0.72 & \cite{BNL04} \\
&&&&&\\[-2mm]
 & Average & & 11\,659\,208.0(6.3) & 0.54 & \cite{BNLfinal}\\
\noalign{\smallskip}\hline
\end{tabular}
\end{table} 
Measured in the experiments is the ratio of the muon precession frequency
$\omega_a=\omega_s-\omega_c$ and the proton precession frequency from
the magnetic field calibration
$\omega_p$: $R=\omega_a/\omega_p$ which together with the ratio of the
magnetic moment of the muon to the one of the proton 
$\lambda=\mu_\mu/\mu_p$
determines the \amm via Eq.~(\ref{amuextraction}).
The CERN determination of $\amu$ was based on the value 
$\lambda=3.1833437(23)$.

The BNL muon $g-2$ experiment has been able to improve and perfect the
method of the last CERN experiments in several respects and was able to
achieve an impressive 14--fold improvement in precision.  
The measurements are $R_{\mu^-}=0.0037072083(26)$ and
$R_{\mu^+}=0.0037072048(25)$ the difference being 
$\Delta R=(3.5\pm3.4) \power{-9}$. Together with
$\lambda=3.18334539(10)$ \cite{PDG04,PDG06} one obtains the new values
\ba
a_{\mu^-}&=&11659214(8)(3) \power{-10} \cs \crn
a_{\mu^+}&=&11659204(7)(5) \power{-10} \epo
\ea 
Assuming CPT symmetry, as valid in any QFT, and taking into account
correlations between systematic errors between the various data sets,
the new average $R=0.0037072063(20)$ was obtained. 
The new average value is then given by~\cite{BNLfinal}
\ba
a_{\mu}&=&11659208.0(5.4)(3.3)[6.3] \power{-10}~~\mathrm{[0.54~ppm]} \epo 
\label{amuBNL}
\ea 
The two uncertainties given are the statistical and the systematic ones. The
total error in square brackets follows by adding in quadrature the statistical
and systematic errors. In Table~\ref{tab:allamudata} all results from CERN
and E821 are collected.
\begin{figure}[h]
\centering
\IfFarbe{%
\includegraphics[height=7cm]{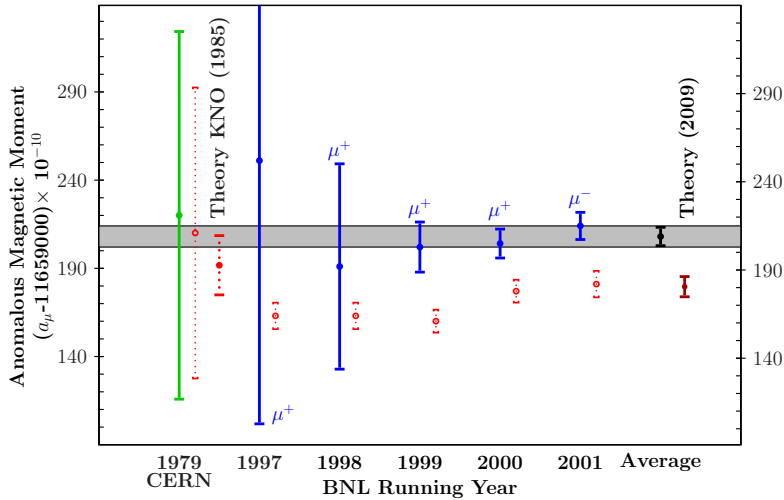}}{%
\includegraphics[height=7cm]{gm2resBNL06.eps}}
\caption{Results for the individual E821 measurements, together with the new
world average and the theoretical prediction. The CERN result is shown
together with the theoretical prediction by Kinoshita et al. 1985, at about
the time when the E821 project was proposed.The dotted vertical bars 
indicate the theory values quoted by the experiments.}
\label{fig:gm2resBNL06} 
\end{figure}
The new average is completely dominated by the BNL results. The
individual measurements are shown also in Fig.~\ref{fig:gm2resBNL06}.
The comparison with the theoretical result including predictions from
SM extensions will be discussed later in Sect.~\ref{sec:thevsexp}. In the following
sections we first review the SM prediction of $\amu$.

\section{QED Prediction of $g-2$}
\label{sec:gm2inQED}
Any precise theoretical prediction requires a precise knowledge of the
fundamental parameters. In QED these are the fine structure
constant $\alpha$ and the lepton masses. As the leading order
result is $\frac{\alpha}{2\pi}$ and since we want to determine  $a_\ell$
with very high precision, the most important basic
parameter for calculating $a_\mu$ is the fine structure
constant. Its most precise value is determined using
of the electron anomalous magnetic moment
\begin{equation}
a_{e}^{\rm exp}= 0.001\, 159\, 652\, 180\, 73(28)[0.24\, \mathrm{ppb}]\cs
\label{new_a_e}
\end{equation}
which very recently~\cite{aenew,aenew08} has been obtained with extreme precision.
Confronting the experimental value with the theoretical prediction as a series in
$\alpha$  (see Sect.~\ref{sec:aele} below) determines~\cite{alnew,Aoyama07,aenew08}
\begin{eqnarray}
\alpha^{-1}(a_e)&=&137.035999084(51)[0.37\, \mathrm{ppb}]\;.
\label{alphainv_a_e}
\end{eqnarray}
This new value has an uncertainty 20 times smaller
than any preceding independent determination of $\alpha$ and we will use it
throughout in the calculation of $a_\mu$.

Starting at 2--loops, higher order corrections include contributions
from lepton loops in which different leptons can circulate and results
depend on the corresponding mass ratios. Whenever needed, we will use
the following values for the muon--electron and muon--tau mass ratios,
and lepton masses~\cite{CODATA00,CODATA05,PDG04,PDG06}
\begin{eqnarray} \begin{array}{c}
m_{\mu}/m_e =  206.768\,2838\,(54) \,,~
m_{\mu}/m_\tau =  0.059\,4592\,(97)\,, \\
m_e     =  0.510\,9989\,918(44) \mathrm{MeV}\,,~
m_{\mu}     =  105.658\,3692\,(94) \mathrm{MeV}\,,~
m_{\tau}     = 1776.99\,(29) \mathrm{MeV}\epo
\end{array}
\label{leptonmasses}
\end{eqnarray}
The primary determination of the electron and muon masses come from
measuring the ratio with respect to the mass of a nucleus and the
masses are obtained in atomic mass units (amu). Therefore the ratios
are known more precisely, than the numbers we get by inserting lepton
masses given in MeV. In fact, the conversion factor to MeV is more
uncertain than the mass of the electron and muon in amu.

Note that the mass--dependent contributions in fact differ for $a_e$,
$\amu$ and $a_\tau$, such that lepton universality is broken: $a_e
\neq \amu \neq a_\tau$.

More SM parameters will be needed for the evaluation of weak and
hadronic contributions. We have collected them in Appendix~\ref{sec:appA}
together with known polylogarithmic functions needed for the representation 
of analytic results of the QED calculations.

Until recently the electron anomaly $a_e$ and, until before the advent
of the Brookhaven muon $g-2$ measurements, also $a_\mu$ were
considered to provide the most clean and precise tests of QED. In fact
the by far largest contribution to the \amm is of pure QED origin, and
with the new determination of $a_e$ by the Harvard electron $g-2$
experiment~\cite{aenew,aenew08} $a_e$ together with its
QED prediction~\cite{Aoyama07} allows for the most precise
determination of the electromagnetic fine structure constant.  The
dominance of just one type of interaction in the electromagnetic
vertex of the leptons, historically, was very important for the
development of QFT and QED, as it allowed to test QED as a model
theory under simple unambiguous conditions. How important such
experimental tests were we may learn from the fact that it took about
20 years from the invention of QED (Dirac 1928 [$g_e=2$]) until the
first reliable results could be established (Schwinger 1948
[$a^{(2)}_e=\alpha/2\pi$]) after a covariant formulation and
renormalization was understood and settled in its main aspects.  

When the precision of experiments improved, the QED part by itself
became a big challenge for theorists, because higher order corrections
are sizable, and as the order of perturbation theory increases, the
complexity of the calculations grows dramatically. Thus experimental
tests were able to check QED up to 7 digits in the prediction which
requires to evaluate the perturbation expansion up to 5 terms (5
loops).  The \amm as a dimensionless quantity exhibits contributions
which are just numbers expanded in powers of $\alpha$, what one would
get in QED with just one species of leptons, and contributions
depending on the mass ratios if different leptons come into play. Thus
taking into account all three leptons we obtain functions of the
ratios of the lepton masses $m_e$, $m_\mu$ and $m_\tau$.  Considering
$a_\mu$, we can cast it into the following form~\cite{KNO90,KM90}
\be
a_\mu^{\rm QED}=A_1+A_2(m_\mu/m_e)+A_2(m_\mu/m_\tau)
+A_3(m_\mu/m_e,m_\mu/m_\tau)\epo
\label{amuQEDform}
\ee
Here $A_1$ denotes the universal term common for all leptons. Also
closed fermion loops contribute to this term provided the fermion is
the muon (=external lepton). The term $A_2$ depends on one scale and gets
contributions from diagrams with closed fermion loops where the
fermion differs from the external one. Such contributions start at the
two loop level: for the muon as the external lepton we have two
possibilities: an additional electron--loop (light--in--heavy)
$A_2(m_\mu/m_e)$ or an additional $\tau$--loop (heavy--in--light)
$A_2(m_\mu/m_\tau)$ two contributions of quite different
character. The first produces large logarithms $\propto \ln
(m_\mu/m_e)^2$ and accordingly large effects while the second, because
of the {\em decoupling} of heavy particles in QED like theories\footnote{The 
Appelquist-Carrazone decoupling--theorem~\cite{AC75}
infers that in theories like QED or QCD, where couplings and masses
are independent parameters of the Lagrangian, a heavy particle of mass
$M$ decouples from physics at lower scales $E_0$ as $E_0/M$ for
$M\to\infty$.},
produces only small effects of order $\propto (m_\mu/m_\tau)^2$. The
two--scale contribution requires a light as well as a heavy extra
loop and hence starts at three loop order.  We will discuss the
different types of contributions in the following.  Each of the terms
is given in renormalized perturbation theory by an appropriate
expansion in $\alpha$:
\ba \begin{array}{cccccccccccccc}
A_1&=&A_1^{(2)} \aldpib &+& A_1^{(4)} \aldpib^2 &+& A_1^{(6)} \aldpib^3 &+&
A_1^{(8)} \aldpib^4 &+& A_1^{(10)} \aldpib^5 &+& \cdots \crn
A_2&=&                & & A_2^{(4)} \aldpib^2 &+& A_2^{(6)} \aldpib^3 &+&
A_2^{(8)} \aldpib^4 &+& A_2^{(10)} \aldpib^5 &+& \cdots \crn
A_3&=&            & &                       & & A_3^{(6)} \aldpib^3 &+&
A_3^{(8)} \aldpib^4 &+& A_3^{(10)} \aldpib^5 &+& \cdots \crn
\end{array} \label{defAs}
\ea
and later we will denote by
\ba
C_L=\sum_{k=1}^{3}\,A_{k}^{(2L)}\cs
\label{defCs}
\ea
the total $L$--loop coefficient of the $(\alpha/\pi)^L$ term.
The present precision of the experimental result~\cite{BNL04,BNLfinal}
\ba
\delta a_\mu^\mathrm{exp}=63 \power{-11}\cs
\label{precE821}
\ea
as well as the future prospects of possible improvements~\cite{LeeRob03}, 
which are expected to be able to reach
\be
\delta a^{\mathrm{fin}}_\mu \sim 10 \power{-11} \cs
\label{precgoal}
\ee
determine the precision at which we need the theoretical prediction.
For the $n$--loop coefficients multiplying $(\alpha/\pi)^n$ the error
Eq.~(\ref{precgoal}) translates into the required accuracies: $\delta
C_{1} \sim 4
\power{-8}$, $\delta C_{2} \sim 1 \power{-5}$, $\delta C_{3} \sim
7 \power{-3}$, $\delta C_{4}\sim 3$ and $\delta C_{5}\sim 1 \power{3~}$.
To match the current accuracy one has to multiply all estimates with a
factor 6, which is the experimental error in units of $10^{-10}$.

\subsection{Universal Contributions}
\noi
$\bullet$ According to Eq.~(\ref{amufullVP}) the leading order contribution
Fig.~\ref{fig:oneloopdia} may be written in the form (see below)
\ba
a_\ell^{(2)~\mathrm{QED}}&=&\frac{\alpha}{\pi} \int\limits_0^1\: \D
x\:(1-x)\:=\:\frac{\alpha}{\pi} \:\ha \cs
\ea
which is trivial to evaluate. This is the famous result of Schwinger
from 1948~\cite{Schwinger48}.
\begin{figure}[h]
\centering
\includegraphics[scale=0.91]{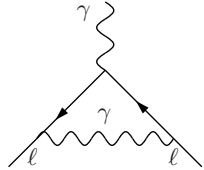}

\caption{The universal lowest order QED contribution to $a_\ell$.}
\label{fig:oneloopdia}
\end{figure}

\begin{figure}[t]
\centering
\includegraphics[scale=0.91]{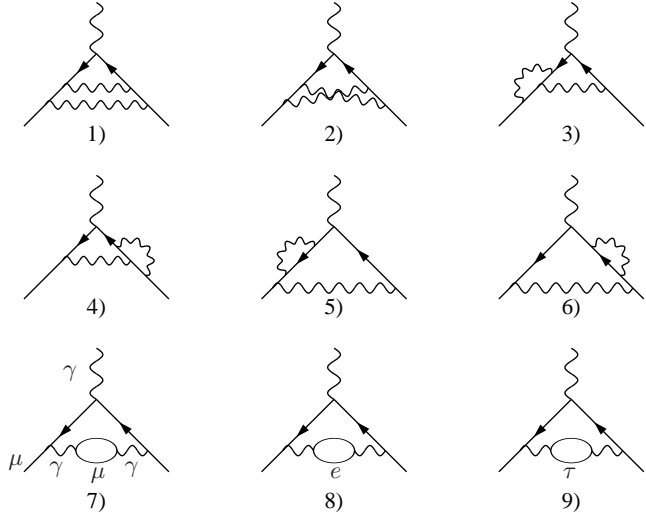}

\caption{Diagrams 1-7 represent the universal second order contribution to
$a_\mu$, diagram 8 yields the ``light'', diagram 9 the ``heavy'' mass
dependent corrections.}
\label{fig:twoloopdia}
\end{figure}

\noi
$\bullet$ At two loops in QED there are the 9 diagrams shown in
Fig.~\ref{fig:twoloopdia} which contribute to $\amu$. The first 6
diagrams, which have attached two virtual photons to the external muon string of lines
contribute to the universal term. They
form a gauge invariant subset of diagrams and yield the result
\bea
A_{1\:\mathrm{\small [1-6]}}^{(4)}= -\frac{279}{144}+ \frac{5\pi^2}{12}-
\frac{\pi^2}{2}\ln 2 + \frac{3}{4} \zeta(3)\epo
\eea
The last 3 diagrams include photon \textit{vacuum polarization} (vap / VP) due
to the lepton loops. The one with the muon loop is also universal in the
sense that it contributes to the mass independent correction

\begin{figure}[t]
\centering
\includegraphics[scale=0.96]{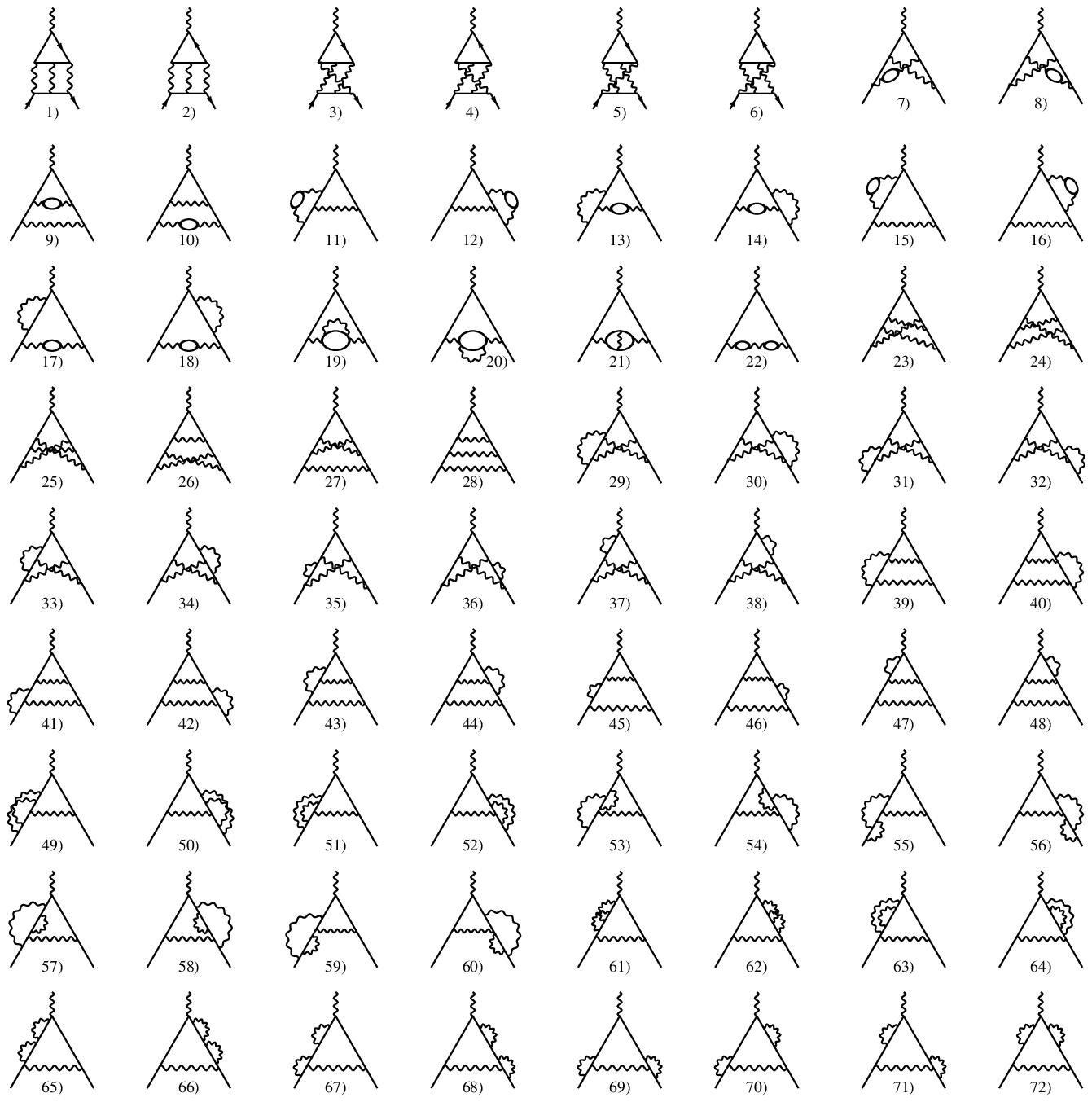}


\caption{The universal third order contribution to $a_\mu$. All
fermion loops here are muon--loops. Graphs 1) to 6) are the
light--by--light scattering diagrams. Graphs 7) to 22) include photon
vacuum polarization insertions. All non--universal contributions follow
by replacing at least one muon in a closed loop by some other fermion.}
\label{fig:threeloopdia1}
\end{figure}

\bea
A_{1\:\mathrm{vap}}^{(4)}(m_\mu/m_\ell=1)= \frac{119}{36}- \frac{\pi^2}{3} \epo
\eea
The complete ``universal'' part yields the coefficient
$A_1^{(4)}$ calculated first by Petermann~\cite{Petermann57} and
by Sommerfield~\cite{Sommerfield57} in 1957:
\be
A_{1\:\mathrm{uni}}^{(4)}=
\frac{197}{144}+ \frac{\pi^2}{12}-
\frac{\pi^2}{2}\ln 2 + \frac{3}{4} \zeta(3)
= -0.328~478~965~579~193~78...
\label{A14univ}
\ee
where $\zeta(n)$ is the Riemann $\zeta$--function of argument $n$ (see
also~\cite{Terentev62}).

\noi
$\bullet$ At three loops in QED there are the 72 diagrams shown in
Fig.~\ref{fig:threeloopdia1} contributing to $g-2$ of the muon. In
closed fermion loops any of the SM fermions may circulate.  The gauge
invariant subset of 72 diagrams where all closed fermion loops are
muon--loops yield the universal one--flavor QED contribution
$A_{1\:\mathrm{uni}}^{(6)}$. This set has been calculated analytically
mainly by Remiddi and his collaborators~\cite{Remiddi}, and Laporta and Remiddi 
obtained the final result in 1996 after finding a trick
to calculate the non--planar ``triple cross'' topology diagram
25) of Fig.~\ref{fig:threeloopdia1}~\cite{LaportaRemiddi96} (see
also~\cite{Ki95}). The
result turned out to be surprisingly compact and reads
\ba
A_{1\:\mathrm{uni}}^{(6)}
&=&\frac{28259}{5184}+ \frac{17101}{810} \pi^2 -
\frac{298}{9}\pi^2 \ln 2 + \frac{139}{18} \zeta(3) 
+\frac{100}{3} \left\{ \mathrm{Li}_4(\frac{1}{2})
+\frac{1}{24}\ln^42-\frac{1}{24}\pi^2 \ln^22 \right\}  \crn && 
-\frac{239}{2160}\pi^4 + \frac{83}{72}\pi^2
\zeta(3)-\frac{215}{24}\zeta(5) 
= 1.181\,241\,456\,587 \ldots 
\label{aell6univ}
\ea
This famous analytical result largely confirmed an earlier numerical
calculation by Kinoshita~\cite{Ki95}. The constants needed for the
evaluation of Eq.~(\ref{aell6univ}) are given in Eqs.~(\ref{Zetanum}) and
(\ref{Linum}).

The big advantage of the analytic result is that it allows a numerical
evaluation at any desired precision. The direct numerical
evaluation of the multidimensional Feynman integrals by Monte Carlo methods
is always of limited precision and an improvement  is always very expensive in
computing power.

\begin{figure}[h]
\centering
\includegraphics[scale=0.91]{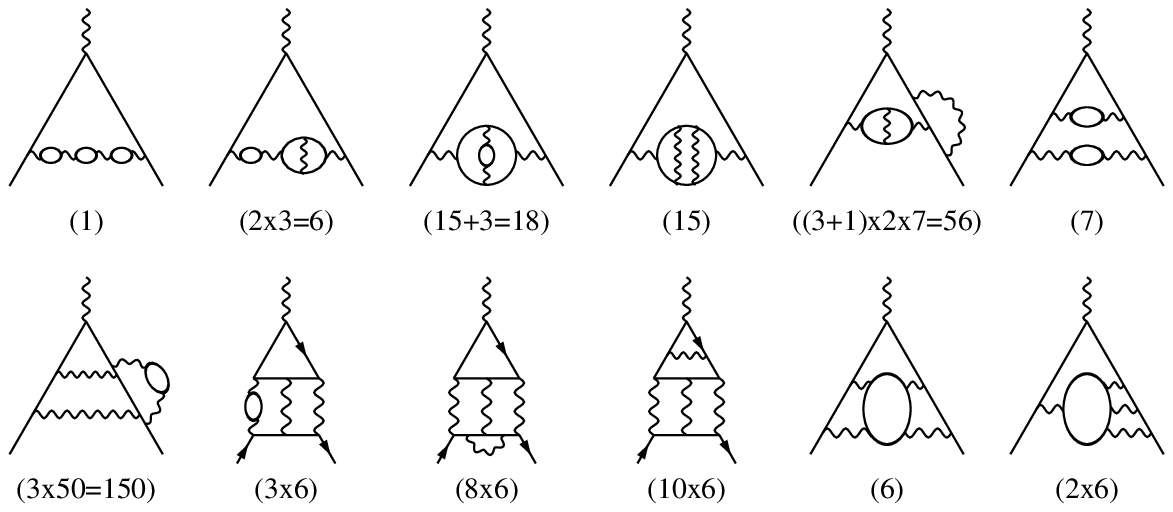}

\caption{Some typical eight order contributions to $a_\ell$ involving
lepton loops. In brackets the number of diagrams of a given type
if only muon loops are considered. The latter contribute to
the universal part.}
\label{fig:fourloopdia1}
\end{figure}

\begin{figure}[h]
\centering
\includegraphics[scale=0.6]{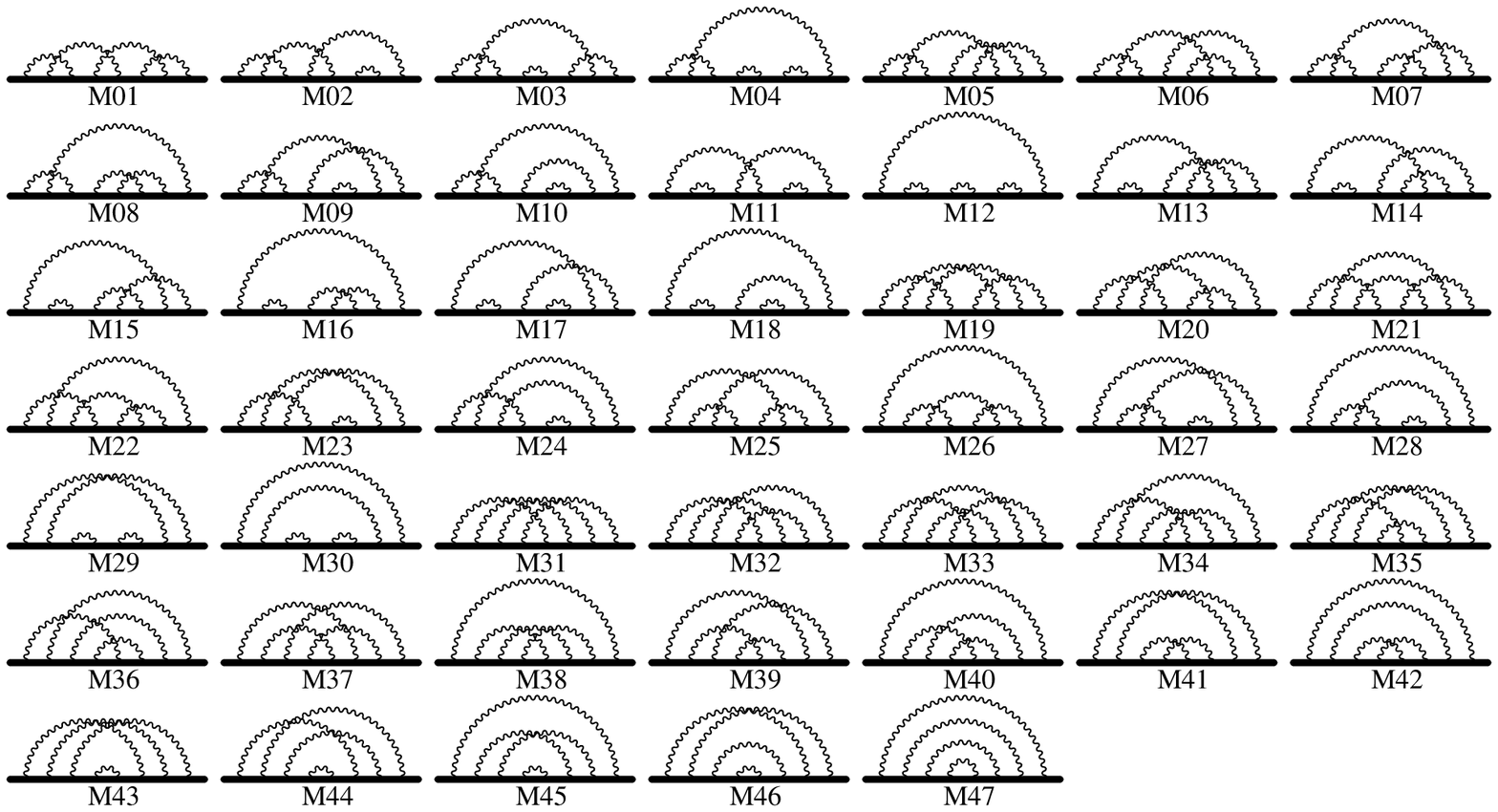}
\caption{4-loop Group V diagrams. 47 self-energy-like diagrams of
$M_{01}$ -- $M_{47}$ represent 518 vertex diagrams [by inserting the external
photon vertex on the virtual muon lines in all possible ways]. 
Reprinted with permission from~\cite{Aoyama07}. Copyright (2007) by
the American Physical Society].}
\label{fig:A18V} 
\end{figure}

\noi
$\bullet$ At four loops there are 891 diagrams [373 have closed lepton
loops (see Fig.~\ref{fig:fourloopdia1}), 518 without fermion
loops=gauge invariant set Group V (see Fig.~\ref{fig:A18V})] with common fermion lines. 
Their contribution has been calculated by numerical methods by Kinoshita and
collaborators. 
The calculation of the 4--loop contribution to $\amu$ is a formidable
task. Since the individual diagrams are much more complicated
than the 3--loop ones, only a few have been calculated analytically
so far~\cite{A8analytic}--\cite{Aguilar08}. In most cases one has to resort to numerical
calculations. This approach has been developed and perfected over the
past 25 years by Kinoshita and his
collaborators~\cite{KL81}--\cite{HughesKinoshita99}
with the very recent recalculations and improvements~\cite{Aoyama07,KinoNio04,Kino05}.
As a result of the enduring heroic effort an
improved answer has been obtained recently by Aoyama, Hayakawa,
Kinoshita and Nio~\cite{Aoyama07} who find
\ba
A^{(8)}_{1}= -1.9144(35)
\label{A8uni}
\ea
where the error is due to the Monte Carlo integration. 
This very recent result is correcting the one published before
in~\cite{KinoNio05} and shifting the coefficient of the
$\left(\frac{\alpha}{\pi} \right)^4$ term by --~0.19 (10\%).
Some error in the cancellation of IR singular terms was found in calculating
diagrams $M_{18}$ ($-0.2207(210)$) and $M_{16}$ ($+0.0274(235)$) in the set of diagrams 
Fig.~\ref{fig:A18V}. The latter 518 diagrams without fermion loops also are responsible for  
the largest part of the uncertainty in Eq.~(\ref{A8uni}).
Note that the universal $O(\alpha^4)$ contribution is sizable, about 6 standard deviations
at current experimental accuracy, and a precise knowledge of this term
is absolutely crucial for the comparison between theory and experiment.

\noi
$\bullet$ The universal 5--loop QED contribution is still largely unknown. 
\begin{figure}[h]
\centering
\includegraphics[scale=0.91]{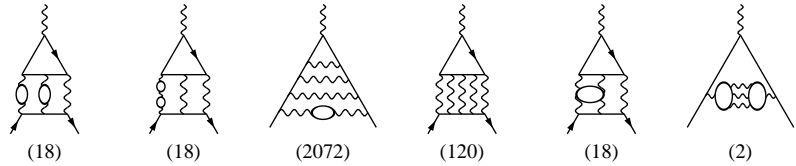}


\caption{Typical tenth order contributions to $a_\ell$ including
fermion loops. In brackets the number of diagrams of the given type.}
\label{fig:fiveloopdia1}
\end{figure}
Using the recipe proposed in Ref.~\cite{CODATA00}, one
obtains the following bound
\be
A^{(10)}_1 =0.0(4.6)\cs
\label{A10uni}
\ee
for the universal part as an estimate for the missing higher order terms.

As a result the universal QED contribution may be written as
\ba
a_\ell^{\rm uni}&=&0.5 \: \left(\frac{\alpha}{\pi} \right)
-0.328\,478\,965\,579\,193\,78 \ldots  \: \left(\frac{\alpha}{\pi} \right)^2
\crn && +1.181\,241\,456\,587 \ldots  \: \left(\frac{\alpha}{\pi} \right)^3
-1.9144(35) \: \left(\frac{\alpha}{\pi} \right)^4
+0.0(4.6)\: \left(\frac{\alpha}{\pi} \right)^5 \crn &=&
0.001\,159\,652\,176\,30(43)(10)(31)[54] \cdots
\label{aellunivesalnum}
\ea
The three errors given are: the error from the uncertainty in
$\alpha$, given in Eq.~(\ref{alphainv_a_e}), the numerical uncertainty
of the $\alpha^4$ coefficient and the error estimated for the missing
higher order terms.

As we already know, the \amm of a lepton is an effect of about 0.12\%,
$g_\ell/2\simeq 1.00116\cdots$. It is remarkable that in spite of the
fact that this observable is so small we know $a_e$ and $a_\mu$ more
precisely than most other precision observables. Note that the first
term $a_\ell^{(2)}\simeq 0.00116141\cdots$ contributes the first three
significant digits of the full result.

\subsection{Electron Anomalous Magnetic Moment and the Fine Structure Constant}
\label{sec:aele}
The universal terms given in Eq.~(\ref{aellunivesalnum}) essentially determine the anomalous
magnetic moment of the electron $a_e$ and therefore allow a precise
determination of the fine structure constant from the experimentally measured
value for $a_e$ by inverting the series in $\alpha$. This is due to the fact
that the effects from heavy leptons (muon, tau) in QED, from hadrons, the
electroweak sector and potential new physics decouple as $(m_e/M)^2$, where
$M$ is some heavy lepton mass or a hadronic, weak or new physics scale.

The electron magnetic moment anomaly likely is the experimentally
most precisely known quantity.  Since recently, a new
substantially improved result
for $a_e$ is available.  It was obtained by Gabrielse et al.~\cite{aenew,aenew08} in
an experiment at Harvard University using a one--electron quantum
cyclotron. The new results from 2006 and 2008 read
\ba
a_{e}^{\rm exp}&=&1.159\, 652\, 180\,85(76) \times 10^{-3}\,[.66\mathrm{ \ ppb}]\,,\crn
a_{e}^{\rm exp}&=&1.159\, 652\, 180\,73(28) \times 10^{-3}\,[.24\mathrm{ \ ppb}]\,,
\label{aeexp}
\ea
the latter with an accuracy 15 times better than  the  earlier result 
\bea
a_{e}^{\rm exp}=1.159\, 652\, 1883(42) \times 10^{-3}\,[3.62\mathrm{ \ ppb}]\,,
\eea
obtained by Dehmelt et al. at Washington University in 
1987~\cite{VanDycketal87,CODATA00}.
The new value is shifting down $a_e$ by 1.8 standard deviations.

The measurements of $a_e$ not only played a key role in the history of
precision tests of QED in particular, and of QFT concepts in general,
today we may use the \amm of the electron to get the most precise
indirect measurement of the fine structure constant $\alpha$. This
possibility of course hangs on our ability to pin down the theoretical
prediction with very high accuracy. Indeed $a_e$ is much saver to
predict reliably than $a_\mu$. The reason is that non--perturbative hadronic
effects as well as the sensitivity to unknown physics beyond the SM
are suppressed by the large factor $m^2_\mu/m^2_e\simeq 42\,753$ in
comparison to $a_\mu$. This suppression has to be put into perspective
with the 2250 times higher precision with which we know $a_e$.
We thus can say that effectively $a_e$ is a factor 19 less
sensitive to model dependent physics than $\amu$.

The prediction is given by a perturbation expansion of the form (see
also Eqs.~(\ref{amuQEDform}), (\ref{defCs}))
\be
a_e^{\rm QED}=\sum_{n=1}^N C_{n} (\alpha/\pi)^n \cs
\label{aeexpansion}
\ee
with terms up to five loops, $N=5$, under consideration.
The experimental precision of $a_e$ requires the knowledge of
the coefficients with accuracies $\delta C_{2}\sim 1 \power{-7}$,
$\delta C_{3}\sim 6 \power{-5}$, $\delta C_{4}\sim 2 \power{-2}$ and
$\delta C_{5}\sim 10$.
For what concerns the universal terms one may conclude by inspecting
the convergence of Eq.~(\ref{aellunivesalnum}) that one would expect the
completely unknown coefficient $C_{5}$ to be $O(1)$ and hence
negligible at present accuracy. In reality it is one of the main
uncertainties, which is already accounted for in
Eq.~(\ref{aellunivesalnum}). Concerning the mass--dependent contributions,
the situation for the electron is quite different from the muon. Since
the electron is the lightest of the leptons a potentially large
``light internal loop'' contribution is absent. For $a_e$ the muon is
a heavy particle $m_\mu \gg m_e$ and its contribution is of the type
``heavy internal loops'' which is suppressed by an extra power of
$m^2_e/m^2_\mu$. In fact the $\mu$--loops tend to decouple and
therefore only yield small terms.  Corrections due to internal
$\mu$--loops are suppressed as $O(2(\alpha/\pi)\, (m_e^2/m_\mu^2))\simeq
1.1\power{-7}$ relative to the leading term and the $\tau$--loops
practically play no role at all. The fact that muons and tau leptons
tend to decouple is also crucial for the unknown 5--loop
contribution, since we can expect that corresponding contributions can
be safely neglected.

The result may be written in the form
\ba
a_e^\mathrm{QED}&=&a_e^{\rm uni}+a_e(\mu)+a_e(\tau)+a_e(\mu,\tau)\cs
\ea
with the universal term given by Eq.~(\ref{aellunivesalnum}) and
\bea
a_e(\mu)&=&5.197\,386\,70(27) \times 10^{-7}
{\left(\frac{\alpha}{\pi} \right)^2}
-7.373\,941\,65(29) \power{-6} {\left(\frac{\alpha}{\pi}
\right)^3}\cs \crn
a_e(\tau)&=&1.83763(60) \times
10^{-9} {\left(\frac{\alpha}{\pi} \right)^2}
 -6.5819(19) \power{-8} {\left(\frac{\alpha}{\pi}
\right)^3}\cs \crn
a_e(\mu,\tau)&=&0.190945(62) \power{-12} {\left(\frac{\alpha}{\pi}
\right)^3} \epo
\nn
\eea
As a result the perturbative expansion
for the QED prediction of $a_e$ is given by
\begin{eqnarray}
a_e^{\rm QED}
 &=& {\alpha\over 2\pi}
 -0.328\,478\,444\,002\,90(60) \left( {\alpha\over \pi}\right)^2
 +1.181\, 234\, 016\, 827(19) \left( {\alpha\over \pi}\right)^3
\nonumber\\&&
-1.9144(35)\left( {\alpha\over \pi}\right)^4
+0.0(4.6)\left({\alpha\over \pi}\right)^5.
\label{aeQED}
\end{eqnarray}
As mentioned before, the completely unknown universal 5--loop term
$C_5 \simeq A^{(10)}_1$ has been estimated to be bounded by the last term. 
The missing 5--loop result represents the largest uncertainty in the 
prediction of $a_e$.

What is missing are the hadronic and weak contributions, which
both are suppressed by the $(m_e/m_\mu)^2$ factor relative to
$a_\mu$. For $a_e$ they are small\footnote{The total
hadronic contribution to $a_e$ is given by 
$a_e^{(4)}(\mathrm{vap,\,had})+a_e^{(6)}(\mathrm{vap,\,had})
+a_e(\mathrm{LbL,\,had}) \sim (1.860\pm 0.015-0.223\mp 0.002+0.039\pm 0.013)
\times 10^{-12}$ (see below).} : $a_e^{\rm had}=1.676(18) \times
10^{-12}$ and $a_e^{\rm weak}=0.039 \times 10^{-12}$,
respectively (see the discussion of the corresponding contributions to
$\amu$ and Sect.~\ref{ssec:elwea_a_e}
below). The hadronic contribution now just starts to be
significant, however, unlike in $a_\mu^{\rm had}$ for the muon,
$a_e^{\rm had}$ is known with sufficient accuracy and is not the
limiting factor here. The theory error is dominated by the missing
5--loop QED term.  As a consequence $a_e$ at this level of
accuracy is theoretically well under control (almost a pure QED
object) and therefore is an excellent observable for extracting
$\alpha$ based on the SM prediction
\begin{eqnarray}
\!\!a_e^{\rm SM} &=& a_e^{\rm QED} \mathrm{[Eq.~(\ref{aeQED})]}
+1.715(18)\times 10^{-12}~
\mbox{(hadronic \& weak)}\,.
\end{eqnarray}

When we compare this result with the very recent extremely precise measurement
of the electron anomalous magnetic moment~\cite{aenew08} given by  Eq.~(\ref{new_a_e})
we obtain
\begin{eqnarray*}
\alpha^{-1}(a_e)&=&137.035999084(33)(12)(37)(2)[51]\;,
\end{eqnarray*}
which is the value Eq.~(\ref{alphainv_a_e})~\cite{aenew08} given
earlier.
The first error is the experimental one of
$a_e^\mathrm{exp}$, the second and third are the numerical
uncertainties of the $\alpha^4$ and $\alpha^5$ terms,
respectively. The last one is the hadronic uncertainty, which is
completely negligible.  The recent correction of the $O(\alpha^4)$
coefficient Eq.~(\ref{A8uni}) (from $-1.7283(35)$ to $-1.9144(35)$)
lead to a 7 $\sigma$ shift in $\alpha(a_e)$. This is the most precise
determination of $\alpha$ at present and we will use it for
calculating $a_\mu$.\\

Of course we still may use $a_e$ for a precision test of QED.
For a theoretical prediction of  $a_e$ we then have to adopt
the best determinations of $\alpha$ which do not depend on $a_e$.
They are~\cite{Cs06,Rb06}
\ba
\alpha^{-1}(\mathrm{Cs})&=&137.03600000(110)[8.0\, \mathrm{ppb}]\;,\label{alphainv_aiCs}\\
\alpha^{-1}(\mathrm{Rb})&=&137.03599884(091)[6.7\, \mathrm{ppb}]\;,
\label{alphainv_aiRb}
\ea
and have been determined by atomic interferometry.
In terms of \label{page:aelexpvsthe}
$\alpha(\mathrm{Cs})$ one gets $a_e= 0.00115965217299(930)$ which
agrees well with the experimental value
$a_e^\mathrm{exp}-a_e^\mathrm{the}=7.74(9.30) \times 10^{-12}$;
and similarly, using the value $\alpha(\mathrm{Rb})$ the prediction is
$a_e= 0.00115965218279(770)$, again in good agreement with experiment
$a_e^\mathrm{exp}-a_e^\mathrm{the}=-2.06(7.70) \times 10^{-12}$.
Errors are completely dominated by the uncertainties in $\alpha$.
The following
Table~\ref{tab:aecontributions} collects the typical contributions to
$a_e$ evaluated in terms of Eqs.~(\ref{alphainv_aiCs},\ref{alphainv_aiRb}).
\begin{table}[h]
\centering
\caption{Contributions to $a_e(h/M)$ in units $10^{-6}$. The three errors 
given in the universal contribution come from the experimental
uncertainty in $\alpha$, from the $\alpha^4$ term and from the
$\alpha^5$ term, respectively.}
\label{tab:aecontributions}
\begin{tabular}{lr@{.}lr@{.}l}
\noalign{\smallskip}\hline\noalign{\smallskip}
contribution &\multicolumn{2}{c}{ $\alpha(h/M_{\rm Cs})$} &\multicolumn{2}{c}{$\alpha(h/M_{\rm Rb})$} \\
\noalign{\smallskip}\hline\noalign{\smallskip}
  universal   & $~~1159$&$652\,16856(929)(10)(31)$& $~~1159$&$652\,17836(769)(10)(31)$ \\
$\mu $--loops & $   0$&$000\,00271~~(0) $        & $   0$&$000\,00271~~(0) $         \\
$\tau$--loops & $   0$&$000\,00001~~(0) $        & $   0$&$000\,00001~~(0) $         \\
 hadronic     & $   0$&$000\,00168~~(2) $        & $   0$&$000\,00168~~(2) $         \\
 weak         & $   0$&$000\,000039~(0) $        & $   0$&$000\,000039~(0) $            \\
\noalign{\smallskip}\noalign{\smallskip}
 theory     & $1159$&$652\,17299(930)$& $1159$&$652\,18279(770)$ \\
 experiment &$1159$&$652\,180\,73~(28)$&$1159$&$652\,180\,73~(28)$ \\
\noalign{\smallskip}\hline
\end{tabular}
\end{table}

\noi
Obviously an improvement of non--$a_e$ determinations of $\alpha$ by a factor 20 
would allow a much more stringent test of QED, and therefore  
would be very important.
At present, assuming that $\left| \Delta a_e^{\rm New\ Physics}\right|
\simeq m_e^2 / \Lambda^2$ where $\Lambda$ approximates the scale of
``New Physics'', the agreement between $\alpha^{-1}(a_e)$ and
$\alpha^{-1}(\mathrm{Rb06})$ probes the scale $\Lambda \lsim
O(\mbox{250 GeV})$. To access the much more interesting range of $\Lambda
\sim O(\mbox{1 TeV})$  would also require
a reliable estimate of the first significant digit of the 5--loop QED
contribution, and an improved calculation of the 4--loop QED
contribution to $a_e^{\rm SM}$.

\subsection{Mass Dependent Contributions}
Since fermions, as demanded by the SM\footnote{Interactions are known
to derive from a local gauge symmetry principle, which implies the
structure of gauge couplings, which must be of vector (V) or
axial--vector (A) type.}, only interact via photons or other spin one
gauge bosons, mass dependent corrections at first show up at the
2--loop level via photon vacuum polarization effects. At
three loops light--by--light scattering loops show up, etc.  As all
fermions have different masses, the fermion-loops give rise to mass
dependent effects, which were calculated at two loops
in~\cite{SWP57,El66} (see
also~\cite{LdeR69}--\cite{LiMeSa93}), at three loops
in~\cite{Ki67}--\cite{FGdR05}, and at four loops
in~\cite{A8analytic}--\cite{Aguilar08},\cite{KinoNio04}. For five loops only partial estimates
exist~\cite{A8:10RG,Aguilar08},\cite{Ka93}--\cite{Baikov:2008si}.

The leading mass dependent effects come from photon vacuum
polarization, which leads to charge screening.
Including a factor $e^2$ and considering the renormalized photon propagator
(wave function renormalization factor $Z_\gamma$) we have
\begin{equation}
\I \:e^2\:D^{' \mu \nu}_\gamma (q)=\frac{- \I g^{\mu \nu} \: e^2\:Z_\gamma}{q^2\:
\left( 1+\Pi'_\gamma(q^2)\right)}+ {\rm \
gauge \ terms \ }\cs
\end{equation}
which in effect means that the charge has to be replaced by an
energy-momentum scale dependent {\em running charge}
\begin{equation}
e^2 \to e^2(q^2)=\frac{e^2Z_\gamma}{1+\Pi'_\gamma(q^2)}=\frac{e^2}{1+(\Pi'_\gamma(q^2)-\Pi'_\gamma(0))} \;,
\label{runninge}
\end{equation}
where $Z_\gamma$ is fixed to obtain the classical charge in the
Thomson limit $q^2 \to 0$. In perturbation theory the lowest order
diagram which contributes to $\Pi'_\gamma(q^2)$ is
\begin{figure}[h]
\centering
\includegraphics{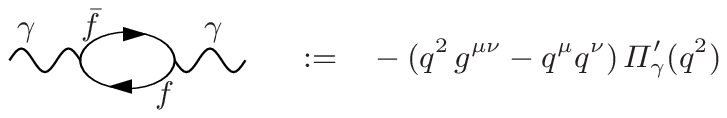}
\end{figure}

\noindent
and describes the virtual creation and re-absorption of fermion pairs
$\gamma^* \rightarrow e^+e^-$, $\mu^+ \mu^-$, $\tau^+ \tau^-$, $u\bar{u}$, $d
\bar{d}$, $\cdots (\mathrm{had})\rightarrow ~~\gamma^*$~.
The photon self--energy function may
also be defined by the time-ordered correlator of two electromagnetic
currents as
\be
\I\:e^2\,\int \D^4 x\: \E^{\I qx} \bra{0}T j^\mu_\mathrm{em}(x)j^\nu_\mathrm{em}(0)\ket{0}
=-(q^2\,g^{\mu \nu}-q^\mu q^\nu)\,\Pi'_\gamma(q^2)\cs
\label{jemspectral}
\ee
which is purely transversal by virtue of electromagnetic current conservation $
\partial_\mu j^\mu_\mathrm{em}(x)=0$.

In terms of the fine structure constant $\alpha=\frac{e^2}{4\pi}$ Eq.~({\ref{runninge})
reads
\begin{equation}
\alpha(q^2)=\frac{\alpha}{1-\Delta \alpha(q^2)}\;\;\;;\;\;\;\Delta \alpha(q^2) =
- {\rm Re}\:\left(\Pi'_\gamma(q^2)-\Pi'_\gamma(0)\right) \;.
\label{dalpdef}
\end{equation}
The various contributions to the shift in the fine structure constant
come from the leptons (lep = $e$, $\mu$ and $\tau$), the 5 light
quarks ($u$, $b$, $s$, $c$, and $b$) and/or the corresponding hadrons
(had). The top quark is too heavy to give a relevant contribution. The
hadronic contributions will be considered later. 

The renormalized photon self--energy is an analytic function and
satisfies the dispersion relation (DR)
\ba
-\frac{\Pi'_{\gamma\:\mathrm{ren}} (k^2)}{k^2} = \int\limits_0^\infty
\frac{\D s}{s}\, \frac{1}{\pi}\Impa \Pi'_\gamma (s)\:\frac{1}{k^2-s} \epo
\label{ppDR}
\ea
Note that the only $k$ dependence under the convolution integral shows
up in the last factor.  Thus, in a generic VP contribution\\
\begin{figure}[h]
\vspace*{-6mm}
\centering
\includegraphics{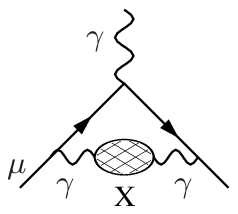}
\vspace*{-4mm}
\end{figure}

\noi where the ``blob'' is the full photon propagator, including all kinds
of contributions as predicted by the SM or beyond, the free photon
propagator in the 1--loop vertex graph in the next higher order is
replaced by
\bea
-\I g_{\mu \nu}/k^2 \to -\I g_{\mu \nu}/(k^2-s)\cs
\eea
which is the exchange of a photon of mass square $s$. This result
then has to be convoluted with the imaginary part of the photon
vacuum polarization. The calculation of the
contribution from the massive photon proceeds exactly
as in the massless case. Again $F_\mathrm{M}(0)$ most
simply may be calculated using the projection method which allows to work at
$q^2=0$.  The result is~\cite{BKK56,BdeR68}
\be
a^{(2)\:\mathrm{heavy}\:\gamma}_\mu \equiv
\frac{\alpha}{\pi}\,K_\mu^{(2)}(s) =
\frac{\alpha}{\pi} \int\limits_0^1\: \D x\: \frac{x^2\:(1-x)}{x^2+(s/m^2_\mu)(1-x)}\cs
\label{vapokernel}
\ee
which is the leading order contribution to $\amu$ from an exchange of a
photon with square mass $s$. For $s=0$ we get the known
Schwinger result. Utilizing this result and Eq.~(\ref{ppDR}), the contribution from the
``blob'' to $g-2$ reads
\be
\amu^{(X)}= \frac{\alpha}{\pi^2} \int\limits_0^\infty \: \frac{\D s}{s} \:
\Impa \Pi^{'(X)}_\gamma (s)\: K_\mu^{(2)}(s) \epo
\label{amuvapo}
\ee
If we exchange integrations and evaluating the DR we arrive at~\cite{LPdR72}
\ba
\amu^{(X)} &=& \frac{\alpha}{\pi} \int\limits_0^1\: \D x\:(1-x)\:
\int\limits_0^\infty \: \frac{\D s}{s} \:
\:\frac{1}{\pi} \:\Impa \Pi^{'(X)}_\gamma (s)\: \frac{x^2}{x^2+(s/m^2_\mu)(1-x)} \crn
&=&\frac{\alpha}{\pi} \int\limits_0^1\: \D x\:(1-x)\: \left[-\Pi^{'(X)}_\gamma(s_x) \right]
\mathrm{\,, \ with \ \ } s_x=-\frac{x^2}{1-x} \: m_\mu^2 \epo
\label{amuxint}
\ea
The last simple representation in terms of $\Pi^{'(X)}_\gamma(s_x)$ follows
using
\bea
\frac{x^2}{x^2+(s/m^2_\mu)(1-x)}=-s_x \: \frac{1}{s-s_x} \epo
\eea
Formally, this means that we may replace the free photon propagator 
by the full transverse propagator in the 1--loop muon vertex~\cite{Lautrup77}:
\ba
\amu^{(X),\mathrm{resummed}} &=&
\frac{\alpha}{\pi} \int\limits_0^1\: \D x\:(1-x)\:\left( 1-\Pi_{\gamma\:\mathrm{ren}}^{'(X)}(s_x)
+(\Pi_{\gamma\:\mathrm{ren}}^{'(X)}(s_x))^2+\cdots \right) \crn
&=&\frac{\alpha}{\pi} \int\limits_0^1\: \D x\:(1-x)\:
\left(\frac{1}{1+\Pi_{\gamma\:\mathrm{ren}}^{'(X)}(s_x)}\right)\epo 
\label{amufullVP}
\ea
By Eq.~(\ref{runninge}) this is equivalent to the contribution
of a free photon interacting with dressed charge (effective fine
structure constant).  However, since
$\Pi_{\gamma\:\mathrm{ren}}'(k^2)$ is negative and grows
logarithmically with $k^2$ the full photon propagator develops a so
called \textit{Landau pole} where the effective fine structure
constant becomes infinite. Thus resumming the
perturbation expansion under integrals produces a problem and one
better resorts to the order by order approach, by expanding the full
propagator into its geometrical progression. In this case 
Eq.~(\ref{amufullVP}) may be considered as a very useful bookkeeping device, collecting
effects from different contributions and different orders.

The running of $\alpha$ caused by vacuum polarization effects is
controlled by the renormalization group (RG). The latter 
systematically takes care of the terms enhanced by large short--distance
logarithms of the type $\ln m_\mu/m_e$ in the case of $\amu$. Since in
QED one usually adopts an on shell renormalization scheme the RG
for $\amu$ is actually the Callan-Symanzik (CS)
equation~\cite{LdeR74}, which in the limit $m_e \ll m_\mu$, i.e. neglecting
power corrections in $m_e/m_\mu$, takes the homogeneous
form
\bea
\left(m_e \frac{\partial}{\partial m_e}+\beta(\alpha)\:\alpha
\frac{\partial}{\partial \alpha}
\right)\:a_\mu^{(\infty)}\left(\frac{m_\mu}{m_e},\alpha \right)=0\cs
\eea
where $a_\mu^{(\infty)}(\frac{m_\mu}{m_e},\alpha)$ is the
corresponding asymptotic form of $a_\mu$ and $\beta(\alpha)$ is the
QED $\beta$--function. The latter governs the charge screening of the
electromagnetic charge. To leading order the charge is running
according to
\ba
\alpha(\mu)&=& \frac{\alpha}{1-\frac{2}{3}\frac{\alpha}{\pi}\:\ln
\frac{\mu}{m_e}}\simeq \alpha\,\left(1+\frac{2}{3}\frac{\alpha}{\pi}\:\ln
\frac{\mu}{m_e} +\cdots\right) \epo
\ea
The solution of the CS equation amounts to
replace $\alpha$ by the running fine structure constant
$\alpha(m_\mu)$ in $a_\mu^{(\infty)}(\frac{m_\mu}{m_e},\alpha)$, which
implies taking into account the leading logs of higher orders. If we
replace in the 1--loop result $\alpha \to \alpha(m_\mu)$ we obtain
\ba
\amu=\frac{1}{2}\frac{\alpha}{\pi}\:
(1+\frac{2}{3}\frac{\alpha}{\pi}\:\ln \frac{m_\mu}{m_e})\cs
\label{RG1loop}
\ea
which reproduces precisely the leading term of the 2--loop result
given below.
Since $\beta$ is known
to four loops~\cite{GKLS90} and also $a_\mu$ is known analytically at three loops,
it is possible to obtain the important higher leading logs quite
easily. For more elaborate RG estimates of contributions 
to $a_\mu^{(8)}$ and $a_\mu^{(10)}$ we refer to Ref.~\cite{A8:10RG}.

\subsubsection{2--loop Vacuum Polarization Insertions}
The leading mass dependent non--universal contribution is due to the
last two diagrams of Fig.~\ref{fig:twoloopdia}.  The coefficient now
is a function of the mass $m_\ell$ of the lepton forming the closed
loop. For actually calculating the VP contributions the 
1--loop photon vacuum polarization is needed. It is given by
\ba
\Pi'_{\gamma\:\mathrm{ren}}(q^2)&=&-\frac{\alpha}{\pi} \int\limits_0^1
\D z\:2z\:(1-z)\:\ln (1-z\:(1-z)\:q^2/m_\ell^2) \crn
&=&\frac{\alpha}{\pi} \int\limits_0^1
\D t\:t^2\:(1-t^2/3)\:\frac{1}{4m_\ell^2/q^2-(1-t^2)} \cs
\label{Pigammaint}
\ea
and performing the integral yields
\ba
\Pi'_{\gamma\:\mathrm{ren}}(q^2)
&=&-\frac{\alpha}{3\pi}\left\{\frac{8}{3}-\beta_\ell^2+\frac{1}{2}\beta_\ell\,(3-\beta_\ell^2)\,\ln
\frac{\beta_\ell-1}{\beta_\ell+1}\right\}
\cs
\label{Pigammaana}
\ea
where  $\beta_\ell=\sqrt{1-4m_\ell^2/q^2}$ is the lepton velocity.
The imaginary part is given by
the simple formula
\ba
\Impa \Pi'_\gamma(q^2)&=&\frac{\alpha}{3} 
\left(1+\frac{2m_\ell^2}{q^2} \right)\,\beta_\ell \epo
\label{vapoimag}
\ea
For $q^2<0$ the amplitude $\Pi'_{\gamma\:\mathrm{ren}}(q^2)$ is negative
definite and what is needed in Eq.~(\ref{amuxint}) is $-\Pi^{'(\ell)}_\gamma(-\frac{x^2}{1-x}\,m_\mu^2)$ 
or Eq.~(\ref{Pigammaana}) with $\beta_\ell=\sqrt{1+4\,x_\ell^2\,(1-x)/x^2}$,
where $x_\ell=m_\ell/m_{\mu}$ and $m_\ell$ is the mass of the virtual
lepton in the vacuum polarization subgraph.

Using the representation Eq.~(\ref{amuxint}) together
with Eq.~(\ref{Pigammaint}) the VP insertion
was computed in the late 1950s~\cite{SWP57} for $m_\ell= m_e$ and
neglecting terms of $O(m_e/m_{\mu})$. Its exact expression was
calculated in 1966~\cite{El66} and may be written in compact form as~\cite{Passera04}
\ba
A_{2\:\mathrm{vap}}^{(4)}(1/x)   \! &=& \!
-\frac{25}{36} - \frac{\ln x}{3}
+x^2 \left(4+3\ln x \right)
+x^4 \left[ \frac{\pi^2}{3} -2\ln x \, \ln \left(\frac{1}{x}-x\right)
-{\rm Li}_2(x^2)\right]
\nonumber \\   \!&&\! 
+ \, \frac{x}{2} \left(1-5 x^2\right) \!\left[\frac{\pi^2}{2}
  - \ln x \, \ln \left( \frac{1-x}{1+x} \right)
  - {\rm Li}_2(x) + {\rm Li}_2(-x) \right] \crn
  \! &=& \! 
-\frac{25}{36} - \frac{\ln x}{3}
+x^2 \left(4+3\ln x \right)
+x^4 \left[2 \ln^2(x) -2\ln x \, \ln \left(x-\frac{1}{x}\right)
+{\rm Li}_2(1/x^2)\right]
\nonumber \\   \!&&\! 
+ \, \frac{x}{2} \left(1-5 x^2\right) \!\left[
  - \ln x \, \ln \left( \frac{x-1}{x+1} \right)
  + {\rm Li}_2(1/x) - {\rm Li}_2(-1/x) \right]~~~~ (x>1) \epo
\label{A2vapexact}
\ea
\noi
The first form is valid for arbitrary $x$.
For $x>1$ some of the logs as well as $\Liz{x}$ develop a cut and a
corresponding imaginary part like the one of $\ln(1-x)$. Therefore,
for the numerical evaluation in terms of a series expansion, it is an
advantage to rewrite the $\Liz{x}$'s in terms of $\Liz{1/x}$'s,
according to Eq.~(\ref{spencerel}), which leads to the second form. 

There are two different regimes for the mass dependent
effects, the light electron loops and the heavy tau loops~\cite{SWP57,El66}:

\noi
$\bullet$ {\bf Light} internal masses give rise to potentially large logarithms of
mass ratios which get singular in the limit $m_\mathrm{light} \to 0$\\
\begin{figure}[h]
\vspace*{-6mm}
\hspace*{2.5cm}
\includegraphics[scale=0.91]{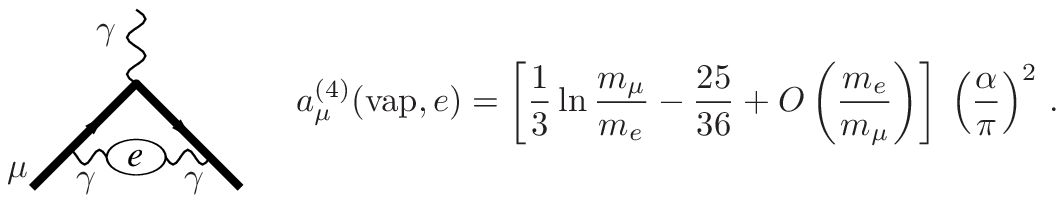}
\end{figure}

\noi
Here we have a typical result for a light field which produces a large
logarithm $\ln \frac{m_\mu}{m_e}\simeq 5.3$, such that the first term
$\sim 2.095$ is large relative to a typical constant second term $-0.6944$.
Here the exact 2--loop result is
\ba
a^{(4)}_\mu(\mathrm{vap},e)\simeq  1.094\,258\,3111(84)\:\left(
\frac{\alpha}{\pi}\right)^2=5.90406007(5)\times 10^{-6} \epo 
\ea
The error is due to the uncertainty in the mass ratio $(m_e/m_\mu)$.
The leading term as we have shown is due to the charge screening
according to the RG.

For comparison we next consider the

\noi  
$\bullet$ {\bf equal} internal mass case, which yields the pure number
\\

\begin{figure}[h]
\vspace*{-4mm}
\hspace*{1cm}
\includegraphics[scale=0.91]{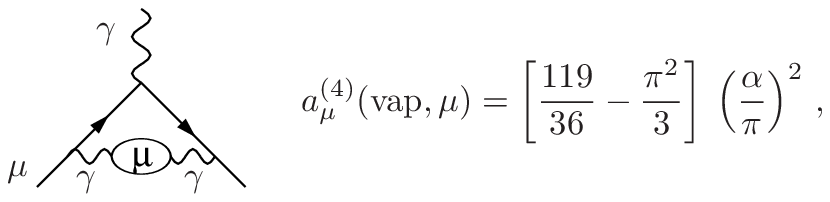}
\end{figure}

\noi
and is already included in the universal part Eq.~(\ref{A14univ}). 
The result is typical for these kind of radiative correction
calculations: a rational term of size $3.3055...$ and a
transcendental $\pi^2$ term of very similar magnitude $3.2899...$ but of
opposite sign largely cancel. The result is only 0.5\% of
the individual terms:
\be
a^{(4)}_\mu(\mathrm{vap},\mu)\simeq 0.015\,687\,4219 \:\left(
\frac{\alpha}{\pi}\right)^2 =  8.464\,13320 \times 10^{-8}\epo
\label{a4_vapm}
\ee

\noi
$\bullet$ {\bf Heavy} internal masses decouple in
the limit $m_\mathrm{heavy} \to \infty$ and thus only yield small
power corrections\\[-4mm]
\begin{figure}[h]
\hspace*{1cm}
\includegraphics[scale=0.91]{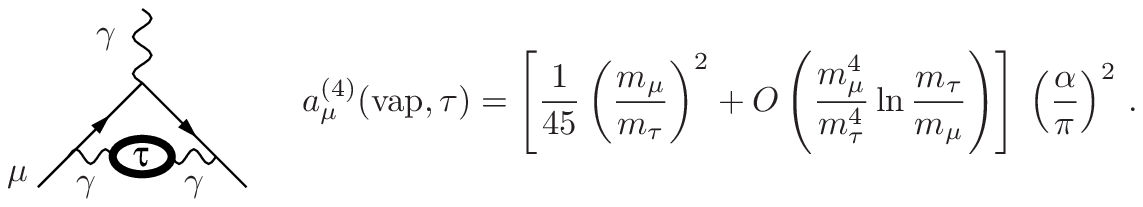}
\end{figure}

\noi
Here we have a typical ``heavy physics'' contributions, from a state
of mass $M \gg m_\mu$, yielding a term proportional to
$m_\mu^2/M^2$. This means that besides the order in $\alpha$ there is
an extra suppression factor, e.g.  $O(\alpha^2) \to Q(\alpha^2
\frac{m_\mu^2}{M^2})$ in our case. To unveil new heavy states thus
requires a corresponding high precision in theory and experiment.  For
the $\tau$ the contribution is relatively tiny
\ba
a^{(4)}_\mu(\mathrm{vap},\tau)\simeq 0.000\,078\,064(25)\:\left(
\frac{\alpha}{\pi}\right)^2=4.2120(13)\times 10^{-10}\,,
\ea
with the error from the mass ratio $(m_\mu/m_\tau)$.
Note that at the level of accuracy reached by the Brookhaven
experiment ($63 \power{-11}$), the contribution is non--negligible.
At the 2--loop level a $e-\tau$ mixed
contribution is not possible, and hence
$A_3^{(4)}(m_{\mu}/m_e,m_{\mu}/m_{\tau}) = 0$.

The complete 2--loop QED contribution from the diagrams displayed in
Fig.~\ref{fig:twoloopdia} is given by
\bea
C_2 =  A_{1\:\mathrm{uni}}^{(4)} +
A_{2\:\mathrm{vap}}^{(4)}(m_{\mu}/m_e) +
A_{2\:\mathrm{vap}}^{(4)}(m_{\mu}/m_{\tau}) = 0.765 \, 857 \, 410 \,(27) \cs
\eea
and we have
\ba
\amu^{(4)~\mathrm{QED}}&=& 0.765 \, 857 \, 410 \,(27)\:
\left(\frac{\alpha}{\pi}\right)^2 \simeq 413217.620(14)
\power{-11}
\label{QED2all}
\ea
for the complete 2--loop QED contribution to $\amu$.
The errors of $A_2^{(4)}(m_{\mu}/m_e)$ and
$A_2^{(4)}(m_{\mu}/m_{\tau})$ have been added in quadrature as the
errors of the different measurements of the lepton masses may be
treated as independent. The combined error $\delta
C_2 = 2.7 \times 10^{-8}$ is negligible by the
standards $1 \times 10^{-5}$ estimated after Eq.~(\ref{precgoal}).

\subsubsection{3--loop: Light-by-Light Scattering and Vacuum
Polarization Insertions}
At three loops, in addition to photon vacuum polarization corrections, 
a new kind of contributions shows up exhibiting the so called
\textit{light--by--light scattering}
(LbL) insertions: closed fermion loops with four photons
attached. Note that the
physical process $\gamma \gamma \to \gamma \gamma$ of light--by--light
scattering involves real on--shell
photons. There are 6 diagrams which follow from the first one in
Fig.~\ref{fig:LbLinsertions}, by permutation of the photon vertices on
the external muon line, 
\begin{figure}[t]
\centering
\includegraphics[scale=0.91]{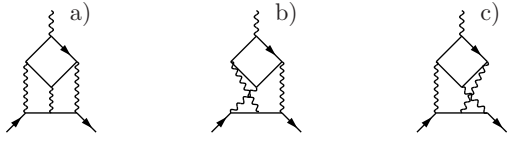}
\vspace*{-4mm}
\caption{Light--by--light scattering insertions in the electromagnetic
vertex.}
\label{fig:LbLinsertions}
\end{figure}
plus the ones obtained by reversing the direction of the fermion loop.
Remember that closed fermion loops with three photons vanish by
Furry's theorem. Again, besides the equal mass case $m_\mathrm{loop}=m_\mu$
there are two different regimes for electron and tau loops~\cite{LR93,KOPV03}, respectively:

\noi
$\bullet$ {\bf Light} internal masses also in this case give rise to potentially
large logarithms of mass ratios which get singular in the limit
$m_\mathrm{light} \to 0$\\[-4mm]
\begin{figure}[h]
\hspace*{1cm}
\includegraphics{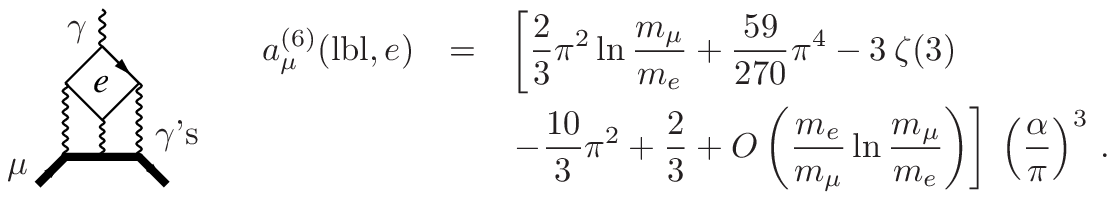}
\end{figure}

\noi
This again is a light loop which yields an unexpectedly large
contribution 
\be
a^{(6)}_\mu(\mathrm{lbl},e)\simeq 20.947\,924\,89(16)
\:\left(
\frac{\alpha}{\pi}\right)^3= 2.625\,351\,02(2)\times 10^{-7} \cs 
\ee
with the error from the $(m_e/m_\mu)$ mass ratio. Historically, it was calculated
first numerically by Aldins et al.~\cite{LBLlep}, after a 1.7
$\sigma$ discrepancy with the CERN measurement~\cite{CERN68} in 1968
showed up.

Again, for comparison we also consider the

\noi
$\bullet$ {\bf equal} internal masses case, which yields a pure number\\[-4mm]


\begin{figure}[h]
\hspace*{1cm}
\includegraphics{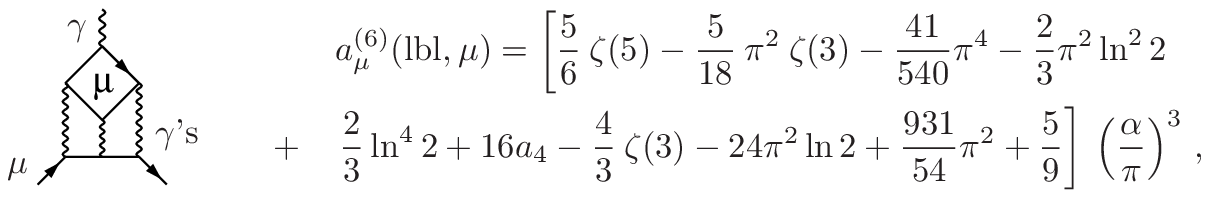}
\end{figure}

\noi
and has been included in the universal part Eq.~(\ref{aell6univ}) already.
The constant $a_4$ is defined in Eq.~(\ref{Linum}).
The single scale QED contribution is much smaller
\be
a^{(6)}_\mu(\mathrm{lbl},\mu)\simeq  0.371005293\:\left(
\frac{\alpha}{\pi}\right)^3=4.64971652 \times 10^{-9}\cs
\label{a6_lblm}
\ee
but is still a substantial contributions at the required level of
accuracy.

\noi
$\bullet$ {\bf Heavy} internal masses again decouple in the limit
$m_\mathrm{heavy} \to \infty$ and thus only yield small power
corrections\\[-4mm]
\begin{figure}[h]
\hspace*{1cm}
\includegraphics{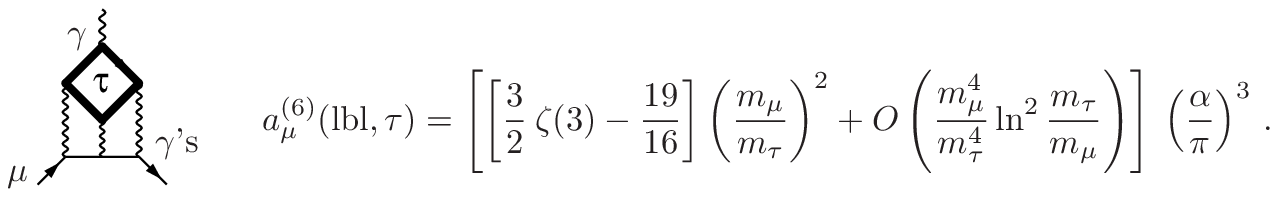}
\end{figure}

\noi
Numerically we obtain
\be
a^{(6)}_\mu(\mathrm{lbl},\tau)\simeq 0.002\,142\,83(69) \:\left(
\frac{\alpha}{\pi}\right)^3= 2.685\,56(86) \times 10^{-11}\epo 
\ee
This contribution could play a role for a next generation precision experiment only.
The error indicated is from the  $(m_\mu/m_\tau)$ mass ratio.

All other corrections follow from Fig.~\ref{fig:threeloopdia1} by
replacing at least one muon in a loop by another lepton or quark. The
corresponding mass dependent corrections are of particular interest
because the light electron loops yield contributions which are
enhanced by large logarithms. Results for $A_2^{(6)}$ have been
obtained in~\cite{Ki67,A26early,La93,LR93,KOPV03}, for $A_3^{(6)}$
in~\cite{SaLi91,LiMeSa93,CS99,FGdR05,Aguilar08}.  For the
light--by--light contribution, graphs 1) to 6) of
Fig.~\ref{fig:threeloopdia1}, the exact analytic result is
known~\cite{LR93}, but only the much simpler asymptotic expansions
have been published. At present the following series expansions are
sufficient to match the requirement of the precision needed: for
electron LbL loops we have
\ba
A_{2\:\mathrm{lbl}}^{(6)}(m_\mu/m_e)&=&
\frac{2}{3}\pi^2 \ln \frac{m_\mu}{m_e} +\frac{59}{270}\pi^4 -3
\zeta(3) -\frac{10}{3}\pi^2 +\frac{2}{3} \crn
&& \hspace*{-2.4cm} +\,\left(\frac{m_e}{m_\mu}\right)^{~}\,
\bigg[\frac{4}{3}\pi^2 \ln
\frac{m_\mu}{m_e} -\frac{196}{3}\pi^2 \ln 2 + \frac{424}{9}\pi^2 \bigg] \crn
&& \hspace*{-2.4cm} +\,\left(\frac{m_e}{m_\mu}\right)^2\,
\bigg[-\frac{2}{3}\ln^3 \frac{m_\mu}{m_e}
+\left(\frac{\pi^2}{9}-\frac{20}{3}\right)\:\ln^2 \frac{m_\mu}{m_e}
-\left(\frac{16}{135}\pi^4 +4 \zeta(3)
-\frac{32}{9}\pi^2
+\frac{61}{3}\right)\:\ln \frac{m_\mu}{m_e}
 \crn && \hspace*{0cm}
+\frac{4}{3} \pi^2 \zeta(3)-\frac{61}{270}\pi^4 + 3\:\zeta(3)
+\frac{25}{18}\pi^2 -\frac{283}{12}\bigg] \crn
&& \hspace*{-2.4cm} +\,\left(\frac{m_e}{m_\mu}\right)^3\,
\bigg[\frac{10}{9}\pi^2 \ln \frac{m_\mu}{m_e} -\frac{11}{9}\pi^2
\bigg] \crn
&& \hspace*{-2.4cm} +\,\left(\frac{m_e}{m_\mu}\right)^4\,
\bigg[\frac{7}{9}\ln^3 \frac{m_\mu}{m_e} + \frac{41}{18}\ln^2 \frac{m_\mu}{m_e}
+ \left( \frac{13}{9}\pi^2 + \frac{517}{108}\right) \ln \frac{m_\mu}{m_e}
+ \frac{1}{2}\zeta(3)
+ \frac{191}{216}\pi^2 + \frac{13283}{2592} \bigg]
\crn &&
\, + \, O\left(\left(m_e/m_\mu\right)^5\right)\, 
=20.947\,924\,89(16)\cs
\label{a6_2lble}
\ea
\noi where here and in the following we use $m_e/m_\mu$ as given in
Eq.~(\ref{leptonmasses}). The leading term in the $(m_e/m_\mu)$ expansion
turns out to be surprisingly large. It has been calculated first
in~\cite{LautrupSamuel77}. Prior to the exact calculation
in~\cite{LR93} good numerical estimates $20.9471(29)$~\cite{Kinoshita88}
and $20.9469(18)$~\cite{Samuel92} have been available. For $\tau$ LbL
loops one obtains
\ba
A_{2\:\mathrm{lbl}}^{(6)}(m_\mu/m_\tau)&=&
\frac{m_\mu^2}{m_\tau^2}\left[ \frac{3}{2}\zeta_3-\frac{19}{16}\right]
\nonumber \\
&&\hspace{-2.1cm}
+\frac{m_\mu^4}{m_\tau^4}\left[\frac{13}{18}\zeta_3
-\frac{161}{1620}\zeta_2-\frac{831931}{972000}
-\frac{161}{3240}L^2-\frac{16189}{97200}L\right]\nonumber \\
&&\hspace{-2.1cm}
+\frac{m_\mu^6}{m_\tau^6}\left[\frac{17}{36}\zeta_3-\frac{13}{224}\zeta_2
-\frac{1840256147}{3556224000}
-\frac{4381}{120960}L^2-\frac{24761}{317520}L
\right]\nonumber \\
&& \hspace{-2.1cm}
+\frac{m_\mu^8}{m_\tau^8}\left[\frac{7}{20}\zeta_3 -
\frac{2047}{54000}\zeta_2 - \frac{453410778211}{1200225600000}
-\frac{5207}{189000}L^2-\frac{41940853}{952560000}L
\right] \nonumber \\
&& \hspace{-2.1cm}
+\frac{m_\mu^{10}}{m_\tau^{10}}\!\left[\frac{5}{18}\zeta_3 -
\frac{1187}{44550}\zeta_2- \frac{86251554753071}{287550049248000}
- \frac{328337}{14968800}L^2
-\frac{640572781}{23051952000}L \right]
\crn
&& \hspace{-2.1cm}
+ O\left((m_\mu/m_\tau)^{12}\right) = 0.002\,142\,833(691)\cs
\label{a6_2lblt}
\ea
where $L = \ln(m_\tau^2/m_\mu^2)$, $\zeta_2=\zeta(2)=\pi^2/6$ and
$\zeta_3=\zeta(3)$. The expansion given in~\cite{LR93} in place of
the exact formula has been extended in~\cite{KOPV03} with the result
presented here.

Vacuum polarization insertions contributing to $a^{(6)}$ may origin
from one or two internal closed fermion
loops. The vacuum polarization 
insertions into photon lines again yield mass dependent effects if one
or two of the $\mu$ loops of the universal contributions are
replaced by an electron or a $\tau$. Here we first give the
numerical results for the coefficients of
$\left(\frac{\alpha}{\pi}\right)^3$~\cite{La93,CS99,FGdR05}:
\begin{figure}[h]
\vspace*{-4mm}
\hspace*{1cm}
\includegraphics{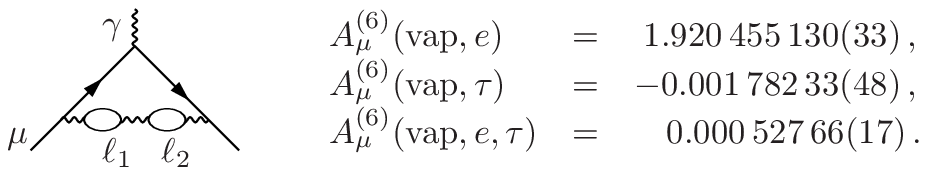}

\end{figure}

\noi
Again the exact results are available~\cite{La93} but the following
much simpler asymptotic expansions are adequate at present precision:
for electron loops replacing muon loops in
Fig.~\ref{fig:threeloopdia1} one finds
\ba
A_{2\:\mathrm{vap}}^{(6)}(m_\mu/m_e)&=&
\frac{2}{9} \ln^2 \frac{m_\mu}{m_e} +\left(\zeta(3)-\frac{2}{3}\pi^2 \ln 2 +
\frac{1}{9}\pi^2 +\frac{31}{27}\right)\: \ln \frac{m_\mu}{m_e} \crn
&& \hspace*{-0.4cm} + \frac{11}{216} \pi^4
-\frac{2}{9} \pi^2 \ln^2 2 - \frac{8}{3}a_4 -\frac{1}{9}\ln^4 2 -3
\zeta(3) +\frac{5}{3}\pi^2 \ln 2 -\frac{25}{18}\pi^2+\frac{1075}{216} \crn
&& \hspace*{-2.4cm}  +\,\left(\frac{m_e}{m_\mu}\right)^{~}\,
\bigg[-\frac{13}{18}\pi^3 -\frac{16}{9}\pi^2 \ln 2 +
\frac{3199}{1080}\pi^2\bigg] \crn
&&  \hspace*{-2.4cm} +\,\left(\frac{m_e}{m_\mu}\right)^2\,
\bigg[\frac{10}{3}\ln^2 \frac{m_\mu}{m_e} -\frac{11}{9}\ln \frac{m_\mu}{m_e}
-\frac{14}{3}\pi^2 \ln 2 -2 \zeta(3)
+\frac{49}{12}\pi^2-\frac{131}{54} \bigg] \crn
&& \hspace*{-2.4cm} +\,\left(\frac{m_e}{m_\mu}\right)^3\,
\bigg[\frac{4}{3}\pi^2 \ln \frac{m_\mu}{m_e} + \frac{35}{12}\pi^3
-\frac{16}{3}\pi^2 \ln 2 -\frac{5771}{1080}\pi^2 \bigg] 
\crn && \hspace*{-2.4cm} 
+\,\left(\frac{m_e}{m_\mu}\right)^4\,
\bigg[- \frac{25}{9}\ln^3\left(\frac{m_\mu}{m_e}\right) -
\frac{1369}{180}\ln^2\left(\frac{m_\mu}{m_e}\right)
+ \left( -2\zeta(3) + 4\pi^2\ln 2
- \frac{269}{144}\pi^2
- \frac{7496}{675}\right) \ln \frac{m_\mu}{m_e}
\crn && \hspace*{-0.4cm}
- \,\frac{43}{108}\pi^4 \,+\, \frac{8}{9}\pi^2\ln^2 2 \,+\,\frac{80}{3}a_4
\,+\, \frac{10}{9}\ln^4 2 \, 
+\,\frac{411}{32}\zeta(3) \,+\, \frac{89}{48}\pi^2\ln 2
\,-\, \frac{1061}{864}\pi^2 \,
\crn && \hspace*{-0.4cm}
-\, \frac{274511}{54000}
\bigg]
+ \, O\left(\left(m_e/m_\mu\right)^5\right)\, 
=1.920\,455\,130(33)\epo
\label{a6_2vape}
\ea
\noi The leading and finite terms were first given
in~\cite{BarbieriRemiddi75}, the correct $(m_e/m_\mu)$ terms have been
given in~\cite{SaLi91}.  In contrast to the LbL contribution the
leading logs of the VP contribution may be obtained relatively easy by
renormalization group considerations using the running fine structure
constant~\cite{LdeR74,BarbieriRemiddi75a}. In place of the known but
lengthy exact result only the expansion shown was presented
in~\cite{La93}. Despite the existence of large leading logs the VP
contribution is an order of magnitude smaller than the one from the
LbL graphs. Replacing muon loops in Fig.~\ref{fig:threeloopdia1} by
tau loops in all possible ways one obtains
\ba
A_{2\:\mathrm{vap}}^{(6)}(m_\mu/m_\tau)&=& \left( \frac{m_\mu}{m_\tau}\right)^2\,
\bigg[-\frac{23}{135}\ln \frac{m_\tau}{m_\mu} -\frac{2}{45}\pi^2 +\frac{10117}{24300}\bigg]
 \crn && \hspace*{-2.4cm}
+ \left(\frac{m_\mu}{m_\tau}\right)^4\,
\bigg[\frac{19}{2520}\ln^2 \frac{m_\tau}{m_\mu}
-\frac{14233}{132300} \ln \frac{m_\tau}{m_\mu} +
\frac{49}{768}\zeta(3)-\frac{11}{945}\pi^2+\frac{2976691}{296352000}\bigg]
 \crn && \hspace*{-2.4cm}
+ \left(\frac{m_\mu}{m_\tau}\right)^6\,
\bigg[\frac{47}{3150}\ln^2 \frac{m_\tau}{m_\mu}
-\frac{805489}{11907000} \ln \frac{m_\tau}{m_\mu} +
\frac{119}{1920}\zeta(3)-\frac{128}{14175}\pi^2
+\frac{102108163}{30005640000}\bigg]
\crn &&
+O\left( \left(m_\mu/m_\tau\right)^8\right) = -0.001\,782\,327(484)\epo
\label{a6_2vapt}
\ea
\noi Also in this case, in place of exact result obtained in~\cite{La93} only
the expansion shown was given in the paper.  As has been cross checked 
recently against the exact results  in~\cite{Passera04}, all the expansions
presented are sufficient for numerical evaluations at the present
level of accuracy.

Starting at three loops, a contribution to
$A_3(m_\mu/m_e,m_\mu/m_\tau)$, depending on two mass ratios, shows 
up. The relevant term is due to diagram 22) of Fig.~\ref{fig:threeloopdia1}
with one fermion loop an electron--loop and the other a
$\tau$--loop. According to Eq.~(\ref{amufullVP})
we may write
\ba
\hspace*{-6mm}\left. a_{\mu}^{(6)}(\mathrm{vap},e,\tau)\right|_{\mathrm{dia}\:22)} &=&
\frac{\alpha}{\pi}\int_{0}^{1} dx (1-x)\ 2\
\left[-\Pi_{\gamma\:\mathrm{ren}}^{'\:e}\left(\frac{-x^2}{1-x}m_{\mu}^2 \right) \right]
\left[-\Pi_{\gamma\:\mathrm{ren}}^{'\:\tau}
\left(\frac{-x^2}{1-x}m_{\mu}^2 \right) \right],
\label{mixedVP}
\ea

\noi
which together with Eq.~(\ref{Pigammaint}) leads to a three--fold
integral representation. However, since
$\Pi_{\gamma\:\mathrm{ren}}^{'\:\ell}$ given by Eq.~(\ref{Pigammaint})
is analytically known, Eq.~(\ref{mixedVP}) represents a 1--dimensional
integral. It has been calculated as an expansion in the
two mass ratios in~\cite{SaLi91,CS99} and was extended to
$O((m^2_\mu/m^2_\tau)^{5})$ recently in~\cite{FGdR05}. The result
reads
\ba
A_{3\:\mathrm{vap}}^{(6)}(m_\mu/m_e,m_\mu/m_\tau)&=&
\left(\frac{m_{\mu}^2}{m_{\tau}^2}\right)\,
\bigg[\frac{2}{135}\ln\frac{m_{\mu}^2}{m_e^2}-\frac{1}{135}\bigg] \crn
& & \hspace*{-3.5cm}+ \left(\frac{m_{\mu}^2}{m_{\tau}^2}\right)^2
\bigg[-\frac{1}{420}\ln\frac{m_{\tau}^2}{m_{\mu}^2}\ln\frac{m_{\tau}^2\ m_{\mu}^2}{m_e^4}
-\frac{37}{22050}\ln\frac{m_{\tau}^2}{m_e^2} +\frac{1}{504}\ln\frac{m_{\mu}^2}{m_e^2}
+\frac{\pi^2}{630}
-\frac{229213}{12348000}\bigg] \crn
& & \hspace*{-3.5cm} + \left(\frac{m_{\mu}^2}{m_{\tau}^2}\right)^3
\bigg[-\frac{2}{945}\ln\frac{m_{\tau}^2}{m_{\mu}^2}\ln\frac{m_{\tau}^2\ m_{\mu}^2}{m_e^4}
-\frac{199}{297675}\ln\frac{m_{\tau}^2}{m_e^2} -\frac{1}{4725}\ln\frac{m_{\mu}^2}{m_e^2}
+\frac{4\pi^2}{2835}
-\frac{1102961}{75014100}\bigg] \crn
& & \hspace*{-3.5cm} + \left(\frac{m_{\mu}^2}{m_{\tau}^2}\right)^4
\bigg[-\frac{1}{594}\ln\frac{m_{\tau}^2}{m_{\mu}^2}\ln\frac{m_{\tau}^2\ m_{\mu}^2}{m_e^4}
-\frac{391}{2058210}\ln\frac{m_{\tau}^2}{m_e^2} -\frac{19}{31185}\ln\frac{m_{\mu}^2}{m_e^2}
+\frac{\pi^2}{891}
-\frac{161030983}{14263395300}\bigg] \crn
& & \hspace*{-1.0cm}
+\frac{2}{15}\frac{m_e^2}{m_{\tau}^2}-\frac{4\pi^2}{45}\frac{m_e^3}{m_{\tau}^2 m_{\mu}}
+
\cO\left[\left(\frac{m_{\mu}^2}{m_{\tau}^2}\right)^5 \ln\frac{m_{\tau}^2}{m_{\mu}^2}
\ln\frac{m_{\tau}^2\ m_{\mu}^2}{m_e^4} \right]+\cO\left(
\frac{m_e^2}{m_{\tau}^2} \frac{m_{\mu}^2}{m_{\tau}^2}\right) \crn &=&
0.00052766(17) \epo
\label{vapeletau}
\ea
\noi
The result is in agreement with the numerical evaluation~\cite{La93}.
The $\tau$--lepton mass uncertainty determines the error. The
leading--logarithmic term of this expansion corresponds to simply
replacing $\alpha(q^2=0)$ by $\alpha(m_\mu^2)$ in the 2--loop diagram
with a $\tau$ loop. The last term, with odd powers of $m_e$ and
$m_\mu$, has been included although it is not relevant numerically.
It illustrates typical contributions of the eikonal expansion, the
only source of terms non--analytical in masses squared.

With Eqs.~(\ref{aell6univ}) and (\ref{a6_2lble}) to (\ref{vapeletau}) the
complete 3--loop QED contribution to $a_\mu$ is now known
analytically, either in form of a series expansion or exact. The mass
dependent terms may be summarized as follows:
\ba \begin{tabular}{lcr@{.}l}
$A^{(6)}_{2}(m_\mu/m_e)$                            &=& $22$&$868\,380\,02(20)$,\\
$A^{(6)}_{2}(m_\mu/m_\tau)$                           &=& $ 0$&$000\,360\,51(21)$,\\
$A^{(6)}_{3\:\mathrm{vap}}(m_\mu/m_e,m_\mu/m_\tau)$ &=& $ 0$&$000\,527\,66(17)$.
\end{tabular}
\label{A36num}
\ea
As already mentioned above, the $A^{(6)}_{2}(m_\mu/m_e)$ contribution
is surprisingly large and predominantly from light--by--light
scattering via an electron loop. The importance of this term was
discovered in~\cite{LBLlep}, improved by numerical calculation
in~\cite{KM90} and calculated analytically in~\cite{LR93}.
Adding up the relevant terms we have
\bea
C_3=24.050 \, 509 \, 64 \,(46)
\eea
or
\ba
\amu^{(6)~\mathrm{QED}}&=& 24.050 \, 509 \, 64 \,(46)\: \left(\frac{\alpha}{\pi}\right)^3
\simeq 30141.902(1) \power{-11}
\ea
as a result for the complete 3--loop QED contribution to $\amu$.
We have combined the first two errors of Eq.~(\ref{A36num}) in quadrature and
the last linearly, as the latter depends on the same errors in the
mass ratios.


\subsubsection{4--loop: Light Lepton Insertions}
Also at four loops, the light internal electron loops, included in $A^{(8)}_{2}(m_\mu/m_e)$, 
give the by far largest contribution. Here
469 diagrams contribute which may be divided into four gauge invariant
(\gi) groups:\\

\noi
\textit{Group I:} 49 diagrams obtained from the 1--loop muon vertex
by inserting 1--, 2-- and 3--loop lepton VP subdiagrams, i.e.,
the internal photon line of Fig.~\ref{fig:oneloopdia} is replaced by the full
propagator at three loops. The group is subdivided into four \gis
subclasses I(a), I(b), I(c) and I(d) as illustrated in Fig.~\ref{fig:a8I}.
\begin{figure}[h]
\centering
\includegraphics{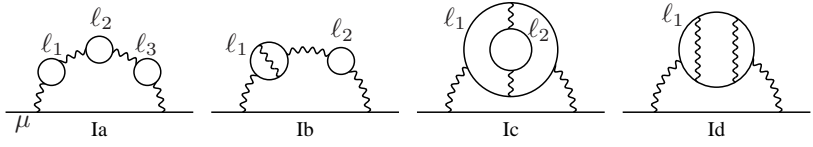}

\caption{Typical diagrams of subgroups Ia (7 diagrams), Ib (18 diagrams),
Ic (9 diagrams) and Id (15 diagrams). The lepton lines represent
fermions propagating in an external magnetic field. $\ell_i$ denote VP
insertions.}
\label{fig:a8I}
\end{figure}
Results for this group have been obtained by numerical and analytic
methods~\cite{KinoNio04,A8analytic}. The numerical result~\cite{KinoNio04}
\bea
A^{(8)}_{2\,I}= 16.720\,359 \:(20) \cs
\eea
has been obtained by using simple integral representations.

\noi
\textit{Group II:} 90 diagrams generated from the 2--loop muon
vertex by inserting 1--loop and/or 2--loop lepton VP subdiagrams as
shown in Fig.~\ref{fig:a8II}.
\begin{figure}[h]
\centering
\includegraphics{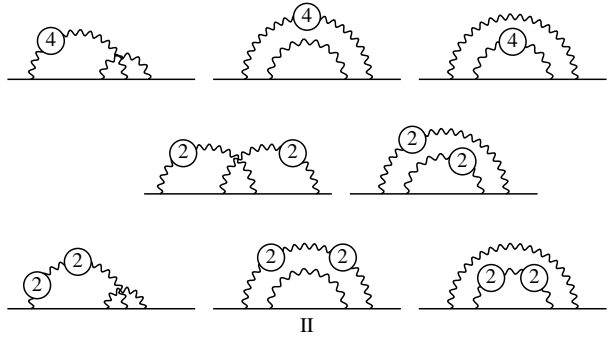}

\caption{Typical diagrams of group II (90 diagrams). The lepton lines
as in Fig.~\ref{fig:a8I}. 2 and 4, respectively, indicate second
(1--loop subdiagrams) and
fourth (2--loop subdiagrams) order lepton--loops.}
\label{fig:a8II}
\end{figure}
As for the previous case, results for this group have been obtained by numerical and analytic
methods~\cite{KinoNio04,A8analytic}. The result here is~\cite{KinoNio04}
\bea
A^{(8)}_{2\,II}= -16.674\,591 \:(68) \epo
\eea

\noi
\textit{Group III:} 150 diagrams generated from the 3--loop muon vertex
Fig.~\ref{fig:threeloopdia1} by inserting one 1--loop electron VP
subdiagrams in each internal photon line in all possible
ways. Examples are given in Fig.~\ref{fig:a8III}.
\begin{figure}[h]
\centering
\includegraphics{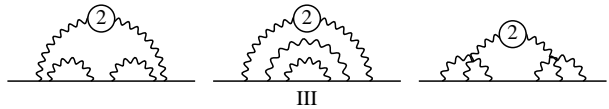}

\caption{Typical diagrams of group III (150 diagrams). The lepton lines
as in Fig.~\ref{fig:a8I}.}
\label{fig:a8III}
\end{figure}
This is a group which has been calculated numerically only. The result
found in~\cite{KinoNio04} reads
\bea
A^{(8)}_{2\,III}= 10.793\,43 \:(414) \epo
\eea

\noi
\textit{Group IV:} 180 diagrams with muon vertex containing
LbL subgraphs deco\-rated with additional radiative corrections.  This
group is subdivided as shown in Fig.~\ref{fig:a8IV} into \gis subsets
IV(a), IV(b), IV(c) and IV(d).
\begin{figure}
\centering
\includegraphics{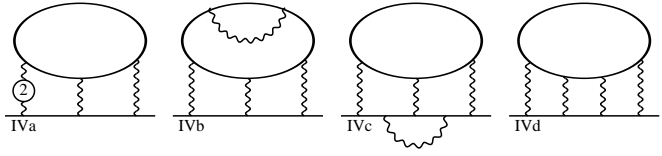}

\caption{Typical diagrams of subgroups IVa (54 diagrams), IVb (60
diagrams), IVc (48 diagrams) and IVd (18 diagrams). The lepton lines
as in Fig.~\ref{fig:a8I}.}
\label{fig:a8IV}
\end{figure}
\noi The calculation of the corresponding contribution is at the limit of present
possibilities. The result has been evaluated by two independent methods in~\cite{KinoNio04}
and reads
\bea
A^{(8)}_{2\,IV}= 121.8431\:(59) \epo
\eea
The sum of the results from the different groups thus reads
\be
A^{(8)}_{2}(m_\mu/m_e) = 132.6823(72)\epo
\label{A8ele}
\ee

A small contribution to $A^{(8)}_{3}$ comes from the diagrams
which depend on 3 masses. There are 102 diagrams
containing two or three closed loops of VP and/or LbL type,
defined above as the classes I (30 diagrams), II (36 diagrams)
and IV (36 diagrams).  The results found in~\cite{KinoNio04} read
\ba \begin{tabular}{lcr@{.}l}
$A^{(8)}_{3I }(m_\mu/m_e,m_\mu/m_\tau)$ &=&$0$&$007\:630\:(01)\,,$
\\
$A^{(8)}_{3II}(m_\mu/m_e,m_\mu/m_\tau)$ &=&$ -0$&$053\:818\:(37)\,,$
\\
$A^{(8)}_{3IV}(m_\mu/m_e,m_\mu/m_\tau)$ &=&$  0$&$083\:782\:(75)\,,$
\end{tabular}
\ea
and are adding up to the value
\be
A^{(8)}_{3}(m_\mu/m_e,m_\mu/m_\tau) =  0.037\:594\:(83)\epo
\label{A8mixed}
\ee
A rough estimate of the $\tau$--loops contribution
performed in~\cite{KinoNio04} yields
\be
A^{(8)}_{2}(m_\mu/m_\tau) = 0.005(3) \epo
\label{A8tau}
\ee

\noi
Note that all mass dependent as well as the mass independent
$O(\alpha^4)$ QED contributions to $\amu$ have been recalculated by
different methods by Kinoshita and collaborators~\cite{KinoNio04,KinoNio05,Aoyama07}.
There is also some progress in analytic calculations~\cite{LMR03}.
Adding the $A^{(8)}$ terms discussed above we obtain
\bea
C_4=130.8105(85)\;,
\eea
which yields
\ba
\amu^{(8)~\mathrm{QED}}&=& 130.810 \, 5 \,(85)\: \left(\frac{\alpha}{\pi}\right)^4
\simeq 380.807(25) \power{-11}\;,
\ea
the result for the complete 4--loop QED contribution to $\amu$.

\subsubsection{5--loop and Summary of QED Contributions}
Also at five loops electron loop insertions are the leading contributions
(see Fig.~\ref{fig:fiveloopdia1}).
Here the number of diagrams is 9080, a very discouraging number as
Kinoshita~\cite{Kino05} remarks.
This contribution originally was evaluated using renormalization group
(RG) arguments in~\cite{KM90,Ka93}. 
The new estimate by Kinoshita and Nio~\cite{Kino05,KinoNio06} is
\be
A^{(10)}_{2}(m_\mu/m_e) = 663(20),\crn
\label{A10_2e}
\ee
and was obtained by numerically evaluating all Feynman diagrams, which are
known or likely to be enhanced. The error estimate should cover all
remaining subleading contributions. The number in Eq.~(\ref{A10_2e}) was
subsequently cross--checked by Kataev~\cite{Kataev05}. 
A very recent calculation from the class of leading tenth order
contributions (singlet (SI) VP insertion diagrams which
includes the last diagram of Fig.~\ref{fig:fiveloopdia1} with two
electron LbL loops) yields $A^{(10)}_2(m_\mu/m_e)=-1.263 44
(14)$~\cite{Aoyama:2008gy}. This result has been reproduced at the 3\%
level by an asymptotic expansion in~\cite{Baikov:2008si}, where also
the much larger 4--loop non-singlet (NS) VP insertion with electron
loops has been calculated: $A^{(10),{\rm as}}_2(m_\mu/m_e)=63.481_\mathrm{NS}
-1.21429_\mathrm{SI}=62.2667$. Since the leading terms are included in 
Eq.~(\ref{A10_2e}) already and subleading terms are unknown in general,
we will stay with the above result in the following.

Thus, taking into account  Eq.~(\ref{A10uni}), we arrive at
\bea
C_5 \sim 663.0(20.0)(4.6)
\eea
or\\[-8mm]
\ba
\amu^{(10)~\mathrm{QED}}& \sim & 663(20)(4.6)\: \left(\frac{\alpha}{\pi}\right)^5
\simeq 4.483(135)(31) \power{-11}
\ea
as an estimate of the 5--loop QED contribution.\\

In Table~\ref{tab:amuqedcontributions2} we collect the results of the
QED calculations. In spite of the fact that the expansion coefficients
$C_i$ multiplying $(\alpha/\pi)^i$ grow rapidly with the order, the
convergence of the perturbative expansion of $a_\mu^{\mathrm{QED}}$ is
good. This suggests that the perturbative truncation error is
well under control at the present level of accuracy.
\begin{table}[ht]
\centering
\caption{The QED contributions to $a_\mu$.}
\label{tab:amuqedcontributions2}
\begin{tabular}{lr@{.}llr@{.}l}
&&&&&\\[-3mm]
\noalign{\smallskip}\hline\noalign{\smallskip}
 & \multicolumn{2}{c}{$C_i$} & &
\multicolumn{2}{c}{$a_\mu^{(2i)\:\mathrm{QED}}\power{11}$} \\
\noalign{\smallskip}\hline\noalign{\smallskip}
$C_1$     & $\, 0$&$5    $&
$~a^{(2)}$ & $\,         116140973$&$289(43)$ \\
$C_2$     & $\, 0$&$765 \, 857 \, 410 \,(27)    $&
$~a^{(4)}$ & $\, 413217$&$620(14)    $ \\
$C_3$     & $\, 24$&$050 \, 509 \, 64 \,(46)    $&
$~a^{(6)}$ & $\, 30141$&$902(1)    $ \\
$C_4$     & $\, 130$&$8105(85)    $&
$~a^{(8)}$ & $\, 380$&$807(25)    $ \\
$C_5$     & $\,663$&$0(20.0)(4.6)$&
$~a^{(10)}$ & $\, 4$&$483(135)(31)    $ \\
\noalign{\smallskip}\hline
\end{tabular}
\end{table}

The universal QED terms have been given in
Eq.~(\ref{aellunivesalnum}) and together with the mass dependent QED terms of
the 3 flavors ($e$, $\mu$, $\tau$) we obtain
\be
a_\mu^\mathrm{QED}=116\, 584\, 718.104(.044)(.015)(.025)(.139)[.148] \power{-11}\epo
\label{QEDfin}
\ee
The errors are given by the uncertainties in
$\alpha_\mathrm{input}$, in the mass ratios,
the numerical error on $\alpha^4$ terms and the guessed
uncertainty of the $\alpha^5$ contribution, respectively.

Now we have to address the question what happens beyond QED. What is
measured in an experiment includes effects from the real world and we
have to include the contributions from all known particles and
interactions such that from a possible deviation between theory and
experiment we may get a hint of the yet unknown physics.

\section{Hadronic Vacuum Polarization Corrections}
\label{sec:hadvap}
On a perturbative level we may obtain the hadronic vacuum polarization
contribution by replacing internal lepton loops in the QED VP
contributions by quark loops, adapting charge, color multiplicity and
the masses accordingly. Since quarks are, however, confined inside
hadrons, a quark mass cannot be defined in the same natural way as a
lepton mass and quark mass values depend in various ways on the
physical circumstances. Moreover, the running strong coupling
``constant'' $\alpha_s(s)$ becomes large at low energies
$E=\sqrt{s}$. Therefore perturbative QCD (pQCD) fails to ``converge''
in any practical sense in this region and pQCD may only be trusted
above about 2 GeV and away from thresholds and resonances. The low
energy structure of QCD with confinement and the spontaneous breaking
of chiral symmetry in the chiral limit (on a Lagrangian level
characterized by vanishing (current) quark masses) is completely
beyond the scope of pQCD.  Low energy QCD is characterized by its
typical spectrum of low lying hadronic states, the pseudoscalar pions,
the Kaons and the $\eta$ as quasi Goldstone bosons (true ones in the
chiral limit), the pseudoscalar singlet $\eta'$, the spin-1 vector
bosons $\rho$, $\omega$, $\phi$ and by the order parameters of chiral
symmetry breaking, like the quark condensates $\braket{\bar{q}q}\neq 0$
($q=u,d,s$). For the calculation of the hadronic contributions $\amuh$
to the $g-2$ of the muon, baryons like proton and neutron do not play a
big role.

Quarks contribute to the electromagnetic current according to their charge
\ba
j^{\mu\:\mathrm{had}}_\mathrm{em}=\sum_c \left(\frac{2}{3}\bar{u}_c\gamuc
u_c-\frac{1}{3}\bar{d}_c\gamuc d_c-\frac{1}{3}\bar{s}_c\gamuc s_c
+\frac{2}{3}\bar{c}_c\gamuc c_c -\frac{1}{3}\bar{b}_c \gamuc b_c +
\frac{2}{3}\bar{t}_c \gamuc t_c\right) \epo
\label{emhadcurrent} 
\ea
The hadronic electromagnetic current
$j^{\mu\:\mathrm{had}}_\mathrm{em}$ is a color singlet and hence
includes a sum over colors indexed by
$c$. Its contribution to the electromagnetic current correlator
Eq.~(\ref{jemspectral}) defines $\Pi^{'\:\mathrm{had}}_\gamma(s)$, which
enters the calculation of the leading order hadronic contribution to
$\amuh$, diagrammatically given by Fig.~\ref{fig:gmudia}.  
\begin{figure}
\centering
\includegraphics[height=3.5cm]{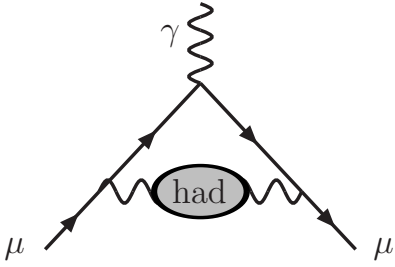}
\caption{Leading hadronic contribution to $g-2$.}
\label{fig:gmudia}
\end{figure}

Perturbative QCD fails to be a reliable tool for estimating $\amuh$ and known
approaches to low energy QCD like chiral perturbation theory as well
as extensions of it which incorporate spin-1 bosons or
lattice QCD  are far from being able to make precise predictions. We therefore
have to resort to a semi-phenomenological approach using dispersion
relations together with the optical theorem and experimental data.

The basic relations are 
\bit
\item analyticity (deriving from causality), which allows to write the
DR 
\be
\Pi'_{\gamma}(k^2)-\Pi'_{\gamma}(0)=
\frac{k^2}{\pi}\int\limits_{0}^{\infty} \D s\,
\frac{{\rm Im} \Pi'_{\gamma}(s)}{s\:(s-k^2-\ie)}\;\;.
\label{subDR}
\ee
\item optical theorem (deriving from unitarity), which relates the imaginary part of the vacuum
polarization amplitude to the total cross section in $\epm$--annihilation
\be
{\rm Im} \Pi'_{\gamma}(s)=\frac{s}{4\pi
\alpha(s)}\:\sigma_\mathrm{tot}(\epm \to \mathrm{anything}):=\frac{\alpha(s)}{3}\,R(s)\,,
\label{opttheo}
\ee
\eit
with
\be
R(s)=\sigma_{\rm tot}/\frac{4 \pi \alpha(s)^2}{3s} \epo
\label{Rsdef}
\ee
The normalization factor is the point cross section (tree level)
$\sigma_{\mu \mu}(\epm \to \gamma^* \to \mu^+ \mu^-)$ in the limit $s
\gg 4 m_\mu^2$. 
We obtain the hadronic contribution  if we restrict ``anything'' to
hadrons. The complementary leptonic part may be calculated reliable in
perturbation theory and the production of a lepton pair at lowest order is given by
\be
R_\ell(s)=
\sqrt{1-\frac{4m_\ell^2}{s}}\left(1+\frac{2m_\ell^2}{s}\right) \cs
(\ell=e,\mu,\tau), 
\label{Rell}
\ee
which may be read off from the imaginary part given in Eq.~(\ref{vapoimag}).
This result provides an alternative way to calculate the renormalized vacuum
polarization function Eq.~(\ref{Pigammaint}), namely, via the DR
Eq.~(\ref{ppDR}) which now takes the form
\be
\label{PiRell}
\Pi^{'\ell}_{\gamma\:\mathrm{ren}}(q^2)=\frac{\alpha q^2}{3\pi}\int_{4 m_\ell^2}^\infty\!\!
\D s\, \frac{R_\ell(s)}{s(s-q^2-\ie)}\,,
\ee
yielding the vacuum polarization due to a lepton--loop.

In contrast to the leptonic part, the hadronic contribution cannot be
calculated analytically as a perturbative series, but it can be
expressed in terms of the cross section of the reaction
$e^+e^-\rightarrow\mbox{hadrons}$, which is known from
experiments.\label{page:vaphadhow1} Via
\be
R_\mathrm{had}(s)=
\sigma(e^+e^-\rightarrow\mbox{hadrons})/\frac{4 \pi \alpha(s)^2}{3s}\,,  
\label{Rshaddef}
\ee
we obtain the relevant hadronic vacuum polarization
\be
\label{Piphad}
\Pi^{'\mathrm{had} }_{\gamma\:\mathrm{ren}}(q^2)=\frac{\alpha q^2}{3\pi}\int_{4 m_\pi^2}^\infty\!\!
\D s\, \frac{R_\mathrm{had}(s)}{s(s-q^2-\ie)} \epo
\ee
At low energies, where the dominating final state consists of two
charged pions\footnote{A much smaller contribution is due to $\gamma^*
\to \pi^0 \gamma$, the hadronic final state with the lowest threshold $s>m^2_{\pi^0}$.}, 
the cross section is given by the square of the electromagnetic form factor of the pion 
$\Fem^{(0)}(s)$ (effective $\pi^+\pi^-\gamma$ vertex undressed from VP
effects, see below),
\be\label{Rh}
R_\mathrm{had}(s)=\frac{1}{4}\left(1-\frac{4m_\pi^2}{s}\right)^{\frac{3}{2}}\,
|\Fem^{(0)}(s)|^2\co\hspace{2em} 4\,m_\pi^2<s<9\,m_\pi^2\co
\ee
which directly follows from the corresponding imaginary part
\bea
\Impa \Pi^{'\:(\pi)}_{\gamma}(q^2)=\frac{\alpha}{12}\:(1-4m_\pi^2/s)^{3/2}
\eea
of a pion loop in the photon vacuum polarization. At $s=0$ we have
$\Fem^{(0)}(0) = 1$, i.e.,  $\Fem^{(0)}(0) $ measures the classical pion charge in units of
$e$. For point--like pions we would have $\Fem^{(0)}(s) \equiv 1$. 
There are three differences between the pionic loop integral and
those belonging to the lepton loops:
\begin{itemize}
\item the masses are different
\item the spins are different
\item the pion is composite -- the Standard Model leptons are
  elementary
\end{itemize}
The compositeness manifests itself in the occurrence of the form
factor $\Fem(s)$, which generates an enhancement: at the $\rho$ peak,
$|\Fem(s)|^2$ reaches values of about 45, while the quark parton model
would give about 7. The remaining difference in
the expressions for the quantities $R_\ell(s)$ and $R_\mathrm{had}(s)$ in
Eqs.~(\ref{Rell}) and (\ref{Rh}), respectively, originates in the fact that the
leptons carry spin $\frac{1}{2}$, while the spin of the pion
vanishes. Near threshold, the angular momentum barrier suppresses the
function $R_\mathrm{had}(s)$ by three powers of momentum, while $R_\ell(s)$ is
proportional to the first power. 
The suppression largely compensates
the enhancement by the form factor -- by far the most important
property is the mass, which sets the relevant scale.

  \subsection{Lowest Order Vacuum Polarization Contribution}
\label{ssec:amuhadlo}

Using Eq.~(\ref{amuvapo}) together with Eq.~(\ref{opttheo}), the $O(\alpha^2)$
contributions to $\amuh$ may be
directly evaluated in terms of $R_\mathrm{had} (s)$ defined in
Eq.~(\ref{Rshaddef}).  More precisely we may write
\ba
\amuhlo = \left(\frac{\alpha m_\mu}{3\pi}
\right)^2 \bigg(\!
\int\limits_{m_{\pi^0}^2}^{E^2_{\rm cut}}\! \D s\,
\frac{R^{\mathrm{data}}_\mathrm{had}(s)\,\hat{K}(s)}{s^2}
+\!\! \int\limits_{E^2_{\rm cut}}^{\infty}\! \D s\,
\frac{R^{\mathrm{pQCD}}_\mathrm{had}(s)\,\hat{K}(s)}{s^2} \bigg), 
\label{AM}
\ea
with a cut $E_{\rm cut}$ in the energy, separating the
non--perturbative part to be evaluated from the data and the
perturbative high energy tail to be calculated using pQCD.
The kernel $K(s)$ is represented by Eq.~(\ref{vapokernel}) discarding the
factor $\alpha/\pi$. This
integral can be performed analytically. Written in terms of the
variable
\bea
x=\frac{1-\beta_\mu}{1+\beta_\mu}\;,\;\;\beta_\mu=\sqrt{1-4m^2_\mu/s}, 
\eea
the result reads\footnote{The representation Eq.~(\ref{KS}) of $K(s)$ is
valid for the muon (or electron) where we have $s>4m^2_\mu$ in the
domain of integration $s>4m^2_\pi$, and $x$ is real, and $0 \leq x
\leq 1$. For the $\tau$ Eq.~(\ref{KS}) applies for $s>4m^2_\tau$. In the
region $4m^2_\pi < s < 4 m^2_\tau$, where $0<r=s/m^2_\tau<4$, we may
use the form
\be
K(s)=\half-r+\half r\:(r-2)\:\ln(r)-\left(1-2r+\half
r^2\right)\:\varphi/w, 
\ee
with $ w=\sqrt{4/r-1}$ and $\varphi=2 \arctan (w)$.}~\cite{BdeR68}
\ba
K(s)&=&\frac{x^2}{2}\:(2-x^2)
    +\frac{(1+x^2)(1+x)^2}{x^2}
       \left(\ln(1+x)-x+\frac{x^2}{2} \right)
    +\frac{(1+x)}{(1-x)}\:x^2 \ln(x)\;\;. 
\label{KS}
\ea
We have written the integral Eq.~(\ref{AM}) in terms of the rescaled function
\ba 
\hat{K}(s)=\frac{3 s}{m^2_\mu}K(s), 
\label{KShat}
\ea 
which is only slowly varying
in the range of integration. It increases monotonically from 0.63... at
the $\pi\pi$ threshold $s=4m^2_\pi$ to 1 at $s=\infty$. The graph is shown in
Fig.~\ref{fig:gm2kernel}.
\begin{figure}[h]
\centering
\includegraphics[height=4cm]{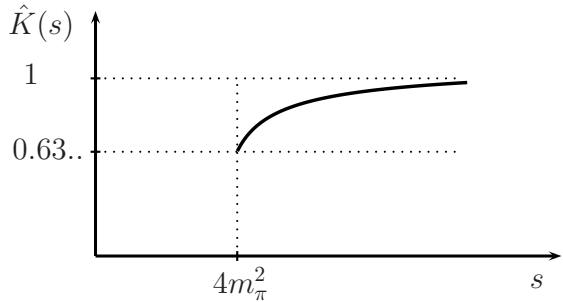}
\caption{Graph of weight function $\hat{K}(s)$ of the $g-2$ dispersion integral.}
\label{fig:gm2kernel}
\end{figure}

%
Note the $1/s^2$--enhancement of
contributions from low energies in $\amu$. Thus the $g-2$ kernel gives
very high weight to the low energy range, in particular to the lowest
lying resonance, the $\rhon$.
Thus, this $1/E^4$ magnification of the low energy region by the
$\amu$ kernel--function together with the existence of the pronounced
$\rhon$ resonance in the $\pi^+\pi^-$ cross--section are responsible
for the fact that pion pair production $\epm \to \ppm$ gives the by
far largest contribution to $\amuh$. The $\rho$ is the lowest lying
vector--meson resonance and shows up in $\pi^+\pi^- \to \rhon$ at
$m_\rho\sim 770$ MeV.  This dominance of
the low energy hadronic cross--section by a single simple two--body
channel is good luck for a precise determination of $\amu$, although a
very precise determination of the $\pi^+\pi^-$ cross--section is a
rather difficult task. The experimental data for the low energy region are
shown in Fig.~\ref{fig:epemdata}. Below about 810 MeV
$\sigma_\mathrm{tot}^\mathrm{had}(s) \simeq
\sigma_{\pi\pi}(s)$ to a good approximation
but at increasing energies more and more channels open and
``measurements of $R$'' get more difficult. In the light sector of
$u,d,s$ quarks, besides the $\rho$ there is the $\omega$, which is
mixing with the $\rho$, and the $\phi$ resonance, essentially a
$\bar{s}s$ bound system. In the charm region we have the pronounced
$\bar{c}c$--resonances, the $J/\psi_{\rm 1S},\psi_{\rm 2S}, \cdots$
resonance series and in the bottom region the $\bar{b}b$--resonances
$\Upsilon_{\rm 1S},\Upsilon_{\rm 2S},\cdots$. Many of the resonances
are very narrow as indicated in Fig.~\ref{fig:rhadron}.
\begin{figure}[t]
\centering
\IfFarbe{%
\includegraphics[height=6cm]{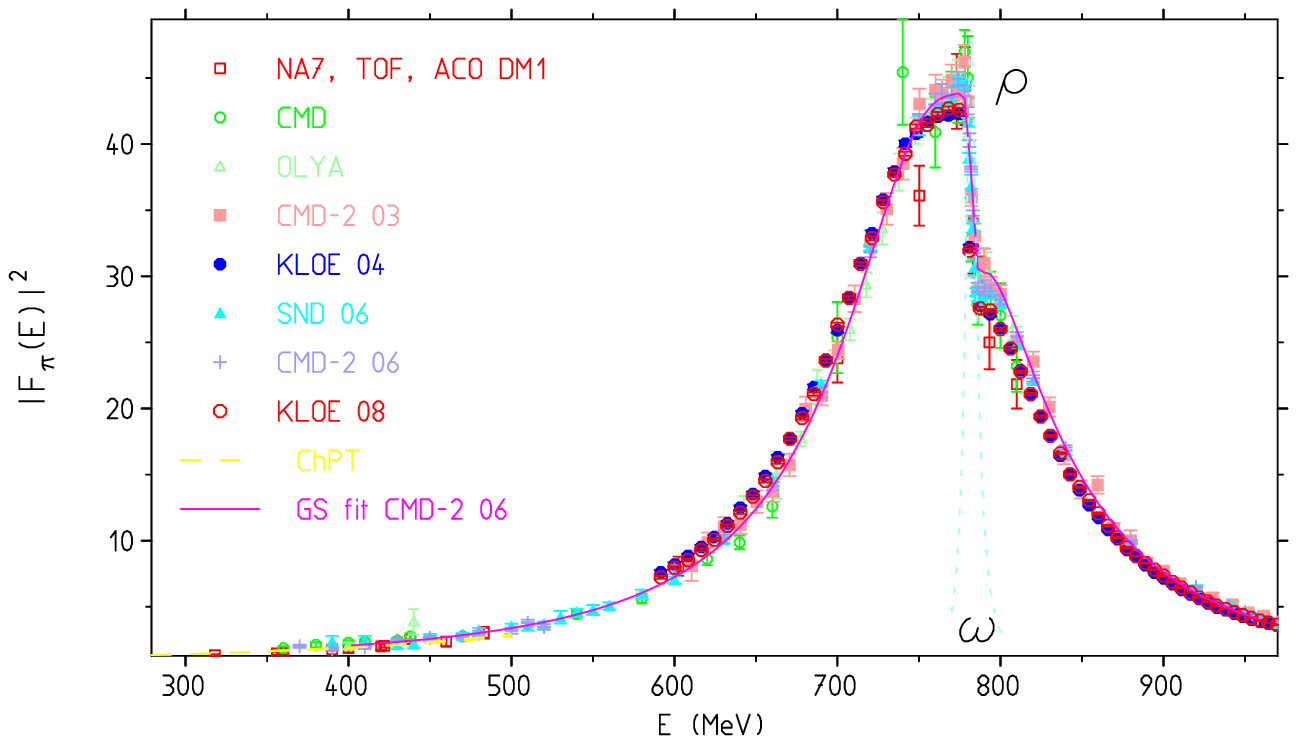}}{%
\includegraphics[height=6cm]{ffpp08all.eps}}

\IfFarbe{%
\includegraphics[height=6cm]{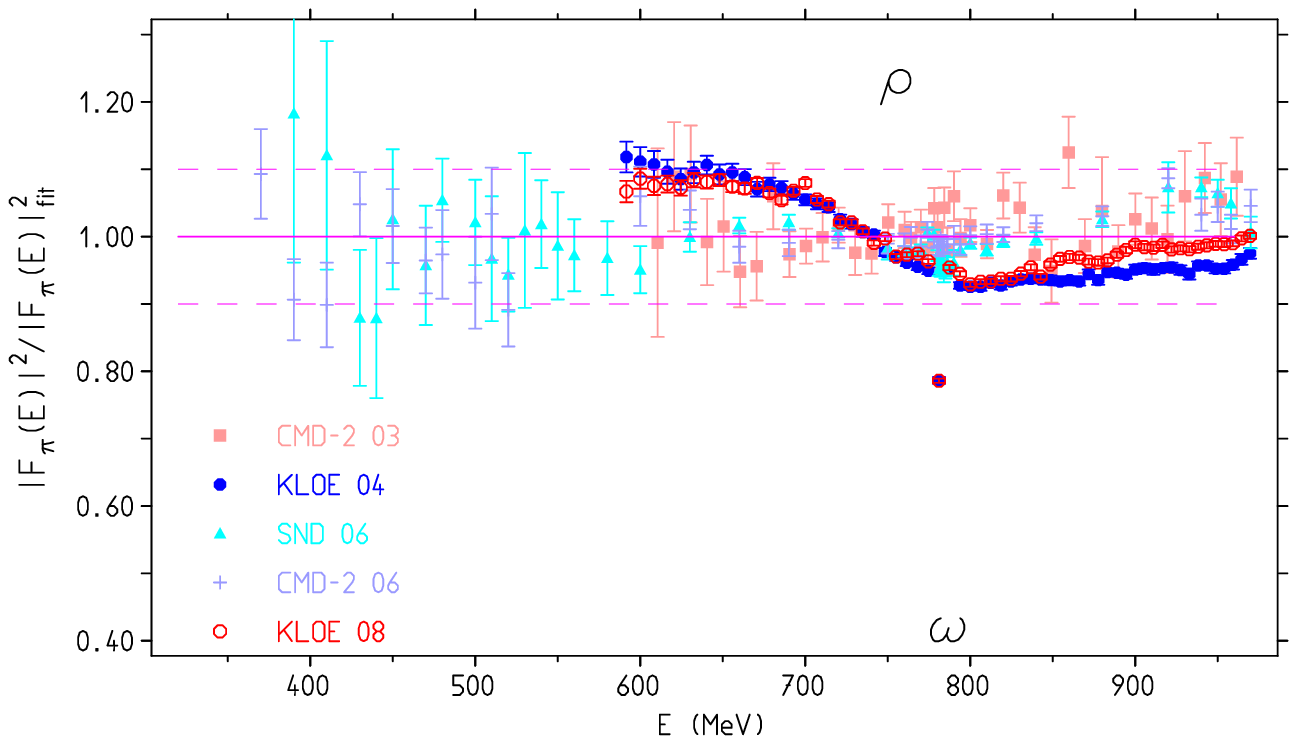}}{%
\includegraphics[height=6cm]{ffpp08rat.eps}}
\caption{\small
The dominating low energy domain is given by the channel $\epm \ra
\pi^+\pi^-$ which exhibits the $\rho$--resonance. The $\rho-\omega$ mixing,
due to isospin breaking by $m_u \neq m_d$, is distorting the ideal Breit-Wigner
resonance shape of the $\rho$. The ratio $|F_\pi(E)|^2/|F_\pi(E)|^2_\mathrm{fit}$  
shows the fairly good compatibility of the newer measurements relative to a CMD-2 fit.
Dashed horizontal lines mark $\pm$ 10\%.}
\label{fig:epemdata}
\end{figure}

A collection of $\epm$--data at energies $>$ 1 GeV is shown in
Fig.~\ref{fig:rhadron}~\cite{FJ06}. The compilation is an up--to--date version of earlier
ones~\cite{EJ95},\cite{AY95}-\cite{WhalleyR03} by different groups. For
detailed references and comments on the data we refer to~\cite{EJ95}
and the more recent experimental publications by MD-1~\cite{MD196},
BES~\cite{BES02}, CMD-2~\cite{CMD203,CMD206}, KLOE~\cite{KLOE04,KLOE08},
SND~\cite{SND06}, BaBar~\cite{BaBar05} and~\cite{BaBar07}. A
list of experiments and references till 2003 is given
in~\cite{WhalleyR03}, where
the available data are collected.\\ 

\begin{figure}[t]
\centering
\IfFarbe{%
\includegraphics[height=5cm]{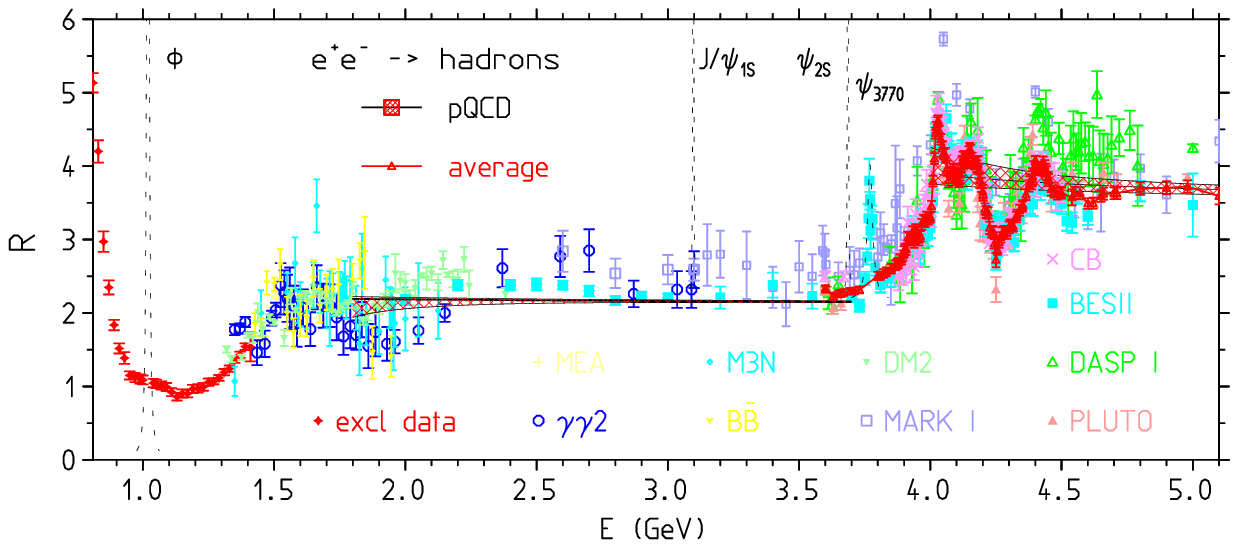}}{%
\includegraphics[height=5cm]{ree0105.eps}}

\IfFarbe{%
\includegraphics[height=5cm]{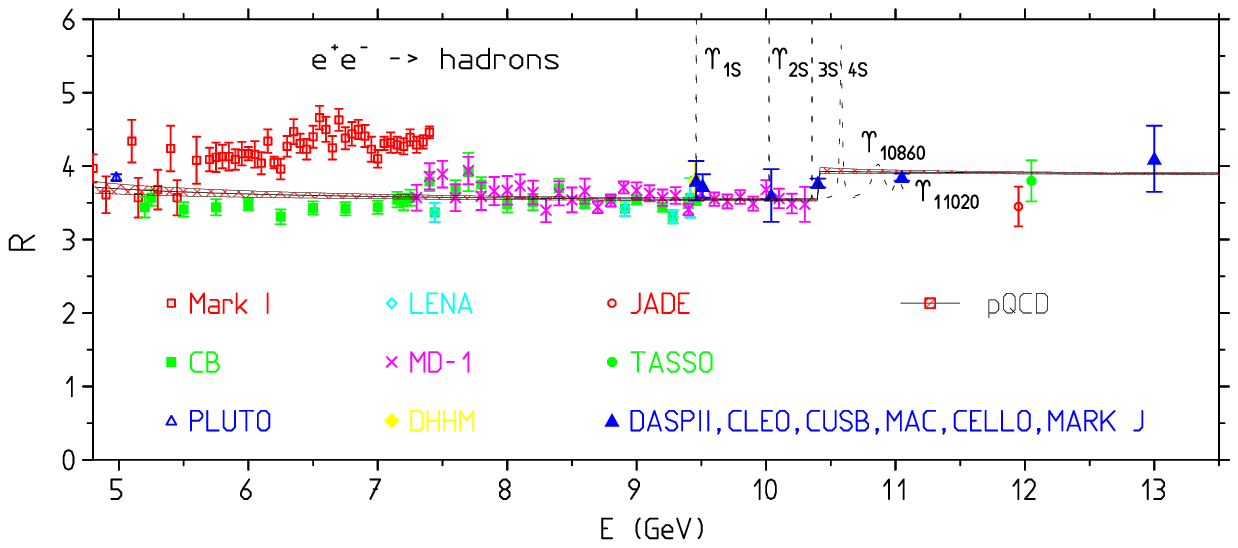}}{%
\includegraphics[height=5cm]{ree0513.eps}}
\caption{Experimental results for $R^\mathrm{had}_\gamma(s)$ in the
range 1 GeV $<E=\sqrt{s}<$ 13 GeV, obtained at the various $\epm$ storage
rings. The perturbative
quark--antiquark production cross--section is also displayed
(pQCD). Parameters: $\alpha_s(M_Z)=0.118 \pm 0.003$, $M_c=1.6 \pm
0.15~\gv$, $M_b=4.75 \pm 0.2~\gv$ and the \MSb scale varied in the
range $\mu \in (\sqrt{s}/2,2\sqrt{s})$.}
\label{fig:rhadron}
\end{figure}

For the evaluation of the basic integral Eq.~(\ref{AM}) we
take $R(s)$ data up to $\sqrt{s}=E_{cut}=5.2$ GeV and for the
$\Upsilon$ resonance--region between 9.46 and 13 GeV and apply
perturbative QCD from 5.2 to 9.46 GeV and for the high energy tail
above 13 GeV. The result obtained is~\cite{FJ08} (update including~\cite{KLOE08})
\be
\amuhlo=(690.30 \pm 5.26)[(692.37 \pm 5.58)] \times 10^{-10}
\label{amuhadLO}
\ee
and is based on a direct integration of all relevant 
$\epm$--data available\footnote{The corresponding contribution to
$a_e$ reads $a_e^{(4)}(\mathrm{vap,\,had})=(1.860\pm 0.015) \times
10^{-12}\,.$ Since the kernel Eq.~(\ref{KShat}) of the
integral Eq.~(\ref{AM}) also depends on the lepton mass, the result
does not scale with $(m_e/m_\mu)^2$ but is about 15\% larger.}. 
In braces the value before including the new KLOE
result~\cite{KLOE08}.

Note that the different data sets shown in Fig.~\ref{fig:epemdata}
exhibit systematic deviations in the distribution which are not yet
understood.  In contrast, the $\amuh$ integrals in general are in good
agreement\footnote{In the common KLOE energy range (591.6,969.5) MeV
individual contributions for $\amuhlo$ based on the latest [2004/2008]
data are: $387.20(0.50)(3.30)$ [KLOE], $392.64(1.87)(3.14)$ [CMD-2]
and $390.47(1.34)(5.08)$ [SND].}. A very recent \textit{preliminary}
precision measurement of the $\epm \to \ppm(\gamma)$
cross section with the ISR method by BaBar~\cite{Davier08} once more reveals 
large (in comparison with the claimed experimental errors) discrepancies
with respect to previous results. The integrated result yields a
shift $\delta \amuh (\pi\pi) \simeq + 13.5 \power{-10}$ and seems to be in
much better agreement with corresponding results obtained from the $\tau$
spectral--functions (see below), however, the spectrum is much steeper
(-10\% at 0.5 GeV up to +10\% at 1 GeV) than the one from ALEPH, for example. The
new $\epm$--based result agrees better (at $\pm$ 5\% level) with the more recent 
Belle $\tau$ results. We will say more about the possibility to use
$\tau$ data for the calculation of $\amuh$, below.

For the $\epm$--based result (\ref{amuhadLO}),
the size of contributions and squared errors from different
energy regions are illustrated in Fig.~\ref{fig:gmusta}. 

\begin{figure}[ht]
\vspace*{-7mm}
\centering
\IfFarbe{%
\includegraphics[height=5cm]{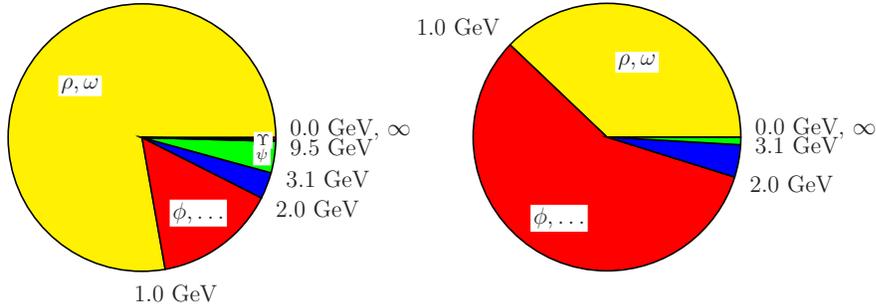}}{%
\includegraphics[height=5cm]{gmdist.eps}}
\caption{The distribution of contributions (left) and errors (right)
in \% for $\amuhlo\;$ from different energy regions. The error of a
contribution $i$ shown is
$\delta^2_{i\:{\rm tot}}/\sum_i \delta^2_{i\:{\rm tot}}$ in \%. The
total error combines statistical and systematic errors in quadrature.}
\label{fig:gmusta}
\end{figure}

Some other recent evaluations are collected in Table~\ref{tab:otheramuhad1}.
Differences in errors
\begin{table}[t]
\caption{Some recent evaluations of $\amuhlo$.}
\label{tab:otheramuhad1}
\centering
\begin{tabular}{clcc}
\hline\noalign{\smallskip}
$\amuhlo \times 10^{10}$ & data & Ref. \\
\noalign{\smallskip}\hline\noalign{\smallskip}
$ 696.3[7.2]$ &$e^+e^-$&\cite{DEHZ03}  \\
$ 711.0[5.8]$ &$e^+e^-+\tau$  &\cite{DEHZ03} \\
$ 694.8[8.6]$ &$e^+e^-$&\cite{GJ03} \\
$ 684.6[6.4]$ & $e^+e^-$ TH&\cite{SN03} \\
$ 699.6[8.9]$ &$e^+e^-$&\cite{ELZ03} \\
$ 692.4[6.4]$ &$e^+e^-$&\cite{HMNT04} \\
\noalign{\smallskip}\hline
\end{tabular} \hspace*{6mm}
\begin{tabular}{clcc}
\hline\noalign{\smallskip}
$\amuhlo \times 10^{10}$ & data & Ref. \\
\noalign{\smallskip}\hline\noalign{\smallskip}
$ 693.5[5.9]$ &$e^+e^-$&\cite{TY04}\\        
$ 701.8[5.8]$ &$e^+e^-+\tau$&\cite{TY04}\\   
$ 690.9[4.4]$ &$e^+e^-$$^{**}$&\cite{DEHZ06}\\
$ 689.4[4.6]$ &$e^+e^-$$^{**}$&\cite{HMNT06}\\
$ 692.1[5.6]$ &$e^+e^-$$^{**}$&\cite{FJ06} \\
$ 690.3[5.3]$ &$e^+e^-$$^{**}$&\cite{FJ08} \\
\noalign{\smallskip}\hline
\end{tabular}
\end{table}
come about mainly by utilizing more ``theory--driven'' concepts : use
of selected data sets only, extended use of perturbative QCD in place
of data [assuming local duality], sum rule methods, low energy
effective methods~\cite{LeCo02}. The last four ($^{**}$)
results include the recent data from SND, CMD-2, and
BaBar. The last update also includes the most recent data 
from BaBar~\cite{BaBar07} and KLOE~\cite{KLOE08}.

\begin{figure}[t]
\centering
\parbox{1.2in}{%
\hspace*{-3.5cm}%
\includegraphics[height=5cm]{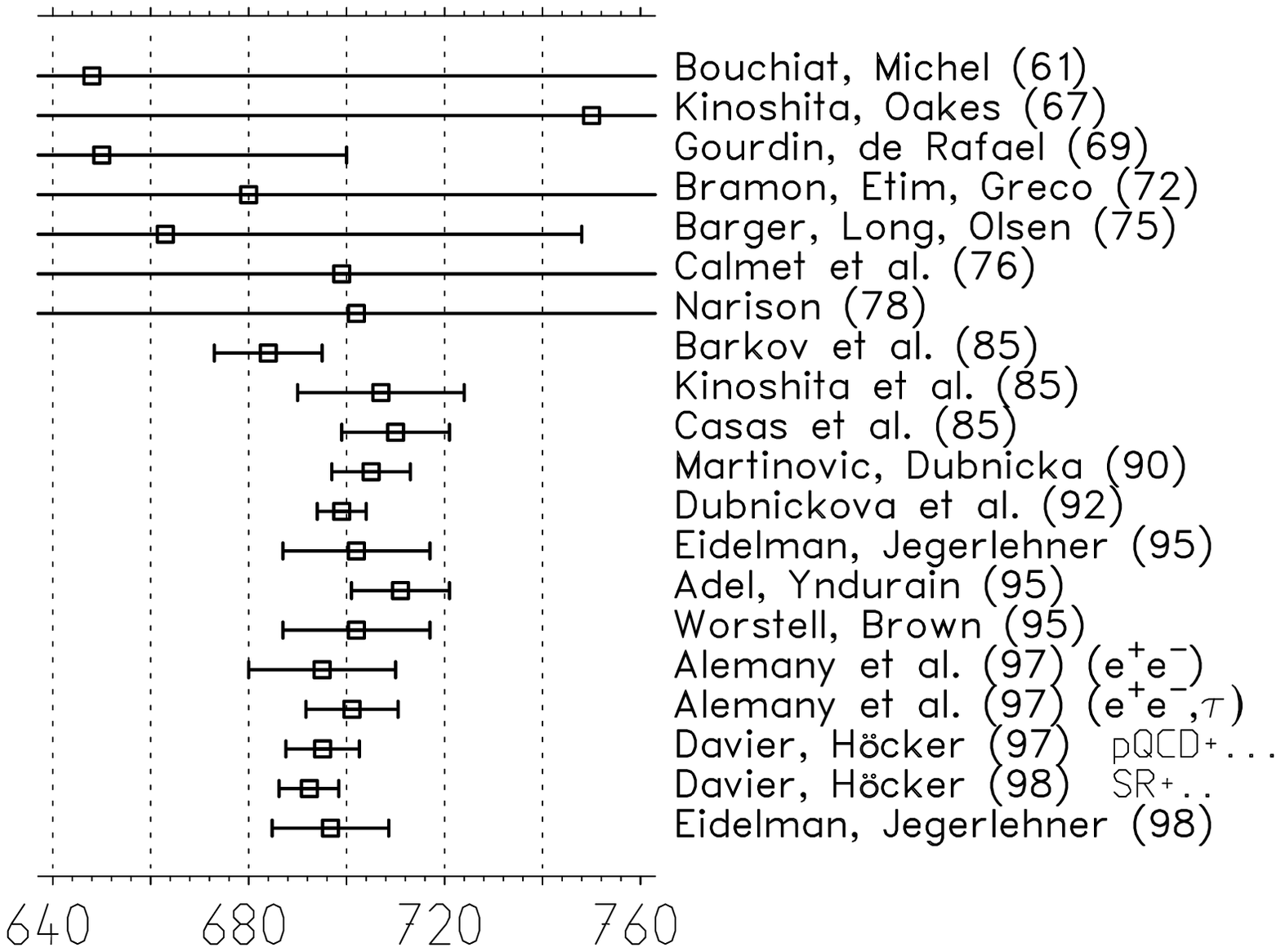}}%
\qquad
\begin{minipage}{1.2in}%
\hspace*{-0.5cm}%
\includegraphics[height=5cm]{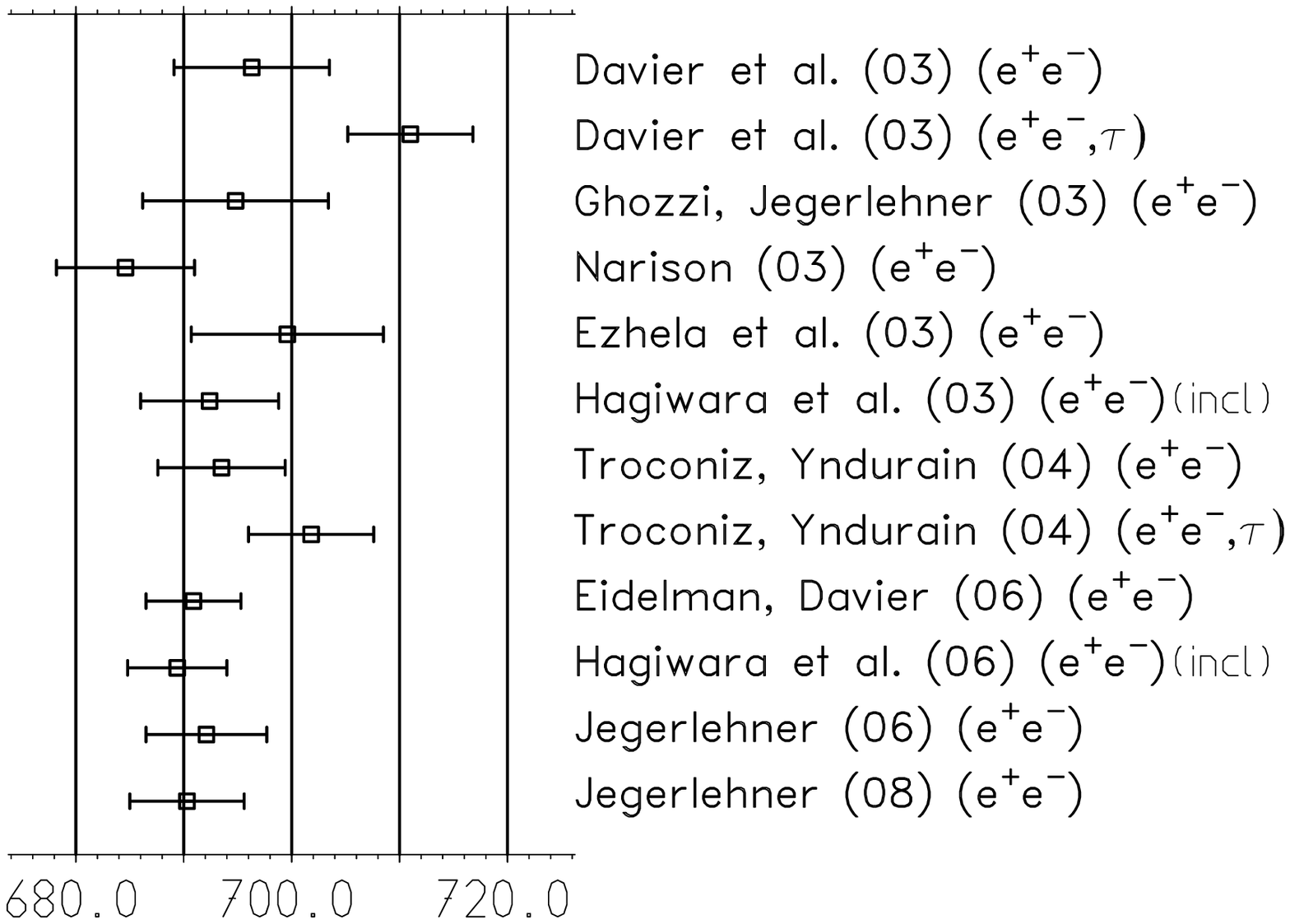}
\end{minipage}%
\caption{History of evaluations before 2000
(left)~\cite{BM}--\cite{GdeR},\cite{BEG72}--\cite{Narison78},\cite{KNO84}--\cite{CLY85},\cite{MD,DDS,EJ95},
\cite{AY95}--\cite{MOR00},
and some more recent ones
(right)~\cite{DEHZ03}--\cite{HMNT06},~\cite{FJ06,FJ08}; ($\epm$) =
$\epm$--data based, ($\epm$,$\tau$) = in addition include data from
$\tau$ spectral functions (see Sect.~\ref{sssec:taudecayHSF}).}
\label{fig:gmuhist}
\end{figure}

There have been many independent evaluations of $\amuhlo$ in the
past\footnote{The method how to calculate hadronic vacuum polarization
effects in terms of hadronic cross--sections was developed long time
ago by Cabibbo and Gatto~\cite{CabibboGatto61}. First estimations were
performed in~\cite{BM}--\cite{GdeR},\cite{BEG72,Narison78}. As cross--section measurements
made further progress much more precise estimates became possible in
the mid 80's~\cite{KNO84}--\cite{CLY85}. A more detailed analysis
based on a complete up--to--date collection of data followed about 10
years later~\cite{EJ95}.}
and some of the more recent ones are listed in
Table~\ref{tab:otheramuhad1}. Fig.~\ref{fig:gmuhist} gives a fairly
complete history of the evaluations based on $\epm$--data.

Before we will continue with a discussion of the higher order hadronic
contributions, we first present additional details about what
precisely goes into the DR Eq.~(\ref{AM}) and briefly discuss some issues concerning the
determination of the required hadronic cross--sections.

     \subsubsection{Dispersion Relations and Hadronic $e^+e^-$--Annihilation Cross Sections}
To leading order in $\alpha$, the hadronic ``blob'' in
Fig.~\ref{fig:gmudia} has to be identified with the photon
self--energy function $\Pi^{'\:\mathrm{had}}_\gamma(s)$. The latter we
may relate to the cross--section $\epm \to$ hadrons by means of the DR Eq.~(\ref{subDR})
which derives from the correspondence Fig.~\ref{fig:opticalthx}
\begin{figure}[t]
\centering
\includegraphics{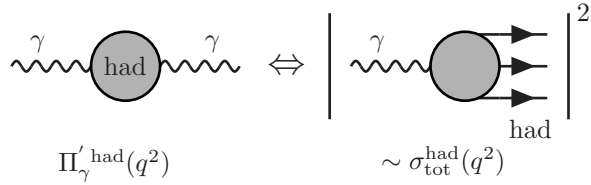}
\caption{Optical theorem for the hadronic contribution to the photon propagator.}
\label{fig:opticalthx} 
\begin{picture}(10,10)(0,0)
\end{picture}
\end{figure}
based on unitarity (optical theorem) and
causality (analyticity), as elaborated earlier. Note that
$\Pi^{'\:\rm had}_\gamma(q^2)$ is a one particle irreducible (1PI)
object, represented by diagrams which cannot be cut into two
disconnected parts by cutting a single photon line.  At low energies
the imaginary part is related to intermediate hadronic states like
$\pi^0\gamma,\rho,\omega,\phi,\cdots,$
$\pi\pi,3\pi,4\pi,\cdots,$ $\pi\pi\gamma,,\cdots,KK,KK\pi\cdots$ which in the DR
correspond to the states produced in $\epm$--annihilation via
a virtual photon. At least one hadron plus any
strong, electromagnetic or weak interaction contribution counts. 
$\epm$--data in principle may be used up to energies where
$\gamma-Z$ interference comes into play above about 40 GeV.

Experimentally, what is determined is of the form
\bea
R^{\mathrm{exp}}_\mathrm{had}(s)=\frac{N_{\rm had}\:{ (1+\delta_{\rm RC})}}{N_{\rm
norm}\:\varepsilon}\:{
\frac{\sigma_{\rm norm}(s)}{\sigmmbp (s)}}\; ,
\eea
where $N_{\rm had}$ is the number of observed hadronic events, $N_{\rm
norm}$ is the number of observed normalizing events, $\varepsilon$ is
the detector efficiency--acceptance product of hadronic events while
$\delta_{\rm RC}$ are radiative corrections to hadron production.
$\sigma_{\rm norm}(s)$ is the physical cross--section for normalizing
events, including all radiative corrections integrated over the
acceptance used for the luminosity measurement, and $\sigmmbp (s)$
$=4\pi\al^2/3s$ is the normalization.  This also shows that a
precise measurement of $R(s)$ requires precise knowledge of the relevant
radiative corrections.

Radiation effects may be used to measure $\sigha (s')$ at all energies $\sqrt{s'}$
lower than the fixed energy $\sqrt{s}$ at which an accelerator is
running~\cite{BaierFadin68}. This is possible due to initial state radiation (ISR),
which can lead to huge effects for kinematical reasons.
The relevant \textit{radiative return} (RR) mechanism is illustrated in
Fig.~\ref{fig:RR}: in the radiative process
$\epm \to \ppm \gamma$, photon radiation from the initial state
reduces the invariant mass from $s$ to $s'=s\,(1-k)$ of the produced
final state, where $k$ is the fraction of energy carried away by the
photon radiated from the initial state.
\begin{figure}[b]
\centering
\IfFarbe{%
\includegraphics[height=3.0cm]{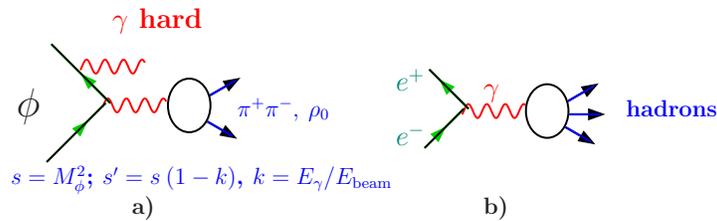}}{%
\includegraphics[height=3.0cm]{RR.eps}}
\caption{a) Principle of the radiative return determination of the $\ppm$ cross--section
by KLOE at the $\phi$--factory DA$\varPhi$NE. At the $B$--factory at SLAC,
using the same mechanism, BaBar has measured many other channels
at higher energies. b) Standard measurement of $\sigha$ in an energy scan,
by tuning the beam energy.}
\label{fig:RR}
\end{figure}
Such RR cross-section measurements are particularly interesting for
machines running on--resonance like the $\phi$-- and $B$--factories,
which have enhanced event rates as they are running on top of a
peak~\cite{RR1,Benayoun99,Phokahra}. The first dedicated RR experiment
has been performed by KLOE at DA$\Phi$NE/Frascati, by measuring the
$\ppm$ cross--section~\cite{KLOE04,KLOE08} (see Fig.~\ref{fig:epemdata}
and Refs.~\cite{Czyz:2007bc,Kluge:2008fb}).

Results for exclusive multi--hadron production channels from BaBar
play an important role in the energy range between 1.4 to 2 GeV. In
fact new data became available for most of the channels of the
exclusive measurements in this region. In contrast the inclusive
measurements date back to the early 1980's and show much larger
uncertainties.

It is important to note that what we need in the DR is the 1PI
``blob'' which by itself is not what is measured, i.e. it is not a
physical observable. In reality the virtual photon lines attached to
the hadronic ``blob'' are dressed photons (full photon propagators
which include all possible radiative corrections) and in order to
obtain the 1PI part one has to undress the cross section by amputation
of the full photon lines. The $e^+e^-
\to \mathrm{hadrons}$ transition amplitude is
proportional to $e^2$ from the $e^+e^- \gamma^*$ and the $\gamma^*
q\bar{q}$ (hadrons) vertices which in the cross section appears in
quadrature ${ \propto}\, e^4$ or $\alpha^2$. Due to the running of
the electromagnetic charge the physical (dressed) cross section is 
${ \propto}\, \alpha^2(s)$.
Undressing requires to replace the running $\alpha(s)$ by the classical $\alpha$:
\be
\sigma^{(0)}_\mathrm{tot}(\epm \to \mathrm{hadrons})=\sigma_\mathrm{tot}(\epm \to \mathrm{hadrons})\,
\left(\frac{\alpha}{\alpha(s)}\right)^2
\label{bearizationofsigma}
\ee
and, using Eq.~(\ref{subDR}) we obtain
\be
\Pi'_{\gamma}(k^2)-\Pi'_{\gamma}(0)=
\frac{k^2}{4\pi^2 \alpha}\int\limits_{0}^{\infty} \D s\,
\frac{\sigma^{(0)}_\mathrm{tot}(\epm \to \mathrm{hadrons})}{(s-k^2-\ie)}\;\;.
\ee
It should be stressed that using the physical cross section in the DR gives a
nonsensical result. In order to get the photon propagator we
have to subtract in any case the effective charge from the external
$\epm \gamma$ vertex at the correct scale.
Thus if we would use
\bea
\frac{k^2}{4\pi^2 \alpha}\int\limits_{0}^{\infty} \D s\,
\frac{\sigma_\mathrm{tot}(\epm \to \mathrm{hadrons})}{(s-k^2-\ie)}
\eea
we would be double counting the VP effects. 
In contrast with the linearly in $\alpha/\alpha(s)$ rescaled cross--section
\be
\frac{k^2}{4\pi^2}\int\limits_{0}^{\infty} \D s\,
\frac{1}{\alpha(s)}\frac{\sigma_\mathrm{tot}(\epm \to \mathrm{hadrons})}{(s-k^2-\ie)}\,,
\ee
we obtain the hadronic shift Eq.~(\ref{amufullVP}) for the full  photon propagator.\\ 

Since what we need is the hadronic blob, in processes as displayed in
Fig.~\ref{fig:RR}, it is evident that a precise extraction of the
desired object requires a subtraction of all radiative corrections not
subsummed in the blob. Thus besides the VP effects, in particular, the
initial state radiation (ISR) has to be subtracted. It is well known
that photon radiation leads to infrared (IR) singularities if not
virtual and real (soft) radiation are included on the same footing
(Bloch-Nordsieck~prescription).  Thereby soft photon radiation has to
be included at least up to energies $0<E<E_\mathrm{cut}$ where
$E_\mathrm{cut}$ is the detection threshold of the detector
utilized. Any charged particle cross--section measurement requires
some detector dependent cuts in photon phase space and the detector
dependent radiation effects must be subtracted in order to obtain a
detector independent meaningful physics cross--section. Besides the
ISR there is final state radiation (FSR) as well as initial--final
state interference effects, where the latter to leading order drop out
in total cross sections for $C$ symmetric cuts. While ISR from the
$\epm$ initial state it calculable in QED to any desired order of
precision, the calculation of the FSR for hadronic final states is not
fully under control.  In addition, experimentally it is not possible
to distinguish ISR from FSR photons. Fortunately the most important
channel contributing to $\amuhlo$ is the two--body $\pi^+\pi^-$ one
and FSR usually is calculated in scalar QED (sQED) which however is
appropriate only for relatively soft photons which see the pions as
point particles. In fact a generalized version of sQED is applied
where the point form--factor $F^\mathrm{point}_\pi=1$ is replaced by
the experimentally determined pion form--factor $F_\pi(s)$. The FSR
contribution will be discussed in Sect.~\ref{ssec:hohad}. On the level
of the quarks the leading order FSR contribution would be given by the
diagrams 19) to 21) of Fig.~\ref{fig:threeloopdia1} and corresponds
to the inclusion of the final state radiation correction of $R(s)$.
The Kinoshita-Lee-Nauenberg (KLN) theorem infers that these fully
inclusive corrections are not enhanced by any logarithms. This also
infers that the model dependence of the FSR contribution is at worst
moderate.  For a more detailed discussion see
Refs.~\cite{Arbuzov97}--\cite{GHJJ03} and references therein.

     \subsubsection{Hadronic $\tau$-Decays and Isospin Violations}
\label{sssec:taudecayHSF}
In principle, the $I=1$ iso--vector part of $e^+e^- \to
\mathrm{hadrons}$ can be obtained in an alternative way by using the
precise vector spectral functions from hadronic $\tau$--decays $\tau
\to \nu_\tau + \mathrm{hadrons}$ which are related by an isospin
rotation~\cite{tsai,ADH98}, like $\pi^0\pi^- \to \pi^+\pi^-$, as
illustrated in Fig.~\ref{fig:isofig} for the most relevant $2\pi$
channel.  After isospin violating corrections, due to photon radiation
and the mass splitting $m_d-m_u \neq 0$, have been applied, there
remains an unexpectedly large discrepancy between the $e^+e^-$- and
the $\tau$-based determinations of $a_\mu$~\cite{DEHZ03}, as may be
seen in Table~\ref{tab:otheramuhad1}. Possible explanations are so far
unaccounted isospin breaking~\cite{GJ03} and/or experimental problems
with the data. For example, FSR corrections in the charged current
$\tau$--channel are expected to be more model dependent than in the
neutral $\epm$--channel as they exhibit a much larger short distance
sensitivity. Since the $e^+e^-$-data are more directly related to what
is required in the dispersion integral, one usually advocates to use
the $e^+e^-$ data only in the evaluation of $\amuhlo$.

\begin{figure}
\centering
\IfFarbe{%
\includegraphics[height=3cm]{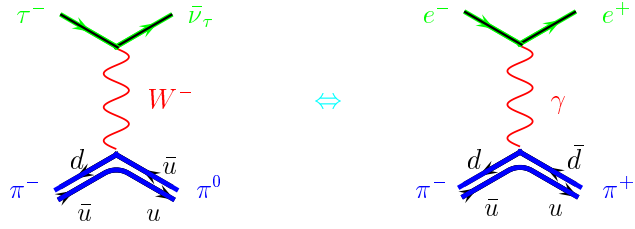}}{%
\includegraphics[height=3cm]{ISOfig.eps}}
\caption{$\tau$--decay vs. $\epm$--annihilation: the involved hadronic
matrix--elements $\bra{{\rm out}~\pi^+\pi^-}j^{I=1}_\mu(0)\ket{0}$ and
$\bra{{\rm out}~\pi^0\pi^-}J^{-}_{{\rm V}\mu}(0)\ket{0}$ are related by isospin.}
\label{fig:isofig}
\end{figure}

Precise $\tau$--spectral functions became available in 1997ff from
ALEPH, OPAL and CLEO~\cite{ALEPH}--\cite{CLEO} and the idea to use the
$\tau$ spectral data to improve the evaluation of the hadronic
contributions $\amuh$ was pioneered by Alemany, Davier and
H\"ocker~\cite{ADH98}.  More recently an new measurement was presented by 
Belle~\cite{BELLE08}. Data sets for $|F_\pi|^2$ are displayed in
Fig.~\ref{fig:taudata}. Taking into account the $\tau$--data increases
the contribution to $\amuh$ by 2 $\sigma$ (see
Table~\ref{tab:otheramuhad1} and Fig.~\ref{fig:gmuhist}). The
unexpectedly large discrepancy between isospin rotated $\tau$--data,
corrected for isospin violations, and the direct $e^+e^-$--data
remains one of the unsolved problems. This on the one hand means that
doubts continue to exist that low energy hadronic cross--sections are
sufficiently well under control, on the other hand a solution of the
problem would contribute to reduce hadronic errors on $g-2$
predictions further.

\begin{figure}[t]
\centering
\IfFarbe{%
\includegraphics[height=5.9cm]{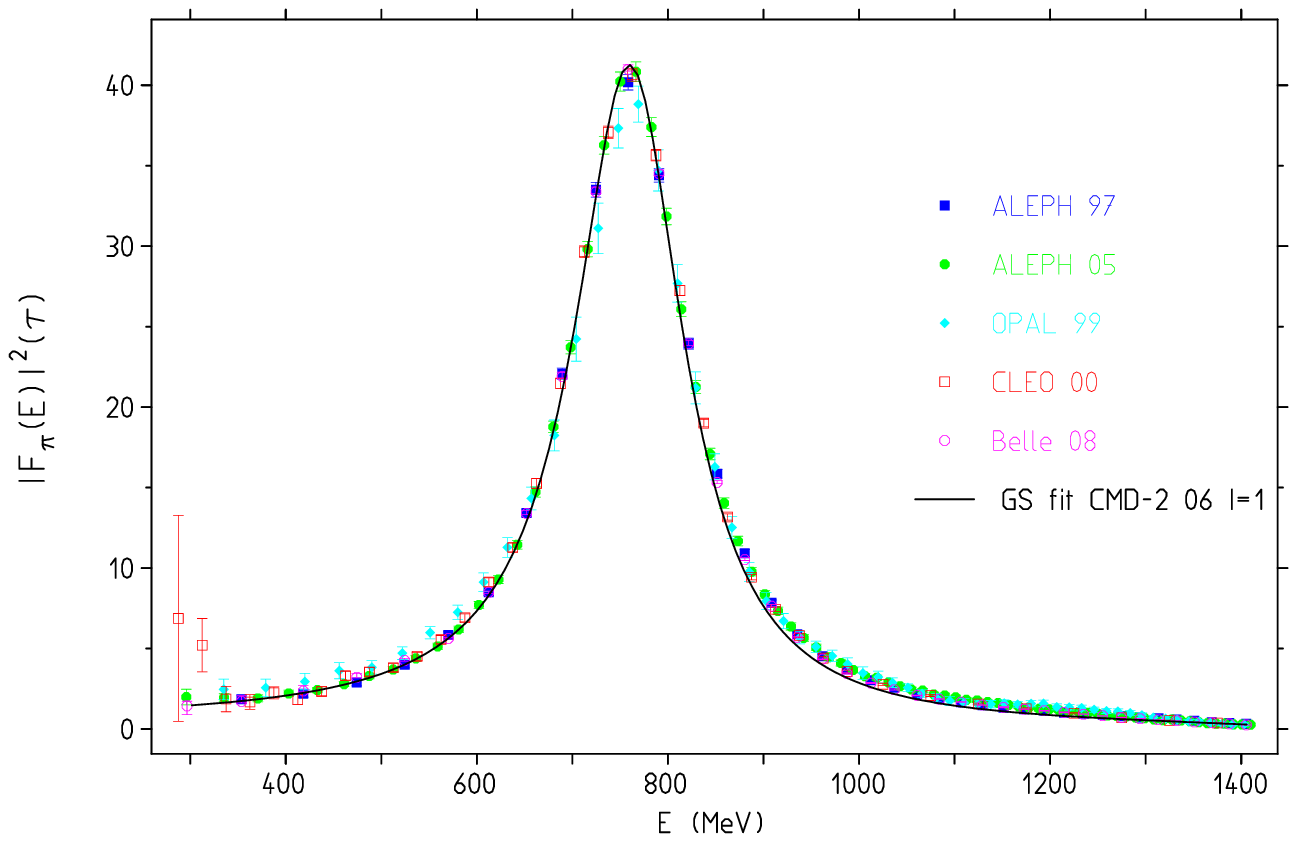}}{%
\includegraphics[height=5.9cm]{ffpptau08.eps}}

\IfFarbe{%
\includegraphics[height=6cm]{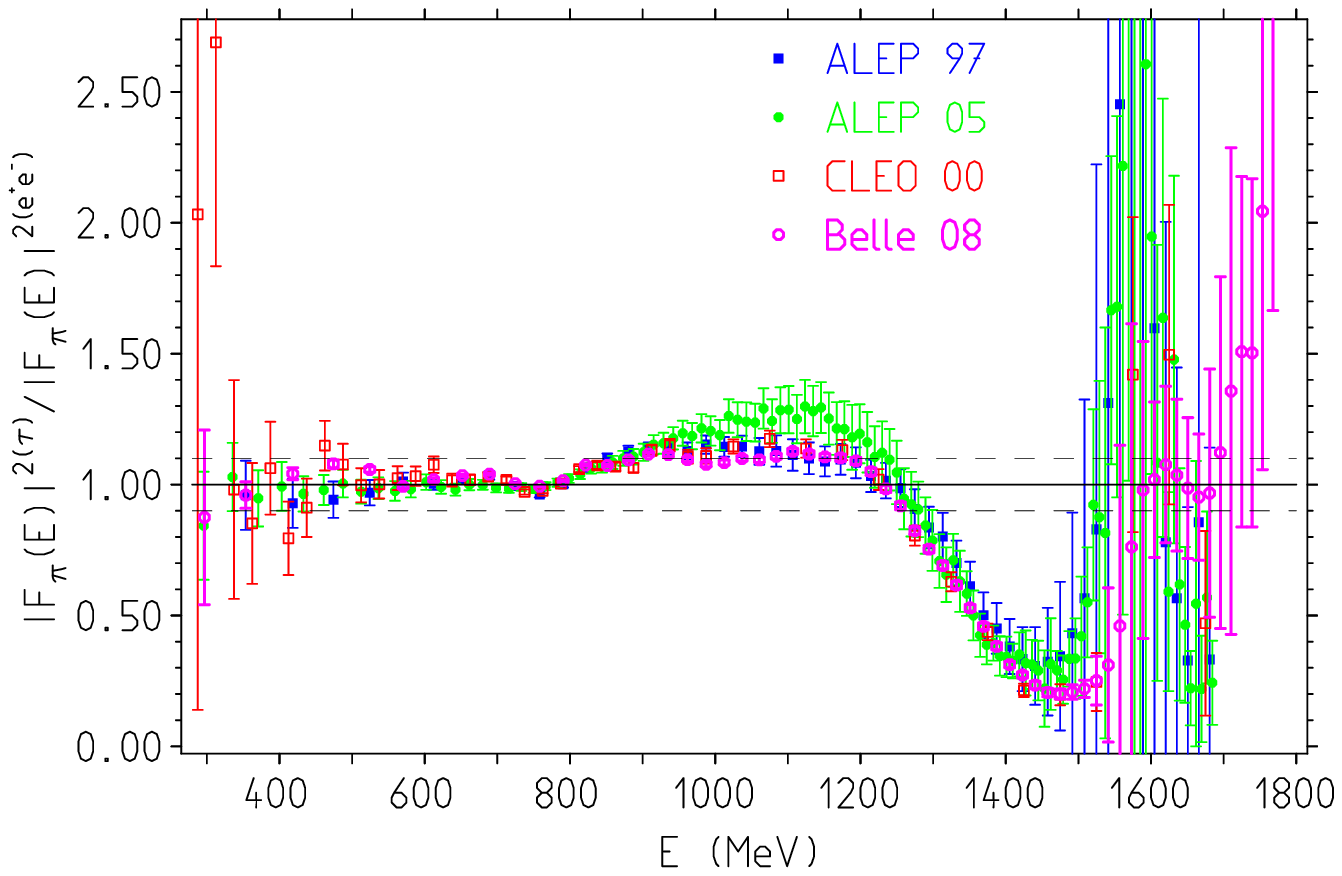}}{%
\includegraphics[height=6cm]{rffpptau08.eps}}
\caption{\small 
Modulus square of the $I=1$ pion form factor extracted from $\tau^\pm \ra
\nu_\tau \pi^\pm \pi^0 $ which shows the $\rho^\pm$--resonance.  
The ratio $|F_\pi(E)|^2(\tau)/|F_\pi(E)|^2_\mathrm{fit}(\epm[I=1])$
illustrates the missing consistency of the $\tau$--data relative to
a CMD-2 fit.  Dashed horizontal lines mark $\pm$ 10\% (see also~\cite{BELLE08,DHZ05}).
Note that the reference fit line represents the $\epm$ data only below about
1 GeV (see Fig.~\ref{fig:epemdata}). At higher energies data for $\epm
\to \ppm$ are rather poor, but old Orsay DM2 data as well as the new
preliminary BaBar radiative return data~\cite{Davier08} also exhibit
the dip at 1.5 GeV, i.e. our ``normalization'' above 1 GeV is to be
considered as arbitrary.}
\label{fig:taudata}
\end{figure}

For the dominating $2\pi$ channel, the precise relation we are talking
about may be derived by comparing the relevant lowest order
diagrams Fig.~\ref{fig:isofig}, which for the $\epm$ case translates
into
\begin{equation}
\sigma^{(0)}_{\pi\pi}\equiv
\sigma_0 (e^+e^- \to \pi^+\pi^-) = \frac{4\pi\alpha^2}{s}\: v_0(s)
\label{eepp}
\end{equation}
and for the $\tau$ case into
\ba
\frac{1}{\Gamma} \frac{d\Gamma}{ds}
 (\tau^- \to \pi^-\pi^0 \nu_\tau) &=&
\frac {6 |V_{ud}|^2 S_{EW}}{m_\tau^2}
\frac{B(\tau^-\rightarrow \nu_\tau\,e^-\,\bar{\nu}_e)}
{B(\tau^-\rightarrow \nu_\tau\,\pi^- \pi^0)}\: 
\left(1-\frac{s}{m_\tau^2}\right)
\left(1 + \frac {2s}{m_\tau^2}\right)\: v_-(s), 
\label{taupp}
\ea
where $|V_{ud}|=0.9746\pm0.0006$~\cite{PDG04} denotes the CKM weak
mixing matrix element and $S_{\mathrm{EW}}=1.0198\pm0.0006$ accounts
for electroweak radiative
corrections~\cite{Marciano:vm}--\cite{CEN},\cite{DEHZ03}.  The \sfs\
are obtained from the corresponding invariant mass distributions. The
$B(i)$'s are branching ratios, $B(\tau^-\rightarrow \nu_\tau\,e^-\,\bar{\nu}_e)=(17.810\pm0.039)\%$,
$B(\tau^-\rightarrow \nu_\tau\,\pi^- \pi^0)=(25.471\pm0.129)\%$. SU(2) symmetry (CVC) would imply
\begin{equation}
v_-(s) =  v_0(s)\;\;.
\label{CVCrel}
\end{equation}
The spectral functions $v_i(s)$ are related to the pion form factors
$F^i_\pi(s)$ by
\begin{equation}
v_i(s)=\frac{\beta_i^3(s)}{12} |F^i_\pi(s)|^2\;\;;\;\;\;(i=0,-), 
\label{sfvsff}
\end{equation}
where $\beta_i (s)$ is the pion velocity. The difference in phase
space of the pion pairs gives rise to the relative factor
$\beta^3_{\pi^-\pi^0}/\beta^3_{\pi^-\pi^+}$.

Before a precise comparison via Eq.~(\ref{CVCrel}) is possible all kinds of
isospin breaking effects have to be taken into account. 
For the $\pi \pi$ channel the most relevant corrections have been 
investigated in~\cite{CEN,FloresBaez:2006gf}. The corrected version of Eq.~(\ref{CVCrel})
may be written in the form
\ba
\sigma_{\pi\pi}^{(0)}=\left[ \frac{K_{\sigma}(s)}{K_\Gamma(s)}\right]
\: \frac{d \Gamma_{\pi\pi [ \gamma ] }}{ds} \times
\frac{R_{\rm IB}(s)}{S_{\rm EW}}, 
\ea
with
\bea
K_\Gamma (s) = \frac{G_F^2\:|V_{ud}|^2\:m_\tau^3 }{384 \pi^3}\:
\left(1-\frac{s}{m_\tau^2}
\right)^2 \left( 1+2\: \frac{s}{m_\tau^2}\right);\; K_\sigma
(s) = \frac{\pi \alpha^2}{3 s},
\label{Ksigma}
\eea
and the isospin breaking correction
\ba
R_{\rm IB}(s) = \frac{1}{G_{\rm EM}(s)} \:
\frac{\beta^3_{\pi^-\pi^+}}{\beta^3_{\pi^- \pi^0}} \:
\left| \frac{F_V(s)}{f_+(s)}\right|^2
\ea
\noi includes the QED corrections to $\tau^- \to \nu_\tau \pi^-
\pi^0$ decay with virtual plus real soft and hard photon radiation integrated over
all phase space.

Originating from Eq.~(\ref{sfvsff}), $\beta^3_{\pi^-\pi^+}/\beta^3_{\pi^-
\pi^0}$ is a phase space correction due to the $\pi^\pm - \pi^0$ mass
difference.  $F_V(s) = F^0_\pi(s)$ is the neutral current (NC) vector form
factor, which exhibits besides the $I=1$ part an $I=0$
contribution. The latter $\rho-\omega$ mixing term is due to the
SU(2) breaking by the $m_d-m_u$ mass difference. Finally, $f_+(s)=F^-_\pi$
is the charged current (CC) $I=1$ vector form factor. One of the leading isospin
breaking effects is the $\rho - \omega$ mixing correction included in
$|F_V(s)|^2$. The form--factor corrections, in principle, also should
include the electromagnetic shifts in the masses and the widths of the
$\rho$'s\footnote{Because of the strong resonance enhancement, 
especially in the $\rho$ region, a small
isospin breaking shift in mass and width between $\rhon$ and
$\rho^\pm$, typically $\Delta m_\rho=m_{\rho^\pm}-m_{\rhon}\sim
2.5~\mathrm{MeV}$ and $\Delta
\Gamma_\rho=\Gamma_{\rho^\pm}-\Gamma_{\rhon}\sim 1.5~\mathrm{MeV}$
and similar shifts for the higher resonances $\rho'$, $\rho''$ and the
mixing amplitudes of these states, causes a large effect in the tails
by the kinematical shift this implies.}. Up to this last mentioned
effect, discussed in~\cite{GJ03} (also see~\cite{MW05}), which was
considered to be negligible in earlier isospin breaking estimates, all
the corrections were applied in~\cite{DEHZ03} but were not able to
eliminate the observed discrepancy between $v_-(s)$ and $v_0(s)$.  The
deviation is starting at the peak of the $\rho$ and is increasing with
energy to about 10-20\%. More precisely, Fig.~\ref{fig:taudata} shows
a good agreement below about 800 MeV, a 10\% enhancement between 800
and 1200 MeV and a pronounced dip around 1500 MeV. The trend shown by
the ALEPH 97 and CLEO data is clearly stressed by the new Belle
measurement~\cite{BELLE08}. The Belle data differ substantially from
the ALEPH 05 data, however, which lie higher by more than 10\% in the
intermediate range [+20\% relative to $\epm$].\\

We should mention here once more that photon radiation by hadrons is
poorly understood theoretically. The commonly accepted recipe is to
treat radiative corrections of the pions by scalar QED, except for the
short distance (SD) logarithm proportional to $\ln M_W/m_{\pi}$ which
is replaced by the quark parton model result and included in $S_{\rm
EW}$ by convention. This SD log is present only in the weak charged
current transition $W^{+*} \to\pi^+\pi^0\:(\gamma)$, while in the
charge neutral electromagnetic current transition $\gamma^* \to
\pi^+\pi^-\:(\gamma)$ this kind of leading log is absent. In any case
there is an uncertainty in the correction of the isospin violations by
virtual and real photon radiation which is hard to quantify. We also
should stress that the possible isospin breaking resonance parameter
shifts, like $\Delta m_\rho$ and $\Delta \Gamma_\rho$, so far have not
been determined unambiguously. Note, however, that the new Belle
results~\cite{BELLE08} precisely confirm the earlier observed shifts
in mass and width of the $\rho$~\cite{Davier03,GJ03}, which could be
part of the source of additional isospin violations.\\
  
A reason for the $\tau$ vs $e^+e^-$ discrepancy could be the way
(incoherent) the pure $I=1$ part (given by $\tau$--data) is combined with
the missing $I=0$ contribution (approximately separated out from the
$\epm$-data). What is needed and what is measured in $\epm$ is the
interference $|A_1(s)+A_0(s)|^2$ which in any case must be smaller
than $|A_1(s)|^2+|A_0(s)|^2$.  In any case, one has to keep in mind
that isospin breaking can only have two origins: the $m_u-m_d$ mass
difference and electromagnetic effects and the latter require a small
\textbf{positive} mass difference $m_{\rho^\pm}-m_{\rhon} \sim
1~\mathrm{MeV}$ and similar for $\rho'$ and $\rho''$. In fact, a fit
of the data for the $\rho$ yields a factor of 2 larger result, which
maybe is a problem. It should be noted that the fits including several
masses, widths and mixings are not very stable. In
Ref.~\cite{Benayoun08} effects of the $\rho$--$\omega$--$\phi$ mixing
on the dipion mass spectrum in $e^+e^-$--annihilation and
$\tau$--decay were analyzed within the HLS effective model and it was
suggested that they could explain the observed isospin breakings.  As
a possibility one also may consider the case that the $\tau$--data
based evaluation of $\amuh$ is the more reliable one. 
The integrated data in the range $m_\pi - 1.8~\gv$ after applying
known isospin violation corrections is given by
\bit
\item Belle ($\tau$)~\cite{BELLE08}
\be
a_\mu^{\pi\pi}=(523.5\pm1.5(\mathrm{exp})\pm 2.6(\mathrm{Br})\pm2.5 (\mathrm{iso})) \power{-10}\cs
\ee
\item ALEPH, CLEO, OPAL ($\tau$)~\cite{DEHZ03}
\be
a_\mu^{\pi\pi}=(520.1\pm2.4(\mathrm{exp})\pm 2.7(\mathrm{Br})\pm2.5 (\mathrm{iso})) \power{-10}\cs
\ee
\eit
which compares to
\bit
\item CMD2, SND ($\epm$)~\cite{DEHZ06}
\be
a_\mu^{\pi\pi}=(504.6\pm3.1(\mathrm{exp})\pm 0.9(\mathrm{rad})) \power{-10}\cs
\ee
\eit
where errors are the experimental ones (exp), from the normalizing
branching fraction (Br), from isospin breaking corrections (iso)
and from radiative corrections (rad). Including $\tau$ data shifts the
theoretical prediction by $\delta \amuh \simeq + 17.9 \power{-10}$
thus would improve
the agreement between theory and experiment for $a_\mu$ to the 1.2
$\sigma$ level, see Sect.~\ref{sec:thevsexp}. However, using the
$\tau$--data would also increase the ``gap'' between a too low value
of indirect Higgs mass determinations in comparison with the known
direct lower bound. This does certainly not support the idea that the
$\tau$--data based evaluation is more likely to be the correct
choice~\cite{Passera:2008jk}.

     \subsubsection{Perturbative QCD Contributions}
\label{subsub:epempQCD}
The high energy tail of the basic dispersion integral Eq.~(\ref{AM})  can 
safely be calculated in pQCD because of the  
{\em asymptotic freedom} of QCD. The latter property infers that the
effective strong interaction constant $\alpha_s(s)$ gets weaker the
higher the energy scale $E=\sqrt{s}$, and we may calculate the hadronic
current correlators in perturbation theory as a power series in
$\alpha_s/\pi$. The object of interest is
\be
\rho(s)=\frac{1}{\pi} \Impa \Pi'_\gamma (s)\;;\;\;
\Pi^{\mu \nu}_\gamma (q)=(q^\mu q^\nu -q^2 g^{\mu \nu})\,\Pi'_\gamma (q^2): \mysymb{0}{2}{20}{20}{prsym1v}~~\epo
\ee
The QCD perturbation expansion diagrammatically is given by
\bea
\mysymb{0}{2}{20}{20}{prsym1v}~~&=& \mysymb{0}{2}{20}{20}{prsym2}~~ +
\mysymb{0}{2}{20}{20}{prsym2ct} \crn
&& \crn
&+& \mysymb{0}{2}{20}{20}{prsym2x}~~ + \mysymb{0}{2}{20}{20}{prsym2y}~~ +
\mysymb{0}{2}{20}{20}{prsym2z} \crn
&& \crn
&+& \mysymb{0}{2}{20}{20}{prsym2a}~~ + \mysymb{0}{2}{20}{20}{prsym2b}~~ +
\mysymb{0}{2}{20}{20}{prsym2c} \crn
&& \crn
&+& \mysymb{0}{2}{20}{20}{prsym2d}~~ + \cdots \crn
\eea
Lines \mysymb{0}{2}{10}{10}{vblinei}~~ show external photons,
\mysymb{0}{2}{10}{10}{felinei}~~ propagating quarks/anti\-quar\-ks and
\mysymb{0}{2}{10}{10}{gllinei}~~ propagating gluons.  The vertices $\otimes$ are marking
renormalization counter term insertions. They correspond to
subtraction terms which render the divergent integrals finite.

Perturbative vacuum polarization effects were first discussed by
Dirac~\cite{Dirac34} in QED (for photons and electrons in place of
gluons and quarks) and
finally unambiguously calculated at the 1--loop level by Schwinger~\cite{Schwingervapo49}
and Feynman~\cite{Feynmanvapo49}. Soon later
Jost and Luttinger~\cite{JostLuttinger50} presented the first 2--loop
calculation.

In zeroth order in the strong coupling $\alpha_s$ we have
\bea
2\, \Impa \mysymb{0}{2}{20}{20}{prsym2}~~=
\mysymb{0}{2}{10}{10}{symb1qed2}~~\raisebox{2ex}{$2$}
\eea
which is proportional to the the free quark--antiquark production
cross--section~\cite{FL74} in the so called
\textit{Quark Parton Model}, describing
quarks with the strong interactions turned off. Because of asymptotic
freedom this picture should be a good approximation
asymptotically in the high energy limit of QCD. In pQCD, of course, we
only can calculate the $q\bar{q}$ ($q=u,d,s,\cdots$) production cross--section
and not the physical hadron production cross--section $\sighad$ itself. 
In this case the $R$ function
corresponding  to Eq.~(\ref{Rshaddef}) is defined by
\be
R(s)^\mathrm{pert} \doteq \frac{\sigqua}{\frac{4\pi\alpha^2}{3s}} =\frac{12 \pi^2}{e^2} \rho(s)^\mathrm{pert}\cs
\ee
which for sufficiently large $s$ can be calculated perturbatively. The result is given 
by~\cite{Rpert1,Rpert2,Rpert3,Rpert4}
\ba
R(s)^\mathrm{pert}&=&N_c\:\sum\limits_q
Q_q^2\:\frac{v_q}{2}\:\left(3-v_q^2 \right)\:\Theta(s-4m_q^2) 
\left\{
1 + a c_1(v_q) + a^2 c_2 + a^3
c_3 +a^4 c_4 \cdots \right\}, 
\label{Rpert}
\ea
where $a=\alpha_s (s)/\pi$ and, assuming $4m_q^2\ll s$, i.e. in the
massless approximation
\bea
c_1&=&1, \crn
c_2&=&C_2(R)\:\left[ - \frac{3}{32}\,C_2(R)-\frac34 \beta_0\,
\zeta(3)-\frac{33}{48}\,N_q+\frac{123}{32}\,N_c\right] \crn
&=&\frac{365}{24}-\frac{11}{12}\,N_q - \beta_0\, \zeta(3) \simeq 1.9857
-0.1153\, N_q, \crn
c_3&=&-6.6368-1.2002\, N_q - 0.0052\, N_q^2  -1.2395\: (\sum_q
Q_q)^2/(3 \sum_q Q_q^2), \crn
c_4&=&-0.010\,N_q^3\,+\,1.88\,N_q^2\,-\,34.4\,N_q\,+\,135.8
-\pi^2\beta_0^2 \left(1.9857-0.1153\, N_q+\frac{5\beta_1}{6\beta_0}\right), 
\eea
in the \MSb scheme. $N_q=\sum_{q:4m_q^2\leq s}1$ is the
number of active quark flavors. The mass dependent threshold factor in front
of the curly brackets in Eq.~(\ref{Rpert}) is a function of the velocity
$v_q=\left(1-\frac{4m_q^2}{s} \right)^{1/2}$ and the exact mass dependence
of the first correction term
\bea
c_1(v_q)=\frac{2\pi^2}{3v_q}-(3+v_q)\left(\frac{\pi^2}{6}-\frac14\right)
\eea
is singular (Coulomb singularity due to soft gluon final state
interaction) at threshold.
The singular terms exponentiate~\cite{Novikov77}:
\bea
1+x &\to& \frac{2x}{1-\E^{-2x}}\;;\;\; x=\frac{2\pi \alpha_s}{3\beta}\,,\\
\left(1+c_1(v_q)\:\frac{\alpha_s}{\pi} +\cdots \right) &\to&
\left(1+c_1(v_q)\:\frac{\alpha_s}{\pi}-\frac{2 \pi \alpha_s}{3v_q}
\right)\:\frac{4 \pi \alpha_s}{3v_q}\: \frac{1}{1-\exp
\left\{-\frac{4\pi \alpha_s}{3v_q}\right\}} \epo
\eea
In the ranges where we apply pQCD the strength of the coupling is
still substantial. This requires renormalization group improvement of
perturbative predictions. Thus, as usual, the coupling $\alpha_s$ and
the masses $m_q$ have to be understood as running parameters:
\bea
R\left(\frac{m_{0q}^2}{s_0},\alpha_s(s_0)\right)=R\left(\frac{m_{q}^2(\mu^2)}{s},\alpha_s(\mu^2)\right)\;;\;\;\mu=\sqrt{s}, 
\eea
where $\sqrt{s_0}$ is a reference energy.
Mass effects are important once one approaches a threshold from the
perturbatively save region sufficiently far above the thresholds. They have been calculated up to
three loops by Chetyrkin, K\"uhn and collaborators~\cite{ChK95} and
have been implemented in the FORTRAN routine {\tt RHAD} by Harlander
and Steinhauser~\cite{HS02}.\\

Where can we trust the perturbative result?  In the complex
$s$--plane, perturbative QCD is supposed to work best in the deep
Euclidean region away from the physical region characterized by a cut
along the positive real axis for $s>s_0=4m^2$ where $m$ is the mass of
the lightest particles which can be pair--produced.  Fortunately, the
physical region to a large extent is accessible to pQCD as well
provided the energy scale is sufficiently large and one looks for the
appropriate observable.

The imaginary part corresponds to the jump of
the vacuum polarization function $\Pi'(q^2)$ across the cut. On the cut
we have the thresholds of the physical states, with lowest lying channels:
$\ppm$, $\pi^0 \ppm$, $\cdots$ and resonances $\rho$, $\omega$,
$\phi$, $J/\psi \cdots$, $\Upsilon \cdots$, $\cdots$. QCD is confining
the quarks inside
hadrons. In any case the quarks {\em hadronize}, a non--perturbative phenomenon which
is poorly understood in detail.
Neither the physical {\em thresholds} nor the {\em resonances} are
obtained with perturbation theory!  In
particular, the perturbative
quark--pair thresholds in Eq.~(\ref{Rpert}) do not nearly approximate the
physical thresholds for the low energy region below about 2 GeV. 
At higher energies pQCD works sufficiently far away from
thresholds and resonances, i.e. in regions where $R(s)$ is a slowly
varying function.  Fig.~\ref{fig:rhadron} shows
the $\epm$--data together with the perturbative QCD
prediction. Less problematic is the space--like (Euclidean) region
$-q^2\: \to \: \infty$, since it is away from thresholds and
resonances. 

The time--like quantity $R(s)$ intrinsically is non-perturbative and 
exhibits bound states, resonances, instanton effects ($\eta'$) and in particular
the hadronization of the quarks. In applying pQCD to describe real
physical cross--sections of hadro--production one needs a ``rule''
which bridges the asymptotic freedom regime with the confinement
regime, since the hadronization of the colored partons produced in the
hard kicks into color singlet hadrons eludes a quantitative
understanding.  The rule is referred to as \textit{quark hadron
duality}\footnote{Quark--hadron duality was first observed
phenomenologically for the structure function in deep inelastic
electron--proton scattering~\cite{BG70}.}~\cite{PQW76,Shifman00},
which states that for large $s$ the average non--perturbative hadron
cross--section equals the perturbative quark cross--section:
\be
\overline{\sigma(e^+e^-\rightarrow\mbox{hadrons})}(s) \simeq
\sum\nolimits_{q}\sigma(e^+e^-\rightarrow q\bar{q},q\bar{q}g,\cdots)(s)\cs
\label{qh-duality}
\ee
where the averaging extends from the hadron production threshold up to
$s$--values which must lie sufficiently far above the quark--pair
production threshold (global duality). Qualitatively, such a behavior
is visible in the data Fig.~\ref{fig:rhadron} above about 2 GeV
between the different flavor thresholds sufficiently above the lower
threshold.  A glance at the region from 4 to 5 GeV gives a good flavor
of duality at work. Note however that for precise reliable predictions it has
not yet been possible to quantify the accuracy of the duality
conjecture. A quantitative check would require much more precise
cross--section measurements than the ones available today. Ideally,
one should attempt to reach the accuracy of pQCD predictions. In
addition, in dispersion integrals the cross--sections are weighted by
different $s$--dependent kernels, while the duality statement is
claimed to hold for weight unity. One procedure definitely is
contradicting duality reasonings: to ``take pQCD plus resonances'' or
to ``take pQCD where $R(s)$ is smooth and data in the complementary
ranges''. Also adjusting the normalization of experimental data to
conform with pQCD within energy intervals (assuming local duality) has
no solid foundation. Nevertheless, the application of pQCD in the
regions advocated in~\cite{HS02} seems to be on fairly solid ground on
a phenomenological level. A more conservative use of pQCD is possible
by going to the Euclidean region and applying the Adler
function~\cite{Adlerfun} method as proposed in Refs.~\cite{EJKV98,FJ98,FJ03}.
As mentioned earlier, the low energy structure of QCD  also exhibits non--perturbative 
quark condensates. The latter also yield contributions to $R(s)$,
which for large energies are calculable by the operator product expansion
of the current correlator Eq.~(\ref{jemspectral})~\cite{SVZ}. The corresponding
$\braket{m_q \bar{q}q}/s^2$ power corrections in fact are small at energies where pQCD
applies~\cite{EJKV98,FJ86} and hence not a problem in our context.

  \subsection{Higher Order Hadronic Vacuum Polarization Corrections}
\label{ssec:hohad}
At order $O(\alpha^3)$ there are several classes of hadronic
VP contributions with typical diagrams shown in
Fig.~\ref{fig:ammhohad}. They have been estimated first
in~\cite{CNPdeR76}. Classes ($a$) to ($c$) involve leading hadronic VP
insertions and may be treated using DRs together with experimental
$\epm$--annihilation data. Class ($d$) involves leading QED
corrections of the charged hadrons and correspond to the inclusion of
hadronic final state radiation (FSR). 

The $O(\alpha^3)$ hadronic contributions
from classes ($a$), ($b$) and ($c$) may be evaluated without
particular problems as described in the following.

\begin{figure}[t]
\centering
\includegraphics[]{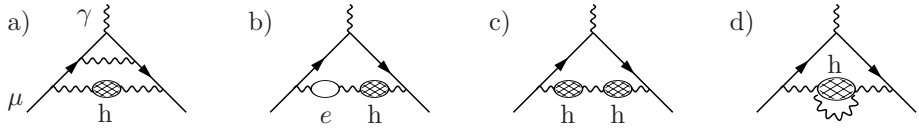}
\caption{Hadronic higher order VP contributions: a)-c) involving LO vacuum
polarization, d) involving HO vacuum polarization (FSR of hadrons).}
\label{fig:ammhohad}
\end{figure}

At the 3--loop level all diagrams of Fig.~\ref{fig:threeloopdia1}
which involve closed muon--loops are contributing to the hadronic
corrections when at least one muon--loop is replaced by a quark--loop
dressed by strong interactions mediated by virtual gluons.

\noi
\underline{Class ($a$)} consists of a subset of 12 diagrams of Fig.~\ref{fig:threeloopdia1}: diagrams 7)
to 18) plus 2 diagrams obtained from diagram 22) by replacing one
muon--loop by a hadronic ``bubble'', and yields a contribution of the
type
\be
a_\mu^{(6)[(a)]}=\left(\frac{\alpha}{\pi}\right)^3\frac{2}{3}
\int\limits_{4m_\pi^2}^{\infty}\frac{\D s}{s}\:R(s)\:K^{[(a)]}\left(s/m^2_\mu
\right), 
\label{hohada}
\ee
where $K^{[(a)]}(s/m^2_\mu)$ is a QED function which was obtained
analytically by Barbieri and Remiddi~\cite{BR74}. The kernel function is the
contribution to $a_\mu$ of the 14 two--loop diagrams obtained from
diagrams 1) to 7) of Fig.~\ref{fig:twoloopdia} by replacing one of the
two photons by a ``heavy photon'' of mass $\sqrt{s}$. The convolution
Eq.~(\ref{hohada}) then provides the insertion of a photon self--energy part
into the photon line represented by the ``heavy photon'' according to
the method outlined after Eq.~(\ref{ppDR}). While the exact expressions
are given in~\cite{BR74} some sufficiently precise handy approximations
have been given by Krause~\cite{Krause96} in form of an expansion up to fourth order
in $m^2/s$ which reads
\ba
K^{[(a)]}(s/m^2)&=& {m^2\over s}\left\{
 \left[{223\over54}-2\zeta(2)-{23\over36}\ln
{s\over m^2}\right]\right. \label{fig1c}\\
&&+ {m^2\over s}\left[{8785\over 1152}-{37\over8}\zeta(2)
-{367\over216}\ln{s\over m^2}
+{19\over144}\ln^2{s\over m^2}\right] \nonumber \\
&&+ {m^4\over s^2}\left[{13072841\over 432000}-{883\over40}\zeta(2)
-{10079\over3600}\ln{s\over m^2}
+{141\over80}\ln^2{s\over m^2}\right] \nonumber \\
&&+ \left.{m^6\over s^3}\left[{2034703\over 16000}-{3903\over40}\zeta(2)
-{6517\over1800}\ln{s\over m^2}
+{961\over80}\ln^2{s\over m^2}\right]\right\} \; .\nonumber
\ea
Here $m$ is the mass of the external lepton, $m=m_\mu$ in our case.
The expanded approximation is more practical for the evaluation of the
dispersion integral, because it is numerically more stable in general.

\noi
\underline{Class ($b$)} consists of 2 diagrams only, obtained from
diagram 22) of Fig.~\ref{fig:threeloopdia1}, and one may write this
contribution in the form
\ba
a_\mu^{(6)[(b)]}&=&\left(\frac{\alpha}{\pi}\right)^3
\frac{2}{3}\int\limits_{4m_\pi^2}^{\infty}\frac{\D
s}{s}R(s)\:K^{[(b)]}(s/m^2_\mu), 
\ea
with
\ba
K^{[(b)]}(s/m^2_\mu)&=&\int\limits_{0}^{1}\D x\:\frac{x^2\:(1-x)}{x^2+(1-x)\:s/m_\mu^2}
\left[-\hat{\Pi}^{'\:e}_\gamma
\left(-\frac{x^2}{1-x}\frac{m_\mu^2}{m_e^2}\right)\right], 
\label{Kbkernel}
\ea
where we have set $\Pi^{'}=\frac{\alpha}{\pi}\hat{\Pi}^{'}$.
Using Eq.~(\ref{Pigammaana}) with
$z=-\frac{x^2}{1-x}\frac{m_\mu^2}{m_e^2}$, we have
\bea
\hat{\Pi}^{'\:e}_\gamma (z)= {8\over 9}
-{\beta^2\over 3}+\left({1\over
2}-{\beta^2\over 6}\right)\beta\ln \frac{\beta-1}{\beta+1}\;\;
\mathrm{ \ with \ }\;\; \beta = \sqrt{1+4{1-x\over x^2}{m_e^2\over m^2_\mu}}\,.
\eea 
Here the kernel function is the
contribution to $a_\mu$ of the 2 two--loop diagrams obtained from
diagram 8) of Fig.~\ref{fig:twoloopdia} by replacing one of the
two photons by a ``heavy photon'' of mass $\sqrt{s}$.

In diagram b) $m_f^2/m^2=(m_e/m_\mu)^2$ is very small and one
may expand $\beta$ in terms of this small parameter. 
The expansion of Eq.~(\ref{Kbkernel}) to fifth order in $m^2/s$ and to first order in
$m_f^2/m^2$ is given by
\ba
K^{[(b)]}(s) &=& {m^2\over s}\left\{   \left(-{1\over 18}
                  + {1\over 9}\ln{m^2\over m_f^2}\right)\right.\nonumber\\
    &&
	     + {m^2\over s}\left(- {55\over 48} +{\pi^2\over 18}
             + {5\over 9}\ln{s\over m_f^2} + {5\over 36}\ln{m^2\over m_f^2}
             - {1\over 6}\ln^2{s\over m_f^2}
             + {1\over 6}\ln^2{m^2\over m_f^2}\right)\nonumber\\
    &&
	     + {m^4\over s^2}\left(- {11299\over 1800} +{\pi^2\over 3}
             + {10\over 3}\ln{s\over m_f^2} -{1\over 10}\ln{m^2\over m_f^2}
             - \ln^2{s\over m_f^2}
             + \ln^2{m^2\over m_f^2}\right)\nonumber\\
    &&
	     - {m^6\over s^3 } \left(
              {6419\over 225}
             - {14\over 9}\pi^2
             + {76\over 45}\ln{m^2\over m_f^2}
             - {14\over 3}\ln^2{m^2\over m_f^2}
             - {140\over 9}\ln{s\over m_f^2}
             + {14\over 3}\ln^2{s\over m_f^2}\right)\nonumber\\
    &&
	     \left. - {m^{8}\over s^4 } \left(
              {53350\over 441}
             - {20\over 3}\pi^2
             + {592\over 63}\ln{m^2\over m_f^2}
             - 20\ln^2{m^2\over m_f^2}
             - {200\over 3}\ln{s\over m_f^2}
             + 20\ln^2{s\over m_f^2}\right)\right\}\nonumber\\
    &&
	     +{m_f^2\over m^2}\left[
               {m^2\over s } - {2\over 3}{m^4\over s^2}
              - {m^6\over s^3 } \left(
              - 2\ln{s\over m^2}
              + {25\over 6}\right)
              - {m^8\over s^4 } \left(
              - 12\ln{s\over m^2}
              + {97\over 5}\right)\right.\nonumber\\
    &&
	      \left.\hspace{1cm} - {m^{10}\over s^5 } \left(
              - 56\ln{s\over m^2}
              + {416\over 5}\right)
               \right]\; .
\label{fig1b}
\ea

If we neglect terms ${\cal O}({m_f^2\over m^2})$ the $x$--integration
in Eq.~(\ref{Kbkernel}) may be performed analytically with the result~\cite{Krause96}
\ba
K^{[(b)]}(s)&=&
-\left({5\over 9}+{1\over3}\ln{m_f^2\over m^2}\right)
\times \biggl\{{1\over2}-(x_1+x_2) \nn \\
&&
+{1\over x_1-x_2}\left[
x_1^2(x_1-1)\ln\left({-x_1\over 1-x_1}\right)
-x_2^2(x_2-1)\ln\left({-x_2\over 1-x_2}\right)\right]\biggr\}-{5\over12}
\nonumber\\
&&
+{1\over3}(x_1+x_2)+{1\over 3(x_1-x_2)}
\left\{
x_1^2(1-x_1)\left[{\rm Li}_2\left({1\over x_1}\right)
-{1\over2}\ln^2\left({-x_1\over 1-x_1}\right)\right]
\right.\nonumber\\
&&
\left.-x_2^2(1-x_2)\left[{\rm Li}_2\left({1\over x_2}\right)
-{1\over2}\ln^2\left({-x_2\over 1-x_2}\right)\right]\right\}\:,
\label{Pipexpand}
\ea
%
with $x_{1,2}={1\over2}(b\pm\sqrt{b^2-4b})$ and $b=s/ m^2$.

\noi
\underline{Class ($c$)} includes the double hadronic VP insertion,
which is given by
\ba
a_\mu^{(6)[(c)]}&=&\left(\frac{\alpha}{\pi}\right)^3
\frac{1}{9}
\int\limits_{4m_\pi^2}^{\infty}\frac{\D s}{s}\frac{\D s'}{s'}R(s)\:
R(s')\:K^{[(c)]}(s,s'), 
\ea
where
\ba
K^{[(c)]}(s,s')=\int\limits_{0}^{1}\D
x\:\frac{x^4\:(1-x)}{[x^2+(1-x)\:s/m_\mu^2][x^2+(1-x)\:s'/m_\mu^2]}\epo
\nn
\ea
This integral may be performed analytically.
Setting $b=s/m^2$ and
$c=s'/m^2$ one obtains for $b\neq c$
\ba
K^{[(c)]}(s,s')&=&
{1\over 2} - b - c - {{\left( 2 - b \right) \,{b^2}\,\ln (b)}\over
    {2\,\left( b - c \right) }} -
  {{{b^2}\,\left( 2 - 4\,b + {b^2} \right) \,
      \ln ({{b + {\sqrt{- \left( 4 - b \right) \,b  }}}\over
         {b - {\sqrt{- \left( 4 - b \right) \,b  }}}})}\over
    {2\,\left( b - c \right)
     {\sqrt{- \left( 4 - b \right) \,b  }}\,
       }} \nonumber \\
&&\hspace*{-4mm} -
  {{\left( -2 + c \right) \,{c^2}\,\ln (c)}\over
    {2\,\left( b - c \right) }} +
  {{{c^2}\,\left( 2 - 4\,c + {c^2} \right) \,
      \ln ({{c + {\sqrt{- \left( 4 - c \right) \,c }}}\over
         {c - {\sqrt{- \left( 4 - c \right) \,c }}}})}\over
    {2\,\left( b - c \right) \,{\sqrt{- \left( 4 - c \right)  \,c
          }}}} \; ,
\label{form2c1}
\ea
and for $b=c$
\ba
K^{[(c)]}(s,s')&=&
{1\over 2} - 2\,c
+{c\over2}\left( -2 + c - 4\,\ln (c) + 3\,c\,\ln (c) \right)
+ {{c\,\left( -2 + 4\,c - {c^2} \right) }
\over {2(-4 + c)}} \nn \\
&&  +    {{c\,\left( 12 - 42\,c + 22\,{c^2} - 3\,{c^3} \right) \,
          \ln ({{c + {\sqrt{\left( -4 + c \right) \,c}}}\over
             {c - {\sqrt{\left( -4 + c \right) \,c}}}})}\over
        {2\left( -4 + c \right) \,{\sqrt{\left( -4 + c \right)
\,c}}}} \epo
\label{form2c2}
\ea

Results obtained by different groups, for so far unaccounted higher
order vacuum polarization effects, are collected in
Table~\ref{tab:amuHOdet}.
\begin{table}[t]
\centering
\caption{Higher order contributions from diagrams a) - c) (in units
$10^{-11}$). Note that errors between contributions a) and b) are
100\% anticorrelated, and the contribution c) is suppressed. The error of
$\amuhho$ is also close to 100\% anticorrelated to the one of the leading term
$\amuhlo$.} 
\label{tab:amuHOdet}
\begin{tabular}{ccccc}
&&&&\\[-3mm]
\hline\noalign{\smallskip}
 $a_\mu^{(6)[(a)]}$ &$a_\mu^{(6)[(b)]}$ &$a_\mu^{(6)[(c)]}$ &$\amuhho$ & Ref.  \\
\noalign{\smallskip}\hline\noalign{\smallskip}
 -199(4)  & 107(3) & 2.3(0.6) & -90(5) &\cite{KNO84}\\
 -211(5)  & 107(2) &2.7(0.1) & -101(6) &\cite{Krause96} \\
 -209(4)  & 106(2)&  2.7(1.0)& -100(5) & \cite{ADH98}\\
 -207.3(1.9)  &106.0(0.9) &3.4(0.1) &-98(1) & \cite{HMNT04,HMNT06} \\
 -207.5(2.0) & 104.2(0.9)& 3.0(0.1) & -100.3 (1.1) &\cite{FJ06}\\
\noalign{\smallskip}\hline
\end{tabular}
\end{table}
We will adopt the estimate\footnote{Our evaluation of the contribution to $a_e$ 
is $a_e^{(6)}(\mathrm{vap,\,had})=(-0.223\pm 0.002) \times
10^{-12}\,.$ The result is dominated by the diagram
Fig.~\ref{fig:ammhohad}a) which now includes the electron loop, while
diagram b) includes the muon loop and is suppressed  by a factor $(m_e/m_\mu)^2$.
A similar suppression factor applies for the other diagrams
(see also~\cite{Krause96}).}
\begin{equation}
\amuhho=(-100.3\pm 1.1)\:\times 10^{-11}\;
\label{amuhadHO}
\end{equation}
obtained with the compilation~\cite{FJ06}.

\noi \label{page:FSRcorr}
\underline{Class ($d$)} exhibits 3 diagrams  (diagrams 19) to 21) of
Fig.~\ref{fig:threeloopdia1} and corresponds to the leading hadronic
contribution with $R(s)$ corrected for final state radiation.
We thus may write this correction by replacing
\be
R(s)\to R(s)\:\eta(s)\:\frac{\alpha}{\pi}
\label{FSRetacorr}
\ee
in the basic integral Eq.~(\ref{AM}). This correction is particularly
important for the dominating two pion channel for which
$\eta(s)$ may be calculated in scalar QED. The result reads~\cite{Sch89,Melnikov01}
\ba
\eta(s)&=&\frac{1+\bpi^2}{\bpi} \Biggl\{
4 {\rm Li}_2 \left(\frac{1-\bpi}{1+\bpi} \right)+
2 {\rm Li}_2 \left(-\frac{1-\bpi}{1+\bpi} \right) \nn \\ &&
-3 \log \left(\frac{2}{1+\bpi} \right) \:
\log \left(\frac{1+\bpi}{1-\bpi} \right) -
2 \log ( \bpi ) \: \log \left(\frac{1+\bpi}{1-\bpi} \right)
\Biggr\} \nn \\ &&
-3 \log \left(\frac{4}{1-\bpi^2} \right) -
4 \log ( \bpi )  \nn \\ &&
+ \frac{1}{\bpi^3} \left[ \frac{5}{4}(1+\bpi^2)^2-2 \right]\:
\log \left(\frac{1+\bpi}{1-\bpi} \right)+
\frac{3}{2} \frac{1+\bpi^2}{\bpi^2}\,,
\label{etaFSR}
\ea
and provides a good measure for the dependence of
the FSR on the pion mass. Neglecting the pion mass is
obviously equivalent to taking the high energy limit $\eta(s\to\infty) = 3\;.$
As sQED treats the pions as point--like particles the hard part of the
spectrum, where photons couple to quarks rather than to the hadron,
is certainly not taken into account properly. Since we are not able to
unambiguously calculate radiation from strongly bound systems one
should focus much more on direct measurements of the 
spectrum~\cite{GHJJ03}.
In Fig.~\ref{fig:qcal} the sQED correction $\eta(s)$ is plotted
as a function of the center of mass energy. We observe that
for energies below 1 GeV the pion mass leads to a considerable
enhancement of the FSR corrections.  Regarding the desired precision,
ignoring the pion mass would therefore lead to wrong results.
As usual, close to the threshold of charged particle production 
the Coulomb force between the two final state particles leads to
substantial corrections.  In this limit ($s \simeq 4m_{\pi}^2$) the factor 
$\eta(s)$ becomes sin\-gular
[$\eta(s)\to\pi^2/2\beta_{\pi}$] which means that the $O(\alpha)$
result for the FSR correction cannot be trusted anymore.  Fortunately,
the singular terms are known to all orders of perturbation theory and can
be resummed. In fact the leading terms
exponentiate and one obtains~\cite{Sch89}:
\ba
R^{(\gamma)}(s) &=& R(s) \;\left(1+\eta(s)\frac{\alpha}{\pi}-
\frac{\pi\alpha}{2\beta_{\pi}}
\right)\;\frac{\pi\alpha}{\beta_{\pi}}
\times \left[1-\exp\left(-\frac{\pi\alpha}{\beta_\pi}\right)\right]^{-1}\eqp
\ea
While only the exponentiated correction
yields the correct answer close to the threshold, the deviation from the
non--exponentiated one is below $1\,\%$ above  $\sqrt{s}=0.3\;\gv$.
\begin{figure}[t]
\centering
\IfFarbe{%
\includegraphics[height=4.5cm]{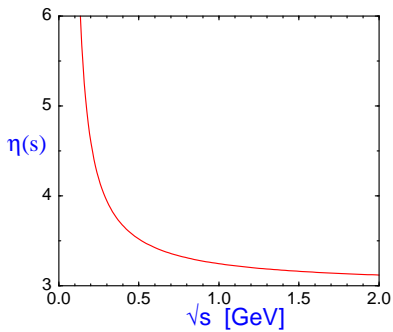}}{%
\includegraphics[height=4.5cm]{FSReta.eps}}
\caption{The FSR correction factor $\eta(s)$ as a function of the
c.m. energy $\sqrt{s}$.}
\label{fig:qcal}
\end{figure}
The $O(\alpha)$ $\pi^+\pi^-\gamma$ correction calculated in sQED yields
\be
\delta^{\gamma} \amuhlo=a_\mu^{(6)[(d)]}=(38.6 \pm 1.0) \times 10^{-11}\cs
\ee
as a contribution to $\amu$. Here, we added a guesstimated error which of course is not the true
model error, the latter remaining unknown\footnote{One could expect
that due to $\gamma -\rhon$ mixing (VMD type models~\cite{VMDmodel},
see below) the sQED
contribution gets substantially reduced. However, due to the low
scales $\sim m_\mu,m_\pi$ involved here, in relation to $M_\rho$, the
photons essentially behave classically in this case. Also, the bulk of
the VP contribution at these low scales comes from the neutral
$\rhon$--exchange, while the FSR is due to the dissociated charged $\pi^+\pi^-$ intermediate state
as assumed in sQED. Fig.~\ref{fig:qcal} shows that the main contribution comes from very low
energies.}. In the inclusive region above
typically 2 GeV, the FRS corrections are well represented by the inclusive
photon emission from quarks. However, since in inclusive measurements
experiments commonly do not subtract FSR, the latter is included
already in the data and no additional contribution has to be taken
into account.  In more recent analyses this contribution is usually
included as the $\pi^+\pi^-\gamma$ channel in the leading hadronic 
VP contribution, in particular in the value given in Eq.~(\ref{amuhadLO}) .

\section{Hadronic Light-by-Light Scattering Contribution}
\label{sec:lbl}

The most problematic set of hadronic corrections are those related
to hadronic light--by--light scattering, which for the first
time show up at order $O(\alpha^3)$ via the diagrams of
Fig.~\ref{fig:hadronicLbL}.  
\begin{figure}[h]
\centering
\includegraphics{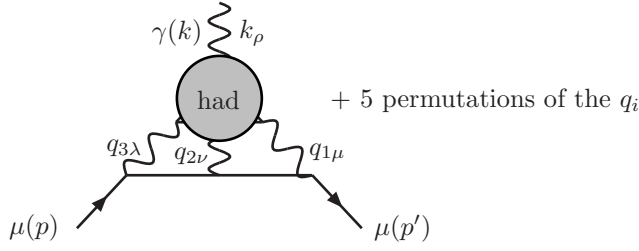}

\caption{Assignment of momenta for the calculation of the hadronic contribution
of the light--by--light scattering to the muon electromagnetic vertex.}
\label{fig:hadronicLbL}
\end{figure} 
We already know from the leptonic
counterpart Fig.~\ref{fig:LbLinsertions} that such contributions can be
dramatically  
enhanced and thus represent an important contribution which has to be
evaluated carefully. The problem is that even for
real--photon light--by--light scattering, perturbation theory is far
from being able to describe reality, as the reader may convince
himself by a glance at Fig.~\ref{fig:LBLfacts}, showing sharp spikes of
$\pi^0$, $\eta$ and $\eta'$ production, 
\begin{figure}[t]
\centering
\includegraphics[height=6cm]{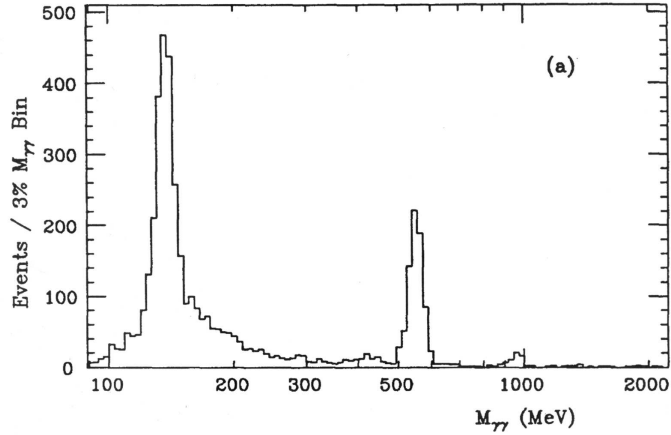}
\caption{The invariant $\gamma\gamma$ mass spectrum obtained with
the Crystal Ball detector~\cite{LBLfacts}. The three spikes seen
represent the $\gamma\gamma \to$ pseudoscalar (PS)
$\to \gamma\gamma$ excitations: PS=$\pi^0,\eta,\eta'$.}
\label{fig:LBLfacts}
\end{figure}
while pQCD predicts a smooth continuum (see Fig.~\ref{fig:LbLviahadrons}).  
\begin{figure}[h]
\centering
\includegraphics{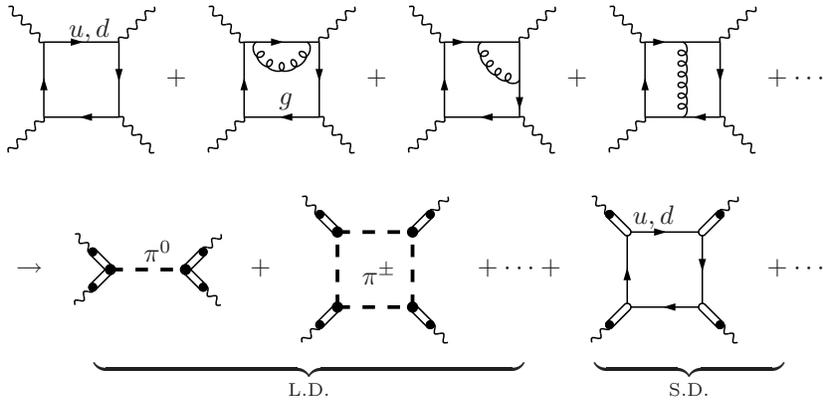}

\caption{Hadronic light--by--light scattering is dominated by $\pi^0$--exchange
in the odd parity channel, pion loops etc. at long distances
(L.D.) and quark loops including hard gluonic corrections at short
distances (S.D.). The photons in the effective theory couple to
hadrons via $\gamma-\rho^0$ mixing.}
\label{fig:LbLviahadrons}
\end{figure}

As a contribution to the \amm three of the four photons in
Fig.~\ref{fig:hadronicLbL} are virtual and to be integrated over all
four--momentum space, such that a direct experimental input for the
non--perturbative dressed four--photon correlator is not available. In this
case one has to resort to the low energy effective descriptions of QCD like
\textit{chiral perturbation theory} (CHPT) extended to include vector--mesons.
Note that early evaluations assumed that the main contribution to hadronic
light-by-light scattering comes from momentum regions around the muon mass. It
was later observed in Refs.~\cite{HKS95,BijnensLBL} that the higher momentum
region, around $500 - 1000~\mbox{MeV}$, also gives important
contributions. Therefore, hadronic resonances beyond the Goldstone bosons of
CHPT need to be considered as well. The Resonance Lagrangian Approach (RLA) is
realizing vector--meson dominance model (VMD) ideas in accord with the low
energy structure of QCD~\cite{EckerCPT}.  Other effective theories are the
extended Nambu-Jona-Lasinio (ENJL) model~\cite{BijnensLBL} (see
also~\cite{deRafaelENJL94}) or the very similar hidden local symmetry (HLS)
model~\cite{HKS95,HK98}; approaches more or less accepted as a framework for
the evaluation of the hadronic LbL effects. The amazing fact is that the
interactions involved in the hadronic LbL scattering process are the parity
conserving QED and QCD interactions while the process is dominated by the
parity odd pseudoscalar meson--exchanges. This means that the effective
$\pi^0\gamma\gamma$ interaction vertex exhibits the parity violating $\gafi$
coupling, which of course in $\gamma\gamma \to \pi^0 \to \gamma\gamma$ must
appear twice (an even number of times).  The process indeed is induced by the
parity odd $O(p^4)$ Wess-Zumino-Witten (WZW) effective Lagrangian
term~\cite{WessZumino71,Witten83} 
\be
\cL^{(4)}_\mathrm{WZW}=\frac{\alpha}{\pi} \frac{N_c}{12
F_\pi}\:\left(\pi^0+\frac{1}{\sqrt{3}}\:\eta_8+2
\sqrt{\frac{2}{3}}\:\eta_0  \right)\:\tilde{F}_{\mu \nu} F^{\mu \nu}\epo
\label{WZW}
\ee
The latter reproduces the ABJ anomaly~\cite{ABJanomaly} on the level of the
hadrons.  $\pi^0$ is the neutral pion field, $F_\pi$ the pion decay constant
($F_\pi = 92.4\,\mbox{MeV}$). The pseudoscalars $\eta_8,\eta_0$ are mixing
into the physical states $\eta,\eta'$.  However, the constant WZW form factor
yields a divergent result, applying a cut--off $\Lambda$ one obtains the
leading term 
\begin{figure}[h]
\centering
\includegraphics{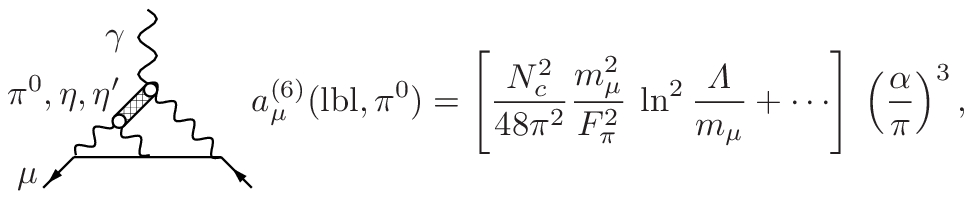}
\end{figure}

\noi with an universal coefficient ${\cal C}=N_c^2m_\mu^2/(48 \pi^2
F_\pi^2)$~\cite{KnechtNyffeler01,KNPdeR01}; in the VMD dressed cases $M_V$
represents the cut--off $\Lambda \to M_V$ if $M_V \to \infty$\footnote{Since
the leading term is divergent and requires UV subtraction, we expect this term
to drop from the physical result, unless a physical cut--off tames the
integral, like the physical $\rho$ in effective theories which implement the
VMD mechanism.}. For the case of $\pi^0$-exchange, a two-dimensional integral
representation for $a_\mu^{\mathrm{LbL};\pi^0}$ has been derived in
Ref.~\cite{KnechtNyffeler01} (in terms of the moduli of the Euclidean loop
momenta $|Q_1|$ and $|Q_2|$) for a certain class of form factors including the
VMD dressed case. The universal weight functions multiplying the
model-dependent form factors clearly show the relevance of momenta of order
$500 - 1000~\mbox{MeV}$.

A new quality of the problem encountered here is the fact that the integrand
depends on 3 invariants $q_1^2$, $q_2^2$, $q_3^2$, where $q_3=-(q_1+q_2)$.  In
contrast, the hadronic VP correlator, or the VVA triangle with an external zero
momentum vertex (which enters the electroweak contribution, see
Sect.~\ref{sec:weak}), only depends on a single invariant $q^2$. In the
latter case, the invariant amplitudes (form factors) may be separated into a
low energy part $q^2\leq \Lambda^2$ (soft) where the low energy effective
description applies and a high energy part $q^2 > \Lambda^2$ (hard) where pQCD
works. In multi--scale problems, however, there are mixed soft--hard regions,
where no answer is available in general, unless we have data to constrain the
amplitudes in such regions. In our case, only the soft region
$q_1^2,q_2^2,q_3^2 \leq \Lambda^2$ and the hard region $q_1^2,q_2^2,q_3^2 >
\Lambda^2$ are under control of either the low energy effective field theory
(EFT) and of pQCD, respectively. In the other domains operator product
expansions and/or soft versus hard factorization ``theorems'' \`a la
Brodsky-Farrar~\cite{BrodskyFarrar73} may be applied.

Another problem of the RLA is that the low energy effective
theory is non--renormalizable and thus has unphysical UV behavior,
while QCD is renormalizable and has the correct UV behavior (but
unphysical IR behavior). As a consequence of the mismatch of the
functional dependence on the cut--off, one cannot match the two
pieces in a satisfactory manner and one obtains a cut--off dependent
prediction.  Unfortunately, the cut--off dependence of the sum is not
small even if one varies the cut--off only within ``reasonable''
boundaries around about 1 or 2 GeV, say. Of course the resulting
uncertainty just reflects the model dependence and so to say
parametrizes our ignorance. An estimate of the real model dependence
is difficult as long as we are not knowing the true solution of the
problem. In CHPT and its extensions, the low energy constants
parametrizing the effective Lagrangian are accounting for the
appropriate S.D. behavior, usually. Some groups however prefer an
alternative approach based on the fact that the weakly coupled
large--$N_c$ QCD, i.e., $SU(N_c)$ for $N_c \to \infty$ under the
constraint $\alpha_s N_c$=constant, is theo\-reti\-cally better known than
true QCD with $N_c=3$. It is thus tempting to approximate QCD as an
expansion in $1/N_c$~\cite{tHooft74,Manohar01,LargeNcCHPT}.

Of course, also applying a large--$N_c$ expansion one has to respect
the low energy properties of QCD as encoded by CHPT. In CHPT the
effective Lagrangian has an overall factor $N_c$, while the $U$
matrix, exhibiting the pseudoscalar fields, is $N_c$ independent. Each
additional meson field has a $1/F_\pi \propto 1/\sqrt{N_c}$. In the
context of CHPT the $1/N_c$ expansion thus is equivalent to a
semiclassical expansion. The chiral Lagrangian can be
used at tree level, and loop effects are suppressed by powers of
$1/N_c$. Note, however, that for instance the low-energy constants $L_i$ which
appear at order $p^4$ in the chiral Lagrangian have different weights in the
$N_c$ counting.  

The various hadronic LbL contributions in the effective theory are shown in
Fig.~\ref{fig:lbldiagabc} and the corresponding $1/N_c$ and chiral $O(p)$
counting is given in Table~\ref{tab:LEeffcounting}~\cite{deRafaelENJL94}. Note
that the chiral counting refers to the contribution to the 4-point function
$\langle VVVV\rangle$ and not to $a_\mu$ itself. Based on this classification
it was argued in Ref.~\cite{deRafaelENJL94} that the (constituent) quark-loop
represents the irreducible part of the 4-point function and should be included
as a separate contribution (although maybe with dressed couplings of the
constituent quarks to the photons, which arises naturally in the ENJL model
employed in Ref.~\cite{deRafaelENJL94}), in addition to the exchanges or loops
of resonances. Within the CHPT approach, this irreducible part can be viewed
as a local counterterm contribution $\bar \psi \sigma^{\mu\nu} \psi
F_{\mu\nu}$ to $a_\mu$.  In particular, it was argued in
Ref.~\cite{deRafaelENJL94} that the (constituent) quark-loop should not be
used as a substitute for the hadronic contributions (exchanges and loops with
resonances), as was done in earlier evaluations of the hadronic light-by-light
scattering contribution to $g-2$ in Refs.~\cite{CNPdeR76,KNO84}.

\begin{figure}
\centering
\includegraphics[height=2.75cm]{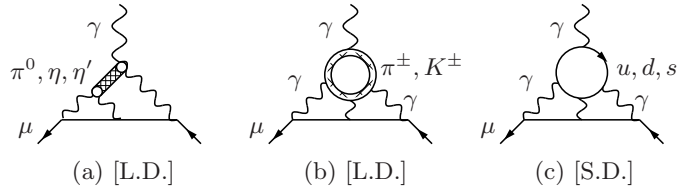}
\caption{Hadronic light--by--light scattering diagrams in a low energy
effective model description. Diagrams (a) and (b) represent the long
distance [L.D.] contributions at momenta $p\leq \Lambda$, diagram (c)
involving a quark loop which yields the leading short distance [S.D.] part at
momenta $p \geq \Lambda$ with $\Lambda\sim 1~\mathrm{to}~2~\gv $ an UV
cut--off. Internal photon lines are dressed by $\rho-\gamma$ mixing.}
\label{fig:lbldiagabc}
\end{figure}

\begin{table}[t]
\centering
\caption{Orders with respect to $1/N_c$  and chiral expansion of typical
  leading contributions shown in Fig.~\ref{fig:lbldiagabc}.}  
\label{tab:LEeffcounting}
\begin{tabular}{lccl}
&&&\\[-3mm]
\hline\noalign{\smallskip}
 Diagram &  $1/N_c$ expansion & $p$ expansion & type\\
\noalign{\smallskip}\hline\noalign{\smallskip}
 Fig.~\ref{fig:lbldiagabc}(a) & $N_c$ & $p^6$ & $\pi^0,\eta,\eta'$
exchange \\
 Fig.~\ref{fig:lbldiagabc}(a) & $N_c$ & $p^8$ & $a_1,\rho,\omega$ exchange \\
 Fig.~\ref{fig:lbldiagabc}(b) & $1$ & $p^4$ & meson loops ($\pi^\pm,~K^\pm$) \\
 Fig.~\ref{fig:lbldiagabc}(c) & $N_c$ & $p^8$ & quark loops  \\
\noalign{\smallskip}\hline
\end{tabular}
\end{table}

Based on refined effective field theory models, two major efforts in
evaluating the full $a_\mu^{\mathrm{LbL}}$ contribution were made by Hayakawa,
Kinoshita and Sanda (HKS 1995)~\cite{HKS95}, Bijnens, Pallante and Prades (BPP
1995)~\cite{BijnensLBL} and Hayakawa and Kinoshita (HK 1998)~\cite{HK98} (see
also Kinoshita, Nizic and Okamoto (KNO 1985)~\cite{KNO84}). Although the
details of the calculations are quite different, which results in a different
splitting of various contributions, the results are in good agreement and
essentially given by the $\pi^0$-pole contribution, which was taken with the
wrong sign, however. In order to eliminate the cut--off dependence in
separating L.D. and S.D. physics, more recently it became favorable to use
quark--hadron duality, as it holds in the large $N_c$ limit of
QCD~\cite{tHooft74,Manohar01}, for modeling of the hadronic
amplitudes~\cite{deRafaelENJL94}. The infinite series of narrow vector states
known to show up in the large $N_c$ limit is then approximated by a suitable
lowest meson dominance (LMD) ansatz~\cite{LMD98}, assumed to be saturated by
known low lying physical states of appropriate quantum numbers. This approach
was adopted in a reanalysis by Knecht and Nyffeler (KN
2001)~\cite{KnechtNyffeler01,KNPdeR01}, in which they discovered a sign
mistake in the dominant $\pi^0,\eta,\eta'$ exchange contribution (see
also~\cite{BCM02,RMW02}), which changed the central value by $+167 \times
10^{-11}$, a 2.8 $\sigma$ shift, and which reduced a larger discrepancy
between theory and experiment.  More recently Melnikov and Vainshtein (MV
2004)~\cite{MV03} found additional problems in previous calculations, this
time in the short distance constraints (QCD/OPE) used in matching the high
energy behavior of the effective models used for the $\pi^0,\eta,\eta'$
exchange contribution.  Most evaluations have adopted the pion-pole
approximation which, however, violates four-momentum conservation at the
external $\pi^0 \gamma^* \gamma$ vertex, if used too naively, as pointed out
in Refs.~\cite{MV03,FJ07,Jegerlehner:2008zz}.  In the following we will
attempt an evaluation which avoids such manifest inconsistencies.  Maybe some
of the confusion in the recent literature was caused by the fact that the
distinction between off-shell and on-shell (pion-pole) form factors was not
made properly.

Let us start now with a setup of what one has to calculate
actually. We will closely follow Ref.~\cite{KnechtNyffeler01} in the
following. The hadronic light--by--light scattering contribution to
the electromagnetic vertex is represented by the diagram
Fig.~\ref{fig:hadronicLbL}.
According to the diagram, a complete discussion of the hadronic
light--by--light contributions involves the full rank--four hadronic
vacuum polarization tensor
\be
\Pi_{\mu\nu\lambda\rho}(q_1,q_2,q_3) =
\int \D^4x_1\: \D^4x_2\: \D^4x_3
\,\E^{\I\,(q_1x_1 + q_2x_2 + q_3 x_3)}\,
\langle\,0\,\vert\,T
\{j_{\mu}(x_1)j_{\nu}(x_2)j_{\lambda}(x_3)j_{\rho}(0)\}
\,\vert\,0\,\rangle \epo
\label{lblcorrelator}
\ee
The external photon momentum $k$ is incoming, the $q_i$'s of the virtual
photons are outgoing from the hadronic ``blob''.  Here $j_{\mu}(x) \equiv
(\bar{\psi}\:\hat{Q} \gamma_\mu \psi)(x)$ ($\bar{\psi} =
(\bar{u},\bar{d},\bar{s})$, $\hat{Q}=\mathrm{diag}(2,-1,-1)/3$ the charge
matrix) denotes the light quark part of the electromagnetic current.  Since
$j_{\mu}(x)$ is conserved, the tensor $\Pi_{\mu\nu\lambda\rho}(q_1,q_2,q_3)$
satisfies the Ward-Takahashi identities
$\{q_1^{\mu};q_2^{\nu};q_3^{\lambda};k^{\rho}\}
\Pi_{\mu\nu\lambda\rho}(q_1,q_2,q_3)\,=\,0 \,,$ with $k=(q_1+q_2+q_3)$ which
implies
\be
\Pi_{\mu\nu\lambda\rho}(q_1,q_2,k-q_1-q_2) = -
k^\sigma (\partial / \partial k^\rho)\:
\Pi_{\mu\nu\lambda\sigma}(q_1,q_2,k-q_1-q_2)\cs
\label{WTrelation}
\ee
and thus tells us that the object of interest is linear in $k$ when we
go to the static limit $k^\mu \to 0$ in which the \amm is defined.  As
a consequence the electromagnetic vertex amplitude takes the form ${
\Pi}_{\rho}(p\,',p) = k^{\sigma}{ \Pi}_{\rho\sigma}(p\,',p)$ and the
hadronic light--by--light contribution to the muon anomalous magnetic
moment is given by (see also~\cite{BarbieriRemiddi75})
\be
F_{\rm M}(0)\,=\,
\frac{1}{48m_\mu}\,
\Tr \left\{(\not\! p + m_\mu)[\gamma^{\rho},\gamma^{\sigma}](\not\! p + m_\mu)
\Pi_{\rho\sigma}(p,p)\right\}
\,.
\label{F2trace}
\ee
The required vertex tensor amplitude is determined by
\ba
{\Pi}_{\rho\sigma}(p\,',p)
&=& -\I e^6 
\,\int\frac{\D^4q_1}{(2\pi)^4}\frac{\D^4q_2}{(2\pi)^4}\,
\frac{1}{q_1^2\,q_2^2\,(q_1+q_2-k)^2}\,
\frac{1}{(p\,'-q_1)^2-m_\mu^2}\,
\frac{1}{(p-q_1-q_2)^2-m_\mu^2}
\nonumber\\
&& 
\qquad \quad \times ~~
\gamma^{\mu}\:(\not\! p\,'- \not\!q_1+m_\mu)\:
\gamma^{\nu}\:(\not\! p\, - \not\! q_1- \not\! q_2+m_\mu)\:
\gamma^{\lambda}
\nonumber\\
&& 
\qquad \quad \times ~~
\frac{\partial}{\partial k^{\rho}}\,
\Pi_{\mu\nu\lambda\sigma}(q_1,q_2,k-q_1-q_2)\,,
\label{Gamma2}
\ea
where now $k=0$ such that $p'=p$ and $q_3=-(q_1+q_2)$.  After performing
the trace (see below) we have what we actually need to calculate. The
integral to be performed is 8 dimensional. Thereof 3 integrations can
be done analytically. In general, one has to deal with a 5 dimensional
non--trivial integration over 3 angles and 2 moduli.

The hadronic tensor $\Pi_{\mu\nu\lambda\sigma}(q_1,q_2,k-q_1-q_2)$ in
Eq.~(\ref{lblcorrelator}) or (\ref{Gamma2}) is a very complicated object,
because it has an unexpectedly complex structure as we will see, in no way
comparable with the leptonic counterpart. The general covariant decomposition
involves 138 Lorentz structures of which 32 can contribute to
$g-2$~\cite{BijnensLBL}. Fortunately, this tensor is dominated by the
pseudoscalar exchanges $\pi^0,\eta,\eta',...$ (see Fig.~\ref{fig:LBLfacts}),
described by the WZW effective Lagrangian (\ref{WZW}) at lowest order in the
chiral expansion. This fact rises hope that a half--way reliable estimate
should be possible. Generally, the perturbative QCD expansion only is useful
to evaluate the short distance tail, while the dominant long distance part
must be evaluated using some low energy effective model which 
includes the pseudoscalar Goldstone bosons as well as the vector mesons as
shown in Fig.~\ref{fig:LbLviahadrons}.

\subsection{Pseudoscalar--exchange Contribution}
\label{sec:pionpole}

Here we discuss the dominating hadronic contributions which are due to
the neutral pseudoscalar--exchange diagrams shown in
Fig.~\ref{fig:LbLpionpole}. 
\begin{figure}[h]
\centering
\includegraphics{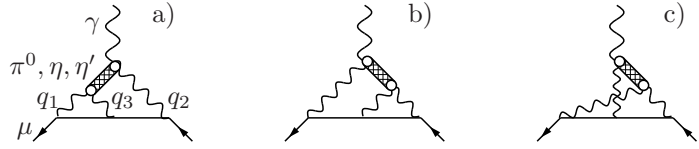}

\caption{Leading hadronic light--by--light scattering diagrams.
In accord with Eq.~(\ref{a_pion_2}), here all photon momenta are chosen
incoming to the pion transition form factors. Internal photon lines are
dressed by $\rho-\gamma$ mixing.}  
\label{fig:LbLpionpole}
\end{figure}

We first concentrate on the exchange of the neutral pion.  The key object
which enters the Feynman diagrams is the {\bf off-shell} $\pi^0 \gamma \gamma$
form factor $\FF((q_1+q_2)^2,q_1^2, q_2^2)$ which is defined, up to small
mixing effects with the states $\eta$ and $\eta^\prime$, via the Green's
function $\VVP$ in QCD
\ba
\lefteqn{ \int \D^4 x\, \D^4 y \, \E^{\I (q_1 \cdot x + q_2 \cdot y)} \, 
\langle\,0 | T \{ j_\mu(x) j_\nu(y) P^3(0) \} | 0 \rangle \, } \crn 
&=&  \varepsilon_{\mu\nu\alpha\beta} \, q_1^\alpha q_2^\beta \, 
 {\I \langle{\overline\psi}\psi\rangle \over F_\pi} \, {\I \over (q_1 +
   q_2)^2 - m_\pi^2} \, \FF((q_1 + q_2)^2,q_1^2,q_2^2) \cs
\label{FFoffshellpi}
\ea
where $P^3 = \bar\psi i \gamma_5 {\lambda^3 \over 2} \psi = \left( \bar u
i\gamma_5 u - \bar d i \gamma_5 d \right) / 2$. Note that we denote by
$\langle{\overline\psi}\psi\rangle$ the {\bf single flavor} bilinear quark
condensate. The form factor is of course Bose symmetric
$\FF((q_1+q_2)^2,q_1^2,q_2^2) = \FF((q_1+q_2)^2,q_2^2,q_1^2)$, as the two 
photons are indistinguishable. We will later also use the following notation
\ba
\cA_{\mu\nu}(\pi^0 \to \gamma^* \gamma^*) &=&\I \int \D^4 x\, \E^{\I q \cdot x}
\langle\,0 | T \{ j_\mu(x) j_\nu(0) 
\} | \pi^0(p) \rangle \, \crn &=&  \varepsilon_{\mu\nu\alpha\beta} \, q^\alpha
p^\beta \, \FFa(m_\pi^2,q^2,(p-q)^2) \cs   
\label{FFonshellpi}
\ea
where now the pion is on-shell, but the photons are in general off-shell. 

We would like to stress that the identification of the pion-exchange
contribution in the full hadronic light-by-light scattering amplitude in $g-2$
according to Fig.~\ref{fig:LbLpionpole} only makes sense, if the pion is
on-shell (or nearly on-shell). If one is (far) off the mass shell of the
exchanged particle (here the pion), it is not possible to separate different
contributions to the $g-2$, unless one uses some particular model where for
instance elementary pions can propagate. In this sense, only the result for
the pion-pole contribution to $g-2$ with on-shell form factors is
model-independent. On the other hand, the pion-pole contribution is only a
part of the full result, since in general the form factors will enter the
calculation with off-shell momenta. Although the contribution in a particular
channel will then be model-dependent, the sum of all off-shell contributions
in all channels will again lead, at least in principle, to a model-independent
result.

Apart from the $\pi^0 \gamma \gamma$ form factor everything is known and may
be worked out (see e.g.~\cite{KnechtNyffeler01}) with the result
\ba
a_{\mu}^{\mathrm{LbL};\pi^0}& = & - e^6
\int {d^4 q_1 \over (2\pi)^4} {d^4 q_2 \over (2\pi)^4}
\,\frac{1}{q_1^2 q_2^2 (q_1 + q_2)^2[(p+ q_1)^2 - m_\mu^2][(p - q_2)^2 -
    m_\mu^2]} 
\nonumber \\
&& \hspace*{-5mm} \quad \quad \times \left[
{\FF(q_2^2, q_1^2, q_3^2) \ \FFc(q_2^2, q_2^2, 0) \over q_2^2 -
m_{\pi}^2} \ T_1(q_1,q_2;p) \nonumber \right. \\
&& \hspace*{-5mm} \quad \quad \quad + \left. {\FF(q_3^2, q_1^2,  q_2^2) \
  \FFc(q_3^2, q_3^2, 
0) \over q_3^2 - m_{\pi}^2} \ T_2(q_1,q_2;p) \right] ,
\label{a_pion_2}
\ea
with
\ba
T_1(q_1,q_2;p) & = & {16 \over 3}\, (p \cdot q_1) \, (p \cdot q_2) \,
(q_1 \cdot q_2)
\,-\, {16 \over 3}\, (p \cdot q_2)^2 \, q_1^2 \nonumber \\
&& \!\!\!\!\!
-\, {8 \over 3}\, (p \cdot q_1) \, (q_1 \cdot q_2) \, q_2^2
\,+\, 8 (p \cdot q_2) \, q_1^2 \, q_2^2
\,-\,{16 \over 3} (p \cdot q_2) \, (q_1 \cdot q_2)^2 \nonumber
\\
&&\!\!\!\!\!
+\, {16 \over 3}\, m_\mu^2 \, q_1^2 \, q_2^2
\,-\, {16 \over 3}\, m_\mu^2 \, (q_1 \cdot q_2)^2 \, , \nonumber \\
T_2(q_1,q_2;p) & = & {16 \over 3}\, (p \cdot q_1) \, (p \cdot q_2) \,
(q_1 \cdot q_2) \,-\,{16 \over 3}\, (p \cdot q_1)^2 \, q_2^2
\nonumber \\
&&\!\!\!\!\!
 +\, {8 \over 3}\, (p \cdot q_1) \, (q_1 \cdot q_2) \, q_2^2
\,+\, {8 \over 3}\, (p \cdot q_1) \, q_1^2 \, q_2^2
\,\nonumber \\
&&\!\!\!\!\!
 +\, {8 \over 3}\, m_\mu^2 \, q_1^2 \, q_2^2
\,-\, {8 \over 3}\, m_\mu^2 \, (q_1 \cdot q_2)^2 \, .
\ea
The first and the second graphs in Fig.~\ref{fig:LbLpionpole} give rise to
identical contributions, leading to the term with $T_1$, whereas the third
graph gives the contribution involving $T_2$.  The factor $T_2$ has been
symmetrized with respect to the exchange $q_1\leftrightarrow -q_2$.  Note that
now the external photon has zero four-momentum ($k^\mu=0$) such that
$q_3=-(q_1+q_2)$.

The result in Eq.~(\ref{a_pion_2}) does not depend on the direction of the
muon momentum vector $p$ such that we may average in Euclidean space over the
directions $\hat{P}$:
\begin{equation}
\braket{\cdots}=\frac{1}{2\pi^2}\int \D \Omega(\hat{P})\, \cdots
\end{equation}
using the technique of Gegenbauer polynomials (hyperspherical approach), see
Ref.~\cite{hyperspherical_approach}. Since all $p$ dependent terms are
independent of the pseudoscalar form factors one may perform the integrations
in general. After reducing numerators of the amplitudes $T_i$ against the
denominators of the propagators one is left with the following integrals
($(4)\equiv(P+Q_1)^2+m_\mu^2$ and $(5)\equiv (P-Q_2)^2+m_\mu^2$ with
$P^2=-m_\mu^2$)
\ba
\braket{\frac{1}{(4)}\frac{1}{(5)}}&=&\frac{1}{m_\mu^2 R_{12}}\arctan
\left(\frac{zx}{1-zt}\right), \crn 
\braket{(P\cdot
Q_1)\,\frac{1}{(5)}}&=&-\,(Q_1\cdot Q_2)\,
\frac{\left(1-R_{m2}\right)^2}{8m_\mu^2}, \crn 
\braket{(P\cdot
Q_2)\frac{1}{(4)}}&=&~\,(Q_1\cdot Q_2)\,
\frac{\left(1-R_{m1}\right)^2}{8m_\mu^2}, \crn  
\braket{\frac{1}{(4)}}&=&-\,\frac{1-R_{m1}}{2m_\mu^2}, \crn 
\braket{\frac{1}{(5)}}&=&-\,\frac{1-R_{m2}}{2m_\mu^2} , 
\ea 
where $R_{mi}=\sqrt{1+4m_\mu^2/Q_i^2}$ and $(Q_1\cdot Q_2)=Q_1\, Q_2\, t$ with
$t=\cos \theta$, $\theta$ the angle between the two Euclidean four--vectors
$Q_1$ and $Q_2$.  Denoting $x=\sqrt{1-t^2}$, we have $R_{12}=Q_1\, Q_2\, x$
and
\bea z=\frac{Q_1 Q_2}{4
m_\mu^2}\left(1-R_{m1}\right)\left(1-R_{m2}\right)\epo 
\eea 
We have thus eliminated all momentum dependences up to the three which also
show up in the hadronic form factors $Q_1^2$, $Q_2^2$, and $Q_3^2$ or
equivalently on $(Q_1\cdot Q_2)=Q_1 Q_2 \cos \theta$ and end up with a
3--dimensional integral over $Q_1=|Q_1|$, $Q_2=|Q_2|$ and $t= \cos \theta$:
\be
a_{\mu}^{\mathrm{LbL};\pi^0} = -\frac{2\alpha^3}{3\pi^2}
\int_0^\infty\,\D Q_1 \D Q_2 \int_{-1}^{+1} \D t\,\sqrt{1-t^2}\, Q_1^3\,
Q_2^3 \ 
\left[ F_1\,P_6\,I_1(Q_1,Q_2,t)+F_2\,P_7\,I_2(Q_1,Q_2,t)\right] , 
\label{pp3drep}
\ee
where $P_6=1/(Q_2^2+m_\pi^2),$ and $P_7=1/(Q_3^2+m_\pi^2)$ denote the Euclidean
single particle exchange propagators. The integration kernels $I_1$ and $I_2$,
which factorize from the dependence on the hadronic form-factors in $F_1$ and
$F_2$, are given by 
\ba
I_1(Q_1,Q_2,t) & = & X(Q_1,Q_2,t)\,\biggl(
      8\,P_1\,P_2 \, (Q_1\cdot Q_2) \crn
      &&-2\, P_1\,P_3 \, ( Q_2^4/m_\mu^2 - 2\,Q_2^2 )
       -2\, P_1 \, ( 2 - Q_2^2/m_\mu^2 + 2\,(Q_1\cdot Q_2)\,/m_\mu^2 )\crn
      &&+4\, P_2\,P_3 \, Q_1^2  - 4\, P_2 
       -2\, P_3 \, (   4 + Q_1^2/m_\mu^2 - 2\,Q_2^2/m_\mu^2 ) + 2/m_\mu^2
       \biggr)\crn 
      &&-2\, P_1\,P_2 \, (  1 + (1-R_{m1})\,(Q_1\cdot Q_2)\,/m_\mu^2 )\crn
      &&+ P_1\,P_3 \, ( 2 - (1-R_{m1})\,Q_2^2/m_\mu^2 )
       + P_1 \, (1-R_{m1})/m_\mu^2\crn
      &&+ P_2\,P_3 \, ( 2 + (1-R_{m1})^2\,(Q_1\cdot Q_2)\,/m_\mu^2 )
       +3\, P_3 \, (1-R_{m1})/m_\mu^2 \nonumber , \crn  
I_2(Q_1,Q_2,t) & = & X(Q_1,Q_2,t)\,\biggl(
       4\,P_1\,P_2 \, (Q_1\cdot Q_2)\crn
      &&+2\, P_1\,P_3 \, Q_2^2 - 2\, P_1
       +2\, P_2\,P_3\, Q_1^2
       -2\, P_2 - 4 P_3 - 4/m_\mu^2 \biggr) \crn
      && -2\, P_1\,P_2 
      -3\, P_1 \, (1-R_{m2})/(2 m_\mu^2)
      -3\, P_2 \, (1-R_{m1})/(2 m_\mu^2)\crn
      &&+ P_1\,P_3 \, ( 2 + 3\,(1-R_{m2})\,Q_2^2/(2 m_\mu^2)
           + (1-R_{m2})^2\,(Q_1\cdot Q_2)\,/(2 m_\mu^2))\crn
      &&+ P_2\,P_3 \, ( 2 + 3\,(1-R_{m1})\,Q_1^2/(2 m_\mu^2)
	   + (1-R_{m1})^2\,(Q_1\cdot Q_2)\,/(2 m_\mu^2))\crn
      &&- P_3 \, (2 - R_{m1} - R_{m2})/(2 m_\mu^2), \label{I2} 
\ea
where we used the notation $P_1=1/Q_1^2,\, P_2=1/Q_2^2,$ and
$P_3=1/Q_3^2$ for the Euclidean propagators and introduced the auxiliary
function 
\begin{equation}
X(Q_1,Q_2,t)=\frac{1}{Q_1 Q_2\,x} \arctan \left(\frac{zx}{1-zt}\right)\;,
\end{equation}
which has the following asymptotic expansion for small $x$, near the forward
and backward points: 
\bea
X(Q_1,Q_2,t)=\frac{1}{Q_1 Q_2}\,\left\{ \begin{tabular}{ccc} 
$\frac{z}{1-z}\left(1 + \frac16 \frac{z\,(z-3)}{(1-z)^2}\,x^2\right)+O \left(
  {x}^{3} \right)  
$&$\mathrm{ \ for \ } $&$ t > 0$\\
$\frac{z}{1+z}\left(1 + \frac16 \frac{z\,(z+3)}{(1+z)^2} x^2\right)+O \left(
  {x}^{3} \right)  
$&$\mathrm{ \ for \ } $&$ t < 0$\\
\end{tabular} \right. \epo
\eea

Equation~(\ref{pp3drep}) provides the general set up for studying any type of
single particle exchange contribution as a 3--dimensional integral
representation. The non-per\-tur\-bative factors according to
Eq.~(\ref{a_pion_2}) are given by
\ba
F_1&=&\FF(-Q_2^2, -Q_1^2, -Q_3^2) \ \FFc(-Q_2^2, -Q_2^2, 0)\:,\; \nonumber \\ 
F_2&=&\FF(-Q_3^2, -Q_1^2, -Q_2^2) \ \FFc(-Q_3^2, -Q_3^2,0), 
\ea
and will be considered next. Note that $F_2$ is symmetric under the exchange
$Q_1 \leftrightarrow Q_2$. We used this property to write $I_2(Q_1, Q_2, t)$
in Eq.~(\ref{I2}) in a symmetric way.

\subsubsection{The $\pi^0\gamma\gamma$ Transition Form Factor: Experimental
  and Theoretical Constraints}
\label{sec:formfactors}

Above we have formally reduced the problem of calculating the
$\pi^0$--exchange contribution diagrams in Fig.~\ref{fig:LbLpionpole} to the
problem of calculating the integral Eq.~(\ref{pp3drep}). The non--perturbative
aspect is now confined in the form--factor function $\FF(q_3^2,q_1^2,q_2^2)$
defined in Eq.~(\ref{FFoffshellpi}), which is largely unknown.  For the time
being we have to use one of the hadronic models mentioned above together with
pQCD as a constraint on the high energy asymptotic behavior. Fortunately some
experimental data are also available. The constant~$\FFabc(m_\pi^2,0,0)$ is
well determined by the $\pi^0 \to \gamma \gamma$ decay rate.  The on--shell
transition amplitude in the chiral limit follows from the WZW--Lagrangian
Eq.~(\ref{WZW}), and is given by
\be
M_{\pi^0\gamma\gamma}=
-e^2\:\FFabc(0,0,0)=\frac{e^2 N_c}{12 \pi^2 F_\pi}
=\frac{\alpha}{\pi F_\pi} \approx 0.025~\gv^{-1}\cs
\ee
and with $F_\pi \sim 92.4~\mv$ and quark color number $N_c=3$, rather
accurately predicts the experimental result
\be
|M^\mathrm{exp}_{\pi^0\gamma\gamma}|=\sqrt{64 \pi
  \Gamma_{\pi^0\gamma\gamma}/m_\pi^3} 
=0.025\pm0.001~\gv^{-1}\epo
\ee
Note that the amplitude $M_{\pi^0\gamma\gamma}$, defined to be finite in
the chiral limit, in terms of the conventional amplitude
$\cM_{\pi^0\gamma\gamma}=-e^2\,
\cA_{\mu\nu}(\pi^0 \to \gamma
\gamma)\,\veps^{*\mu}({q_1,\lambda_1})\,\veps^{*\nu}({q_2,\lambda_2})$ 
follows (up to a phase) via
$\sum_{\lambda_1,\lambda_2}\left|\cM_{\pi^0\gamma\gamma}\right|^2=
\frac{m_\pi^4}{2}|M_{\pi^0\gamma\gamma}|^2\,.$ 

Additional experimental information is available for $\FFac(m_\pi^2,-Q^2,0)$
coming from experiments $\epm \to \epm \pi^0$.  Note that the production of an
on--shell pion at large $-q_1^2=Q^2$ is only possible if the real photon is
highly energetic, i.e., $q_2^0=|\vec{q}_2|$ large. This is different from the
$g-2$ kinematical situation at the external photon vertex, where the external
photon has zero four--momentum. By four--momentum conservation thus only
$\FFc(-Q^2,-Q^2,0)$ and \textbf{not} $\FFc(m_\pi^2,-Q^2,0)$ can enter at the
\textbf{external} vertex.

For the \textbf{internal} vertex both photons are virtual, and luckily,
experimental data on $\FFac(m_\pi^2,-Q^2,0)$ is available from
CELLO~\cite{CELLO90} and CLEO~\cite{CLEO98}, which provides a crucial
constraint on this form factor.  Fortunately, this constrains the border
domain of \textbf{one} of the problematic mixed soft--hard regions at the
\textbf{internal} vertex. Experiments fairly well confirm the
Brodsky-Lepage~\cite{LepageBrodsky80} evaluation of the large $Q^2$ behavior
\be \label{Brodsky_Lepage} 
\lim\limits_{Q^2 \to \infty} \: \FFac(m_\pi^2,-Q^2,0) \sim - \frac{2
F_\pi}{Q^2}.
\ee
In this approach the transition form factor is represented as a
convolution of a hard scattering amplitude (HSA) and the soft
non--perturbative meson wave function and the asymptotic behavior follows from
a pQCD calculation of the HSA.  Together with the constraint from $\pi^0$
decay, $\lim\limits_{Q^2 \to 0} \: \FFac(m_\pi^2,-Q^2,0)=\frac{-1}{4\pi^2
F_\pi}$, an interpolating formula
\ba
\FFac(m_\pi^2,-Q^2,0) \simeq \frac{-1}{4\pi^2 F_\pi}\frac{1}{1+(Q^2/8\pi^2
  F_\pi^2)} 
\label{FFacfit}
\ea
was proposed, which in fact gives an acceptable fit to the data.  Refinements
of form factor calculations/models were discussed and compared with the data
in~\cite{CLEO98} (see also~\cite{EfrRad80,PIGGother,PIGGpQCD,Khod99}).

Apart from these experimental constraints, any satisfactory model for the
off-shell form factor $\FF((q_1 + q_2)^2, q_1^2, q_2^2)$ should match at large
momentum with short-distance constraints from QCD that can be calculated using
the OPE. In Ref.~\cite{KN_EPJC_01} the short-distance properties for the
three-point function $\VVP$ in Eq.~(\ref{FFoffshellpi}) in the chiral limit
and assuming octet symmetry have been worked out in detail (see also
Ref.~\cite{VVP_earlier} for earlier partial results). At least for the pion
the chiral limit should be a not too bad approximation\footnote{As pointed out
in Ref.~\cite{Nyffeler_RADCOR_2002}, the integrals in Eq.~(\ref{a_pion_2}) are
infrared safe for $m_\pi \to 0$. This can also be seen within the EFT approach
to light-by-light scattering proposed in Refs.~\cite{KNPdeR01,RMW02} to be
discussed later in Sect.~\ref{sec:LbLEFT}.}, however, for the $\eta$ and, in
particular, for the non-Goldstone boson $\eta^\prime$ further analysis will be
necessary.

It is important to notice that the Green's function $\VVP$ is an order
parameter of chiral symmetry. Therefore, it vanishes to all orders in
perturbative QCD in the chiral limit, so that the behavior at short distances
is smoother than expected from naive power counting arguments.  Two limits are
of interest. In the first case, the two momenta become simultaneously large,
which in position space describes the situation where the space-time arguments
of all the three operators tend towards the same point at the same rate. To
leading order and up to corrections of order $\order{\alpha_s}$ one obtains
the following behavior for the form factor\footnote{In the chiral limit, the
relation between the off-shell form factor and the single invariant function
${\cal H}_V$ which appears in $\VVP$ is given by $\FF((q_1 + q_2)^2, q_1^2,
q_2^2) = - (2/3) (F_0 / \langle{\overline\psi}\psi\rangle_0) (q_1 + q_2)^2
{\cal H}_V(q_1^2, q_2^2, (q_1 + q_2)^2)$, see Ref.~\cite{KN_EPJC_01} for
details.} 
\be
\lim_{\lambda \to \infty} \FF((\lambda q_1 + \lambda q_2)^2, (\lambda
  q_1)^2, (\lambda q_2)^2) =  {F_0 \over 3} \, {1 \over \lambda^2}
{q_1^2 + q_2^2 + (q_1+q_2)^2 \over q_1^2 q_2^2}  \nonumber + \order{{1\over
    \lambda^4}} \, . \label{FF_OPE_1} 
\ee

The second situation of interest corresponds to the case where the relative
distance between only two of the three operators in $\VVP$ becomes small.  It
so happens that the corresponding behaviors in momentum space involve, apart
from the correlator $\langle A P\rangle$ which, in the chiral limit, is
saturated by the single-pion intermediate state,
\be
\int d^4 x e^{ip \cdot x}
\langle 0 \vert T \{ A_\mu^a(x) P^b(0) \} \vert 0 \rangle =
\delta^{ab} \langle{\overline\psi}\psi\rangle_0 \, {p_\mu \over p^2} \, , 
\ee   
(we denote by $\langle{\overline\psi}\psi\rangle_0$ the {\bf single
  flavor} bilinear quark condensate in the chiral limit) the two-point
  function $\langle VT\rangle$ of the vector current and the antisymmetric
  tensor density, 
\be \label{Pi_VT}
\delta^{ab}(\Pi_{\rm VT})_{\mu\rho\sigma}(p)\,=\, 
\int d^4x e^{ip \cdot x}
\langle 0 \vert T \{ V_\mu^a(x) 
({\overline\psi}\,\sigma_{\rho\sigma}\frac{\lambda^b}{2}\,\psi)(0)\}\vert
0\rangle \, ,  
\ee
with $\sigma_{\rho\sigma}={i\over 2}[\gamma_{\rho},\gamma_{\sigma}]$
(the similar correlator between the axial current and the tensor
density vanishes as a consequence of invariance under charge conjugation). 
Conservation of the vector current and invariance under parity then give
\be 
(\Pi_{\rm VT})_{\mu\rho\sigma}(p)\,=
\,(p_{\rho}\eta_{\mu\sigma}-p_{\sigma}\eta_{\mu\rho})\,\Pi_{\rm VT}(p^2)
\,. 
\ee
The leading short-distance behavior of this two-point
function is given by (see also \cite{Craigie:1981jx})
\be \label{VT_OPE} 
\lim_{\lambda \to \infty}\Pi_{\rm VT}((\lambda p)^2)\,=\,
-\,\frac{1}{\lambda^2}
\,\frac{\langle{\overline\psi}\psi\rangle_0}{p^2}
\,+\,{\cal O}\left(\frac{1}{\lambda^4}\right)\,. 
\ee
The short-distance behavior of the form factor then reads 
\be
\lim_{\lambda \to \infty} \FF(q_2^2, (\lambda q_1)^2, (q_2-\lambda q_1)^2) 
= {2 F_0 \over 3} {1 \over \lambda^2} {1 \over q_1^2} 
+ \order{{1\over \lambda^3}} \, , \label{FF_OPE_2}
\ee
when the space-time arguments of the two vector currents in $\VVP$ approach
each other and 
\be
\lim_{\lambda \to \infty} \FF((\lambda q_1 + q_2)^2, (\lambda q_1)^2,q_2^2)
= - {2 \over 3} {F_0 \over \langle{\overline\psi}\psi\rangle_0}
\Pi_{\rm VT}(q_2^2) 
+ \order{{1\over \lambda}} \, , \label{FF_OPE_3} 
\ee
when the space-time argument of one of the vector currents approaches the one
of the pseudoscalar density.  

In particular, at the external vertex in light-by-light scattering in
Eq.~(\ref{a_pion_2}), the following limit is relevant~\cite{Nyffeler:2009tw}  
\be
\lim_{\lambda \to \infty} \FFc((\lambda q_1)^2, (\lambda q_1)^2,0)
= - {2 \over 3} {F_0 \over \langle{\overline\psi}\psi\rangle_0}
\Pi_{\rm VT}(0) 
+ \order{{1\over \lambda}} \, .
\label{FF_OPE_3_zeromomentum}  
\ee
Note that there is no fall-off in this limit, unless $\Pi_{\rm VT}(0)$
vanishes. As pointed out in Ref.~\cite{Belyaev_Kogan}, the value of $\Pi_{\rm
  VT}(p^2)$ at zero momentum is related to the quark condensate magnetic 
susceptibility $\chi$ of QCD in the presence of a constant external
electromagnetic field, introduced in Ref.~\cite{Ioffe_Smilga} 
\be
\langle 0 | \bar{q} \sigma_{\mu\nu} q | 0 \rangle_{F} = e \, e_q \, \chi
\, \langle{\overline\psi}\psi\rangle_0 \, F_{\mu\nu}, 
\ee
with $e_u = 2/3$ and $e_d = -1/3$. With our definition of
$\Pi_{\rm VT}$ in Eq.~(\ref{Pi_VT}) one then obtains the relation (see also
Ref.~\cite{Mateu_Portoles})  
\be \label{Pi_VT0_Chi}
\Pi_{\rm VT}(0) = - {\langle{\overline\psi}\psi\rangle_0 \over 2} \chi. 
\ee

Unfortunately there is no agreement in the literature what the actual value of
$\chi$ should be. In comparing different results one has to keep in mind that
$\chi$ actually depends on the renormalization scale $\mu$. In
Ref.~\cite{Ioffe_Smilga} the estimate $\chi(\mu = 0.5~\mbox{GeV}) = -
(8.16^{+2.95}_{-1.91})~\mbox{GeV}^{-2}$ was given in a QCD sum rule evaluation
of nucleon magnetic moments. This value was confirmed by the recent
reanalysis~\cite{Narison:2008jp} which yields $\chi = - (8.5 \pm
1.0)~\mbox{GeV}^{-2}$. A similar value $\chi = - N_c / (4 \pi^2 F_\pi^2) = -
8.9~\mbox{GeV}^{-2}$ was obtained by Vainshtein~\cite{Vainshtein03}. From the
explicit expression of $\chi$ it is not immediately clear what should be the
relevant scale $\mu$. Since pion dominance was used in the matching with the
OPE below some higher states, it was argued in Ref.~\cite{Vainshtein03} that
the normalization point is probably rather low, $\mu \sim
0.5~\mbox{GeV}$. Calculations within the instanton liquid model yield
$\chi^{\rm ILM}(\mu \sim 0.5-0.6~\mbox{GeV}) =
-4.32~\mbox{GeV}^{-2}$~\cite{Chi_ILM_1}, where the scale is set by the inverse
average instanton size $\rho^{-1}$. The value of $\chi
\langle{\overline\psi}\psi\rangle_0 = 42~\mbox{MeV}$ at the same scale
obtained in Ref.~\cite{Chi_ILM_1} agrees roughly with the result
$35-40~\mbox{MeV}$ from Ref.~\cite{Chi_ILM_2} derived in the same model. On
the other hand, assuming that $\Pi_{\rm VT}(q^2)$ is well described by the
multiplet of the lowest-lying vector mesons (LMD) and satisfies the OPE
constraint from Eq.~(\ref{VT_OPE}), leads to the ansatz~\cite{Balitsky_Yung,
Belyaev_Kogan, KN_EPJC_01}
\be \label{Pi_VT_LMD}
\Pi_{\rm VT}^{\rm LMD}(q^2) \,=\, -\,\langle{\overline\psi}\psi\rangle_0\,
\frac{1}{q^2-M_V^2} \, .  
\ee
Using Eq.~(\ref{Pi_VT0_Chi}) then leads to the estimate $\chi^{\rm LMD} = - 2
/ M_V^2 = -3.3~\mbox{GeV}^{-2}$~\cite{Balitsky_Yung}. Again, it is not obvious
at which scale this relation holds. In analogy to estimates of low-energy
constants in chiral Lagrangians~\cite{EckerCPT}, it might be at $\mu =
M_V$. This LMD estimate was soon afterwards improved by taking into account
higher resonance states ($\rho^\prime, \rho^{\prime\prime}$) in the framework
of QCD sum rules, with the results $\chi(0.5~\mbox{GeV}) = - (5.7 \pm
0.6)~\mbox{GeV}^{-2}$~\cite{Belyaev_Kogan} and $\chi(1~\mbox{GeV}) = - (4.4
\pm 0.4)~\mbox{GeV}^{-2}$~\cite{Balitsky_etal}. A more recent
analysis~\cite{Ball_etal} yields, however, a smaller absolute value
$\chi(1~\mbox{GeV}) = - (3.15 \pm 0.30)~\mbox{GeV}^{-2}$, close to the
original LMD estimate.   For a
quantitative comparison of all these estimates for $\chi$ we would have to
run them to a common scale, for instance 1 GeV, which can obviously not be
done within perturbation theory starting from such low scales as $\mu =
0.5~\mbox{GeV}$.\footnote{A further complication arises in comparisons with
papers from the early 1980's because not only $\mu = 0.5~\mbox{GeV}$ was
frequently used, but also 1-loop running with a low $\Lambda_{\rm QCD}^{n_f=3}
= 100-150~\mbox{MeV}$, whereas more recent estimates yield
$\Lambda_{\overline{\rm MS}}^{n_f=3} = 346~\mbox{MeV}$ (at
4-loop)~\cite{Bethke_2000}.} Finally, even if the RG running could be
performed non-perturbatively, it is not clear what would be the relevant scale
$\mu$ in the context of hadronic light-by-light scattering.

Further important information on the (on-shell) pion form factor in
Eq.~(\ref{FFonshellpi}) has been obtained in Ref.~\cite{ShuVai82} based on
higher-twist terms in the OPE and worked out in~\cite{NSVVZ84}. We consider
the $\pi^0 \to \gamma \gamma $ transition amplitude Eq.~(\ref{FFonshellpi})
with $j_\mu=\frac23 \bar{u}\gamma_\mu u- \frac13\bar{d}\gamma_\mu d$ the
relevant part of the electromagnetic current. In the chiral limit, the first
two terms of $\FFa(0,-Q^2,-Q^2)$ for large Euclidean momentum
$Q^2 \to \infty$ read~\cite{ShuVai82}
\ba 
\cA^{\mu\nu} (\pi^0 \to \gamma^* \gamma^*)&=&\I \frac{1}{3} \,
\varepsilon^{\mu\nu\alpha\beta} \langle\,0 |
\frac{{q}_\alpha}{{q}^2}\,j^{(3)}_{5\beta} -\frac{8}{9}
\frac{{q}_\alpha}{{q}^4}\,\tilde{j}^{(3)}_\beta | \pi^0(p) \rangle, 
\label{FFonshellpiope}
\ea
where $j^{(3)}_{5\mu}=\bar{\psi}\lambda^3 \gamma_\mu \gamma_5\psi$ and
$\tilde{j}^{(3)}_{\mu}= g_s\, \bar{\psi}\lambda^3 \gamma^\rho \, T^a
\tilde{G}^{a}_{\rho\mu}\psi$. The matrix elements are parametrized as follows:
$\langle\,0 |j^{(3)}_{5\mu}(0) | \pi^0(p) \rangle =2\I \,F_0 p_\mu $ and
$\langle\,0 |\tilde{j}^{(3)}_{\mu}(0) | \pi^0(p) \rangle =-2\I \,F_0 p_\mu \,
\delta^2$. For the pion form factor this implies
\ba
\frac{\FFa(0,-Q^2,-Q^2)}{\FFabc(0,0,0)}=
\frac{8}{3}\pi^2 F_0^2\left\{\frac{1}{Q^2}-\frac{8}{9}
\frac{\delta^2}{Q^4}+\cdots \right\}
\label{NSVVZ}
\ea
and the sum rule estimate performed in~\cite{NSVVZ84} yields $\delta^2 =(0.2
\pm 0.02)~\gv^2$.

\subsubsection{Hadronic Light-by-light Scattering and the Triangle Anomaly}

In this subsection we present the additional QCD short-distance constraints
which have been derived by Melnikov and Vainshtein in Ref.~\cite{MV03},
closely following their notations. The light-by-light scattering
amplitude is written as follows
\ba
{\cal M} & = & \alpha^2 N_c \mbox{Tr}[\hat Q^4] \, {\cal A} 
= \alpha^2 N_c \mbox{Tr}[\hat Q^4] \, {\cal A}_{\mu_1 \mu_2 \mu_3 \gamma 
  \delta} \, \epsilon_1^{\mu_1} \epsilon_2^{\mu_2} \epsilon_3^{\mu_3}
\, f^{\gamma\delta} \nonumber \\  
& = & - e^3 \int d^4x d^4y e^{-i q_1 x - i q_2 y}
\, \epsilon_1^{\mu_1} \epsilon_2^{\mu_2} \epsilon_3^{\mu_3} \, \langle 0 |
T\{ j_{\mu_1}(x) j_{\mu_2}(y) j_{\mu_3}(0) \} | \gamma \rangle , 
\label{VVVgamma}
\ea
with the photon momenta $q_i$ (incoming, $\sum q_i = 0$) and the photon
polarization vectors $\epsilon_i$. The first three photons are 
virtual, while the fourth one represents the external magnetic field and can
be regarded as a real photon with vanishingly small momentum $q_4$. 
The field strength tensor of the external soft photon is denoted by
$f^{\gamma\delta} = q_4^\gamma \epsilon_4^\delta - q_4^\delta
\epsilon_4^\gamma$. Since for $a_\mu$ only terms linear in $q_4$
are needed, see Eq.~(\ref{WTrelation}), one can set $q_4 = 0$ in the amplitude
${\cal A}_{\mu_1 \mu_2 \mu_3 \gamma \delta}$.  

The authors Ref.~\cite{MV03} then consider the momentum region $q_1^2 \approx
q_2^2 \gg q_3^2$, with the well-known OPE result (see also
Eq.~(\ref{FFonshellpiope}))  
\be
i \int d^4x d^4y e^{-i q_1 x - i q_2 y} T\{ j_{\mu_1}(x) j_{\mu_2}(y)\} = \int
d^4z e^{-i (q_1 + q_2) z} {2 i \over \hat q^2} \, \varepsilon_{\mu_1 \mu_2 \delta
\rho} \, \hat q^\delta j_5^\rho(z) + \ldots , 
\ee
where $j_5^\rho = {\overline \psi} \hat Q^2 \gamma^\rho \gamma_5 \psi$ is the
axial current and $\hat q = (q_1 - q_2)/2 \approx q_1 \approx - q_2$. Only the
leading term for large Euclidean $\hat q$ has been retained in the OPE. The
momentum $q_1 + q_2 = - q_3$ flowing through $j_5^\rho$ is assumed to much
smaller than $\hat q$. In this way the matrix element in Eq.~(\ref{VVVgamma})
can be related in the particular kinematical limit $q_1^2 \approx q_2^2 \gg
q_3^2$ to the amplitude  
\be
T_{\mu_3 \rho}^{(a)} = i \int d^4z e^{i q_3 z} \, \langle 0 | T \{
j_{5\rho}^{(a)}(z) j_{\mu_3}(0) \} | \gamma
\rangle ,  
\ee
if the current $j_5^\rho$ is expressed as a linear combination of the
isovector, $j_{5\rho}^{(3)} = {\overline \psi} \lambda_3 \gamma_\rho \gamma_5
\psi$, hypercharge $j_{5\rho}^{(8)} = {\overline \psi} \lambda_8 \gamma_\rho
\gamma_5 \psi$, and the $SU(3)$ singlet, $j_{5\rho}^{(0)} = {\overline \psi}
\gamma_\rho \gamma_5 \psi$, currents. 
%
%
The amplitude $T_{\mu_3 \rho}^{(a)}$ involves the axial current
$j_{5\rho}^{(a)}$ and two electromagnetic currents, one with momentum $q_3$
and the other one (the external magnetic field) with vanishing momentum. The
triangle amplitude for such kinematics was studied in
Ref.~\cite{Vainshtein03}, see also Refs.~\cite{CMV03,CMV03ext,KPPdR04}. It was
found that $T_{\mu_3 \rho}^{(a)}$ can be written in terms of two independent
functions $w_{L,T}^{(a)}(q_3^2)$
\be
T_{\mu_3 \rho}^{(a)} = - {i e N_c \mbox{Tr}[\lambda_a \hat Q^2] \over 4 \pi^2}
\left\{ w_L^{(a)}(q_3^2) q_{3\rho} q_3^\sigma \tilde f_{\sigma \mu_3} 
+ w_T^{(a)}(q_3^2) \left( - q_3^2 \tilde f_{\mu_3 \rho} + q_{3\mu_3}
q_3^\sigma \tilde f_{\sigma\rho} - q_{3\rho} q_3^\sigma \tilde f_{\sigma\mu_3}
\right)  \right\} , 
\ee
where $\tilde f_{\sigma\mu_3} = {1 \over 2} \epsilon_{\sigma \mu_3 \alpha
  \beta} f^{\alpha\beta}$. The first (second) amplitude is related to the
  longitudinal (transversal) part of the axial current, respectively. In terms
  of hadrons, the invariant function $w_{L(T)}$ describes the exchanges of the
  pseudoscalar (axial vector) mesons.

In perturbation theory these invariant functions are completely fixed by the
ABJ anomaly and one obtains for massless quarks from the triangle diagram 
\be \label{wL_wT}
w_L^{(a)}(q^2) = 2 w_T^{(a)}(q^2) = - {2 \over q^2} . 
\ee
In the chiral limit, the result for $w_L^{(3,8)}$ is exact to all orders in
perturbation theory~\cite{ABtheorem69} and there are also no nonperturbative
contributions~\cite{'tHooft79}. As shown in Ref.~\cite{Vainshtein03}, in the
chiral limit, the relation (\ref{wL_wT}) is true to all orders, however,
$w_T^{(3,8)}$ receives nonperturbative corrections. The poles in $w_L^{(3,8)}$
at $q^2 = 0$ are identified with the poles of the Goldstone bosons, $\pi^0$ in
$w_L^{(3)}$ and $\eta$ in $w_L^{(8)}$. 

For $q_1^2 \approx q_2^2 \gg q_3^2$, one can therefore write the hadronic
light-by-light scattering amplitude as follows 
\ba
{\cal A}_{\mu_1 \mu_2 \mu_3 \gamma \delta} f^{\gamma\delta} & = & {8 \over
  \hat q^2} \epsilon_{\mu_1 \mu_2 \delta \rho} \hat q^\delta \sum_{a=3,8,0}
W^{(a)} \left\{ w_L^{(a)}(q_3^2) q_3^\rho q_3^\sigma \tilde f_{\sigma \mu_3}
\right.  \nonumber \\
& & \qquad \qquad \qquad \qquad \qquad \left. 
+ w_T^{(a)}(q_3^2) \left( - q_3^2 \tilde f^\rho_{~\mu_3} + q_{3\mu_3}
q_3^\sigma \tilde f^\rho_{~\sigma} - q_3^\rho 
q_3^\sigma \tilde f_{\sigma\mu_3} \right)  \right\} + \ldots, 
\label{LbL_wL_wT}
\ea
where the weights $W^{(a)}$ are given by $ W^{(3)}=\frac{1}{4}$,
$W^{(8)}=\frac{1}{12}$ and $W^{(0)}=\frac{2}{3}$.   

The expression in Eq.~(\ref{LbL_wL_wT}) is then extrapolated to arbitrary
values of $q_1^2, q_2^2$ by writing ${\cal A} = {\cal A}_{\rm PS} + {\cal
A}_{AV} + \mbox{permutations}$, with the ansatz\footnote{From now on we only 
  consider the pseudoscalar exchanges and use Euclidean space notation
  as in Ref.~\cite{MV03}.} 
\be 
{\cal A}_{\rm PS} = \sum_{a=3,8,0} W^{(a)} \, \phi_L^{(a)}(q_1^2, q_2^2)
\, w_L^{(a)}(q_3^2) \, (f_2^{\mu\nu} \tilde f_1^{\nu\mu}) (\tilde
f^{\rho\sigma} f_3^{\sigma\rho}), 
\ee 
where $f_i^{\mu\nu} = q_i^\mu \epsilon_i^\nu - q_i^\nu
\epsilon_i^\mu$ denote the field strength tensors. The form
factors $\phi_L^{(a)}(q_1^2, q_2^2)$ account for the dependence of the
amplitude on $q_{1,2}^2$, i.e.\ the internal interaction vertex in $a_\mu$
with two virtual photons, whereas the meson propagator and the external
interaction vertex form the triangle amplitude described by the functions
$w_L^{(a)}(q_3^2)$. 

For the pion one obtains, outside the chiral limit, 
\be
w_L^{(3)}(q_3^2) = {2 \over q_3^2 + m_\pi^2}, 
\ee
whereas the ABJ anomaly fixes $\phi_L^{(3)}(0,0) = N_c / (4 \pi^2
F_\pi^2)$. Defining the $\pi^0\gamma^*\gamma^*$ form factor as follows  
$F_{\pi^0\gamma^*\gamma^*}(q_1^2, q_2^2) = \phi_L^{(3)}(q_1^2, q_2^2) /
\phi_L^{(3)}(0,0)$, one finally obtains the result 
\be \label{A_pi0_MV}
{\cal A}_{\pi^0} = - {N_c W^{(3)} \over 2 \pi^2 F_\pi^2} \, 
{F_{\pi^0\gamma^*\gamma^*}(q_1^2, q_2^2) \over q_3^2 + m_\pi^2} \,
(f_2^{\mu\nu} \tilde f_1^{\nu\mu}) (\tilde f^{\rho\sigma} f_3^{\sigma\rho})  
+ \mbox{permutations}. 
\ee

By relating the $\langle VVV | \gamma \rangle$ matrix element to the triangle
amplitude $\langle A V | \gamma\rangle$, in particular to the invariant
function $w_L^{(3)}(q_3^2)$, Melnikov and Vainshtein deduce that no form
factor $F_{\pi^0\gamma^*\gamma}(q_3^2, 0)$ should be used at the external
vertex, but only a constant factor, see Eq.~(\ref{A_pi0_MV}). They rightly
point out that such a form factor violates momentum conservation at the
external vertex and criticize the procedure adopted in earlier
works~\cite{BijnensLBL,HKS95,HK98,KnechtNyffeler01}. However, it is obvious
from their expressions (reproduced above), that they only consider the
on-shell pion form factor $\FFa(q_1^2, q_2^2) \equiv \FFa(m_\pi^2, q_1^2,
q_2^2)$ (e.g.\ at the internal vertex) and not the off-shell pion form factor
$\FF(q_3^2, q_1^2, q_2^2)$. Therefore, contrary to the claim in their paper,
they only consider the {\bf pion-pole} contribution to hadronic light-by-light
scattering.  Actually, also a second argument by Melnikov and Vainshtein in
favor of a constant form factor at the external vertex was based on the use of
on-shell form factors. Since $\FFac(q_3^2,0) \equiv \FFac(m_\pi^2, q_3^2,0)
\sim 1/q_3^2$, for large $q_3^2$, according to Brodsky-Lepage, the use of a
(non-constant) on-shell form factor at the external vertex would lead to an
overall $1/q_3^4$ behavior which contradicts Eq.~(\ref{A_pi0_MV}).

Translated into our notation employed in Eq.~(\ref{a_pion_2}),
Refs.~\cite{KnechtNyffeler01,Bijnens_Persson_01} and maybe also earlier
works, considered, e.g.\ for the first diagram of
Fig.~\ref{fig:LbLpionpole}, the form factors in the pion-pole approximation
\be
{\cal F}_{\pi^0 \gamma^* \gamma^*}(m_\pi^2,
q_1^2,q_3^2)\,\cdot {\cal F}_{\pi^0 \gamma^* \gamma}(m_\pi^2,
q_2^2,0)\,.
\ee
Although pole--dominance might be expected to give a reasonable approximation,
it is not correct as it was used in those references. The point is that the
form factor sitting at the external photon vertex in the pole approximation
[read $\FFac(m_\pi^2,q_2^2,0)$] for $q_2^2 \neq m_\pi^2$ violates
four--momentum conservation $k^\mu=0$~\cite{MV03,FJ07,Jegerlehner:2008zz}. 
The latter requires
$\FFc(q_2^2, q_2^2, 0)$. In order to avoid this inconsistency, Melnikov and
Vainshtein proposed to use
\be
{\cal F}_{\pi^0 \gamma^* \gamma^*}(m_\pi^2,
q_1^2,q_3^2)\,\cdot {\cal F}_{\pi^0 \gamma \gamma}(m_\pi^2, m_\pi^2,0)\,,
\ee 
i.e.\ a constant (WZW) form factor at the external vertex. The absence of a
form factor at the external vertex in the pion-pole approximation follows
automatically, if one carefully considers the momentum dependence of the form
factor. This procedure is also consistent with any quantum field theoretical
framework for hadronic light-by-light scattering, for instance, if one uses a
(resonance) Lagrangian to derive the form factors, and where a different
treatment of the internal and external vertex (apart from the kinematics) is
not possible. On the other hand, taking the diagram more literally, would
require
\be 
{\cal F}_{{\pi^0}^* \gamma^* \gamma^*}(q_2^2, q_1^2,q_3^2)\,\cdot {\cal
F}_{{\pi^0}^* \gamma^* \gamma}(q_2^2, q_2^2,0)\,,
\ee 
as the more appropriate amplitude, see Eq.~(\ref{a_pion_2}). In fact, we
advocate the consistent use of off-shell form factors on both vertices as
explained earlier.  As will be shown in more detail in
Sect.~\ref{sec:new_evaluation_pseudoscalars}, the use of appropriate
off-shell form factors within the framework of large-$N_c$ QCD {\bf does} lead
to a short-distance behavior which qualitatively agrees with the OPE
constraints which were derived in Ref.~\cite{MV03}
albeit with a different constant  $c_{\rm MV}=-0.274$
vs. $c_{\rm JN}= -0.092$ (factor 3 lower).\footnote{The large momentum
behavior of the full light-by-light scattering amplitude for other momentum
regions was also derived in Ref.~\cite{MV03} by evaluating exactly the
massless quark loop. Although the ansatz with a constant form factor at the
external vertex in Eq.~(\ref{A_pi0_MV}) does not satisfy all of these
constraints, it was argued in Ref.~\cite{MV03} that the effects of these other
short-distance constraints on the final numerical result is negligible.}

\subsubsection{The $\pi^0\gamma\gamma$ Transition Form Factor in different
  Models}

After the presentation of the experimental and theoretical constraints we now
turn to some of the ans\"atze for the $\pi^0\gamma\gamma$ form factor, which
have been used in the literature to evaluate the pion-exchange (or pion-pole)
contribution and which are based on or are motivated by different models for
low-energy hadrons. All these ans\"atze have certain drawbacks, thus leading
to different results with inherent model-dependent uncertainties which are
difficult to estimate. 

The simplest model is the constant WZW form factor (recall that $q_3 = - (q_1
+ q_2)$) 
\be
\FF^{\rm WZW}(q_3^2, q_1^2,q_2^2) = - \frac{N_c }{ 12 \pi^2 F_\pi} \, ,
\label{FF_WZW} 
\ee
which leads, however, to a divergent result in the integral in
Eq.~(\ref{a_pion_2})\footnote{Actually, the contribution involving the term
$T_2$ in Eq.~(\ref{a_pion_2}) is finite even for a constant form factor, see
Refs.~\cite{HK98,KnechtNyffeler01}. The numerical value is in fact always
much smaller, less than 5\%, than the results obtained for the part with $T_1$
with more realistic form factors.}, since there is no damping at high
energies. One can use some momentum cutoff around $1-2$~GeV, but this
procedure is completely arbitrary. Nevertheless, the WZW form factor serves as
a physical normalization to the $\pi^0 \to \gamma\gamma$ decay rate and all
models satisfy the constraint  
\be \label{FF_normalization_WZW}
\FFabc(m_\pi^2,0,0) = \FFabc^{\rm WZW}(m_\pi^2,0,0) = - {N_c \over 12 \pi^2
  F_\pi}. 
\ee

One way to implement a damping at high momentum is the VMD prescription
($\gamma-\rho$ mixing) which works reasonably well in many applications to
low-energy hadronic physics. It follows automatically in the HLS model which
was used in Refs.~\cite{HKS95,HK98} to evaluate the full hadronic
light-by-light scattering contribution. The HLS models implements VMD in a
consistent way, respecting chiral symmetry and electromagnetic gauge
invariance. It leads to the form factor
\be
\FF^{\rm VMD}(q_3^2,q_1^2,q_2^2) = - \frac{N_c }{ 12 \pi^2 F_\pi} \frac{M_V^2
}{(q_1^2 - M_V^2)} \frac{M_V^2 }{ (q_2^2 - M_V^2)} \, . 
\label{FF_VMD}
\ee
Note that the on- and off-shell VMD form factors are identical, since they
do not depend on the momentum $q_3^2$ which flows through the pion-leg.  The
problem with the VMD form factor is that the damping is now too strong as it
behaves like $\FFa(m_\pi^2, -Q^2, -Q^2) \sim 1/Q^4$, instead of $\sim 1/Q^2$
deduced from the OPE, see Eq.~(\ref{FF_OPE_2}).

Another model for the form factor $\FF$ which was for instance used in
Refs.~\cite{HKS95,HK98,Bartos_etal_02} is the constituent quark model
(CQM). The off-shell form factor is given by a quark triangular loop
\ba
F^\mathrm{CQM}_{\pi^{0*} \gamma^* \gamma^*}(q^2,p_1^2,p_2^2)&=&
2M_q^2\:C_0(M_q,M_q,M_q;q^2,p_1^2,p_2^2)\crn
&\equiv& \int[\D \alpha]\:\frac{2M_q^2}
{M_q^2-\alpha_2 \alpha_3 p_1^2-\alpha_3 \alpha_1 p_2^2-\alpha_1 \alpha_2
  q^2}\cs 
\ea
where $[\D \alpha] = \D \alpha_1\D \alpha_2\D
\alpha_3\:\delta(1-\alpha_1-\alpha_2-\alpha_3)$ and $M_q$ is a constituent
quark mass ($q=u,d,s$). For $p_1^2=p_2^2=q^2=0$ we obtain
$F^\mathrm{CQM}_{\pi^{0*} \gamma^* \gamma^*}(0,0,0)=1$, which is the proper
ABJ anomaly. Note the symmetry of $C_0$ under permutations of the arguments
($p_1^2,p_2^2,q^2$).  For large $p_1^2$ at $p_2^2 \sim 0,~q^2 \sim 0$ or
$p_1^2 \sim p_2^2$ at $q^2 \sim 0$ the asymptotic behavior is given by
\be \label{FFCQMasymptotics}
F^\mathrm{CQM}_{\pi^{0} \gamma^* \gamma}(0,p_1^2,0)\sim
r\:
\ln^2 r\;,\;\;
F^\mathrm{CQM}_{\pi^{0} \gamma^* \gamma^*}(0,p_1^2,p_1^2)\sim
2\:r\: \ln r , 
\ee 
where $r=\frac{M_q^2}{-p_1^2}$. The same behavior follows for $q^2 \sim p_1^2$
at $p_2^2 \sim 0$.  Note that in all cases we have the same power behavior
$\sim 1/p_i^2$ modulo logarithms. However, also this model has some
drawbacks. It is possible to reproduce the correct OPE behavior (up to the
logarithmic factors) by choosing the constituent quark mass as $M_q=2\pi
F_\pi/\sqrt{N_c}\sim 335~\mv$, which is close to $M_u = M_d = 300$~MeV often
used in the literature. However, the same mass leads to a coefficient in the
Brodsky-Lepage limit, which is too small by a factor of 6. Fitting instead the
Brodsky-Lepage behavior would lead to the unrealistic value $M_q = \sqrt{24}
\pi F_\pi / \sqrt{N_c} \sim 820~\mv$. In general, the description of the data
with the CQM form factor is rather poor, since the log$^2$ in
Eq.~(\ref{FFCQMasymptotics}) is distorting the power law for the values of
$p_1^2$ probed in the experiment, see Ref.~\cite{HK98}. Furthermore, the
permutation symmetry of the arguments in $F^\mathrm{CQM}_{\pi^{0*} \gamma^*
\gamma^*}(q_3^2, q_1^2, q_2^2)$ is not based on any symmetry of the original
QCD Green's function $\VVP$ in Eq.~(\ref{FFoffshellpi}). Therefore there is
also a damping in the other OPE limit studied above, $F^\mathrm{CQM}_{\pi^{0*}
\gamma^* \gamma}(q^2, q^2, 0) \sim 1/q^2$, which does not agree with the
result from Eq.~(\ref{FF_OPE_3_zeromomentum}), unless $\Pi_{\rm VT}(0) = 0$. The
vanishing of $\Pi_{\rm VT}(0)$ contradicts, however, the relation between
$\Pi_{\rm VT}(0)$ and the magnetic susceptibility $\chi$ in
Eq.~(\ref{Pi_VT0_Chi}). Finally, it was argued in Ref.~\cite{deRafaelENJL94}
that maybe one has to dress the coupling of the photons to the constituent
quarks \`a la VMD which leads to a further damping at high momenta. Of course,
all of this is very model dependent. A more complicated ansatz for the form
factor, based on the nonlocal chiral quark model, was employed recently in
Ref.~\cite{Dorokhov_Broniowski} to evaluate the pion-exchange
contribution. See that paper and references therein for a description of the
model and the explicit expression for the form factor.

In Ref.~\cite{BijnensLBL} the ENJL model was used to evaluate the pseudoscalar
exchange diagrams. This calculation was cross-checked in Ref.~\cite{HKS95} by
using a simplified version where the momentum dependence of some parameters,
like $F_\pi, M_\rho$, was neglected. The off-shell form factor $\FF$ in the
ENJL model is essentially given by a CQM-like form factor, see
Ref.~\cite{BijnensLBL} and references therein for more details. In the ENJL
model the dressing of the coupling of the constituent quarks to the photon
arises automatically via the summation of chains of quark bubble diagrams. As
for the CQM form factor, not all QCD short-distance constraints are fulfilled
in the ENJL model. In general, the ENJL model is only valid up to some cutoff
of order $800 - 1200~\mbox{MeV}$. Therefore, in Ref.~\cite{BijnensLBL} a
modified version of the ENJL form factor was finally used for the numerical
evaluation of the pion-exchange contribution. In this way some of the
short-distance constraints could be satisfied, in particular to reproduce the
Brodsky-Lepage behavior~(\ref{Brodsky_Lepage}) and the experimental data for
the on-shell form factor $\FFac(m_\pi^2, -Q^2,0)$.

The results for the form factor $\FF$ obtained in different low energy
effective hadronic models as usual do not satisfy all the large momentum
asymptotics required by QCD. Using these form factors in loops thus leads to
cut--off dependent results, where the cut--off is to be varied between
reasonable values ($\sim 1-2~\mbox{GeV}$) which enlarges the model error of
such estimates. Nevertheless it should be stressed that such approaches are
perfectly legitimate and the uncertainties just reflect the lack of precise
understanding of this kind of non-perturbative physics.

In order to eliminate (or at least reduce) this cut--off dependence, other
models for $\FF$ were proposed later in Ref.~\cite{KN_EPJC_01} and then
applied to hadronic light-by-light scattering in Ref.~\cite{KnechtNyffeler01}.
These models are based on the large--$N_c$ picture of QCD, where, in leading
order in $N_c$, an (infinite) tower of narrow resonances contributes in each
channel of a particular Green's function. The low-energy and short-distance
behavior of these Green's functions is then matched with results from QCD,
using CHPT and the OPE, respectively. Based on the experience gained in many
examples of low-energy hadronic physics, and from the use of dispersion
relations and spectral representations for two-point functions, it is then
assumed that with a minimal number of resonances in a given channel one can
get a reasonable good description of the QCD Green's function in the real
world. Often only the lowest lying resonance is considered, lowest meson
dominance, LMD, as a generalization of vector meson dominance VMD. Note that
it might not always be possible to satisfy {\bf all} short-distance
constraints, in particular from the high-energy behavior of form factors, if
only a {\bf finite number} of resonances is included, see
Ref.~\cite{BGLP03}. Ideally, the matching with the QCD constraints and other
informations, e.g.\ from decays of resonances, then determines all the free
parameters in these minimal hadronic ans\"atze (MHA).

In this spirit, on-shell $\FFa(m_\pi^2, q_1^2, q_2^2)$ and off-shell form
factors $\FF(q_3^2, q_1^2, q_2^2)$ were constructed in Ref.~\cite{KN_EPJC_01}
which contain either the lowest lying multiplet of vector resonances (LMD) or
two multiplets, the $\rho$ and the $\rho'$ (LMD+V). Both ans\"atze fulfill
{\bf all} the OPE constraints from Eqs.~(\ref{FF_OPE_1}), (\ref{FF_OPE_2}) and
(\ref{FF_OPE_3}), however, the LMD ansatz does {\bf not} reproduce the
Brodsky-Lepage behavior from Eq.~(\ref{Brodsky_Lepage}). Instead it behaves
like $\FFa^{\rm LMD}(m_\pi^2, -Q^2, 0) \sim \mbox{const}$. The $1/Q^2$
fall-off can be achieved with the LMD+V ansatz with a certain choice of the
free parameters (for more details see below).  The on-shell form factors where
later used in Ref.~\cite{KnechtNyffeler01} to evaluate the pion-pole
contribution, see also Ref.~\cite{Bijnens_Persson_01}. However, as mentioned
earlier, taking on-shell form factors at both vertices violates four-momentum
conservation.

\subsubsection{New Evaluation of the Pseudoscalar-exchange Contribution}
\label{sec:new_evaluation_pseudoscalars}

As stressed above, we advocate to use consistently dressed off-shell form
factors at both vertices, using for our new numerical evaluation of the
pion-exchange contribution the LMD+V {\bf off-shell} form
factor~\cite{KN_EPJC_01}
\ba
\FF^{\rm LMD+V}(p_\pi^2, q_1^2, q_2^2)&=&
 \frac{F_\pi}{3}\,\frac{\cP(q_1^2,q_2^2,p_\pi^2)}{\cQ(q_1^2,q_2^2)} , \crn 
\cP(q_1^2,q_2^2,p_\pi^2)&=&
q_1^2\,q_2^2\,(q_1^2 + q_2^2 + p_\pi^2) + h_1\,(q_1^2+q_2^2)^2
+ h_2\,q_1^2\,q_2^2 + h_3\,(q_1^2+q_2^2)\,p_\pi^2 + h_4\,p_\pi^4 \crn 
&& 
+h_5\,(q_1^2+q_2^2) + h_6\,p_\pi^2 + h_7
, \crn  
\cQ(q_1^2,q_2^2) &=&
(q_1^2-M_{V_1}^2)\,(q_1^2-M_{V_2}^2)\,(q_2^2-M_{V_1}^2)\,(q_2^2-M_{V_2}^2) , 
\label{KNpipioff}
\ea
with $p_\pi^2 \equiv (q_1 + q_2)^2$.  

We would like to point out that using the off-shell LMD+V form factor at the
external vertex leads to a short-distance behavior which qualitatively agrees
with the OPE constraints derived by Melnikov and Vainsthein in
Ref.~\cite{MV03}.  As a matter of fact, taking first $q_1^2 \sim q_2^2 \gg
q_3^2$ and then $q_3^2$ large, one obtains, together with the pion propagator
in Eq.~(\ref{a_pion_2}), an overall $1/q_3^2$ behavior for the pion-exchange
contribution, as expected from Eq.~(\ref{A_pi0_MV}), since, according to
Eq.~(\ref{FF_OPE_3_zeromomentum}), ${\cal F}^{\rm LMD+V}_{{\pi^0}^* \gamma^*
\gamma}(q_3^2,q_3^2,0) \sim \mbox{const}$ for large $q_3^2$. This also
qualitatively agrees with the $1/q_3^2$ fall-off obtained for the quark-box
diagram in light-by-light scattering derived in Ref.~\cite{MV03}.

Before we can apply the above form--factor we have to pin down as far as
possible the additional parameters $h_i$, which come in when the pion is far
off-shell. A detailed analysis of these constraints, as well as a new
calculation of the $\pi^0$--exchange contribution based on the off-shell LMD+V
form factor, has been performed recently by one of the authors
(A.N.)~\cite{Nyffeler:2009tw} and we closely follow the discussion presented
there.

The constants $h_i$ in the ansatz for ${\cal F}^{\rm LMD+V}_{{\pi^0}^*
\gamma^* \gamma^*}$ in Eq.~(\ref{KNpipioff}) are determined as follows. The
normalization with the WZW form factor in Eq.~(\ref{FF_normalization_WZW})
yields $h_7 = - N_c M_{V_1}^4 M_{V_2}^4 / (4 \pi^2 F_\pi^2) - h_6 m_\pi^2 -
h_4 m_\pi^4$. Note that in Refs.~\cite{KN_EPJC_01,KnechtNyffeler01} the small
corrections proportional to the pion mass were dropped, assuming that the
$h_i$ are of order $1-10$ in appropriate units of GeV.  The Brodsky-Lepage
behavior Eq.~(\ref{Brodsky_Lepage}) can be reproduced by choosing $h_1 =
0~\mbox{GeV}^2$. Furthermore, in Ref.~\cite{KN_EPJC_01} a fit to the CLEO data
for the on-shell form factor $\FFac^{\rm LMD+V}(m_\pi^2, -Q^2, 0)$ was
performed, with the result $h_5 = 6.93 \pm 0.26~\mbox{GeV}^4 - h_3
m_\pi^2$. Again, the correction proportional to the pion mass was omitted in
Refs.~\cite{KN_EPJC_01,KnechtNyffeler01}. As pointed out in Ref.~\cite{MV03},
the constant $h_2$ can be obtained from the higher-twist corrections in the
OPE. Comparing with Eq.~(\ref{NSVVZ}) yields the result
$h_2=-4\,(M_{V_1}^2+M_{V_2}^2) + (16/9)\,\delta^2 \simeq -10.63~\gv^2$, where
we used $M_{V_1} = M_\rho = 775.49~\mbox{MeV}$ and $M_{V_2} = M_{\rho^\prime}
= 1.465~\mbox{GeV}$~\cite{PDG_2008}.

Within the LMD+V framework, the vector-tensor two-point function discussed
earlier reads~\cite{KN_EPJC_01} 
\ba
\Pi_{\rm VT}^{\rm LMD+V}(p^2) & = &  -\,\langle{\overline\psi}\psi\rangle_0\, 
{ p^2 + c_{\rm VT} \over (p^2-M_{V_1}^2) (p^2-M_{V_2}^2) } \, ,
\label{VT_LMD+V} \\ 
c_{\rm VT} & = & {M_{V_1}^2 M_{V_2}^2 \chi \over 2} , 
\label{c_VT}
\ea
where we fixed the constant $c_{\rm VT}$ using Eq.~(\ref{Pi_VT0_Chi}). As
shown in Ref.~\cite{KN_EPJC_01} the OPE from Eq.~(\ref{FF_OPE_3}) for
$\FF^{\rm LMD+V}$ leads to the relation
\be \label{constraint_h1_h3_h4} 
h_1 + h_3 + h_4 = 2 c_{\rm VT} . 
\ee
As noted above, the value of the magnetic susceptibility $\chi(\mu)$
and the relevant scale $\mu$ are not precisely known. Adopting the
estimate presented in Ref.~\cite{Nyffeler:2009tw} we will use $\chi =
-(3.3 \pm 1.1)~\mbox{GeV}^{-2}$ in our numerical evaluation, which
implies the constraint $h_3 + h_4 = -(4.3 \pm 1.4)~\mbox{GeV}^{2}$.
We will vary $h_3$ in the range $\pm 10~\mbox{GeV}^2$ and
determine $h_4$ from Eq.~(\ref{constraint_h1_h3_h4}) and vice versa.  

The coefficient $h_6$ is undetermined as well. Direct phenomenological
constraints are not available. Model estimates within
the resonance Lagrangian and/or large-$N_c$ inspired approaches 
are given in~\cite{Nyffeler:2009tw}. In accordance  with these
estimates we will vary $h_6$ in the range $5\pm 5~\mbox{GeV}^4$.

Of course, the uncertainties of the values of the undetermined parameters
$h_3, h_4$ and $h_6$ and of the magnetic susceptibility $\chi(\mu)$ are a
drawback when using the off-shell LMD+V form factor and will limit the
precision of the final result.  Before presenting our estimate we note that as
a check we have reproduced with our 3-dimensional integral representation for
$a_{\mu}^{\mathrm{LbL};\pi^0}$ in Eq.~(\ref{pp3drep}) the results for various
form factors obtained earlier in the literature, e.g.\ $a_{\mu;
\mathrm{VMD}}^{\mathrm{LbL};\pi^0} = 57 \times 10^{-11}$ with the value for
$M_\rho = 775.49~\mbox{MeV}$ given above.

The results for $a_\mu^{\mathrm{LbL};\pi^0}$ for some selected values of
$\chi, h_3, h_4$ and $h_6$, varied in the ranges discussed above, with fixed
$h_1 = 0~\mbox{GeV}^2, h_2 = - 10.63~\mbox{GeV}^2$ and $h_5 =
6.93~\mbox{GeV}^4 - h_3 m_\pi^2$ are collected in Table~\ref{tab:pi0res}. 

\begin{table}[h] 
\centering
\caption{Results for $a_\mu^{\mathrm{LbL};\pi^0}\times 10^{11}$
obtained with the off-shell LMD+V form factor for some selected values of
$\chi, h_3 [ h_4]$ (imposing the constraint (\ref{constraint_h1_h3_h4})
with $h_1=0~\mbox{GeV}^2$) and $h_6$. The values of the other model parameters
are given in the text.}      
\label{tab:pi0res}
\begin{tabular}{llccc}
\hline 
 &  & ~~$h_6 = 0~\mbox{GeV}^4$~~ & ~~$h_6 = 5~\mbox{GeV}^4$~~ & ~~$h_6 =
10~\mbox{GeV}^4$~~ \\  
\hline
$\chi = -4.4~\mbox{GeV}^{-2}$~ & 
~~$h_{3[4]} = -10~\mbox{GeV}^2$   & 69.8~[66.9] & 75.7~[72.5] & 81.9~[78.4] \\
& ~~$h_{3[4]} =~~ 0~\mbox{GeV}^2$ & 67.8~[68.9] & 73.4~[74.7] & 79.4~[80.8] \\ 
& ~~$h_{3[4]} =~10~\mbox{GeV}^2$ & 65.8~[71.0] & 71.2~[76.9] & 77.0~[83.3] \\ 
\hline 
$\chi = -3.3~\mbox{GeV}^{-2}$~ & 
~~$h_{3[4]} = -10~\mbox{GeV}^2$   & 68.4~[65.3] & 74.1~[70.7] & 80.2~[76.5] \\
& ~~$h_{3[4]} =~~ 0~\mbox{GeV}^2$ & 66.4~[67.3] & 71.9~[72.8] & 77.8~[78.8] \\
& ~~$h_{3[4]} =~10~\mbox{GeV}^2$ & 64.5~[69.2] & 69.7~[75.0] & 75.4~[81.2] \\
\hline 
$\chi = -2.2~\mbox{GeV}^{-2}$~ & 
~~$h_{3[4]} = -10~\mbox{GeV}^2$   & 67.1~[63.8] & 72.7~[69.0] & 78.7~[74.7] \\
& ~~$h_{3[4]} =~~ 0~\mbox{GeV}^2$ & 65.2~[65.7] & 70.5~[71.1] & 76.3~[77.0] \\
& ~~$h_{3[4]} =~ 10~\mbox{GeV}^2$ & 63.3~[67.6] & 68.4~[73.3] & 74.0~[79.3] \\
\hline 
\end{tabular}
\end{table}

Varying $\chi$ in the range $-(3.3 \pm 1.1)~\mbox{GeV}^{-2}$ changes the
result for $a_\mu^{\mathrm{LbL};\pi^0}$ by at most $\pm 2.1 \times
10^{-11}$. The uncertainty in $h_6$ affects the result by up to $\pm 6.4
\times 10^{-11}$.  The variation of $a_\mu^{\mathrm{LbL};\pi^0}$ with $h_3$
(with $h_4$ determined from the constraint in Eq.~(\ref{constraint_h1_h3_h4})
with $h_1=0~\mbox{GeV}^2$ or vice versa) is much smaller, at most $\pm 2.5
\times 10^{-11}$. The variation of $h_5$ by $\pm 0.26~\mbox{GeV}^4$ only leads
to changes of $\pm 0.6 \times 10^{-11}$ in the final result.

Within the scanned region, we obtain a minimal value of
$a_\mu^{\mathrm{LbL};\pi^0} = 63.3 \times 10^{-11}$ for $\chi =
-2.2~\mbox{GeV}^{-2}, h_3 = 10~\mbox{GeV}^2, h_6 = 0~\mbox{GeV}^4$ and a
maximum of $a_\mu^{\mathrm{LbL};\pi^0} = 83.3 \times 10^{-11}$ for $\chi =
-4.4~\mbox{GeV}^{-2}, h_4 = 10~\mbox{GeV}^2, h_6 = 10~\mbox{GeV}^4$.  In the
absence of more information on the precise values of the constants $h_3, h_4$
and $h_6$, we take the average of the results obtained with $h_6 =
5~\mbox{GeV}^4$ for $h_3 = 0~\mbox{GeV}^2$, i.e.\ $71.9 \times 10^{-11}$, and
for $h_4 =0~\mbox{GeV}^2$, i.e.\ $72.8 \times 10^{-11}$, as our central value,
$72.3 \times 10^{-11}$.  To estimate the error, we add all the uncertainties
from the variations of $\chi$, $h_3$ (or $h_4$), $h_5$ and $h_6$ linearly to
cover the full range of values obtained with our scan of parameters. Note that
the uncertainties of $\chi$ and the coefficients $h_3,h_4$ and $h_6$ do not
follow a Gaussian distribution. In this way we obtain our final
estimate~\cite{Nyffeler:2009tw} (see also~\cite{FJ08})
\be \label{amupi0LMD+V}
a_\mu^{\mathrm{LbL};\pi^0} = (72 \pm  12) \times 10^{-11}.  
\ee
Unless one can pin down the ranges of $\chi$ and $h_6$ more precisely, we get
a larger error than previous estimates based e.g.\ on the on-shell LMD+V form
factor (which has less free parameters).  We would like to stress that
although the central value of our result in Eq.~(\ref{amupi0LMD+V}) is rather
close to $a_\mu^{\mathrm{LbL};\pi^0} = (76.5 \pm 6.7) \times 10^{-11}$ given
by Melnikov and Vainshtein~\cite{MV03},\footnote{Actually, using the on-shell
LMD+V form factor at the internal vertex with $h_2 = -10~\mbox{GeV}^2$ and
$h_5 = 6.93~\mbox{GeV}^4$ and a constant WZW form factor at the external
vertex, we obtain $79.8 \times 10^{-11}$, close to the value $79.6 \times
10^{-11}$ given in Ref.~\cite{BP07} and $79.7 \times 10^{-11}$ in
Ref.~\cite{Dorokhov_Broniowski}.} this is {\bf pure coincidence}. We have used
off-shell LMD+V form factors at both vertices, whereas Melnikov and Vainshtein
evaluated the {\bf pion-pole} contribution using the on-shell LMD+V form
factor at the internal vertex and the constant WZW form factor at the external
vertex.\footnote{Note that at the external vertex
$$\frac{3}{F_\pi}\,\FFc^{\rm LMD+V}(q_1^2, q_1^2, 0)\stackrel{q_1^2 \to
\infty}{\to}\frac{h_1+h_3+h_4}{M_{V_1}^2 M_{V_2}^2}=\frac{2c_{\rm
    VT}}{M_{V_1}^2 M_{V_2}^2}=\chi \,,$$ 
while $\FFc^{\rm VVA \ triangle}(q_1^2, q_1^2, 0)|_{m_q=0}= \FFabc^{\rm
  WZW}(0, 0, 0)=\FFabc^{\rm LMD+V}(0, 0, 0)$ utilized in Ref.~\cite{MV03}
  means  
$$\frac{3}{F_\pi}\,\FFabc^{\rm LMD+V}(0, 0, 0)=
\frac{h_7}{M_{V_1}^4M_{V_2}^4}=-\frac{N_c}{4\pi^2F_\pi^2}\simeq
-8.9~\gv^{-2}\,,$$ 
i.e. with Vainshtein's~\cite{Vainshtein03} value of $\chi$ we would precisely
satisfy the Melnikov-Vainshtein~\cite{MV03} short-distance constraint.}

As far as the contribution to $a_\mu$ from the exchanges of the other light
pseudoscalars, $\eta$ and $\eta^\prime$, is concerned, it is not so
straightforward to apply the above analysis within the LMD+V framework to
these resonances. In particular, the short-distance analysis in
Ref.~\cite{KN_EPJC_01} was performed in the chiral limit and assumed octet
symmetry. For the $\eta$ the effect of nonzero quark masses has definitely to
be taken into account. Furthermore, the $\eta^\prime$ has a large admixture
from the singlet state and the gluonic contribution to the axial anomaly will
play an important role. We therefore resort to a simplified approach which was
also adopted in other recent
works~\cite{HKS95,HK98,BijnensLBL,KnechtNyffeler01,MV03} and take the VMD form
factor Eq.~(\ref{FF_VMD}), normalized to the experimental decay width
$\Gamma(\mbox{P} \to \gamma \gamma)$, $\mbox{P} = \eta, \eta^\prime$. We can
fix the normalization by adjusting the pseudoscalar decay constant. Using the
latest values $\Gamma(\eta \to \gamma \gamma) = 0.510 \pm 0.026~\mbox{keV}$
and $\Gamma(\eta^\prime \to \gamma \gamma) = 4.30 \pm 0.15~\mbox{keV}$ from
Ref.~\cite{PDG_2008}, one obtains $F_{\eta, {\rm eff}} = 93.0~\mbox{MeV}$ with
$m_\eta = 547.853~\mbox{MeV}$ and $F_{\eta^\prime, {\rm eff}} =
74.0~\mbox{MeV}$ with $m_{\eta^\prime} = 957.66~\mbox{MeV}$.  We have seen
above that the pion exchange contribution evaluated with off-shell form
factors is not far from the pion-pole contribution. Of course, only a more
detailed analysis will show, whether this approximation works well for $\eta$
and $\eta^\prime$. It should also be kept in mind that the VMD form factor has
a too strong damping for large momenta. From the experience with the pion
contribution, it seems, however, more important to have a good description of
the relevant form factors at small and intermediate energies below 1~GeV,
e.g., by reproducing the slope of the form factor ${\cal
F}_{\mathrm{P}\gamma^*\gamma}(m_{\mathrm{P}}^2, -Q^2, 0)$, at the origin. The
CLEO Collaboration~\cite{CLEO98} has made a fit of the (on-shell) form factors
${\cal F}_{\eta\gamma^*\gamma}(m_\eta^2, -Q^2, 0)$ and ${\cal F}_{\eta^\prime
\gamma^*\gamma}(m_{\eta^\prime}^2, -Q^2, 0)$ using an interpolating formula
similar to Eq.~(\ref{FFacfit}) with an adjustable vector meson mass
$\Lambda_{\mathrm{P}}$.  Taking their values $\Lambda_\eta = 774 \pm
29~\mbox{MeV}$ or $\Lambda_{\eta^\prime} = 859 \pm 28~\mbox{MeV}$ as the
vector meson mass $M_V$ in the expression of the VMD form factor in
Eq.~(\ref{FF_VMD}), we get $a_{\mu}^{\mathrm{LbL};\eta} = (14.5 \pm 4.8)
\times 10^{-11}$ and $a_{\mu}^{\mathrm{LbL};\eta^\prime} = (12.5 \pm 4.2)
\times 10^{-11}$, where we assumed a relative error of 33\%. Note that these
values are somewhat smaller than $a_{\mu}^{\mathrm{LbL};\eta-\mathrm{pole}} =
18 \times 10^{-11}$ and $a_{\mu}^{\mathrm{LbL};\eta^\prime-\mathrm{pole}} = 18
\times 10^{-11}$ given in Ref.~\cite{MV03}, where the constant WZW form factor
was used at the external vertex. Adding up all contributions from the
pseudoscalars, we finally obtain the estimate
\be
a_{\mu}^{\mathrm{LbL;PS}} = (99 \pm 16) \times 10^{-11}\cs
\ee
given in~\cite{Nyffeler:2009tw} (see also~\cite{FJ08}).

\subsection{Summary of the Light-by-Light Scattering Results}
\label{sec:LbLsummary}

We are now ready to summarize the results obtained by the different groups for
the hadronic light-by-light scattering contribution. A comparison of the
different results also sheds light on the difficulties and the model
dependencies in the theoretical estimations achieved so far. 
Very recently, a joint effort to
summarize the results obtained by various groups has been presented by Prades,
de Rafael and Vainshtein [PdRV]~\cite{Prades:2009tw}. The values advocated by
them are included in our tables.

\vspace*{0.3cm} 
\noindent
{\bf Pseudoscalar exchanges} 

According to Table~\ref{tab:LEeffcounting} the diagram
Fig.~\ref{fig:lbldiagabc}(a) with the exchange of pseudoscalars yields the most
important contribution in the large-$N_c$ counting, but requires a model for
the $\FF$ form factor for its evaluation. Although it is subleading in the
chiral expansion in comparison to the loop with charged pions and Kaons
(Fig.~\ref{fig:lbldiagabc}(b)), it turns out that this is the numerically
dominating contribution, see the numbers collected in
Table~\ref{tab:LbLdiaga}.  The dominance of the pseudoscalar exchange
contributions in view of Fig.~\ref{fig:LBLfacts} after all is an experimental
fact. 

\begin{table}[h]
\centering
\caption{Results for the $\pi^0,\eta$ and $\eta'$ exchange contributions.} 
\label{tab:LbLdiaga}
\begin{tabular}{lr@{(}c@{)}lr@{(}c@{)}l}
&\multicolumn{3}{c}{~}&\multicolumn{3}{c}{~}\\[-3mm]
\hline\noalign{\smallskip}
 Model for ${\cal F}_{P^*\gamma^*\gamma^*}$
 &\multicolumn{3}{c}{$a_\mu(\pi^0)\power{11}$}~~& 
\multicolumn{3}{c}{$a_\mu(\pi^0,\eta,\eta')\power{11}$}\\
\noalign{\smallskip}\hline\noalign{\smallskip}
Point coupling & \ttc{~~$+\infty$} & & \ttc{~~~$+\infty$} \\
ENJL (modified) [BPP]~\cite{BijnensLBL,BP07} &~~~~ 59& 9 & &~~~~~~ 85&13& \\
VMD / HLS [HKS,HK]~\cite{HKS95,HK98}  & 57&4& & 83&6& \\
nonlocal $\chi$QM (off-shell)~\cite{Dorokhov_Broniowski} & 65 & 2 & &
\ttc{~~~~~$-$} \\  
LMD+V [KN] (on-shell, $h_2=0~\gv^2$)~\cite{KnechtNyffeler01}& 58&10& & 83&12&
\\ 
LMD+V [KN] (on-shell, $h_2=-10~\gv^2$)~\cite{KnechtNyffeler01}& 63&10& &
88&12& \\ 
LMD+V [MV] (on-shell, constant FF at external vertex)~\cite{MV03}& 77&7& &
114&10& \\ 
LMD+V [PdRV] (on-shell, constant FF at external
vertex)~\cite{Prades:2009tw}& \ttc{~~~~~$-$} & &
114&13& \\ 
LMD+V [N] (off-shell, $\chi = -(3.3\pm 1.1)~\mbox{GeV}^{-2}, h_6 = (5 \pm
5)~\mbox{GeV}^4$)~\cite{Nyffeler:2009tw}
 & 72 & 12 & & 99 & 16 & \\
\noalign{\smallskip}\hline
\end{tabular}
\end{table}

BPP~\cite{BijnensLBL} work within the context of the ENJL model, however, they
take the model only seriously for scales below a few hundred MeV. At higher
momenta, they modify the corresponding ENJL form factor with VMD dressing or
consider a pure VMD form factor. In particular, they try to find a
phenomenological parametrization that interpolates between the ENJL form
factor, which works well below 0.5 GeV, and the measured (on-shell) form
factor ${\cal F}_{\pi^0\gamma^*\gamma}(m_\pi^2, -Q^2,0)$ for Euclidean momenta
above 0.5 GeV and with its asymptotic behavior predicted by QCD
(Brodsky-Lepage). The results for $\eta$ and $\eta'$ are obtained by using a
VMD form factor normalized to the experimental decay rate $\mbox{P} \to
\gamma\gamma$ and rescaled with the ENJL result for $\pi^0$.  HKS,
HK~\cite{HKS95,HK98} work with the HLS model which leads to a VMD form factor,
but also studied the effects of various other kinds of form factors: (dressed)
CQM, ENJL-like\footnote{By ENJL-like we denote the fact that the authors of
Refs.~\cite{HKS95,HK98} took for a cross-check of the evaluation in
Ref.~\cite{BijnensLBL}, the expressions for the form factor in the ENJL model,
however, they neglected the momentum dependence of $f_\pi, M_V$ and the
parameter $g_A$ in this model.}, mixed versions. At the end, they choose a VMD
model where the normalization is fixed by the experimental two-photon decay
width. Furthermore, the rather small error estimate is derived from fitting
the (on-shell) form factor ${\cal F}_{P\gamma^*\gamma}(m_P^2, -Q^2,0)$ to the
available data. This procedure might, however, underestimate the intrinsic
model dependence, in particular, for off-shell values of the form factor. In
Ref.~\cite{Dorokhov_Broniowski} off-shell form factors were used at both
vertices, following the suggestion in Ref.~\cite{FJ07}. These authors use the
nonlocal chiral quark model which shows a strong, exponential suppression for
large pion virtualities. This is very different from what is observed in all
the other models. Finally, we note that all the LMD+V estimates in
Table~\ref{tab:LbLdiaga} only apply to the pion. For the $\eta$ and $\eta'$
a VMD form factor is used, normalized to the experimental decay width.

\vspace*{0.3cm} 
\noindent
{\bf Axial-vector and scalar exchanges}

Next in Table~\ref{tab:LEeffcounting} are the exchanges of other resonances,
like axial-vectors and scalars in a diagram analogous to
Fig.~\ref{fig:lbldiagabc}(a). They are also leading in $N_c$, but of higher
order in the chiral counting, compared to the pseudoscalars. The results for
the axial-vector contribution are collected in Table~\ref{tab:LbLdiagax} and
those for the scalars in Table~\ref{tab:LbLdiagsc}. Since the masses of these
resonances are higher\footnote{Apart from the contribution of a potentially
light, broad $\sigma$-meson $f_0(600)$.} in comparison with the
pseudoscalars, in particular the $\pi^0$, the corresponding suppression by the
propagator leads in general to smaller results, unless the coupling to photons
is extraordinary large.

\begin{table}[h]
\centering
\caption{Results for the axial-vector ($a_1,f_1$ and $f_1'$) exchange
contributions.}
\label{tab:LbLdiagax}
\begin{tabular}{lr@{(}c@{)}lr@{(}c@{)}l}
&\multicolumn{3}{c}{~}&\multicolumn{3}{c}{~}\\[-3mm]
\hline\noalign{\smallskip}
 Model for ${\cal F}_{A^*\gamma^*\gamma^*}$
 &\multicolumn{3}{c}{$a_\mu(a_1)\power{11}$}~~& 
\multicolumn{3}{c}{$a_\mu(a_1,f_1,f_1')\power{11}$}\\
\noalign{\smallskip}\hline\noalign{\smallskip}
ENJL-VMD [BPP] (nonet symmetry)~\cite{BijnensLBL} &~~~~ 2.5&1.0& 
\multicolumn{3}{r}{$-$~~~} \\  
ENJL-like [HKS,HK] (nonet symmetry)~\cite{HKS95,HK98}  & 1.7&1.7&
\multicolumn{3}{r}{$-$~~~} \\ 
LMD [MV] (ideal mixing)~\cite{MV03}& \multicolumn{2}{l}{~~~~~5.7} & &
~~~~~~~~22&5& \\ 
LMD [PdRV]~\cite{Prades:2009tw}& \multicolumn{2}{l}{~~~~~$-$} & &
~~~~~~~~15&10& \\ 
\noalign{\smallskip}\hline
\end{tabular}
\end{table}

In Ref.~\cite{MV03} it was argued that again a constant form factor should be
used at the external vertex to reproduce QCD short-distance constraints,
similarly to the procedure adopted for the pseudoscalars in the same
reference. Using a simple VMD ansatz they also derive the first term in a
series expansion in powers of $m_\mu / M$, where $M$ denotes the axial-vector
mass (assuming nonet-symmetry were they are all treated as equal as done in
Refs.~\cite{BijnensLBL,HKS95,HK98}). They observe that the result strongly
depends on the exact choice of this mass, e.g.\ with $M = M_\rho$ they obtain
$28 \times 10^{-11}$.  The results shown in Table~\ref{tab:LbLdiagax} have
been obtained by using a more sophisticated ansatz for the form factor at the
external vertex which was first proposed in Ref.~\cite{CMV03}. Note that the
form factor now includes a dressing with respect to the one off-shell photon
at the external vertex. In this way they treat the resonances $a_0, f_1,
f_1^\prime$ separately. Since the dressing leads to lower effective
axial-vector masses and since the states $f_0$ and $f_0^\prime$ have an
enhanced coupling to photons (similarly to $\eta$ and $\eta^\prime$), the
final result is a factor of 10 larger than those obtained earlier in
Refs.~\cite{BijnensLBL,HKS95,HK98}. The result for the sum of all resonances
in Ref.~\cite{MV03} does not depend too much on the value of the mixing angle
between $f_0$ and $f_0^\prime$ (treating $f_1$ as pure octet and $f_1^\prime$
as pure $SU(3)$ singlet, they obtain the result $17 \times 10^{-11}$). We
think the procedure adopted in Ref.~\cite{MV03} is an important improvement
over Refs.~\cite{BijnensLBL,HKS95,HK98} and we will therefore take the result
for the axial-vectors from that reference for our final estimate for the full
hadronic light-by-light scattering contribution below, despite the fact that
only on-shell form factors have been used in Ref.~\cite{MV03}. As we argued
above for the pseudoscalar exchanges, we think that one should use
consistently off-shell form factors at the internal and the external vertex.

\begin{table}[h]
\centering
\caption{Results for the scalar exchange contributions.}
\label{tab:LbLdiagsc}
\begin{tabular}{lc}
& \\[-3mm]
\hline\noalign{\smallskip}
 Model for ${\cal F}_{S^*\gamma^*\gamma^*}$ & 
$a_\mu(\mbox{scalars})\power{11}$\\
\noalign{\smallskip}\hline\noalign{\smallskip}
Point coupling & $-\infty$ \\ 
ENJL[BPP]~\cite{BijnensLBL,BP07} & -7(2) \\
ENJL[PdRV]~\cite{Prades:2009tw} & -7(7) \\
\noalign{\smallskip}\hline
\end{tabular}
\end{table}

The contribution from scalar resonances with masses around 1 GeV was first
studied in Ref.~\cite{KNO84}, but found to be negligible ($0.1 \times
10^{-11}$), compared to the dominating $\pi^0$ exchange contribution. Within
the ENJL model used in Ref.~\cite{BijnensLBL}, this scalar exchange
contribution is related via Ward identities to the (constituent) quark
loop. In fact, Ref.~\cite{HK98} argued that the effect of the exchange of
(broad) scalar resonances below several hundred MeV might already be included
in the sum of the (dressed) quark loop and the dressed pion and Kaon
loop. Such a potential double-counting is definitely an issue for the broad
sigma meson $f_0(600)$. Furthermore there is some ongoing debate in the
literature, see the PDG~\cite{PDG06} and references therein, whether the
scalar resonances $f_0(980)$ and $a_0(980)$ are two-quark or four-quark states
(meson molecules). 

The parameters of the ENJL model used in Ref.~\cite{BijnensLBL} have been
determined in Refs.~\cite{BBdeR93,Bijnens:1993ap,BP94}. In particular,
in~\cite{BBdeR93} a fit was performed to various low-energy observables and
resonance parameters, among them a scalar multiplet with mass $M_S =
983~\mbox{MeV}$. However, with those fitted parameters, the ENJL model
actually predicts a rather low mass of $M_S^{\rm ENJL} = 620~\mbox{MeV}$. This
would then correspond more to the light sigma meson $f_0(600)$.  We note that
within a very simple model of a scalar meson $S$ coupled to photons via a term
$S \, F_{\mu\nu} F^{\mu\nu}$, together with a simple VMD-dressing, there
arises again a leading log$^2$ term for $M_\rho \to \infty$. If the coupling
in the above Lagrangian is the same as for the $\pi^0$ in the WZW term, i.e.\
$\alpha N_c / (12 \pi F_\pi)$, then the coefficient of the log-square term is
identical to the universal coefficient found for the pion, except for the
negative sign, see Ref.~\cite{BCM02}. The question is whether the usually
broad scalar resonances can really be described by such a simple resonance
Lagrangian which works best in the large-$N_c$ limit, i.e.\ for very narrow
states.

\vspace*{0.3cm} 
\noindent
{\bf Charged pion and Kaon loops}

Third in Table~\ref{tab:LEeffcounting} are the charged pion- and Kaon-loops 
Fig.~\ref{fig:lbldiagabc}(b) which yield the leading contribution in chiral
counting, but are subleading in $N_c$. The results are given in
Table~\ref{tab:LbLdiagb}.

\begin{table}[h]
\centering
\caption{Results for the (dressed) $\pi^\pm,K^\pm$ loops.}
\label{tab:LbLdiagb}
\begin{tabular}{lll}
&&\\[-3mm]
\hline\noalign{\smallskip}
 Model $\pi^+\pi^-\gamma^*(\gamma^*)$ ~&~~ $a_\mu(\pi^\pm)\power{11}$~&~~
$a_\mu(\pi^\pm,K^\pm)\power{11}$ \\
\noalign{\smallskip}\hline\noalign{\smallskip}
Point coupling (sQED)              & ~~~~~~~$-45.3$   & ~~~~~~~~~$-49.8$  \\
VMD~[KNO, HKS]~\cite{KNO84,HKS95} & ~~~~~~~$-16$     & ~~~~~~~~~~~~$-$ \\
full VMD~[BPP]~\cite{BijnensLBL}   & ~~~~~~~$-18(13)$ & ~~~~~~~~~$-19(13)$ \\
HLS~[HKS,HK]~\cite{HKS95,HK98}    & ~~~~~~~$-4.45$   & ~~~~~~~~~$-4.5(8.1)$ \\
{}[MV]~\cite{MV03}    & ~~~~~~~$~~~-$   & ~~~~~~~~~$~~~0(10)$ \\
full VMD~[PdRV]~\cite{Prades:2009tw}    & ~~~~~~~$~~~-$   & ~~~~~~~~~$-19(19)$
\\ 
\noalign{\smallskip}\hline
\end{tabular}
\end{table}

The result without dressing (scalar QED) is finite, see the EFT analysis
discussed in Sect.~\ref{sec:LbLEFT}, in contrast to the pseudoscalar
exchanges. There was some debate between the authors of
Refs.~\cite{HKS95,HK98} (using the HLS model) and \cite{BijnensLBL} (using
full VMD), on how the dressing of the point vertex has to be implemented
without violating gauge and chiral invariance. This explains the numerical
difference between the two evaluations. The difference of the two values also
indicates the potential model dependence of the result. The most important
fact, however, is that the dressing leads to a rather huge suppression of the
final result compared to the undressed case, so that the final result is much
smaller than the one obtained for the pseudoscalars. This effect was studied
for the HLS model in Ref.~\cite{MV03}, in an expansion in $(m_\pi / M_\rho)^2$
and $\delta = (m_\mu - m_\pi) / m_\pi$, with the result ($L = \ln(M_\rho /
m_\pi)$) 
\ba
a_\mu^{\mathrm{LbL};\pi^\pm} & = & \left( {\alpha \over \pi} \right)^3
\sum_{i=0}^\infty f_i(\delta, L) \left( {m_\pi^2 \over M_\rho^2} \right)^i  
 =  \left( {\alpha \over \pi} \right)^3 (-0.0058)  \nonumber \\ 
& = & \left( -46.37 + 35.46 + 10.98 - 4.70 - 0.3 + \ldots \right) \times
  10^{-11} 
 =  -4.9(3) \times 10^{-11} , 
\ea
where the functions $f_i(\delta,L)$ have been calculated for $i=0, \ldots, 4$
in Ref.~\cite{MV03} and are explicitly given for $i=0,1,2$ there. The
subsequent terms in the last line correspond to the terms in the expansion in
the first line. As one can see, there occurs a large cancellation between the
first three terms in the series and the expansion converges only very
slowly. The main reason is that typical momenta in the loop integral are of
order $\mu = 4 m_\pi \approx 550~\mbox{MeV}$ and the effective expansion
parameter is $\mu / M_\rho$. The authors of Ref.~\cite{MV03} took this as an
indication that the final result is very likely suppressed, but also very
model dependent and that the chiral expansion looses its predictive power. The
pion and Kaon loop contribution is then only one among many potential
contributions of ${\cal O}(1)$ in $N_c$ and they lump all of these into the
guesstimate $a_\mu^{\mathrm{LbL}; N_c^0} = (0 \pm 10) \times
10^{-11}$. However, since this estimate does not even cover the explicit,
although model-dependent, results for the pion and Kaon loops given in
Refs.~\cite{BijnensLBL,HKS95}, we think this procedure is not very
appropriate.

\vspace*{0.3cm} 
\noindent
{\bf Dressed quark loop}

Finally, the last entry in Table~\ref{tab:LEeffcounting} is the (constituent)
quark loop Fig.~\ref{fig:lbldiagabc}(c) which appears as short-distance
complement of the ENJL and HLS low-energy effective models used in
Refs.~\cite{BijnensLBL,HKS95,HK98}. It is again leading order in $N_c$ and of
the same chiral order as the axial-vector and scalar exchanges in $\langle
VVVV \rangle$. As argued in Ref.~\cite{deRafaelENJL94} such a quark-loop can
be interpreted as an irreducible contribution to the 4-point function and
should be added to the other contributions, although a dressing of the
coupling of the constituent quarks with the photons might occur. According to
quark-hadron duality, the (constituent) quark loop also models the
contribution to $a_\mu$ from the exchanges and loops of heavier resonances,
like $\pi^\prime, a_0^\prime, f_0^\prime, p, n, \ldots$ which have not been
included explicitly so far. It also ``absorbs'' the remaining cutoff
dependences of the low-energy effective models. This is even true for the
modeling of the pion-exchange contribution within the large $N_c$ inspired
approach (LMD+V), since not all QCD short-distance constraints in the 4-point
function $\langle VVVV \rangle$ are reproduced with those ans\"atze. Some
estimates for the (dressed) constituent quark loop are given in
Table~\ref{tab:LbLdiagquark}.

\begin{table}[h]
\centering
\caption{Results for the (dressed) quark loops.}
\label{tab:LbLdiagquark}
\begin{tabular}{lc}
&\\[-3mm]
\hline\noalign{\smallskip}
 Model &~~ $a_\mu(\mathrm{quarks})\power{11}$\\
\noalign{\smallskip}\hline\noalign{\smallskip}
Point coupling & $62(3)$   \\
VMD~[HKS, HK]~\cite{HKS95,HK98}   &  $~~~~~~~9.7(11.1)$   \\
ENJL + bare heavy quark~[BPP]~\cite{BijnensLBL} & $21(3)$ \\
Bare $c$-quark only~[PdRV]~\cite{Prades:2009tw} & $2.3$ \\
\noalign{\smallskip}\hline
\end{tabular}
\end{table}

We observe again a large, very model-dependent effect of the dressing of the
photons.  HKS, HK~\cite{HKS95,HK98} used a simple VMD-dressing for the
coupling of the photons to the constituent quarks as it happens for instance
in the ENJL model. On the other hand, BPP~\cite{BijnensLBL} employed the ENJL
model up to some cutoff $\mu$ and then added a bare quark loop with a
constituent quark mass $M_Q = \mu$. The latter contribution simulates the
high-momentum component of the quark loop, which is non-negligible. The sum of
these two contributions is rather stable for $\mu = 0.7, 1, 2$ and 4 GeV and
gives the value quoted in Table~\ref{tab:LbLdiagquark}. A value of $2 \times
10^{-11}$ for the $c$-quark loop is included by BPP~\cite{BijnensLBL}, but not
by HKS~\cite{HKS95,HK98}.

\vspace*{0.3cm} 
\noindent
{\bf Summary}

The totals of all contributions to hadronic light-by-light scattering reported
in the most recent estimations are shown in Table~\ref{tab:LbLrecent}. We have
also included some ``guesstimates'' for the total value. Note that the number
$a_\mu^{\mathrm{LbL;had}} = (80 \pm 40) \times 10^{-11}$ written in the fourth
column in Table~\ref{tab:LbLrecent} under the heading KN was actually not
given in Ref.~\cite{KnechtNyffeler01}, but represents estimates used mainly by
the Marseille group before the appearance of the paper by MV~\cite{MV03}.
Furthermore, we have included in the sixth column the estimate
$a_\mu^{\mathrm{LbL;had}} = (110 \pm 40) \times 10^{-11}$ given recently in
Refs.~\cite{talk_LbL_JP,BP07,Miller07}.  Note that PdRV~\cite{Prades:2009tw}
(seventh column) do not include the dressed light quark loops as a separate contribution. They
assume them to be already covered by using the short-distance constraint from
MV~\cite{MV03} on the pseudoscalar-pole contribution. PdRV add, however, a
small contribution from the bare $c$-quark loop.

\begin{table}[h]
\centering
\caption{Summary of the most recent results for the various contributions to
  $a_\mu^{\mathrm{LbL; had}} \times 10^{11}$. The last column is our estimate
  based on our new evaluation for the pseudoscalars and some of the other
  results.}   
\label{tab:LbLrecent}
\vspace*{0.2cm} 
\begin{tabular}{cr@{$\pm$}lr@{$\pm$}lr@{$\pm$}lr@{$\pm$}lr@{$\pm$}lr@{$\pm$}lr@{$\pm$}l}
\hline\noalign{\smallskip}
Contribution & \ttc{BPP} & \ttc{HKS} & \ttc{KN} & \ttc{MV}
& \ttc{BP} & \ttc{PdRV} &\ttc{N/JN} \\ 
\noalign{\smallskip}\hline\noalign{\smallskip}
 $\pi^0,\eta,\eta'$ & $85$ & $13~$ & $~82.7 $&$ 6.4$ & $~83 $&$ 12~$
& $\,114 $&$ 10\,$ & \ttc{$-$} & $\,114 $&$ 13\,$ & $99$&$16$ \\
$\pi,K$ loops & $-19$ &$ 13~$ &  $-4.5$&$ 8.1$ &\ttc{$-~~$} & \ttc{$-$}
& \ttc{$-$} & $-19$ &$ 19~$ &$-19 $&$ 13~$ \\ 
$\pi,K$ loops + other subleading in $N_c$ & \ttc{$-$} & \ttc{$-~~$} &
\ttc{$-~~$} & $0$&$ 10$ & \ttc{$-$}& \ttc{$-$} & \ttc{$~-$} \\ 
axial vectors & $2.5 $&$ 1.0$ & $1.7 $&$ 1.7$ &\ttc{$-~~$} & $22 $&$
\,5$& \ttc{$-$} & $15 $&$
10$ & $22 $&$ \,5$  \\ 
scalars      & $-6.8$&$ 2.0$ & \ttc{$-~~$} & \ttc{$-~~$} & \ttc{$-$}&
\ttc{$-$} & $-7$&$\,7~$& $-7$&$\,2~$ \\ 
quark loops  ~~& $21 $&$ \,3~$ & $9.7 $&$ 11.1$ & \ttc{$-~~$} & \ttc{$-$} &
 \ttc{$-$} & \ttc{$2.3\,$} & $21 $&$ \,3~$\\
\noalign{\smallskip}\hline\noalign{\smallskip}
total & $83$&$ 32~$ ~~& $89.6 $&$ 15.4$ ~~~& $80 $&$ 40$ ~~~&
$136 $&$ 25$~~~& $110$ & $40$ & $105 $&$ 26$& $116 $&$ 39$ \\ 
\noalign{\smallskip}\hline
\end{tabular}
\end{table}

As one can see from Table~\ref{tab:LbLrecent}, the different models used by
various groups lead to slightly different results for the individual
contributions.  The final result\footnote{For the electron we obtain
$a_e^{\mathrm{LbL};\pi^0} = (2.98 \pm 0.34) \times 10^{-14}$ with the
off-shell LMD+V form factor. This number supersedes the value given in
Ref.~\cite{KnechtNyffeler01}. Note that the naive rescaling
$a_e^{\mathrm{LbL};\pi^0}(\mathrm{rescaled}) = (m_e / m_\mu)^2
a_\mu^{\mathrm{LbL};\pi^0} = 1.7 \times 10^{-14}$ yields a value which is
almost a factor 2 too small. Estimates for the other pseudoscalars are $(0.49
\pm 0.16) \times 10^{-14}[\eta]$ and $(0.39 \pm 0.13) \times 10^{-14}[\eta']$.
Since the other contributions are smaller and/or largely cancel, we arrive at
an estimate $a_e^{\mathrm{LbL; had}} \sim (3.9 \pm 1.3) \times 10^{-14}\,,$
where we assumed a relative error of 33\% to be conservative.}
\be
a_\mu^{\mathrm{LbL; had}}=(116 \pm 39) \power{-11}  
\label{JNLBL}
\ee
for the hadronic light-by-light scattering contribution is dominated by the
pseudoscalar exchange contribution, which we have recalculated from scratch
and beyond the pion-pole approximation which has been used frequently. The
other contributions are smaller, but not
negligible~\cite{BijnensLBL,BP07}. Furthermore, they cancel out to some
extent. Since the
variation of the results for these individual contributions reflects our
inherent ignorance of strong interaction physics in hadronic light-by-light
scattering, it has become customary to take the difference between those
values as an indication of the model uncertainty and to add the errors in
Table~\ref{tab:LbLrecent} linearly (note, however, that
PdRV~\cite{Prades:2009tw} add the errors in quadrature). Maybe this error
estimate is too conservative. For instance, the sum of the dressed pion and
Kaon loops and the dressed quark loops is almost identical for the two
evaluations by BPP~\cite{BijnensLBL} and HKS~\cite{HKS95,HK98} using different
models. But maybe this is a pure numerical coincidence, since these
contributions have a different counting in $N_c$ and $p^2$, see
Table~\ref{tab:LEeffcounting}. Unless one can obtain for instance a more
precise and reliable determination of the pion-loop contribution, it will be
difficult to claim that we really control this kind of hadronic physics. At
the moment, one cannot argue that either the ENJL / full VMD model employed by
BPP or the HLS model used by HSK is superior compared to the other approach.

We also want to stress again that the identification of individual
contributions in hadronic light-by-light scattering (like pion-exchange or the
pion loop) is model dependent as soon as one uses off-shell form
factors. Keeping this caveat in mind, we think that some progress has been
made in recent years in understanding the pseudoscalar and axial-vector
exchange contributions, following Refs.~\cite{KnechtNyffeler01,MV03,BP07} and
our new evaluation for the pseudoscalar-exchange contribution in
Sect.~\ref{sec:new_evaluation_pseudoscalars}. Also the effective field theory
analysis of Refs.~\cite{KNPdeR01,RMW02}, which yields the leading log terms in
the pion exchange contribution, agrees roughly with the numerical values
obtained with different models, although the EFT approach cannot give a
precise number in the end, see Sect.~\ref{sec:LbLEFT}.  Apart from the
numerical differences between BPP and HKS for the dressed pion loop and the
dressed quark loop, there is the issue of the scalar exchange contribution,
see also the discussion above. We think that a priori such a contribution is
likely to be there and the numerical value given in Ref.~\cite{BijnensLBL}
looks reasonable, therefore we have included it in our final estimate. In view
of the relatively large contribution of the axial-vector mesons with masses
around 1300 MeV, it should finally be kept in mind that other states in that
mass region could also contribute significantly to the final result. It is not
clear at present, whether all of these contributions are appropriately modeled
by the dressed quark loop. We will discuss some prospects for improving the
estimate of the hadronic light-by-light scattering contribution to $a_\mu$ and
for reducing its theoretical error in
Sect.~\ref{sec:Conclusions}.

\subsection{An Effective Field Theory Approach to Hadronic Light-by-Light
  Scattering} 
\label{sec:LbLEFT} 

In Ref.~\cite{KNPdeR01} an EFT approach to hadronic light-by-light scattering
was presented based on an effective Lagrangian that describes the physics of
the Standard Model well below 1~GeV. It includes photons, light leptons, and
the pseudoscalar mesons and obeys chiral symmetry and $U(1)$ gauge
invariance. 

The leading contribution to $a_\mu^\mathrm{LbL;had}$, of order\footnote{Note
  that here we count directly the chiral order of the contribution to $a_\mu$,
  in contrast to the counting used in Table~\ref{tab:LEeffcounting} and
  Ref.~\cite{deRafaelENJL94}. In the EFT approach of
  Ref~\cite{KNPdeR01}, the chiral power counting is generalized by treating
  $e,m$ and fermion bilinears as order $p$.} $p^6$, is given by a finite loop
  of charged pions, see Fig.~\ref{fig:lbldiagabc}(b), however, with point-like
  electromagnetic vertices, i.e.\ without the dressing of the photons (scalar
  QED). The numerical value is
  $a_{\mu;\mathrm{sQED}}^{\mathrm{LbL}}(\pi^\pm,K^\pm-\mathrm{loops}) \simeq
  -48 \power{-11}$. Since this contribution involves a loop of hadrons, it is
  subleading in the large-$N_c$ expansion, see Table~\ref{tab:LEeffcounting}.

At order $p^8$ and at leading order in $N_c$, we encounter the divergent
pion-pole contribution, diagrams (a) and (b) of Fig.~\ref{fig:EFT}, involving
two WZW vertices.
\begin{figure}[h]
\centering
\includegraphics[height=3cm]{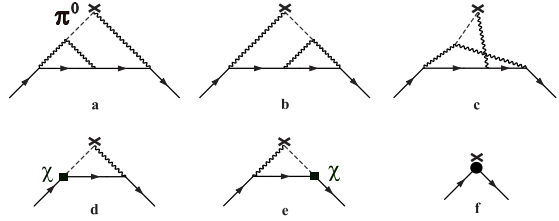}
\caption{The graphs contributing to $a_{\mu}^{\mbox{\tiny{LbyL;$\pi^0$}}}$ at
  lowest order in the effective field theory. 
\label{fig:EFT}}
\end{figure}
The diagram (c) is actually finite. The divergences of the triangular
subgraphs in the diagrams (a) and (b) are removed by inserting the counterterm
$\chi$ from the Lagrangian\footnote{The low-energy constant $\chi$ in this
effective Lagrangian should not be confused with the magnetic susceptibility
discussed earlier.} $\lag^{(6)} = (\alpha^2 / 4 \pi^2 F_\pi) \ \chi \
{\overline\psi} \gamma_\mu \gamma_5 \psi \, \partial^\mu \pi^0 + \cdots$,
leading to the one-loop diagrams (d) and (e). Finally, there is an overall
divergence of the two-loop diagrams (a) and (b) that is removed by a magnetic
moment type counterterm, diagram (f).  Since the EFT involves such a local
contribution, we will not be able to give a precise numerical prediction for
$a_\mu^\mathrm{LbL;had}$.  Nevertheless, it is interesting to consider the
leading and next-to-leading logarithms that are in addition enhanced by a
factor $N_c$ and which can be calculated using the renormalization
group~\cite{KNPdeR01}. The EFT and large-$N_c$ analysis tells us that
\ba
a_\mu^\mathrm{LbL;had} & = & 
\left( {\alpha \over \pi} \right)^3  \Bigg\{
f\left({m_{\pi^\pm} \over m_\mu}, {m_{K^\pm} \over m_\mu}\right)
 + N_c \left( {m_\mu^2 \over 16 \pi^2
F_\pi^2} {N_c \over 3} \right)
\left[ \ln^2 {\mu_0 \over m_\mu} + c_1 \ln {\mu_0 \over m_\mu} + c_0
\right]  \nonumber \\ 
&&\qquad\quad + {\cal O} \left(\!{m_\mu^2 \over \mu_0^2} \times
\mbox{log's}\!\right) + {\cal O} \left(\!{m_\mu^4 \over \mu_0^4} 
N_c \times \mbox{log's}\!\right)\!\!\Bigg\},  \label{a_mu_EFT_N_C}
\ea
where $f(m_{\pi^\pm} / m_\mu, m_{K^\pm} / m_\mu) = -0.038$ represents the
charged pion and Kaon-loop that is formally of order one in the chiral and
$N_c$ counting and $\mu_0$ denotes some hadronic scale, e.g.\ $M_\rho$.  The
coefficient ${\cal C} = (N_c^2 m_\mu^2) / (48 \pi^2 F_\pi^2) \simeq 0.025$ for
$N_c=3$ of the log$^2$ term is universal~\cite{KnechtNyffeler01,KNPdeR01} and
of order $N_c$, since $F_\pi = {\cal O}(\sqrt{N_c})$.

Unfortunately, although the logarithm is sizeable,
$\ln(M_\rho/m_\mu)\simeq1.98$, in $a_\mu^\mathrm{LbL;had}$ there occurs a
cancellation between the log-square and the log-term. In
Ref.~\cite{Nyffeler_Montpellier_02} the result for the VMD form factor for
large $M_\rho$ was fitted to an expression as given in
Eq.~(\ref{a_mu_EFT_N_C}), with the outcome (taking only the diagrams in
Fig.~\ref{fig:lbldiagabc}(a) and (b) into account, which diverge for $M_\rho
\to \infty$)
\ba
a_{\mu;\mathrm{VMD}}^{\mathrm{LbL};\pi^0}
& \doteq & \left( {\alpha \over \pi} \right)^3 {\cal C} 
\ \ \left[ \ln^2 {M_\rho \over m_\mu} + c_1 \ln {M_\rho \over m_\mu} + c_0
\right]   \nonumber \\
& \stackrel{\mbox{\tiny{Fit}}}{=} & \left( {\alpha \over  \pi}
\right)^3 {\cal C}  
\ \ \left[ 3.94 - 3.30 + 1.08 \right] = \left[ 123 - 103 + 34 \right] \times
10^{-11} = 54 \times 10^{-11} \, . 
\ea
This behavior is confirmed by the analytical result derived in
Ref.~\cite{BCM02} in terms of a series expansion in $\delta = (m_\pi^2 -
m_\mu^2) / m_\mu^2$ and $m_\mu^2 / M_\rho^2$. Collecting all terms
proportional to log-square and log, separately, one obtains
$a_{\mu;\mathrm{VMD}}^{\mathrm{LbL};\pi^0} = [136 - 112 + 30] \times 10^{-11}
= 54 \times 10^{-11}$. Note that the coefficient of $\ln^2 (M_\rho / m_\mu)$
in the expansion given in Ref.~\cite{BCM02} also contains corrections of order
$m_\mu^2 / M_\rho^2$, which are not included in the universal term
proportional to ${\cal C}$ in Eq.~(\ref{a_mu_EFT_N_C}). This cancellation
between the different logarithmically enhanced contributions is also visible
in Ref.~\cite{RMW02}. In that paper the remaining parts of $c_1$ have been
calculated: $c_1 = - 2 \chi(\mu_0) / 3 + 0.237 = -0.93^{+0.67}_{-0.83}$, with
our conventions for $\chi$ and $\chi(M_\rho)_{{\rm exp}} =
1.75^{+1.25}_{-1.00}$~\cite{Ametller_01}.

Finally, the EFT analysis shows that the modeling of hadronic light-by-light
scattering by a constituent quark loop, as suggested in 
Refs.~\cite{Pivovarov:2001mw,Erler:2006vu} (see also~\cite{BRinDafneHB95}), 
is not consistent with QCD.\footnote{In any case, any kind of quark loop
fails to explain the observation reproduced in Fig.~\ref{fig:LBLfacts},
which requires an effective description in terms of hadrons as
illustrated in Fig.~\ref{fig:LbLviahadrons}.} The latter
has a priori nothing to do with the full ``quark loop'' in QCD which is dual
to the corresponding contribution in terms of hadronic degrees of freedom.
Equation~(\ref{a_mu_EFT_N_C}) tells us that at leading order in $N_c$ any
model of QCD has to show the behavior $a_\mu^\mathrm{LbL;had} \sim
(\alpha/\pi)^3 N_c [N_c m_\mu^2 / (48 \pi^2 F_\pi^2)] \ln^2\Lambda$, with a
universal coefficient, if one sends the cutoff $\Lambda$ to infinity. From the
analytical result given in Ref.~\cite{LR93}, one obtains the result
$a_\mu^\mathrm{LbL;CQM} \sim (\alpha/\pi)^3 N_c (m_\mu^2 / M_Q^2) + \ldots$,
for $M_Q \gg m_\mu$, if we interpret the constituent quark mass $M_Q$ as a
hadronic cutoff.  Even though one may argue that $N_c / (48 \pi^2 F_\pi^2)$
can be replaced by $1/M_Q^2$, the log-square term is not correctly reproduced
with this model. Therefore, the constituent quark model (CQM) cannot serve as
a reliable description for the dominant contribution to
$a_\mu^\mathrm{LbL;had}$, in particular, its sign. Note that the contribution
of the quark-loop (within the CQM) to $a_\mu$ starts at order $p^8$, i.e.\ it
is of the same chiral order as the pseudoscalar-exchanges and not ${\cal
O}(p^2)$ higher as suggested by the counting in Table~\ref{tab:LEeffcounting}
based on Ref.~\cite{deRafaelENJL94}.

\section{Electroweak Corrections}
\label{sec:weak}
The contribution of weak virtual processes to $g-2$ has been of
interest long before one was actually able to unambiguously calculate
them and before they were playing a role in a comparison with the
experiment. After the renormalizability of the electroweak SM had been
established by 't~Hooft in 1971~\cite{tHooft71} it was possible to make
convincing predictions for $\amu$ beyond QED~\cite{EW1Loop}. The
sensitivity of the last CERN experiment was far from being able to
check the prediction and the weak contribution actually was one of the
motivations to think about a new muon $g-2$ experiment. The test of
the weak contribution is actually one of the milestones achieved by
the Brookhaven experiment E821. The weak contribution
now is almost three standard deviations, and without it the deviation
between theory and experiment would be at the 6 $\sigma$ level.

 \subsection{1-loop Contribution}
\label{weak1}
The leading weak contribution diagrams are shown in
Fig.~\ref{fig:oneloopweak} in the unitary gauge. As $\amu$ is a
physical observable one can calculate it directly in the
non-renormalizable unitary gauge. In the latter only physical
particles are present and diagrams exhibiting Higgs ghosts and
Faddeev-Popov ghosts are absent. The first diagram of
Fig.~\ref{fig:oneloopweak} might be of particular interest as it
exhibits a triple gauge vertex. The coupling of the photon to the charged 
$W$ boson is of course dictated by electromagnetic gauge invariance.
\begin{figure}[h]
\centering
\includegraphics{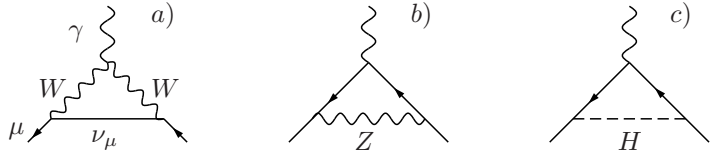}

\caption{The leading weak contributions to $a_\mu$; diagrams in the
physical unitary gauge.}
\label{fig:oneloopweak}
\end{figure}

\noi
In the approximation where tiny terms $O(m_\mu^2/M_{W,Z}^2)$ are neglected, 
the gauge boson contributions are given by~\cite{EW1Loop}
\ba
a^{(2)\:\mathrm{EW}}_\mu(W)&=&\frac{\wz G_\mu m_\mu^2}{16 \pi^2}\:\frac{10}{3}\simeq
+388.70(0) \power{-11}\cs \crn
a^{(2)\:\mathrm{EW}}_\mu(Z)&=&\frac{\wz G_\mu m_\mu^2}{16 \pi^2}\:\frac{(-1+4\,s_W^2)^2-5}{3}\simeq
-193.89(2) \power{-11} \epo
\label{oneloopbosonic}
\ea
For the Higgs exchange one finds
\ba
a^{(2)\:\mathrm{EW}}_\mu(H)&=&\frac{\wz G_\mu m_\mu^2}{4 \pi^2}\: \int\limits_{0}^{1}
dy\:\frac{(2-y)\:y^2}{y^2+(1-y)(m_H/m_\mu)^2} \crn
&\simeq& \frac{\wz G_\mu m_\mu^2}{4 \pi^2} \left\{\begin{array}{lcc}
\frac{m_\mu^2}{m_H^2}\:\ln \frac{m_H^2}{m_\mu^2 } &~~\mathrm{for}~~& m_H \gg m_\mu \\
\frac{3}{2} &~~\mathrm{for}~~& m_H \ll m_\mu \end{array} \right. \crn
 &\leq& 5 \power{-14}~~\mathrm{for}~~ m_H \geq  114~ \gv \cs
\label{SMhiggs}
\ea
and taking into account the LEP bound Eq.~(\ref{Hmassbound}), this is a negligible contribution.
Using the SM parameters given in Eqs.~(\ref{gmu}) and (\ref{sin2W}) we obtain
\ba
\amu^{(2)\:\mathrm{EW}} = (194.82 \pm 0.02) \times 10^{-11}\cs
\label{EW1l}
\ea
where the error is due to the uncertainty in $s_W^2$.

 \subsection{2-loop Contribution}
\label{ssec:weak2}
Typical electroweak 2--loop corrections are the electromagnetic
corrections of the 1--loop diagrams Fig.~\ref{fig:oneloopweak} (part
of the bosonic corrections) or fermionic loop insertions as shown in
Fig.~\ref{fig:twoloopweak}. All these corrections are
proportional to  
\be
\cK_2=\frac{\wz G_\mu\:m_\mu^2}{16 \pi^2}\:\frac{\alpha}{\pi}\simeq
2.70866237\power{-12}\epo
\label{EW2Coeff}
\ee
A first incomplete calculation was presented
by Kukhto, Kuraev, Schiller and Silagadze~\cite{KKSS92} in 1992.
Corrections found turned out to be enhanced
by very large logarithms $\ln M_Z/m_f$, which mainly come from fermion
triangular--loops like in Fig.~\ref{fig:twoloopweak}a. Note that due
to Furry's theorem in QED loops with three photons attached do not contribute 
and the $\gamma \gamma \gamma$--amplitude vanishes. This is different
if the parity violating weak interactions come into play. Contributions from the
two orientations of the closed fermion loops do not cancel and
the $\gamma \gamma Z$, $\gamma Z Z$ and $\gamma W W$ amplitudes do not
vanish. In fact in the $\gamma W W$ triangle charge conservation only
allows one orientation of the fermion loop.

\begin{figure}[h]
\centering
\includegraphics{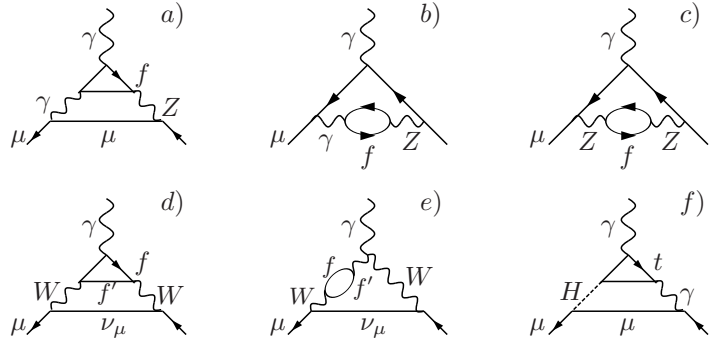}

\caption{Electroweak two--loop diagrams exhibiting fermion loops in the
unitary gauge, $f=(\nu_e,\nu_\mu,\nu_\tau,)\:e,\mu,\tau,u,c,t,d,s,b$ with
weak doublet partners $f'=(e,\mu,\tau,)\:\nu_e,\nu_\mu,\nu_\tau,d,s,b,u,c,t$.
The neutrinos (in brackets) do not couple directly to the
photon and hence are absent in the triangular subgraphs.}
\label{fig:twoloopweak}
\end{figure}

\noi
Diagrams  $a)$ and $b)$, with an internal photon, appear enhanced by a
large logarithm. While contributions from diagram $b)$ mediated by
$\gamma-Z$--mixing are suppressed by small vector coupling 
coefficients\footnote{The $Zf\bar{f}$
vector and axial--vector neutral current (NC) coupling coefficients are given by
\bea
v_f=T_{3f}-2 Q_f\, s_W^2 \;,\;\;\; a_f=T_{3f}\cs
\label{NCcouplings}
\eea
where $T_{3f}$ is the weak isospin $(\pm \frac{1}{2})$ of the fermion $f$\,.}
$(1-4\,s_W^2) \sim 0.1$ (quark loops) or $(1-4\,s_W^2)^2\sim 0.01$ (lepton loops), the
individual triangle fermion loops contributing to the $\gamma
\gamma Z$--vertex of diagram $a)$ lead to un-suppressed corrections
\ba
\amu^{(4)\:\mathrm{EW}}([f])\simeq \cK_2\:2T_{3f}N_{cf}Q^2_f\:
\biggl[3\ln \frac{M_Z^2}{m_{f'}^2} +C_f  \biggr]\cs
\label{KKSSeq}
\ea
in which $m_{f'}=m_\mu$ if $m_f \leq m_\mu$ and
$m_{f'}=m_f$ if $m_f > m_\mu$ and
\bea
C_f=\left\{ \begin{array}{ccl}
5/2                      &\mathrm{ \ for \ }& m_f < m_\mu \cs   \\
11/6-8/9\:\pi^2 &\mathrm{ \ for \ }& m_f = m_\mu \cs  \\
-6                        &\mathrm{ \ for \ }& m_f > m_\mu\epo
\end{array} \right.
\eea
The individual fermion $f$ contribution is proportional to
$N_{cf}Q_f^2a_f$. This is the coefficient of the triangular subdiagram
which exhibits the Adler-Bell-Jackiw (ABJ) or VVA
anomaly~\cite{ABJanomaly}, which must cancel if all fermions are
included~\cite{ABJcancel}. As we know, the \textit{anomaly
cancellation} enforces the known lepton--quark family structure of the
SM. In~\cite{KKSS92} only lepton loops were taken into account and
thus terms due to the quarks of similar size and structure were missed. 
 The anomaly cancellation condition of the SM reads
\be
{\sum}_f N_{cf} Q_f^2a_f=0\:,
\label{ABJCC}
\ee
and hence the leading short distance logarithms proportional to $\ln
M_Z$ are expected to cancel as well. This has been checked to happen
on the level of the quark parton model (QPM) for the 1st and 2nd fermion
family~\cite{PPdeR95,CKM95F,DG98}.

Assuming that we may use the not very well defined \textit{constituent quark
masses} from Eq.~(\ref{CQMasses}) with $M_u,M_d>m_\mu$, the QPM result for the
first family reads~\cite{CKM95F}
\ba
\amu^{(4)\:\mathrm{EW}}([e,u,d])_\mathrm{QPM}&\simeq&
-\cK_2\:\left[
\ln \frac{M_u^8}{m_\mu^6 M_d^2}+\frac{17}{2}\right]
\simeq -4.00 \power{-11},
\label{weak2QPM1st}
\ea
for the second family, with $M_s,M_c>m_\mu$, we obtain
\ba
\amu^{(4)\:\mathrm{EW}}([\mu,c,s])_\mathrm{QPM}&\simeq&
-\cK_2\:\left[
\ln \frac{M_c^8}{m_\mu^6 M_s^2}+\frac{47}{6}-\frac{8\pi^2}{9}\right]
\simeq -4.65 \power{-11}\epo
\label{weak2QPM2nd}
\ea

For the heavy quarks of the third family perturbation theory is applicable and 
the straight forward calculation yields the result~\cite{D'Hoker92,PPdeR95,CKM95F,KPPdeR02} 
\ba
\amu^{(4)\:\mathrm{EW}}([\tau,b,t])&=&
-\cK_2\:
\left[\frac{8}{3}
\ln \frac{m_t^2}{M^2_Z}-\frac{2}{9}\frac{M_Z^2}{m_t^2}\:\left(\ln
\frac{m_t^2}{M^2_Z}+\frac{5}{3}\right) 
+\ln \frac{M^2_Z}{m^2_b}+3 \ln
\frac{M^2_Z}{m^2_\tau}-\frac{8}{3}+~\cdots ~~\right] \crn
&\simeq&-\cK_2\:\times
30.3(3)\simeq-8.21(10) \power{-11} \epo
\label{QPMtbt}
\ea
Terms of order $m_\mu^2/m^2_\tau$, $m_b^2/M_Z^2$, $M_Z^4/m_t^4$
are small and have been neglected.

While the QPM results presented above, indeed confirmed the complete
cancellation of the $\ln M_Z$ terms for the 1st and 2nd family, in the
third family, with the given mass hierarchy, the corresponding terms 
$\ln M_Z/m_\tau$ and $\ln M_Z/m_b$ remain unbalanced as $m_t$ is
larger than $M_Z$ as first pointed out in Ref.~\cite{D'Hoker92}.

We want to stress that the fermionic loops with light-quarks ($u,d,s$) in
Fig.~\ref{fig:twoloopweak} are only meant as a symbolic representation of
another, genuinely non-perturbative hadronic contribution to the muon $g-2$,
similar to the hadronic vacuum polarization and the hadronic light-by-light
scattering contributions considered earlier. Below we will discuss the more
and more sophisticated approaches that have been used over the years in the
literature to control these hadronic uncertainties, going beyond the naive
QPM shown in Fig.~\ref{fig:twoloopweak}.

Improving on the constituent quark model used above, one can also look at the
$Z\gamma\gamma$ contribution in Fig.~\ref{fig:twoloopweak}a) from a purely
hadronic point of view, using chiral perturbation theory (CHPT) as the
low-energy effective field (EFT) theory of QCD. This approach was proposed in
Ref.~\cite{PPdeR95} and the corresponding Feynman diagrams are shown in
Fig.~\ref{fig:XPTuds}(a) and (b). To lowest order in the chiral expansion, the
hadronic $Z\gamma\gamma$ interaction is dominated by the pseudoscalar meson
(the quasi Goldstone bosons) exchange. The corresponding effective
couplings are given by
\be
\cL^{(2)}=- \frac{e}{2\sin \Theta_W \cos \Theta_W} F_\pi \partial_\mu
\left(\pi^0+\frac{1}{\sqrt{3}}\:\eta_8-\frac{1}{\sqrt{6}}\:\eta_0
\right)\:Z^\mu \cs
\ee
which is the relevant part of the $O(p^2)$ chiral effective Lagrangian,
and the effective $O(p^4)$ WZW Lagrangian Eq.~(\ref{WZW}).

\begin{figure}[t]
\centering
\includegraphics[height=2.75cm]{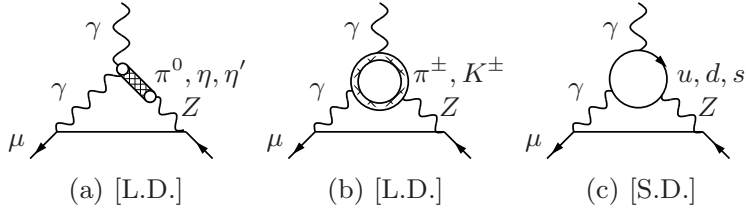}
\caption{The two leading EFT diagrams (L.D.) and the
QPM diagram (S.D.). The charged pion loop is subleading in $N_c$ and will
be discarded.}
\label{fig:XPTuds}
\end{figure}
The $[u,d,s]$ contribution is obtained with a long distance (L.D.) 
part ($E<M_\Lambda$) evaluated in the EFT and a short distance (S.D.) 
part ($E>M_\Lambda$) from Fig.~\ref{fig:XPTuds}(c) evaluated in the
QPM.  The cut--off for matching L.D. and S.D. part typically is
$M_\Lambda=m_P \sim 1~\gv$ to $M_\Lambda=M_\tau \sim 2~\gv$. The
diagrams from Fig.~\ref{fig:XPTuds}, together with their crossed
versions in the unitary gauge, yield, in the chiral limit and up to
terms suppressed by $m_\mu^2/M_\Lambda^2\,,$\footnote{The simplest way to
implement the cut--off $M_\Lambda$ of the low energy effective field
theory is to write for the $Z$--propagator
\bea
\frac{1}{M_Z^2+Q^2}=\underbrace{\frac{1}{M_\Lambda^2+Q^2}}_{EFT}
+ \underbrace{\frac{1}{M_Z^2+Q^2}-\frac{1}{M_\Lambda^2+Q^2}}_{QPM}
\eea
and by using the QPM for the second term. The first term corresponds
to Eq.~(20) of~\cite{PPdeR95}, later corrected in the constant term. In the first term $M_Z$ is
replaced by $M_\Lambda$, in the second term constant terms drop out in
the difference as the quark masses in any case have values far below
the cut--offs $M_\Lambda$ and $M_Z\,.$ For the actual calculation we
may use Eq.~(\ref{VVAtoamu}) below.}
\bea
\amu^{(4)\:\mathrm{EW}}([u,d,s];p<
M_\Lambda)_{\mathrm{EFT}}&=&
\cK_2\:\left[
\frac{4}{3} \ln \frac{M_\Lambda^2}{m_\mu^2}+\frac{2}{3}\right]
\simeq 2.10 \power{-11} \cs
\crn
\amu^{(4)\:\mathrm{EW}}([u,d,s];p>
M_\Lambda)_{\mathrm{QPM}}&=&
\cK_2\:\left[
2 \ln \frac{M_Z^2}{M_\Lambda^2}
\right]
\simeq 4.45 \power{-11} \epo
\eea
In Sect.~\ref{sssec:hadew} below, we will learn that the last diagram
of Fig.~\ref{fig:XPTuds} in fact takes into account the leading term
of Eq.~(\ref{wTgeneral}) which is protected by Vainshtein's relation
Eq.~(\ref{nonrentrans}).

Above a divergent term has been dropped, as the latter cancels against
corresponding terms from the complementary contributions from $e$,
$\mu$ and $c$ fermion--loops. Including the finite contributions from
$e$, $\mu$ and $c$~:
\bea
\amu^{(4)\:\mathrm{EW}}([e,\mu,c])_{QPM}&=&
-\cK_2\:\left[
6 \ln \frac{M_Z^2}{m_\mu^2}-4\ln \frac{M_Z^2}{M_c^2}+\frac{37}{3}
-\frac{8}{9}\pi^2\right]
\crn &  \simeq  & - \cK_2\:\times 51.83 \simeq -14.04 \power{-11}\cs
\eea
the complete answer for the 1st plus 2nd family reads~\cite{PPdeR95}
\ba
\amu^{(4)\:\mathrm{EW}}\left(\left[\begin{tabular}{c}$e,u,d$ \\
$\mu,c,s$ \end{tabular}\right]\right)&=&
-\cK_2\:\left[
\frac{14}{3} \ln \frac{M_\Lambda^2}{m_\mu^2}-4\ln \frac{M_\Lambda^2}{M_c^2}
+\frac{35}{3}
-\frac{8}{9}\pi^2\right]
\crn & \simeq  & - \cK_2\:\times 27.58(46) \simeq -7.47(13) \power{-11} \epo
\label{CHPTeudmsc}
\ea
In Eq.~(\ref{CHPTeudmsc}) the error comes from varying the cut--off
$M_\Lambda$ between 1 GeV and 2 GeV. Below about 1 GeV, the
calculation within the EFT can be trusted, above 2 GeV we can use
pQCD. Fortunately the result is not very sensitive to the choice of
the cut--off. Nevertheless, the mismatch of the cut--off dependencies
of the L.D. and the S.D. parts is a problem and gives raise to worries
about the reliability of the estimate. Therefore, a refined treatment
of these effects is discussed in the following, beginning with a
consideration of the general structure of these contributions.

\subsubsection{Hadronic Effects in Weak Loops and the Triangle Anomaly}
\label{sssec:hadew}
In order to discuss the contribution from VVA
triangle fermions loops, following~\cite{CMV03},
one has to consider the $Z^*\gamma\gamma^*$ amplitude
\ba
T_{\nu \lambda}=\I \int\: \D^4x \:\E^{\I q x}
\bra{0}T\{j_\nu(x)\:j_{5\lambda}(0)\} \ket{\gamma(k)}\cs
\label{ammCCme}
\ea
which  we need for small $k$ up to quadratic terms. The corresponding
covariant decomposition
\ba
T_{\nu \lambda}&=&-\frac{\I\:e}{4\pi^2}\:\left[
w_T(q^2)\:(-q^2\tilde{f}_{\nu \lambda}+q_\nu q^\alpha
\tilde{f}_{\alpha \lambda}-q_\lambda q^\alpha \tilde{f}_{\alpha \nu})+
w_L(q^2)\:q_\lambda q^\alpha
\tilde{f}_{\alpha \nu}
\right]\cs
\label{twotensordec}
\ea
exhibits two terms, a transversal amplitude $w_T$ and a longitudinal one
$w_L$, with respect to the axial current index $\lambda$.

The $g-2$ contribution $a_\mu^{(4)\:\mathrm{EW}}([f])_{\mathrm{VVA}}$
of a fermion $f$ in the $Z^*\gamma \gamma^*$ amplitude, in the unitary
gauge with $Z$ propagator $\I\:(-g_{\mu \nu} +q_\mu
q_\nu/M_Z^2)/(q^2-M_Z^2)$, is given by\footnote{ Since the result does
not depend on the direction of the external muon momentum $p$ we may
average over the 4-dimensional Euclidean sphere which yields the exact
1--dimensional integral representation given.}
\ba
a_\mu^{(4)\:\mathrm{EW}}([f])_{\mathrm{VVA}}&=&
\cK_2\:~\I~
\int\:\D^4 q \:\frac{1}{q^2+2qp}\:\biggl[
\frac{1}{3}\:\left(1+ \frac{2(qp)^2}{q^2m_\mu^2} \right) 
\left(w_L- \frac{M_Z^2}{M_Z^2-q^2}\:w_T \right) +
\frac{M_Z^2}{M_Z^2-q^2}\:w_T \biggr] \crn 
&=&\cK_2
\int_{0}^{\Lambda^2}\! \D Q^2\,\frac16\, \frac{Q^2}{m_\mu^2}\biggl\{
    w_L(Q^2) \, \left((Q^2/m_\mu^2-2)\,(1-R_m) + 2 \right)
\crn
&&\hspace*{25mm} - w_T(Q^2)\; \frac{M_Z^2}{M_Z^2+Q^2}\,
\left( (Q^2/m_\mu^2+4)\,(1-R_m) + 2\right)
\biggr\}\cs
\label{VVAtoamuexact}
\ea
in terms of the two scalar amplitudes $w_{L,T}(q^2)$.  $\Lambda$ is a
cutoff to be taken to $\infty$ at the end, after summing over a
family. We have performed a Wick rotation to Euclidean space with
$Q^2=-q^2$ and $R_m=\sqrt{1+4m_\mu^2/Q^2}$.  For leading estimates we
may expand in $m_\mu^2/Q^2\ll1$.  For contributions from the heavier
states it is sufficient to set $p=0$ except in the phase space where
it would produce an IR singularity. Including the leading corrections
the result takes the simple form
\ba
 a_\mu^{(4)\:\mathrm{EW}}([f])_{\mathrm{VVA}}&\simeq&
\cK_2\:
\int_{m_\mu^2}^{\Lambda^2}\!\!\! \D Q^2\:
\biggl\{
 w_L(Q^2) \, \left( 1 - \frac43\, m_\mu^2/Q^2 +\cdots \right)
\crn && \hspace*{16mm}+ 
w_T(Q^2)\, \frac{M_Z^2}{M_Z^2+Q^2}\,\left( 1 - \frac23\,
m_\mu^2/Q^2 +\cdots \right)
\biggr\}\epo
\label{VVAtoamu}
\ea
Interestingly, the Adler-Bardeen non--renormalization theorem valid
for the anomalous amplitude $w_L$ in full QCD (considering the quarks
$q=u,d,s,c,b,t$ only):
\ba
\left. w_L(Q^2)\right|_{m=0}=
\left. w^{1-\mathrm{loop}}_L(Q^2)\right|_{m=0}=  \sum_q (2T_qQ^2_q)\:  \frac{2 N_c}{Q^2}\cs
\label{wLexact}
\ea
carries over to the perturbative part of the transversal amplitude. In
fact, Vainshtein~\cite{Vainshtein03} has shown that in the chiral limit the relation
\ba
\left. w_T(Q^2)_\mathrm{pQCD}\right|_{m=0}=\ha
\left. w_L(Q^2)\right|_{m=0}
\label{nonrentrans}
\ea
is valid actually to all orders of perturbative QCD in the kinematical
limit relevant for the $g-2$ contribution.
This means that in the chiral limit the perturbative QPM result for $w_T$ is
exact in pQCD. This looks puzzling, since in low energy
effective QCD, which specifies the non--perturbative strong interaction
dynamics, this kind of term seems to be absent. The
non--renormalization theorem has been proven independently
in~\cite{KPPdR04} and was extended to the full off--shell triangle
amplitude to 2--loops in~\cite{JT05}. Note that corrections to Vainshtein's
relation Eq.~(\ref{nonrentrans}) must be of non--perturbative origin. 

A simple heuristic proof of Vainshtein's theorem proceeds by first
looking at the imaginary part of Eq.~(\ref{ammCCme}) and the covariant
decomposition Eq.~(\ref{twotensordec}). In accordance with the
Cutkosky rules the imaginary part of an amplitude is always more
convergent than the amplitude itself. The imaginary part of the
one--loop result is finite and one does not need a regularization to
calculate it unambiguously. In particular, it allows us to use
anti--commuting $\gafi$ to move it from the axial vertex
$\gamma_\lambda \gafi$ to the vector vertex $\gamma_\nu$. In the limit
$m_f=0$, this involves anti--commuting $\gafi$ with an even number of
$\gamma$--matrices, no matter how many gluons are attached to the
quark line joining the two vertices. As a result $\Impa T_{\nu
\lambda}$ must be symmetric under $\nu \leftrightarrow \lambda,~q
\leftrightarrow -q$:
\bea
&&\hspace*{-7mm} \Impa \left[
w_T(q^2)\:(-q^2\tilde{f}_{\nu \lambda}+q_\nu q^\alpha
\tilde{f}_{\alpha \lambda}-q_\lambda q^\alpha \tilde{f}_{\alpha \nu})+
w_L(q^2)\:q_\lambda q^\alpha
\tilde{f}_{\alpha \nu}
\right] \propto q_\nu q^\alpha \tilde{f}_{\alpha \lambda} +
q_\lambda q^\alpha \tilde{f}_{\alpha \nu}\cs
\eea
which, on the r.h.s., requires that $q^2=0$, to get rid of the
antisymmetric term proportional to $\tilde{f}_{\nu \lambda}$, and that
$w_T$ is proportional to $w_L$: $w_L=c\:w_T$; the symmetry follows
when $c=2$.  Thus the absence of an antisymmetric part is possible
only if
\ba
2\, \Impa w_T(q^2)=\Impa w_L(q^2)= \mathrm{constant}\:\times \,\delta(q^2)\cs
\label{ImpawTL}
\ea
where the constant is fixed to be $2\pi \cdot 2T_{3f}N_{cf}Q_f^2$ by the
exact form of $w_L$. Both $w_L$ and $w_T$ are analytic functions
which fall off sufficiently fast at large $q^2$ such that they satisfy
convergent DRs
\bea
w_{T,L}(q^2)=\frac{1}{\pi} \int_0^{\infty}\D s \frac{\Impa w_{T,L}(s)}{s-q^2}\cs
\eea
which together with Eq.~(\ref{ImpawTL}) implies Eq.~(\ref{nonrentrans}). 
According to the Adler-Bardeen non--renormali\-za\-tion theorem and by the
topological nature of the anomaly (see~\cite{Witten83}),
$w_L$ given by Eq.~(\ref{wLexact}) is exact beyond perturbation theory.
Vainshtein's non--renormalization theorem for $w_T$ in the chiral
limit implies
\be
w_T(q^2)=\frac{2T_{3f}N_{cf}Q_f^2}{Q^2}+ \mathrm{non-perturbative \  corrections}\epo
\label{wTgeneral}
\ee

Coming back to the calculation of Eq.~(\ref{VVAtoamu}), we observe that
the contribution from $w_L$ for individual fermions is
logarithmically divergent, but it completely drops for a complete
family due to the vanishing anomaly cancellation coefficient. The
contribution from $w_T$ is convergent for individual fermions due to
the damping by the $Z$ propagator. In fact it is the leading $1/Q^2$
term of the $w_T$ amplitude which produces the $\ln \frac{M_Z}{m}$
terms. However, the coefficient is the same as the one for the
anomalous term and thus for each complete family also the $\ln M_Z$
terms must drop out. Since the leading perturbative contributions
have to cancel the non-perturbative contributions to $w_T$ which are not
constraint by the anomaly cancellation condition require special
attention. Non-perturbative effects are accessible in a systematic
manner via the OPE.

\subsubsection{Non--perturbative Effects via the OPE}
\label{sssec:OPE}
In order to study further the matrix element Eq.~(\ref{ammCCme})
we need to look at the OPE of the two currents
\bea
\hat{T}_{\nu \lambda}&=&\I \int \D^4 x\:\E^{\I qx}\:T\{j_\nu(x)\:j_{5\lambda}(0)\}
=\sum\limits_{i}c^i_{\nu \lambda \alpha_1\ldots \alpha_i}(q)\:\cO_i^{\alpha_1\ldots \alpha_i}\cs
\eea
where the operators $\cO$ are local operators constructed from the
light fields, the photon, light quarks and gluon fields. The
operator matrix elements describe the non--perturbative long range
strong interaction features while the perturbatively calculable Wilson 
coefficients $c^i$ encode the short distance properties. We are
concerned with the matrix element
\ba
T_{\nu \lambda}&=&\bra{0}\hat{T}_{\nu \lambda}\ket{\gamma(k)}
=\sum\limits_{i}c^i_{\nu \lambda \alpha_1\ldots \alpha_i}(q)\:
\bra{0}\cO_i^{\alpha_1\ldots \alpha_i}\ket{\gamma(k)}
\ea
in the classical limit $k \to 0$. The leading contribution
is linear in $\tilde{f}_{\alpha \beta}$ the dual of $f_{\alpha
\beta}=k_\alpha \veps_\beta- k_\beta \veps_\alpha$.  Therefore, only
those operators contribute which have the structure of an
antisymmetric tensor
\ba
\bra{0}\cO_i^{\alpha \beta}\ket{\gamma(k)}= - \I\: \frac{1}{4\pi^2}
\kappa_i\tilde{f}^{\alpha \beta}\epo
\label{MEforkappa3}
\ea
The constants $\kappa_i$ depend on the renormalization scale $\mu$.
Given the tensor structure Eq.~(\ref{twotensordec}),
the operators contributing to $T_{\nu \lambda}$ are of the form
\ba
T_{\nu \lambda}=\sum\limits_i \left\{ c_T^i(q^2)
\:(-q^2\cO^i_{\nu \lambda}+q_\nu q^\alpha
\cO^i_{\alpha \lambda}-q_\lambda q^\alpha \cO^i_{\alpha \nu})+
c^i_L(q^2)\:q_\lambda q^\alpha
\cO^i_{\alpha \nu}
\right\}\epo
\label{optwotensordec}
\ea
Consequently, we may write
\ba
w_{T,L}(q^2)=\sum\limits_i c^i_{T,L}(q^2,\mu^2)\:\kappa_i(\mu^2) \epo
\ea
In this expansion for large $Q^2=-q^2$ the relevance of the terms
is determined by the dimension of the operators, the low dimensional
ones being the most relevant, unless they vanish or are suppressed by
small coefficients due to exact or approximate symmetries, like chiral
symmetry. The functions we expand are analytic in the $q^2$--plane and
the asymptotic expansion for large $Q^2$ is a formal power series in
$1/Q^2$ up to logarithms. This implies that operators of odd dimension
produce terms proportional to the mass $m_f$ of the light fermion
field from which the operator is constructed. Thus, in the chiral limit only
antisymmetric operators of \textit{even} dimensions contribute.\\

In the following discussion of the different terms we will include the
factors $T_{3f}$ at the $Z^\lambda j_{5\lambda}(0)$ vertex (axial
current coefficient), $Q_f$ at the $A^\nu j_{\nu}(x)$ vertex (vector
current coefficient) and the color multiplicity factor $N_{cf}$. An
additional factor $Q_f$ (coupling to the external photon) comes in via
the matrix elements $\kappa_i$ of fermion operators $\bar{\psi}_f \cdots
\psi_f$.  Coefficients $\kappa_i$ which go with helicity flip operators
$\bar{\psi}_{fR}\cdots \psi_{fL}$ or $\bar{\psi}_{fL}\cdots \psi_{fR}$ are proportional to
$m_f$.\\

The leading operator is of dimension $d_{\cO}=2$ and corresponds to
the parity odd dual electromagnetic field strength tensor
\bea
\cO_F^{\alpha \beta}=\frac{1}{4\pi^2}\tilde{F}^{\alpha
\beta}=\frac{1}{4\pi^2}\veps^{\alpha \beta \rho \sigma} \partial_\rho
A_\sigma \epo
\eea
The normalization here is chosen such that $\kappa_F=1$ and hence
$w_{L,T}^F=c_{L,T}^F$.
The coefficient for this leading term is given by
the one--loop triangle diagram and yields
\be
c_L^F[f]=2c_T^F[f]=\frac{4T_{3f}N_{cf}Q_f^2}{Q^2}\:\left[1-\frac{2m_f^2}{Q^2}\:\ln
\frac{Q^2}{\mu^2}+O(\frac{m_f^4}{Q^4})\right]
\label{opeLO}\epo
\ee
Again, the leading $1/Q^2$ term cancels family--wise by quark--lepton
duality. We know that in the chiral limit this is the only contribution to
$w_L$.\\

The next to leading term is the $d_{\cO}=3$ operator given by
\bea
\cO_f^{\alpha \beta}=-\I \bar{f}\sigma^{\alpha \beta}\gafi f\equiv \ha
\veps^{\alpha \beta \rho \sigma}\bar{f}\sigma^{\rho \sigma} f \epo
\eea
Such helicity flip operators only contribute if chiral symmetry is
broken and the coefficients must be of the form
$c^f\propto m_f/Q^4$. The coefficients are determined by tree level
Compton scattering type diagrams and again contribute equally to
both amplitudes
\bea
c^f_L[f]=2c^f_T[f]=\frac{8T_{3f}Q_fm_f}{Q^4} \epo
\eea
For the sake of illustration, not taking into account soft strong
interaction effects, we may calculate the soft photon quark matrix
element in the QPM. The result is UV divergent and in the \MSb scheme
given by
\bea
\kappa_f=-Q_f\:N_{cf}\:m_f\: \ln \frac{\mu^2}{m_f^2} \epo
\eea
When inserted in the $d_{\cO}=3$ contribution to $w_T$ one gets
\bea
\Delta^{(d_{\cO}=3)}w_L=2\Delta^{(d_{\cO}=3)}w_T
=\frac{8}{Q^4}\sum\limits_fT_{3f}\:Q_f\:m_f\:\kappa_f
\eea
and thus recovers precisely the $1/Q^4$ term of
Eq.~(\ref{opeLO}). While this illustrated the use of the OPE we have
just reproduced the pQCD result.  However, in contrast to the leading
$1/Q^2$ term which is not modified by soft gluon interactions, i.e.,
$\kappa_F=1$ is exact, the physical $\kappa_f$ cannot be obtained from
pQCD. So far it is an unknown constant, in fact it is proportional to
the magnetic susceptibility $\chi$ of QCD~\cite{Ioffe_Smilga}, which
we have discussed in Sect.~\ref{sec:formfactors} before.  Here again,
the spontaneous breakdown of the chiral symmetry plays a key role. 
It implies the existence of the quark condensates $\langle
\bar{\psi}\psi \rangle_0\neq0$, which are non--vanishing in the chiral 
limit\footnote{Typically they take values
$\langle\bar{\psi}\psi\rangle_0 ~\simeq~ -(240~\mv)^3$}. Now, unlike
in perturbation theory, $\kappa_f$ need not be proportional to $m_f$.
In fact it is proportional to $\langle
\bar{\psi}\psi \rangle_0$. As the condensate is of dimensionality 3,
another quantity must enter carrying dimension of a mass and which is
finite in the chiral limit. In the $u,d$ quark sector this is either
the pion decay constant $F_0$ or the $\rho$ mass $M_{\rho^0}$. Since 
$\kappa_f$ is given by the matrix element Eq.~(\ref{MEforkappa3})  it must
be proportional to $N_{cf}Q_f$ such that
\bea
\kappa_f=N_{cf}Q_f \frac{\langle \bar{\psi}_f \psi_f \rangle_{0}}{F_0^2}
\eea
and hence~\cite{KPPdeR02,Vainshtein03}
\ba
\Delta^{(d_{\cO}=3)}\,w_L=2\,\Delta^{(d_{\cO}=3)}\,w_T
=\frac{8}{Q^4}\sum\limits_fN_{cf}T_{3f}\:Q^2_f\:
m_f\:\frac{\langle \bar{\psi}_f \psi_f \rangle_{0}}{F_0^2}\epo
\label{deltawL3}
\ea
The overall normalization is chosen such that it reproduces the
expansion of the non--perturbative modification of $w_L$, which
becomes proportional to the pion propagator beyond the chiral limit:
\bea
w_L=\frac{2}{Q^2+m_\pi^2} =\frac{2}{Q^2}-\frac{2m_\pi^2}{Q^4}+\cdots
\eea
We will come back to that point below.

All operators of $d_{\cO}=4$ yield terms 
suppressed by the light quark masses as $m_f^2/Q^4$ and vanish in the
chiral limit. Similarly the dimension $d_{\cO}=5$ operators
are contributing to the $1/Q^6$ coefficient but require a factor $m_f$ and thus again
are suppressed due to close-by chiral symmetry.\\

Interestingly the dimension $d_{\cO}=6$ operators play a more important
role. There is a term which is proportional to the quark condensates
and behaves like $1/Q^6$ and which gives a non--vanishing contribution
in the chiral limit. Such terms only contribute to the transversal
amplitude, and using estimates presented in~\cite{LMD98} one obtains
\ba
\Delta^{(d_{\cO}=6)}\,w_T(Q^2)_\mathrm{NP} \simeq - \frac{16}{9}\pi^2\:\frac{2}{F_0^2}\:
\frac{\alpha_s}{\pi}\frac{\langle \bar{\psi}\psi \rangle^2}
{Q^6}\cs
\ea
for large enough $Q^2$, the $\rho$ mass being the typical scale. This NP
contribution breaks the degeneracy $w_T(Q^2)=\ha w_L(Q^2)$ which
holds for the perturbative part only.

\noi
As a result the consequences of the OPE for the light quarks $u$, $d$ and $s$ in
the chiral limit may be summarized in the relations~\cite{CMV03}:
\ba
\label{wLwTcorrected}
w_L[u,d]_{m_{u,d}=0}&=&-3\:w_L[s]_{m_s=0}=\frac{2}{Q^2}\cs \nn \\
w_T[u,d]_{m_{u,d}=0}&=&-3\:w_T[s]_{m_s=0}=\frac{1}{Q^2}-
\frac{32 \pi \alpha_s}{9\: Q^6}\:\frac{\langle \bar{\psi}\psi
\rangle_0^2}{F_\pi^2}+O(Q^{-8}) \epo 
\ea
The condensates are fixed essentially by the 
Gell-Mann-Oakes-Renner (GOR)
relations
\bea
(m_u+m_d)\:\langle \bar{\psi}\psi
\rangle_0=-F_0^2m_\pi^2 \cs \crn
m_s\langle\bar{\psi}\psi\rangle_0 \simeq
-F_0^2M_K^2 \,,
\eea
and the last term of Eq.~(\ref{wLwTcorrected}) numerically estimates to
$$w_T(Q^2)_\mathrm{NP}\sim-\alpha_s\:(0.772~\gv)^4/Q^6\cs$$ i.e., the
scale is close to the $\rho$ mass. 

As a result non--perturbative corrections to the leading
$\pi^0,\eta,\eta'$ exchange contributions in $w_L$ require the
inclusion of vector--meson exchanges which contribute to $w_T$.
More precisely, for the transversal function the intermediate states
have to be $1^+$ mesons with isospin $1$ and $0$ or $1^-$ mesons with
isospin $1$. The lightest ones are $\rho$, $\omega$ and $a_1$. They
are massive also in the chiral limit.

In principle, the incorporation of vector--mesons, like the $\rho$, in
accordance with the basic symmetries is possible using the Resonance
Lagrangian Approach (RLA)~\cite{EckerCPT}, an extended form
of CHPT.  Like in the light-by-light scattering case discussed before, the
more recent 
analyses are modeling the hadronic amplitudes~\cite{deRafaelENJL94} in
the spirit of large $N_c$ QCD~\cite{tHooft74,Manohar01} where
quark--hadron duality becomes exact. The infinite series of narrow
vector states known to show up in the large $N_c$ limit is then
approximated by a suitable lowest meson dominance (LMD), i.e., amplitudes
are assumed to be saturated by known low lying physical states of
appropriate quantum numbers. This approach was adopted in an analysis
by the Marseille group~\cite{KPPdeR02}.  An analysis which takes into
account the complete structure Eq.~(\ref{wLwTcorrected}) was finalized
in~\cite{CMV03}.  In the narrow width approximation one may write the
ansatz
\ba
\Impa w_T=\pi {\sum}_i\: g_i \:\delta(s-m_i^2)\cs
\ea
where the weight factors $g_i$ satisfy
\ba
{\sum}_i\: g_i=1\;,\;\;{\sum}_i\: g_i m_i^2=0\cs
\ea
in order to reproduce Eq.~(\ref{wLwTcorrected}) in the chiral
limit. Corrections which show up beyond the chiral limit may be
implemented by modifying the second constraint such that they match 
the coefficients of the corresponding terms in the OPE.

While for leptons we have the amplitude
\bea
w_L[\ell]=-\frac{2}{Q^2}\css (\ell=e,\mu,\tau)\cs
\eea
the hadronic counterparts get modified by strong interaction effects as mentioned:
a sufficient number of states with appropriate weight factors has to be included in
order to be able to satisfy the S.D. constraints, obtained via the
OPE. Since the $Z$ does not have fixed parity both vector and axial
vector states couple (see Fig.~\ref{fig:XPTuds}a).
For the 1st family $\pi^0$, $\rho(770)$ and $a_1(1260)$ are taken
into account
\ba
w_L[u,d]&=&\frac{2}{Q^2+m_\pi^2}  \simeq
2\:\left(\frac{1}{Q^2}-\frac{m_\pi^2}{Q^4}+\cdots\right)\cs \crn
w_T[u,d]&=&\frac{1}{M_{a_1}^2-M_\rho^2}\bigg[
 \frac{M_{a_1}^2-m_\pi^2}{Q^2+M_\rho^2}
-\frac{M_\rho^2-m_\pi^2}{Q^2+M_{a_1}^2}
\bigg] \simeq
\left(\frac{1}{Q^2}-\frac{m_\pi^2}{Q^4}+\cdots\right)\cs
\ea
for the 2nd family $\eta'(960)$, $\eta(550)$, $\phi(1020)$ and
$f_1(1420)$ are included
\ba
w_L[s]&=&-\frac{2}{3}\bigg[
\frac{2}{Q^2+M_{\eta'}^2}-\frac{1}{Q^2+m_{\eta}^2}\bigg] \simeq
-\frac{2}{3}\left(\frac{1}{Q^2}-\frac{\tilde{M}_\eta^2}{Q^4}+\cdots\right)\cs \crn
w_T[s]&=&-\frac{1}{3}\frac{1}{M_{f_1}^2-M_\phi^2}\bigg[
 \frac{M_{f_1}^2-m_\eta^2}{Q^2+M_\phi^2}
-\frac{M_\phi^2-m_\eta^2}{Q^2+M_{f_1}^2}
\bigg] \simeq
-\frac{1}{3}\left(\frac{1}{Q^2}-\frac{m_\eta^2}{Q^4}+\cdots\right) \cs
\ea
with $\tilde{M}_\eta^2=2 M_{\eta'}^2-m_\eta^2$. The expanded forms allow
for a direct comparison with the structure of the OPE and reveal that
the residues of the poles have been chosen correctly.

While the contributions to $\amu$ from the heavier states may be
calculated using the simplified integral Eq.~(\ref{VVAtoamu}), for the
leading $\pi^0$ contribution we have to use Eq.~(\ref{VVAtoamuexact}),
which also works for $m_\pi \sim m_\mu$. The
results obtained for the 1st family reads~\cite{CMV03}\footnote{Up to
the common factor $\cK_2$ for pseudoscalar exchanges like  
$w_L(Q^2)=1/(Q^2+m_\pi^2)-1/(Q^2+M_Z^2)$ 
(Pauli-Villars regulated) one obtains the exact result
\bea
F_L(x )&=&\frac16\,\biggl(x \,(x +2) \,f(x)
 -x ^2\ln x +2 x +3\biggr)-\,\ln \frac{m_\mu^2}{M_Z^2}\cs
\eea
where
\bea
f(x)=\left\{\begin{tabular}{ccl} 
$-\sqrt{4/x -1}\,
\left(\arcsin\left(1-\frac{x }{2}\right)+\frac{\pi}{2}\right)$ &
for &
$x < 4$~ ($x=x_\pi$)$\cs$ \\
$\sqrt{1-4/x}\,
\ln \left(-2/(x \sqrt{1-4/x } -x +2)\right)$ &for& $x>4$~ ($x=x_\eta$)$\cs$ 
\end{tabular}\right. 
\eea
with  $x_\pi=m_\pi^2/m_\mu^2$, $x_\eta=m_\eta^2/m_\mu^2$ etc. and
$M_Z$ as a cut-off. For vector exchanges like $w_T(Q^2)=1/(Q^2+M_\rho^2)$ one obtains
\bea
F_T(m_\mu^2/M_\rho^2) &=&-\ln \frac{M_\rho^2}{M_Z^2}-\frac23 \, \frac{m_\mu^2}{M_\rho^2}\,\ln \left(
\frac{M_\rho^2}{m_\mu^2}+1\right) +O\left(\left(m_\mu^2/M_\rho^2\right)^2\right)\epo
\eea
Up to terms $O(m_\mu^2/M_Z^2)$ the result reads
\bea
 F_T(\frac{1}{x_\rho})&=& \frac16 \biggl\{
(x_\rho^2-6\,x_\rho)\,\ln x_\rho-2\,x_\rho-6\,\ln a +9 - x_\rho\,r_\rho\,\ln(-2/(r_\rho-x_\rho+2))
\\ &&\hspace*{-2mm}
-r_\rho\,\ln((x_\rho^4-r_\rho\,
(x_\rho^3-6\,x_\rho^2+10\,x_\rho-4)-8\,x_\rho^3+20\,x_\rho^2-16\,x_\rho+2)/2) \biggr\}\cs
\eea
with $x_\rho=M_\rho^2/m_\mu^2$, $r_\rho=\sqrt{x_\rho^2-4x_\rho}$ and $a=m_\mu^2/M_Z^2\;.$
}
\ba
\amu^{(4)\:\mathrm{EW}}([e,u,d])&\simeq&
-\cK_2\:
\biggl\{
\frac13\,\biggl(r_\pi \,(r_\pi +2) \,
\sqrt{\frac{4}{r_\pi}-1} 
\left[\arcsin \left(1-\frac{r_\pi}{2}\right)+\frac{\pi}{2}\right]
\crn &&
\hspace*{8mm}
 +r_\pi^2\ln r_\pi -2 r_\pi -3\biggr) 
+\ln \frac{M_\rho^2}{m_\mu^2}-\frac{M_\rho^2}{M_{a_1}^2-M_\rho^2}
\ln \frac{M_{a_1}^2}{M_\rho^2} +\frac{5}{2}\biggr\}\crn 
&\simeq&-\cK_2\:\times 8.49(74) = -2.30(20) \power{-11}\cs
\label{LOhad1st}
\ea
with $r_\pi=m_\pi^2/m_\mu^2$.
This may be compared with the QPM result Eq.~(\ref{weak2QPM1st}), which is
about a factor of two larger and once more illustrates the problem of
perturbative calculations in the light quark sector. For the 2nd family
after adding the $\mu$ and the perturbative charm contribution one obtains
\ba
\amu^{(4)\:\mathrm{EW}}([\mu,c,s])&\simeq&
-\cK_2\:\left[
 \frac{2}{3}\:\ln \frac{M_\phi^2}{M_{\eta'}^2}
-\frac{2}{3}\:\ln \frac{M_{\eta'}^2}{m_{\eta}^2}
\right. \crn && \left. 
\hspace*{8mm}
+\frac{1}{3}\frac{M_\phi^2-m_\eta^2}{M_{f_1}^2-M_\phi^2}\ln \frac{M_{f_1}^2}{M_\phi^2}
+4 \ln \frac{M_c^2}{M_\phi^2} + 3 \ln \frac{M_\phi^2}{m_\mu^2} -\frac{8\pi^2}{9}+\frac{59}{6}
 \right] \crn 
&\simeq& -\cK_2\:\times 17.25(1.10)
\simeq -4.67(30) \power{-11}\cs
\label{LOhad2nd}
\ea
which yields a result close to the one obtained with the QPM
Eq.~(\ref{weak2QPM2nd}). For the 2nd family the QPM estimate works
better due to the fact that the non--perturbative light $s$--quark
contribution is suppressed by a factor four relative to the $c$ due to
the different charge.

Altogether, for the 1st plus 2nd family, the large $N_c$ QCD inspired
LMD result is
\ba
\amu^{(4)\:\mathrm{EW}}\left(\left[\begin{tabular}{c}$e,u,d$ \\
$\mu,c,s$ \end{tabular}\right]\right)_{\mathrm{LMD}}
&\simeq&-\cK_2\:\times 25.74
\simeq -6.97 \power{-11}\,,
\label{LOhad12}
\ea
and turns out to be rather close to the very crude
estimate Eq.~(\ref{CHPTeudmsc}) based on separating L.D. and S.D. by a
cut--off in the range 1 to 2 GeV.\\

Note that numerically the differences of the different estimates (QPM,
EFT, large $N_c$) are not substantial.  Following~\cite{CMV03}, we adopt the 
specific forms discussed last in the following.

\subsubsection{Residual Fermion--Loop Effects}
So far unaccounted are sub--leading contributions which come from diagrams
$c),d),e)$ and $f)$ in Fig.~\ref{fig:twoloopweak}. They have been calculated in~\cite{CKM95F,DG98} with
the result
\ba
a^{(4)\:\mathrm{EW}}_{\mu\,\mathrm{NLL}}&=&
-\cK_2\:
\biggl\{\frac{1}{2s_W^2}\left[\frac{5}{8}\frac{m_t^2}{M_W^2}
+\ln \frac{m_t^2}{M_W^2}+\frac{7}{3} \right]
+\Delta C^{tH} \biggr\}\crn
&\simeq& -4.15(11) \power{-11}-(1.1^{-0.1}_{+1.4}) \power{-11}\cs
\label{NLLfermi}
\ea
where $\Delta C^{tH}$ is the coefficient from diagram $f)$
\bea
\Delta C^{tH}=\left\{\begin{tabular}{lc}
$\frac{16}{9}\ln \frac{m_t^2}{m_H^2}+\frac{104}{27}$ & $m_H \ll m_t$ , \\
$\frac{32}{3}\left(1-\frac{1}{\sqrt{3}}\mathrm{Cl}_2\left(\pi/3\right)
\right)$ &  $m_H = m_t$ , \\
$\frac{m_t^2}{m_H^2}\left(8+\frac{8}{9}\pi^2+\frac{8}{3}
\left(\ln \frac{m_H^2}{m_t^2}-1\right)^2 \right)$ & $m_H \gg m_t$ ,
\end{tabular} \right. 
\eea
with typical values $\Delta C^{tH}\!\!=(5.84,4.14,5.66)$ contributing to
Eq.~(\ref{NLLfermi}) by $(-1.58,-1.12,$ $-1.53) \power{-11}$, respectively, for
$m_H=(100,m_t,300)~\gv$.  The first term in Eq.~(\ref{NLLfermi}) is for
$\Delta C^{tH}=0$, the second is the $\Delta C^{tH}$ contribution
for $m_H=m_t$ with uncertainty corresponding to the range $m_H=100~\gv$
to $m_H=300~\gv$. 

\subsubsection{Bosonic Contributions}
In approximate form, the full electroweak bosonic corrections have
been calculated by Czarnecki, Krause and Marciano in
1995~\cite{CKM96B}.  At two loops, in the linear 't~Hooft gauge,
including fermion loops, there are 1678 diagrams to be considered, and
the many mass scales involved complicate the exact calculation
considerably. The calculation~\cite{CKM96B} has been performed by
asymptotic expansions in $(m_\mu/M_V)^2$ and $(M_V/m_H)^2$, where
$M_V=M_W$ or $M_Z$ and $m_H \gg M_V$.  The heavy mass expansion of
course substantially simplifies the calculation. As a further
approximation an expansion in the NC vector couplings was used. The
latter are suppressed like $(1-4\,s_W^2) \sim 0.1$ for quarks and
$(1-4\,s_W^2)^2 \sim 0.01$ for leptons.  As a result a
two--loop electroweak correction
\ba
a^{(4)\:\mathrm{EW}}_\mu(\mathrm{bosonic}) &=&\cK_2\:
\left(
\sum_{i=-1}^2
\left[ a_{2i}\, (s_W^{2})^i + \frac{M_W^2}{m_H^2}b_{2i}\, (s_W^{2})^i\right]
+ O(s_W^6)
\right)\crn &\simeq&
-21.4^{+4.3}_{-1.0} \power{-11}
\label{twoloopbos}
\ea
was found for $M_W=80.392~\gv$ ($s_W^2=1-M_W^2/M_Z^2$) and $m_H=250~\gv$
ranging between $m_H=100~\gv$ and $m_H=500~\gv$.  The expansion
coefficients are given in~\cite{CKM96B}.  The on mass--shell
renormalization prescription has been used and the one--loop
contributions in Eq.~(\ref{oneloopbosonic}) were parametrized in terms
of the muon decay constant $G_\mu$. This means that part of the
two--loop bosonic corrections have been absorbed into the lowest order
result. For the lower Higgs masses the heavy Higgs mass expansion is
not accurate and an exact calculation has been performed by
Heinemeyer, St\"ockinger and Weiglein~\cite{HSW04} and by Gribouk and
Czarnecki~\cite{GC05}. The result has the form
\ba
a^{(4)\:\mathrm{EW}}_\mu(\mathrm{bosonic}) &=&\cK_2\:
\left(c_L^{\mathrm{bos},2L} \ln \frac{m^2_\mu}{M_W^2}+c_0^{\mathrm{bos},2L}\right)\cs
\label{twoloopbosLL}
\ea
where the coefficient of the large logarithm $\ln
\frac{m_\mu^2}{M_W^2}\sim -13.27$ is given by the simple expression
\bea
c_L^{\mathrm{bos},2L}=\frac{1}{18}[107+23\:(1-4s_W^2)^2]\sim 5.96 \epo
\eea
While the leading term is simple, the Higgs mass dependent function
$c_0^{\mathrm{bos},2L}$ in its exact analytic form is rather unwieldy
and therefore has not been published. The numerical result
of~\cite{HSW04} was confirmed in~\cite{GC05}. The 2nd Ref. also
presents a number of semi--analytic intermediate results which give
more insight into the calculation.
\begin{figure}[th]
\vspace*{-2mm}
\centering
\IfFarbe{%
\includegraphics{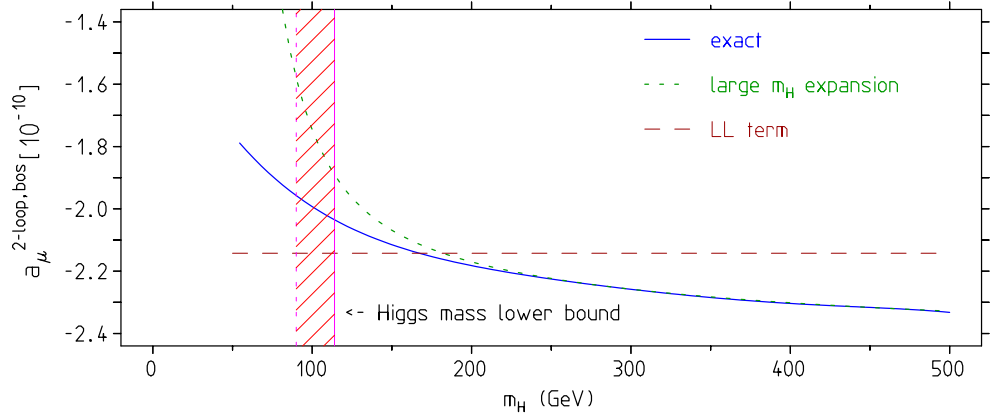}}{%
\includegraphics{amu2lbos.eps}}
\caption{Exact result for the bosonic correction vs.
the asymptotic expansion Eq.~(\ref{twoloopbos}) minus a correction
$0.88 \power{-11}$ and the LL approximation
(first term of Eq.~(\ref{twoloopbosLL})\,).}
\label{fig:amu2lbos}
\end{figure}
Considering a Higgs mass range $m_H=50~\gv$ to $m_H=500~\gv$, say, one may
expand the result as a function of the unknown Higgs
mass in terms of Tschebycheff polynomials defined on
the interval [-1,1].  A suitable variable is $x=(2m_H-550~\gv)/(450~\gv)$
and with the polynomials $t_1=1\cs t_2=x\cs t_{i+2}=2 x t_{i+1}-t_i\cs i=1,\cdots,4$,
we may approximate Eq.~(\ref{twoloopbosLL}) in the given range by
\ba
a^{(4)\:\mathrm{EW}}_\mu(\mathrm{bosonic}) &\simeq& \sum_{i=1}^6\:a_i\:t_i(x) \power{-10}\cs
\label{twoloopbosnumfit}
\ea
with the coefficients given by
$a_1=80.0483$, $a_2=8.4526$, $a_3= -3.3912$, $a_4= 1.4024$, $a_5=-0.5420$
and $a_6=0.2227$. The result is shown in Fig.~\ref{fig:amu2lbos},
which may be translated into
\ba
a^{(4)\:\mathrm{EW}}_\mu(\mathrm{bosonic}) &=& (-21.56^{+1.49}_{-1.05} ) \power{-11}\cs
\label{twoloopbosnum}
\ea
and applies for $m_H=100~\gv$ to $m_H=300~\gv\,.$ The central value is
obtained for $m_H=m_t\,.$ Note that the exact result exhibits a much
more moderate Higgs mass dependence at lower Higgs masses. This also
implies that the uncertainty caused by the unknown Higgs mass is
reduced considerably.\\

\noi
{\bf Summary of the Results for the Weak Contributions}\\
As a rough estimate
the perturbative 2--loop leading logs may be summarized in compact
form by~\cite{CKM96B,PPdeR95,CKM95F}, \cite{DG98,CMV03}
\ba
a^{(4)\:\mathrm{EW}}_{\mu\:{\rm LL}} &=&
-\cK_2
\:\biggl\{
 \left[\frac{215}{9}+\frac{31}{9} (1-4s_W^2)^2\right]\ln\frac{M_Z}{m_\mu}
\nonumber \\
&&  -
\sum_{f\in F} N_f Q_f \left[
12\:T^3_f\, Q_f -\frac{8}{9} \left( T^3_f - 2Q_f s_W^2\right)
\left( 1-4s_W^2\right)\right] \ln\frac{M_Z}{m_f} \biggr\}\:.
\label{eq17}
\ea
Electron and muon loops as  well  as non--fermionic loops produce  the
$\ln(M_Z/ m_\mu)$ terms in this expression  (the first line) while the
sum runs over  $F=\tau,u,d,s,c,b$. The logarithm $\ln(M_Z/m_f)$ in the
sum implies  that the fermion mass $m_f$  is larger  than $m_\mu$. For
the light quarks, such as $u,d$ and $s$, whose current masses are very
small, $m_f$ has a meaning of  an effective hadronic constituent mass.
In    this approximation $a^{(4)\:\mathrm{EW}}_{\mu\:{\rm   LL}}\simeq
-36.72   \power{-11}$,  which is to be compared   with  the full estimate
Eq.~(\ref{EW2l}),  below. Note   that  the $(1-4s_W^2)$  suppressed LL
terms from photonic corrections to diagram Fig.~\ref{fig:oneloopweak}b
[23/9 of the 31/9]  and Fig.~\ref{fig:twoloopweak}b [for $e$ and $\mu$
2$\times$ 4/9 and corresponding terms (2nd term) in the sum  over $f\in F$] only
account  a     negligible   contribution  $-31.54   \power{-13}$.  The
un-suppressed LL terms from Fig.~\ref{fig:twoloopweak}a [2$\times$ 54/9
of the 215/9 for $e$ and $\mu$  plus the corresponding terms (1st
term) in the sum
$f\in F$] in  the above expression cancel for  the 1st and 2nd fermion
family.  What survives   are the  terms  due  to  the virtual   photon
corrections     (bosonic)      of         the     1--loop     diagrams
Fig.~\ref{fig:oneloopweak}a,b  [120/9($W$) - 13/9($Z$) of the 215/9] and
the incomplete  cancellation in the 3rd fermion  family resulting as a
consequence  of the mass separation  pattern $m_\tau,  M_b \ll M_Z \ll
m_t$, relative to the effective cut--off $M_Z$.

The hadronic effects required a much more careful study which takes
into account the true structure of low energy QCD and as leading
logs largely cancel a careful study of the full 2--loop corrections was necessary.
The various weak contributions are collected in Table~\ref{tab:amuweak2}
and add up to the
\begin{table}
\centering
\caption{Summary of weak 2--loop effects in units $10^{-11}$. Fermion triangle loops: 1st, 2nd
and 3rd family LO, fermion loops NLL and bosonic loops (with equation numbers).}
\label{tab:amuweak2}
\begin{tabular}{r@{.}l|r@{.}l|r@{.}l|r@{.}l|r@{.}l}
\multicolumn{2}{c}{~}&\multicolumn{2}{c}{~}&\multicolumn{2}{c}{~}&\multicolumn{2}{c}{~}\\[-3mm]
\hline
 \ttcb{$[e u d]$~LO~Eq.~(\ref{LOhad1st})~}&\ttcb{~$[\mu s c]$~LO~Eq.~(\ref{LOhad2nd})~}& 
\ttcb{~$[\tau b t]$~LO~Eq.~(\ref{QPMtbt})~}& \ttcb{~NLL~Eq.~(\ref{NLLfermi})~} & \ttc{~bosonic~Eq.~(\ref{twoloopbosnum})} \\
\hline
~~ -2&30$\pm0.2$  &~~ -4&67$\pm0.3$ &~~ -8&21$\pm0.1$ &~~ -5&3$^{+0.1}_{-1.4}$ &~~ -21&6$^{+1.5}_{-1.0}$\\
\hline
\end{tabular}
\end{table}
total weak 2--loop contribution
\be
a^{(4)\:\mathrm{EW}}_{\mu}\simeq (-42.08 \pm 1.5[m_H,m_t]\pm 1.0
[\rm had]) \power{-11}\epo
\label{EW2l}
\ee
The high value $-40.98$ corresponds to low $m_H=100~\gv$,
the central value to $m_H=m_t$ and the minimum $-43.47$ to a high $m_H=300~\gv$
(see Eq.~(\ref{Hmassbound})).

Three--loop effects were studied by RG methods first
in~\cite{DG98}. The result
\be
a^{(6)\:\mathrm{EW}}_{\mu\:{\rm LL}}\simeq (0.4 \pm 0.2) \power{-11}
\label{LL3l}
\ee
was later confirmed by~\cite{CMV03}. The error estimates
uncalculated 3--loop contributions.

By adding up Eqs.~(\ref{EW1l}), (\ref{EW2l}) and (\ref{LL3l}) we find the
result\footnote{The result is essentially the same as
\bea
\amu^{\mathrm{EW}} = (154 \pm 1 [\rm had] \pm 2
[m_H,m_t,3-loop])\times 10^{-11}
\eea
of Czarnecki, Marciano and Vainshtein~\cite{CMV03}, which
also agrees numerically  with the one
\bea
\amu^{\mathrm{EW}} = (152 \pm 1 [\rm had])\times 10^{-11}
\eea
obtained by Knecht, Peris, Perrottet and de Rafael~\cite{KPPdeR02}.}
\be
\amu^{\mathrm{EW}} = (153.2 \pm 1.0 [\rm had] \pm1.5
[m_H,m_t,3-loop])\times
10^{-11}\cs
\label{EWfin}
\ee
based on~\cite{CMV03,HSW04,GC05}.

\subsection{2--loop electroweak contributions to $a_e$}
\label{ssec:elwea_a_e}
\noindent
The dominant electroweak 1--loop contributions Eq.~(\ref{oneloopbosonic}) scale 
with high precision with an overall factor $x_{(e\mu)}=(m_e/m_\mu)^2$,
up to terms which are suppressed with higher powers up to logarithms,
like the contribution from the Higgs Eq.~(\ref{SMhiggs}). Thus
\bea
a^{(2)\:\mathrm{EW}}_e=x_{(e\mu)}\,a^{(2)\:\mathrm{EW}}_\mu=45.57(0) \power{-15}\,.
\eea
At two loops various contributions do not scale in this
simple way~\cite{KKSS92,CKM95F,CKM96B}. We therefore present a set of
modified formulae, which allow us to calculate $a^{(4)\:\mathrm{EW}}_e$.
Apart from the overall factor 
\be
\cK_2 \to x_{(e\mu)} \cK_2\simeq
6.3355894\power{-17}\cs
\label{EW2Coeffe}
\ee
the logarithmically enhanced as well as some constant terms change according to
Eq.~(\ref{KKSSeq}), adapted for the electron.
We only present those terms which do not scale trivially.
The QPM results Eqs.~(\ref{weak2QPM1st}) and (\ref{weak2QPM2nd})
are modified to 
\ba
a_e^{(4)\:\mathrm{EW}}([e,u,d])_\mathrm{QPM}&\simeq&
-\cK_2\:\left[
\ln \frac{M_u^8}{m_e^6 M_d^2}+\frac{47}{6}-\frac{8\pi^2}{9}\right]
\simeq -2.36 \power{-15},
\label{weak2QPM1st_e}
\ea
\ba
a_e^{(4)\:\mathrm{EW}}([\mu,c,s])_\mathrm{QPM}&\simeq&
-\cK_2\:\left[
\ln \frac{M_c^8}{m_\mu^6 M_s^2}\right]
\simeq -1.15 \power{-15}\,,
\label{weak2QPM2nd_e}
\ea
for the 1st and 2nd family, respectively. The EFT/QPM estimates used in  Eq.~(\ref{CHPTeudmsc}) now
read
\bea
a_e^{(4)\:\mathrm{EW}}([u,d,s];p<
M_\Lambda)_{\mathrm{EFT}}&=&
\cK_2\:\left[
\frac{4}{3} \ln \frac{M_\Lambda^2}{m_e^2}+\frac{2}{3}\right]
\simeq 1.39 \power{-15} \cs
\crn
a_e^{(4)\:\mathrm{EW}}([u,d,s];p>
M_\Lambda)_{\mathrm{QPM}}&=&
\cK_2\:\left[
2 \ln \frac{M_Z^2}{M_\Lambda^2}
\right]
\simeq 1.04 \power{-15}\,,
\eea
and together with
 \bea
a_e^{(4)\:\mathrm{EW}}([e,\mu,c])_{QPM}&=&
- \cK_2\:\left[
3 \ln \frac{M_Z^2}{m_e^2}+3 \ln \frac{M_Z^2}{m_\mu^2}-4\ln \frac{M_Z^2}{M_c^2}+\frac{23}{6}
-\frac{8}{9}\pi^2\right]
\crn &  \simeq  & - \cK_2\:\times 75.32 \simeq -4.77 \power{-15}\cs
\eea
yield the complete estimate for the 1st plus 2nd family
\ba
a_e^{(4)\:\mathrm{EW}}\left(\left[\begin{tabular}{c}$e,u,d$ \\
$\mu,c,s$ \end{tabular}\right]\right)&=&
-\cK_2\:\left[
\frac{5}{3} \ln \frac{M_\Lambda^2}{m_e^2}+3 \ln \frac{M_\Lambda^2}{m_\mu^2}-4\ln \frac{M_\Lambda^2}{M_c^2}
+\frac{19}{6}
-\frac{8}{9}\pi^2\right]
\crn & \simeq  & - \cK_2\:\times 36.85(46) \simeq -2.33(3) \power{-15} \epo
\label{CHPTeudmsc_e}
\ea
The large $N_c$ QCD inspired LMD result Eq.~(\ref{LOhad1st}) for the
1st family translates into
\ba
a_e^{(4)\:\mathrm{EW}}([e,u,d])&\simeq&
-\cK_2\:
\biggl\{
\frac13\,\biggl(-r_\pi \,(r_\pi +2) \,
\sqrt{1-\frac{4}{r_\pi}} 
\left[\ln \frac{-2}{r_\pi\,\sqrt{1-\frac{4}{r_\pi}}-r_\pi+2}\right]
\crn &&
\hspace*{8mm}
 +r_\pi^2\ln r_\pi -2 r_\pi -3\biggr) 
+\ln \frac{M_\rho^2}{m_e^2}-\frac{M_\rho^2}{M_{a_1}^2-M_\rho^2}
\ln \frac{M_{a_1}^2}{M_\rho^2} -\frac{8\pi^2}{9}+\frac{11}{6}\biggr\}\crn 
&\simeq&-\cK_2\:\times 29.41(2.56) = -1.86(16) \power{-15}\cs
\label{LOhad1st_e}
\ea
with $r_\pi=m_\pi^2/m_e^2$. For the 2nd family Eq.~(\ref{LOhad2nd}) reads
\ba
a_e^{(4)\:\mathrm{EW}}([\mu,c,s])&\simeq&
-\cK_2\:\left[
 \frac{2}{3}\:\ln \frac{M_\phi^2}{M_{\eta'}^2}
-\frac{2}{3}\:\ln \frac{M_{\eta'}^2}{m_{\eta}^2}
\right. \crn && \left. 
\hspace*{8mm}
+\frac{1}{3}\frac{M_\phi^2-m_\eta^2}{M_{f_1}^2-M_\phi^2}\ln \frac{M_{f_1}^2}{M_\phi^2}
+4 \ln \frac{M_c^2}{M_\phi^2} + 3 \ln \frac{M_\phi^2}{m_\mu^2} +2
 \right] \crn 
&\simeq& -\cK_2\:\times 18.19(1.16)
\simeq -1.15(7) \power{-11}\epo
\label{LOhad2nd_e}
\ea

The LL approximation Eq.~(\ref{eq17}) for $a_e$ is given by
\ba
a^{(4)\:\mathrm{EW}}_{e\:{\rm LL}} &=&
-\cK_2
\:\biggl\{
 \left[\frac{161}{9}+\frac{27}{9} (1-4s_W^2)^2\right]\ln\frac{M_Z}{m_e}
\nonumber \\
&&  -
\sum_{f\in F} N_f Q_f \left[
12\:T^3_f\, Q_f -\frac{8}{9} \left( T^3_f - 2Q_f s_W^2\right)
\left( 1-4s_W^2\right)\right] \ln\frac{M_Z}{m_f} \biggr\} 
\nonumber \\
&\simeq&  -14.62\power{-15}\,,
\label{eq17_e}
\ea
where the sum extends over $F=\mu,\tau,u,d,s,c,b$.

Note that the contributions Eqs.~(\ref{QPMtbt}) and (\ref{NLLfermi}) scale with $x_{(e\mu)}$. 
The bosonic contributions only depend on the external fermion mass and
we may use the full 2--loop result Eq.~(\ref{twoloopbosnum}) together with Eq.~(\ref{twoloopbosLL})
to calculate $c_0^{\mathrm{bos},2L}$ which is equal for $\mu$
and $e$ and we obtain 
$a^{(4)\:\mathrm{EW}}_e(\mathrm{bosonic})=-1.02^{+0.35}_{-0.25}
\power{-15}$. Results are collected in Table~\ref{tab:amuweak2_e}.

\begin{table}
\centering
\caption{Summary of weak 2--loop effects contributing to $a_e$ in units $10^{-15}$. Fermion triangle loops: 1st, 2nd
and 3rd family LO, fermion loops NLL and bosonic loops (with equation
numbers, last 3 entries rescaled as described in the text).}
\label{tab:amuweak2_e}
\begin{tabular}{r@{.}l|r@{.}l|r@{.}l|r@{.}l|r@{.}l}
\multicolumn{2}{c}{~}&\multicolumn{2}{c}{~}&\multicolumn{2}{c}{~}&\multicolumn{2}{c}{~}\\[-3mm]
\hline
 \ttcb{$[e u d]$~LO~Eq.~(\ref{LOhad1st_e})~}&\ttcb{~$[\mu s c]$~LO~Eq.~(\ref{LOhad2nd_e})~}& 
\ttcb{~$[\tau b t]$~LO~Eq.~(\ref{QPMtbt})~}& \ttcb{~NLL~Eq.~(\ref{NLLfermi})~} & \ttc{~bosonic~Eqs.~(\ref{twoloopbosLL},\ref{twoloopbosnum})} \\
\hline
~~ -1&86$\pm0.16$  &~~ -1&15$\pm0.07$ &~~ -1&91$\pm0.02$ &~~ -1&09$\pm
0.19$ &~~~~ -1&02$^{+0.35}_{-0.25}$\\
\hline
\end{tabular}
\end{table}
As a result we obtain the total weak 2--loop contribution
\be
a^{(4)\:\mathrm{EW}}_{e}\simeq (-7.03 \pm 0.35[m_H,m_t]\pm 0.23
[\rm had]) \power{-15}\epo
\label{EW2l_e}
\ee
The total weak contribution thus is given by
\be
a^{\mathrm{EW}}_{e}\simeq (38.54 \pm 0.35[m_H,m_t]\pm 0.23
[\rm had]) \power{-15}\epo
\label{EWtot_e}
\ee
Note that the leading log approximation in Eq.~(\ref{eq17_e}) utilizing constituent quarks in this case is quite far
off from the result in Eq.~(\ref{EW2l_e}). Using this approximation we would get the smaller value
$a^{\mathrm{EW}}_{e}\simeq 30.95 \power{-15}$, which was used
frequently in the past. 
  
\section{Muon g-2: Theory versus Experiment}
\label{sec:thevsexp}
A new stage in testing theory and new physics scenarios has been
reached with the BNL muon $g-2$ experiment, which was able to reduce
the experimental uncertainty by a factor 14 to $\sim
63\power{-11}$. We already have summarized the experimental status in
Sect.~\ref{sec:results5}.  The world average experimental muon
magnetic anomaly, dominated by the very precise BNL result, now
is~\cite{BNLfinal}
\be
a_\mu^\mathrm{exp} =1.165 920 80(63) \times
10^{-3}
\label{EXPfinal}
\ee
(relative uncertainty $0.54$ppm), which confronts the SM prediction
(see Table~\ref{tab:theory})
\be
a_\mu^\mathrm{the} =1.165 917 90(65)
\times 10^{-3} \epo
\label{amuTH}
\ee
As ever before, but on a order of magnitude higher level, the \amm of
the muon provides one of the most precise tests of quantum field
theory as a basic framework of elementary particle theory and of QED
and the electroweak SM in particular. But not only that, it also
constrains physics beyond the SM severely.  In fact the $3.2\:\sigma$
deviation between theory and experiment
\be
\delta a_\mu^\mathrm{NP?}=a_\mu^\mathrm{exp}-a_\mu^\mathrm{the}=(290\pm 90)\: \power{-11}\cs
\label{amuNP}
\ee
could be a hint for new physics. Before we discuss possibilities to
explain this deviation assuming it to be \textit{a clear indication of
something missing}, we first will summarize the SM prediction and
recall what are the most relevant effects.

 \subsection{Standard Model Prediction}
In the previous sections, we have discussed in detail the various
contributions which enter the theoretical prediction of $a_\mu$.  We summarize
them in Table~\ref{tab:theory}.  Input parameters were specified in
Sect.~\ref{sec:gm2inQED} and Appendix~\ref{sec:appA}.
\begin{table}[t]
\begin{center}
\caption{Standard model theory and experiment
comparison [in units $10^{-11}$].
\label{tab:theory}}
\centering
\begin{tabular}{lr@{.}lr@{.}lcc}
&\ttc{}~&\ttc{}&&\\[-3mm]
\hline
&\ttc{}~&\ttc{}&&\\[-3mm]
Contribution & \multicolumn{2}{c}{Value} & \multicolumn{2}{c}{Error} &
~~~Equation~~~ & References \\
&\ttc{}~&\ttc{}&&\\[-3mm]
\hline\noalign{\smallskip}
QED incl. 4-loops+LO 5-loops & 116\,584\,718&1 & 0&2 & (\ref{QEDfin})
&  \cite{QEDall} \\
Leading hadronic vacuum polarization & 6\,903&0 & ~~52&6 &
(\ref{amuhadLO}) &  \cite{FJ08}\\
Subleading hadronic vacuum polarization & -100&3 & 1&1 &
(\ref{amuhadHO}) &  \cite{FJ06} \\
Hadronic light--by--light &  116&0 & 39&0 & (\ref{JNLBL}) &\cite{hadLBL}\\
Weak incl. 2-loops & 153&2 & 1&8 & (\ref{EWfin}) & \cite{weak2}\\
&\ttc{}~&\ttc{}&\\ 
Theory        & 116\,591\,790&0 & 64&6 & -- &  \\
Experiment & 116\,592\,080&0 & 63&0 & (\ref{amuBNL}) & \cite{BNLfinal} \\
Exp. - The. ~~  3.2 standard deviations & 290&0 & 90&3 & -- & \\ \noalign{\smallskip}\hline
\end{tabular}
\end{center}
\end{table}
What we notice is that a new quality of ``diving into the depth of
quantum corrections'' has been achieved: the 8th order QED [$\sim 381
\power{-11}$], the weak correction up to 2nd order [$\sim 153 \power{-11}$] and the
hadronic light--by--light scattering [$\sim 116 \power{-11}$] are now in
the focus. The hadronic vacuum polarization effects which played a
significant role already for the last CERN experiment now is a huge
effect of more than 11 SD's. As a non--perturbative effect it still
has to be evaluated largely in terms of experimental data with
unavoidable experimental uncertainties which yield the biggest
contribution to the uncertainty of theoretical predictions.
\begin{figure}[t]
\centering
\IfFarbe{%
\includegraphics[height=9.4cm]{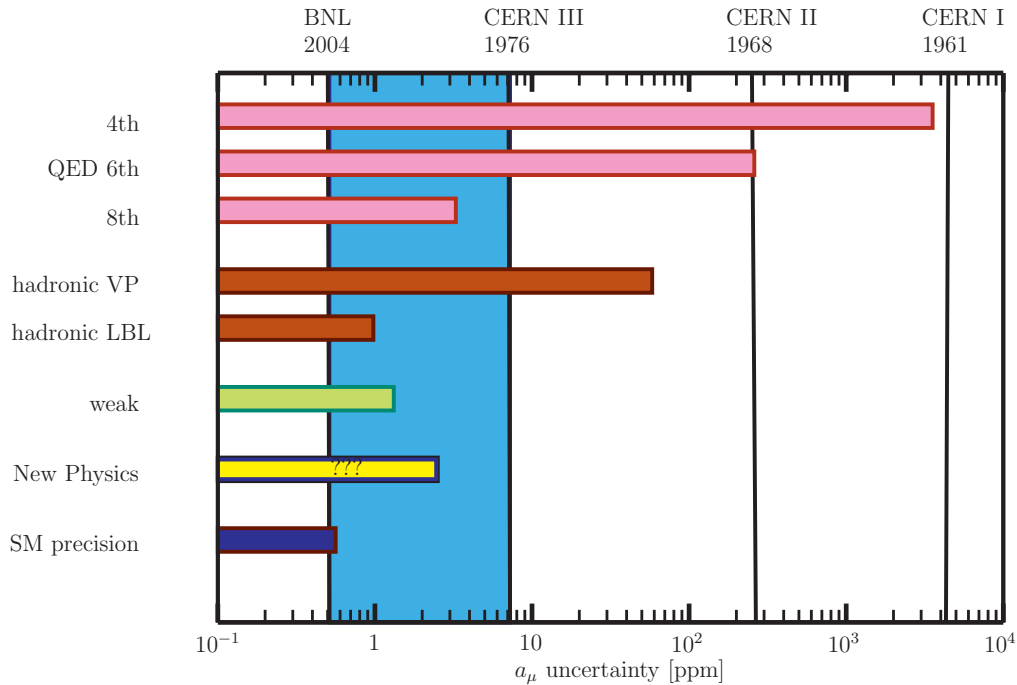}}{%
\includegraphics[height=9.4cm]{amucontrib.eps}}
\caption{Sensitivity of $g-2$ experiments to various contributions. The
increase in precision with the BNL $g-2$ experiment is shown as a gray
(blue) vertical band. New Physics is illustrated by the deviation
$(a_\mu^\mathrm{exp}-a_\mu^\mathrm{the})/a_\mu^\mathrm{exp}\,.$}
\label{fig:amucontrib}
\end{figure}
However, due to substantial progress in the measurement of total hadronic
$\epm$--annihilation cross--sections, the uncertainty from this source
has reduced to a remarkable $\sim 53
\power{-11}$ only. This source of error now is only slightly larger
than the uncertainty in the theoretical estimates of the hadronic
light--by--light scattering contribution [$\sim 39
\power{-11}$]. Nevertheless, we have a solid prediction with a total
uncertainty of $\sim 65 \power{-11}$, which is essentially equal to
the experimental error of the muon $g-2$ measurement. 

Fig.~\ref{fig:amucontrib} illustrates the sensitivity to various
contributions and how it developed in history. The high sensitivity of
$a_\mu$ to physics from not too high scales $M$ above $m_\mu$, which
is scaling like $(m_\mu/M)^2$, and the more than one order of
magnitude improvement of the experimental accuracy has raised many
previously negligible SM contributions to relevance.  We also have reached
an exciting level of sensitivity to New Physics. ``New Physics'' is
displayed in the figure as the ppm deviation of Eq.~(\ref{amuNP})
which is $3.2\:\sigma$. We note that the theory error is somewhat
larger than the experimental one. It is fully dominated by the
uncertainty of the hadronic low energy cross--section data, which
determine the hadronic vacuum polarization and, partially, from the
uncertainty of the hadronic light--by--light scattering
contribution. 

In any case we now have a much more detailed test of the present established 
theory of the fundamental forces and the particle spectrum than we had
before the BNL experiment. At the same time the muon $g-2$ provides insight 
to possible new physics entering at scales below about 1 TeV.
For what concerns the interpretation of the actual deviation between theory and
experiment, we have to remember that such high precision
physics is extremely challenging for both experiment and for theory
and it is not excluded that some small effect has been overlooked or
underestimated at some place. To our present knowledge, it is hard to
imagine that a 3 $\sigma$ shift could be explained by known physics or
underestimated systematic uncertainties, theoretical and/or
experimental. Thus New Physics seems a likely interpretation, if it is
not an experimental fluctuation (0.27\% chance).

It should be noted that the result Eq.~(\ref{amuTH}) is obtained when
relying on the published $\epm$--data for the evaluation of the
hadronic vacuum polarization. If isospin rotated hadronic
$\tau$--decay spectral functions, corrected for known isospin
violations, are included, a substantially larger value for $\amuh$
results: $\delta a_\mu(+\tau) \sim 150 \power{-11}$ and the
``discrepancy'' Eq.~(\ref{amuNP}) reduces to about 1.4 $\sigma$ only,
which would mean that there is agreement between theory and
experiment.  However, as pointed out in Ref.~\cite{Passera:2008jk}
recently, an increase of the hadronic vacuum polarization would also
increase the value of $\alpha(M_Z)$ and as a consequence lower the
indirect Higgs mass bound from LEP precision experiments. In fact the
indirect upper Higgs mass bound, $m_H < 153~\gv$ would move to $m_H<
133~\gv$, closer to be in conflict with the direct exclusion bound of
$m_H > 114~\gv$. This possibility in any case would lead to an
interesting tension for the Standard Model to be in conflict with
experimental facts.

Note that the theoretical predictions obtained by different authors in
general differ by the leading order hadronic vacuum polarization
contribution listed in Tab.~\ref{tab:otheramuhad1} and/or by a
different choice of the hadronic light--by--light scattering
contribution which we have collected in Tab.~\ref{tab:LbLrecent}. The
deviation between theory and experiment then ranges from 0.7 to 4.2
$\sigma$'s. The smallest difference is obtained when including the
isospin rotated $\tau$--data in calculating $\amuh$, as
in~\cite{DEHZ03}, together with the LbL estimate~\cite{MV03}, the
largest using the $\amuh$ estimate~\cite{HMNT06} together with the LbL
estimate~\cite{KnechtNyffeler01} (also see Fig.~7.1 in Ref.~\cite{Jegerlehner:2008zz}).

 \subsection{New Physics Contributions}
\label{ssec:NPcontibution}
Although the SM is very well established as a renormalizable QFT and
describes essentially all experimental data of laboratory and
collider experiments, it is well established that the SM is
\textit{not} able to explain a number of fundamental facts. The SM
fails to account for the existence of non--baryonic cold dark matter (at
most 10\% is normal baryonic matter), the matter--antimatter asymmetry
in the universe, which requires baryon--number $B$ and lepton--number
$L$ violation at a level much higher than in the SM, the
problem of the cosmological constant and so on. Also, a ``complete''
theory should include the 4th force of gravity in a natural way and
explain the huge difference between the weak and the Planck scale
(hierarchy problem). So, new physics is there but how is it realized?
What can the muon $g-2$ tell us about new physics?

New physics contributions, which we know must exist, are part
of any measured number. If we confront an accurately predictable
observable with a sufficiently precise measurement of it, we should be
able to see that our theory is incomplete. New physics is due to
states or interactions which have not been seen by other experiments,
either by a lack of sensitivity or, because the new state was too
heavy to be produced at existing experimental facilities or, because
the signal was still buried in the background.  At the high energy
frontier LEP and the Tevatron have set limits on many species of
possible new particles predicted in a plenitude of models which extend
the SM. The Particle Data Group~\cite{PDG06} includes a long list of
possible states which have not been seen, which
translates into an experimental lower bound for the mass. In contrast
to the direct searches at the high energy frontier, new physics
is expected to change $a_\mu$ indirectly, by virtual 
loop--contributions. In general, assuming Eq.~(\ref{amuNP}) to be a true
effect, the result allows to constrain the parameter space of
extensions of the SM.

The simplest possibility is to add a 4th
fermion family of sequential fermions, where the neutrino has to
have a large mass ($m_{\nu'}> 45$~GeV) as additional light (nearly massless)
neutrinos have been excluded by LEP. The present bounds read 
$m_L> 100$~GeV for a heavy lepton and  $m_{b'}\, \gapprox\, 200$ GeV for a heavy quark.

Similarly, there could exist additional
gauge bosons, like from an extra $U(1)'$. This would imply an
additional $Z$ boson, a sequential $Z'$ which would mix with the SM
$Z$ and the photon.  More attractive are extensions which solve some
real or thought shortcomings of the SM. This includes Grand Unified
Theories (GUT)~\cite{GUT73} which attempt to unify the strong,
electromagnetic and weak forces, which correspond to three different
factors of the local gauge group of the SM, in one big simple local
gauge group
\bea
G_\mathrm{GUT} \supset SU(3)_c\otimes SU(2)_L \otimes U(1)_Y\equiv
G_\mathrm{SM}
\eea
which is assumed to be spontaneously broken in at least two steps
\bea
G_\mathrm{GUT} \to SU(3)_c\otimes SU(2)_L \otimes U(1)_Y \to
SU(3)_c\otimes  U(1)_\mathrm{em}\epo 
\eea
Coupling unification is governed by the renormalization group
evolution of $\alpha_1(\mu)$, $\alpha_2(\mu)$ and $\alpha_3(\mu)$,
corresponding to the SM group factors $U(1)_Y$, $SU(2)_L$ and
$SU(3)_c$, with the experimentally given low energy values, typically
at the $Z$ mass scale, as starting values evolved to very high
energies, the \textit{GUT scale} $M_\mathrm{GUT}$ where couplings should meet.
Within the SM the three couplings do not unify, thus unification
requires new physics as predicted by a GUT
extension~\cite{GUTLEP90}. Also extensions like the left--right ($LR$)
symmetric model are of interest. The simplest possible unifying group
is $SU(5)$ which, however, is ruled out by the fact that it predicts
protons to decay faster than allowed by observation. GUT models like
$SO(10)$ or the exceptional group $E_6$ not only unify the gauge
group, thereby predicting many additional gauge bosons, they also
unify quarks and leptons in GUT matter multiplets. Now quarks and
leptons directly interact via the
\textit{leptoquark} gauge bosons $X$ and $Y$ which carry color,
fractional charge ($Q_X=-4/3$, $Q_Y=-1/3$) as well as baryon and
lepton number. Thus GUTs are violating $B$ as well as $L$, yet with
$B-L$ still conserved. The proton may now decay via $p\to e^+ \pi^0$
or many other possible channels. The experimental proton lifetime
$\tau_{\mathrm{proton}}> 2
\power{29}~\mathrm{years}~~\mathrm{at}~~90\%\,
\mathrm{C.L.}$ requires the extra gauge bosons
to exhibit masses of about $M_\mathrm{GUT} > 10^{16}$ GeV and excludes
$SU(5)$ as it predicts unification at too low 
scales. Note that the stability of the proton requires
$M_\mathrm{GUT}$ to lie not more than a factor 1000
below the Planck scale. In general GUTs
also have additional normal gauge bosons, extra $W'$s and $Z'$s which
mix with the SM gauge bosons. Present bounds here are $M_{Z',W'}> 600-800$~GeV
depending  on the GUT scenario. Contributions from such extra gauge bosons
may be estimated from the weak one--loop contributions by rescaling with
$(M_W/M_{W'_{\rm SM}})^2\sim 0.01$ and hence 1\% of $19.5 \power{-10}$
only, an effect much too small to be of relevance.

In deriving bounds on New Physics it is important to respect
constraints not only from $\amu$ and the direct bounds from the
particle data tables, but also from other precision observables which
are sensitive to new physics via radiative corrections.  Important
examples are the electroweak precision
observables~\cite{LEPEWWG06,EWinPDG06}: $M_W=80.392(29)~\gv\,,$
$\sin^2 \Theta^\ell_\mathrm{eff}=0.23153(16)\,,$ and
$\rho_0=1.0002^{+0.0007}_{-0.0004}\,,$ which are all precisely
measured and precisely predicted by the SM or in extensions of it.
The SM predictions use the very precisely known independent input
parameters $\alpha$, $G_\mu$ and $M_Z$, but also the less precisely
known top mass $m_t=172.6\pm1.4~\gv\,,$~\cite{TopMass08} 
(the dependence on other fermion masses is usually weak, the one on
the unknown Higgs is only logarithmic and already fairly well
constrained by experimental data). The effective weak mixing parameter
essentially determines $m_H=114^{+45}_{-33}~\gv$ at 68\% C.L. (not taking
into account $M_W$). The parameter $\rho_0$ is the tree level (SM
radiative corrections subtracted) ratio of the low energy effective
weak neutral to charged current couplings: $\rho=G_{\rm NC}/G_{\rm
CC}$ where $G_{\rm CC}\equiv G_\mu$. This parameter is rather
sensitive to new physics. Equally important are constraints by 
the $B$--physics branching fractions~\cite{HFAG06}
$\mathrm{BR}(b\to s \gamma)= (3.55\pm0.24^{+0.09}_{-0.10}\pm0.03)\power{-4}\cs$ 
$\mathrm{BR}(B_s\to \mu^+\mu^-)< 1.0 \power{-7}~~(95\%~\mathrm{C.L.})\,.$

Concerning flavor physics, in particular the B factories Belle at KEK
and BaBar at SLAC have set new milestones in confirming the flavor
structure as inferred by the SM. In the latter FCNC are absent at tree
level due to the GIM mechanism and CP-violation and flavor mixing
patterns seem to be realized in nature precisely as implemented by
the three fermion--family CKM mixing scheme. Many new physics models
have serious problems to accommodate this phenomenologically largely
confirmed structure in a natural way. Therefore, the criterion of \textit{Minimal
Flavor Violation} (MFV)~\cite{MFV02} has been conjectured as a
framework for constructing low energy effective theories which include
the SM Lagrangian without spoiling its flavor structure.  The SM
fermions are grouped into three families with two $SU(2)_L$ doublets
($Q_L$ and $L_L$) and three $SU(2)_L$ singlets ($U_R$, $D_R$ and
$E_R$) and the largest group of unitary transformations which commutes
with the gauge group is $G_F=U(3)^5$~\cite{Chivukula87}. The latter
may be written more specifically as
\bea
G_F=SU(3)_q^3 \otimes SU(3)_\ell^2 \otimes U(1)_B\otimes U(1)_L\otimes U(1)_Y\otimes U(1)_{PQ}\otimes U(1)_{E_R} 
\eea
with $SU(3)_q^3=SU(3)_{Q_L} \otimes SU(3)_{U_R} \otimes SU(3)_{D_R}$
and $SU(3)_\ell^2=SU(3)_{L_L} \otimes SU(3)_{E_R}$. The SM Yukawa
interactions break the subgroup $SU(3)_q^3 \otimes SU(3)_\ell^2
\otimes U(1)_{PQ}\otimes U(1)_{E_R}$. However, one may
introduce three dimensionless auxiliary fields
\bea
Y_U \sim (3,\bar{3},1)_{SU(3)^3_q}\css Y_D \sim (3,1, \bar{3})_{SU(3)^3_q}\css
Y_E \sim (3,\bar{3})_{SU(3)^2_\ell} 
\eea  
which provide a convenient bookkeeping for constructing MFV effective
theories. Formally the auxiliary fields allow to write down MFV
compatible interactions as $G_F$ invariant effective interactions. The
MVF criterion requires that a viable dynamics of flavor violation is
completely determined by the structure of the ordinary SM Yukawa
couplings.  Most of the promising and seriously considered new
physics models, which we will consider below, belong to the class of MFV
extensions of the SM. Examples are the R-parity conserving two doublet Higgs
models, the R-parity conserving minimal supersymmetric extension of
the SM~\cite{Altmannshofer07} and the Littlest Higgs model without T-parity.

One important monitor for new physics is the electric dipole moment
which we briefly discussed towards the end of
Sect.~\ref{sec:intro}. The EDM is a direct measure of T--violation,
which in a QFT is equivalent to a CP--violation. Since extensions of
the SM in general exhibit additional sources of CP violation, EDMs are
very promising probes of new physics. An anomalously large EDM of the
muon $d_\mu$ would influence the $\amu$ extraction from the muon
precession data as discussed earlier. We may ask whether $d_\mu$ could
be responsible for the observed deviation in $\amu$. In fact
Eq.~(\ref{EDMomegashift}) tells us that a non--negligible $d_\mu$
would increase the observed $\amu$, and we may estimate
\be
|d_\mu|=\frac{1}{2}\frac{e}{m_\mu}\sqrt{(a^\mathrm{exp}_\mu)^2
-(a^\mathrm{SM}_\mu)^2} = ( 2.42\pm  0.41)\power{-19}\,e \cdot \mathrm{cm}\epo 
\ee
This also may be interpreted as an upper limit $d_\mu < 2.7 \power{-19}\,e\cdot\mathrm{cm}$.
Recent advances in experimental techniques
will allow to perform much more sensitive
experiments for electrons, neutrons and neutral atoms~\cite{Khriplovich97}.
For new efforts to determine $d_\mu$ at much higher precision 
see~\cite{NewEDMexp,Adelmann06}. In the following we will assume that $d_\mu$ is
in fact negligible, and that the observed deviation has
other reasons. As mentioned after Eq.~(\ref{EDMomegashift}), 
in the SM and viable extensions of it $d_\mu$ is expected to be much
smaller that what could be of relevance here (see~\cite{Hoogeveen90,BeSuEDM90}).

As mentioned many times, the general form of contributions from states
of mass $M_\mathrm{NP}\gg m_\mu$ takes the form
\be
a_\mu^\mathrm{NP}=\cC\,\frac{m_\mu^2}{M_\mathrm{NP}^2}
\label{NPgeneric}
\ee
where naturally $\cC=O(\alpha/\pi)$ ($\sim$ lowest order $\amu$), like for the weak contributions
Eq.~(\ref{oneloopbosonic}), but now from interactions and states not
included in the SM. New fermion 
loops may contribute similarly to a
$\tau$--lepton by
\ba
\mysymb{-8}{13}{30}{30}{amm27top}\!\!\!\!\! a^{(4)}_\mu(\mathrm{vap},F)
=\sum_F Q_F^2N_{cF} \left[\frac{1}{45}\left(\frac{m_\mu}{m_F} \right)^2+ \cdots \right]\left(
\frac{\alpha}{\pi}\right)^2,
\label{heavyF}
\ea
which means $\cC=O((\alpha/\pi)^2)$.  Note that the $\tau$
contribution to $\amu$ is $42 \power{-11}$ only, while the 3 $\sigma$
effect we are looking for is $290 \power{-11}$.  As the direct lower
limit for a sequential fermion is about 100 GeV such effects cannot account for the
observed deviation\footnote{It should be noted that heavy sequential fermions
are constrained severely by the $\rho$--parameter (NC/CC effective
coupling ratio), if doublet members are not nearly mass
degenerate. However, a doublet $(\nu_L,L)$ with $m_{\nu_L}=45~\gv$ and
$m_L=100~\gv$ only contributes $\Delta
\rho\simeq 0.0008$ which is within the limit from LEP electroweak 
fits~\cite{LEPEWWG06}. Not yet included is a similar type of contribution from
the 4th family $(t',b')$ doublet mass--splitting, which also would add a
positive term $$\Delta \rho= \frac{\sqrt{2}G_\mu}{16
\pi^2}\,3\,|m_{t'}^{2}-m_{b'}^{2}| + \cdots\,.$$ In this context it
should be mentioned that the so called \textit{custodial symmetry} of
the SM which predicts $\rho_0=1$ at the tree level (independent of any
parameter of the theory, which implies that it is not subject to
subtractions due to parameter renormalization) is one of the severe
constraints to extensions of the SM (see~\cite{CGJZ99})}.  

A rough estimate of the scale $M_{\rm NP}$ required to account for the
observed deviation is given in Table~\ref{tab:MPscales}. An effective
tree level contribution would extend the sensibility to the very
interesting 2 TeV range, however, we know of no compelling scenario
where this is the case.
\begin{table}
\centering
\caption{Typical New Physics scales required to satisfy $\Delta
a_\mu^{\rm NP}=\delta a_\mu$ in Eq.~(\ref{amuNP}).}
\label{tab:MPscales}
\begin{tabular}{cccc}
&&&\\[-3mm]
\hline\noalign{\smallskip}
 $\cC$ & 1 & $\alpha/\pi$ & $(\alpha/\pi)^2$ \\
\noalign{\smallskip}\hline\noalign{\smallskip}
 $M_\mathrm{NP}$ &~~~ $2.0^{+0.4}_{-0.3}~\tv$~~~ &~~~ $100^{+21}_{-13}~\gv$ ~~~&~~~ $5^{+1}_{-1}~\gv$~~~\\
\noalign{\smallskip}\hline
\end{tabular}
\end{table}

   \subsubsection{Generic Contributions from Physics beyond the SM}
\label{sssec:generic}
Common to many of the extensions of the SM are predictions of new
states: scalars S, pseudoscalars P, vectors V or axialvectors A,
neutral or charged. They contribute via one--loop lowest order type
diagrams shown in Fig.~\ref{fig:nplo}. Here, we explicitly assume all
fermions to be Dirac fermions. Besides the SM fermions, $\mu$ in
particular, new heavy fermions $F$ of mass $M$ may be involved, but
fermion number is assumed to be conserved, like in $\Delta \cL_S = f
\bar{\psi}_\mu \psi_FS+{\rm h.c.}$, which will be different in
supersymmetric (SUSY) extensions
discussed below, where fermion number violating Majorana fermions
necessarily must be there.
\begin{figure}[h]
\centering
\IfFarbe{%
\includegraphics[height=1.8cm]{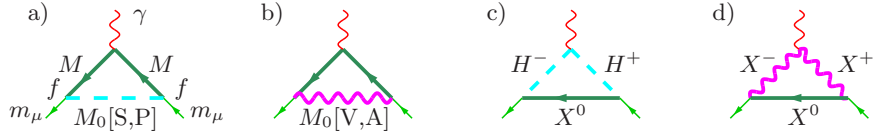}}{%
\includegraphics[height=1.8cm]{NPamvert.eps}}
\caption{Possible New Physics contributions. 
Neutral boson exchange: a) scalar or pseudoscalar and b) vector or
axialvector, flavor changing or not. New charged bosons: c) scalars or
pseudoscalars,  d) vector or axialvector.}
\label{fig:nplo}
\end{figure}
Note that massive spin 1 boson exchange contributions in general have
to be considered within the context of a gauge theory, in order to
control gauge invariance and unitarity. We will present corresponding
contributions in the unitary gauge calculated with dimensional
regularization.  We first discuss neutral boson exchange contributions
from diagrams a) and b). Exotic neutral bosons of mass $M_0$ coupling
to muons with coupling strength $f$ would
contribute~\cite{LPdR72,Leveille77}
\ba
\Delta \amu^{\mathrm{NP}} =\frac{f^2}{4\pi^2}\:\frac{m_\mu^2}{M_0^2}\: L
,~
L=\ha \intzo \! \D x\: \frac{Q(x)}{(1-x)\:(1-\lambda^2\:x)+ (
\epsilon \lambda)^2\:x}\cs 
\label{NPFFgen}
\ea
where $Q(x)$ is a polynomial in $x$ which depends on the type of 
coupling:

\begin{tabular}{lcl}
\centering
Scalar  &:& $Q_\mathrm{S}=\:x^2\:(1+\epsilon -x)$ \\
Pseudoscalar  &:& $Q_\mathrm{P}=\:x^2\:(1-\epsilon -x)$ \\
Vector  &:& $Q_\mathrm{V}=2x\:(1-x)\:(x-2\:(1-\epsilon))\,
+\lambda^2\,(1-\epsilon)^2\:Q_\mathrm{S}$ \\
Axialvector &:& $Q_\mathrm{A}=2x\:(1-x)\:(x-2\:(1+\epsilon))\,
+\lambda^2\,(1+\epsilon)^2\:Q_\mathrm{P}$
\end{tabular}

\noi
with $\epsilon=M/m_\mu$ and $\lambda=m_\mu/M_0$. As an illustration we first consider the
regime of a heavy boson of mass $M_0$ and $m_\mu,M\ll M_0$ for which one gets
\ba \bary{lclcl}
L_\mathrm{S} &=& ~~\frac{M}{m_\mu}\left(\ln \frac{M_0}{M}-\frac{3}{4}
\right)+\frac{1}{6} &\stackrel{M=m_\mu}{=}&~~ \ln
\frac{M_0}{m_\mu}-\frac{7}{12},\\[2mm] 
L_\mathrm{P} &=& -\frac{M}{m_\mu}\left(\ln \frac{M_0}{M}-\frac{3}{4}
\right)+\frac{1}{6}&\stackrel{M=m_\mu}{=}&-\ln
\frac{M_0}{m_\mu}+\frac{11}{12}, \\[2mm] 
L_\mathrm{V} &=&
~~\frac{M}{m_\mu}-\frac{2}{3}&\stackrel{M=m_\mu}{=}&~~\frac{1}{3}, \\[2mm] 
L_\mathrm{A} &=& -\frac{M}{m_\mu}-\frac{2}{3}&\stackrel{M=m_\mu}{=}&-\frac{5}{3}\epo 
\eary
\label{NPGa}
\ea
In accordance with the MFV requirement it is more realistic to assume a flavor conserving neutral
current $M=m_\mu$ as given by the second form. Typical contributions are shown in
Fig.~\ref{fig:NewPhysGen}. Taking the coupling small enough such that
a perturbative expansion in $f$ makes sense, we take $f/(2\pi)=0.1$,
only the scalar exchange could account for the observed deviation with
a scalar mass $480~\gv~< M_0 <~690~\gv$.  Pseudoscalar and
axialvector yield the wrong sign. The vector exchange is too small.
\begin{figure}[t]
\centering
\subfigure[Case: $m_\mu=M \ll M_0$] 
{
    \label{fig:NPGa}
\IfFarbe{%
    \includegraphics[width=5.5cm]{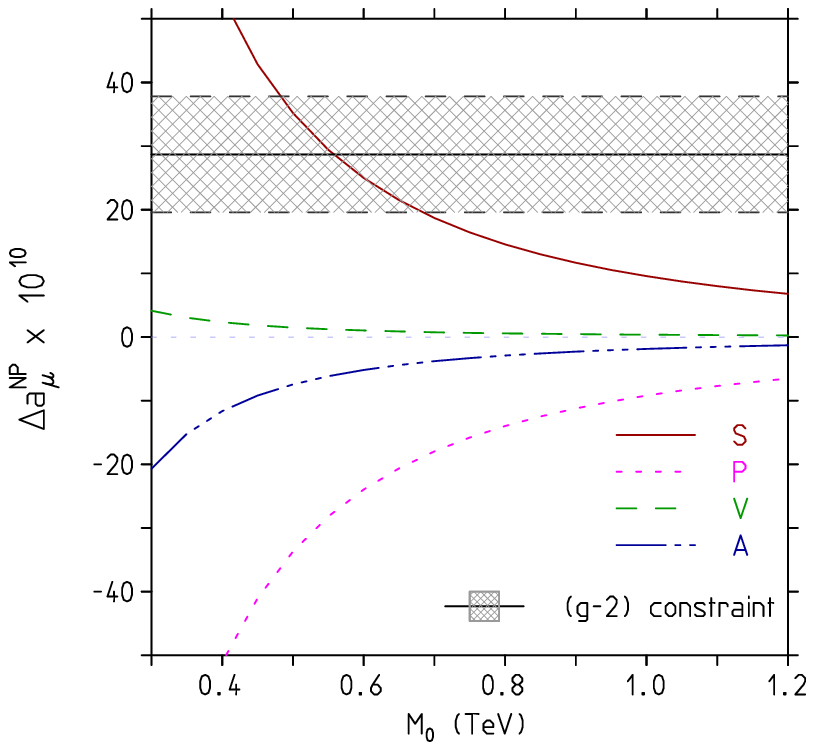}}{%
    \includegraphics[width=5.5cm]{NewPhysGen.eps}}
}
\hspace{1mm}
\subfigure[Case: $m_\mu \ll M_0=M$] 
{
    \label{fig:NPGb}
\IfFarbe{%
    \includegraphics[width=5.0cm]{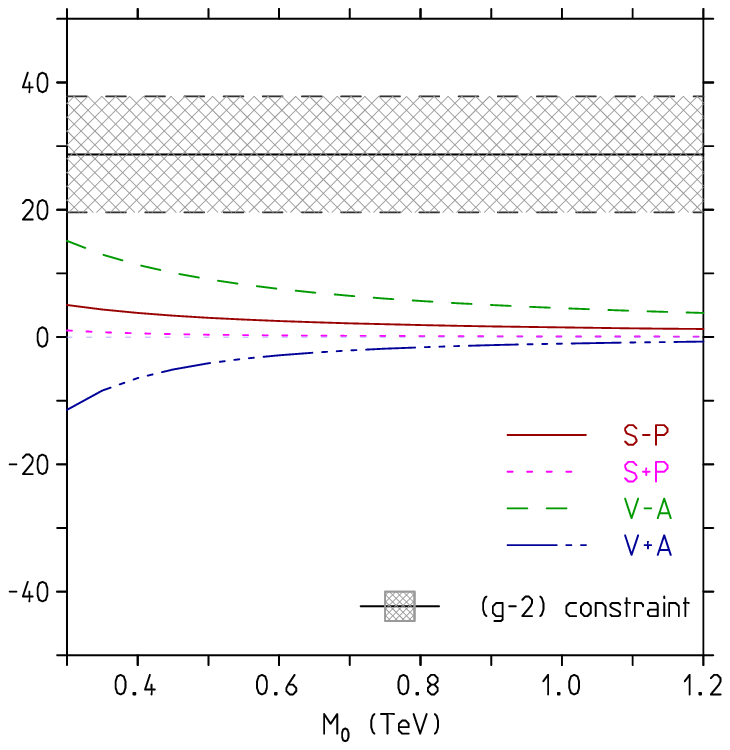}}{%
    \includegraphics[width=5.0cm]{NewPhysGen2.eps}}
\vspace*{-5mm}
}
\caption{Single particle one--loop induced NP effects from Eq.~(\ref{NPFFgen}) for
$f^2/(4\pi^2)=0.01$ (Note, a typical EW SM coupling would be
$e^2/(4\pi^2\,\cos^2 \Theta_W)=0.003$).  S,P,V,A denote scalar,
pseudoscalar, vector and axialvector exchange. Panel (a) uses Eq.~(\ref{NPGa})
for $M=m_\mu$, panel (b) the chiral combinations in Eq.~(\ref{LNPchirala}) for
$M=M_0$, with the large combinations $L_S-L_P$ and $L_V-L_A$
rescaled by the muon Yukawa coupling $m_\mu/v$ in order to compensate for the
huge pre-factor $M/m_\mu$ (see text).}
\label{fig:NewPhysGen}
\end{figure}

As we will see later, in SUSY and littlest Higgs extensions the leading
contributions actually come from the regime $m_\mu \ll M,M_0$ with $M\sim M_0$,
which is of enhanced FCNC type, and thus differs from the case just
presented in Eq.~(\ref{NPGa}). For the combinations of fixed chirality
up to terms of order $O(m_\mu/M)$ one gets
\ba
\!\!\!L_S+L_P&=&\frac{1}{6\,(1-z)^4}\,\left[
2+3z-6z^2+z^3+6z\ln z \right]=\frac{1}{12}\,F_1^C(z),\crn
\!\!\!L_S-L_P&=&\frac{-M}{2m_\mu\,(1-z)^3}\,\left[
3-4z+z^2+2\ln z \right]=\frac{M}{3m_\mu}\,F_2^C(z),\crn
\!\!\!L_V+L_A&=&\frac{-1}{6\,(1-z)^4}\,\left[
8-38z+39z^2-14z^3+5z^4-18z^2\ln z \right]=-\frac{13}{12}\,F^C_3(z),\crn
\!\!\!L_V-L_A&=&\frac{M}{2m_\mu\,(1-z)^3}\,\left[
4-3z-z^3+6z\ln z \right]=\frac{M}{m_\mu}\,F^C_4(z), 
\label{LNPchirala}
\ea
where $z=(M/M_0)^2=O(1)$ and the functions $F^C_i$ are normalized to
$F^C_i(1)=1$.  The possible huge enhancement factors $M/m_\mu$, in
some combination of the amplitudes, typical for flavor changing
transitions, may be compensated due to radiative contributions to the
muon mass (as discussed below) or by a corresponding Yukawa coupling
$f\propto y_\mu=\sqrt{2}\,m_\mu/v$, as it happens in SUSY or little
Higgs extensions of the SM.

The second class of possible new physics transitions
due to charged S,P,V and A modes are represented by the diagrams c) and d)
in Fig.~\ref{fig:nplo}.  It amounts to replace $L$ in Eq.~(\ref{NPFFgen}) according to
\ba
\Delta \amu^{\mathrm{NP}} =\frac{f^2}{4\pi^2}\:\frac{m_\mu^2}{M_0^2}\: L,~
L=\ha \intzo \! \D x\: \frac{Q(x)}{(\epsilon
  \lambda)^2\:(1-x)\:(1-\epsilon^{-2}\:x)+x}\cs  
\label{NPFFgenCC}
\ea
where again $Q(x)$ is a polynomial in $x$ which depends on the type of 
coupling:

\begin{tabular}{lcl}
\centering
Scalar  &:& $Q_\mathrm{S}=-\,x\,(1-x)\:(x+\epsilon)$ \\
Pseudoscalar  &:& $Q_\mathrm{P}=-\,x\,(1-x)\:(x-\epsilon)$ \\
Vector  &:& $Q_\mathrm{V}=-2\,x^2\,(1+x-2\epsilon)+\lambda^2\,(1-\epsilon)^2\:Q_\mathrm{S}$ \\
Axialvector &:& $Q_\mathrm{A}=-2\,x^2\,(1+x+2\epsilon)+\lambda^2\,(1+\epsilon)^2\:Q_\mathrm{P}$
\end{tabular}

\noi
Again, results for V and A are in the unitary gauge calculated with
dimensional regularization.  For a heavy boson of mass $M_0$ and $m_\mu,M\ll M_0$ one finds
\ba \bary{lclcllclcl}
L_\mathrm{S} &=&
-\frac{1}{4}\frac{M}{m_\mu}-\frac{1}{12}&\stackrel{M=m_\mu}{=}&-\frac{1}{3}~,&~~~ 
L_\mathrm{P} &=&
~~\frac{1}{4}\frac{M}{m_\mu}-\frac{1}{12}&\stackrel{M=m_\mu}{=}&~~\frac{1}{6}
\;,  \\[2mm] 
L_\mathrm{V} &=&
~~\frac{M}{m_\mu}-\frac{5}{6}&\stackrel{M=m_\mu}{=}&~~\frac{1}{6}~,&~~~ 
L_\mathrm{A} &=&
-\frac{M}{m_\mu}-\frac{5}{6}&\stackrel{M=m_\mu}{=}&-\frac{11}{6}\epo 
\eary
\label{NPGb}
\ea
The second form given is for a flavor conserving charged current
transition with $M=m_\mu$.

Also for the charged boson exchanges the regime $m_\mu \ll M,M_0$ with
$M\sim M_0$ is of interest in SUSY and littlest Higgs extensions of the SM and we find
\ba
\!\!\!L_S+L_P&=&\frac{-1}{6\,(1-z)^4}\,\left[
1-6z+3z^2+2z^3-6z^2\ln z \right]=-\frac{1}{12}\,F_1^N(z),\crn
\!\!\!L_S-L_P&=&\frac{-M}{2m_\mu\,(1-z)^3}\,\left[
1-z^2+2z\ln z \right]=-\frac{M}{6m_\mu}\,F_2^N(z),\crn
\!\!\!L_V+L_A&=& \frac{-1}{6\,(1-z)^4}\,\left[
10-43\,z+78\,{z}^{2}-49\,{z}^{3}+4\,{z}^{4}+18\,{z}^{3}\ln z \right]
=-\frac{5}{3}\,F^N_3(z),\crn
\!\!\!L_V-L_A&=&\frac{M}{m_\mu\,(1-z)^3}\,
\left[4-15\,z+12\,{z}^{2}-{z}^{3}-6\,{z}^{2}\ln z
  \right]=\frac{2M}{m_\mu}\,F^N_4(z),  
\label{LNPchiralb}
\ea
where $z=(M/M_0)^2=O(1)$ and the functions $F^N_i$ are normalized to
$F^N_i(1)=1$.

At $O((\alpha/\pi)^2)$ new physics may enter via vacuum polarization 
and we may write corresponding contributions
as a dispersion integral Eq.~(\ref{amuvapo}):
\bea
\Delta \amu^{\mathrm{NP}}=\aldpi \intzi \frac{\D s}{s}
\frac{1}{\pi}\:\Impa \Delta \Pi^\mathrm{NP}_\gamma(s)\:K(s) \epo
\eea
Since, we are looking for contributions from heavy yet unknown states
of mass $M \gg m_\mu$, and $\Impa \Delta \Pi^\mathrm{NP}_\gamma(s)
\neq 0$ for $s \geq 4 M^2$ only, we may safely approximate $K(s)\simeq
\frac{1}{3}\frac{m_\mu^2}{s} ~~\mathrm{for}~s \gg m_\mu^2$
such that, with $\frac{1}{\pi}\:\Impa \Delta
\Pi^\mathrm{NP}_\gamma(s)=\frac{\alpha(s)}{\pi}\,R^\mathrm{NP}(s)$
\bea
\Delta \amu^{\mathrm{NP}}=\frac{1}{3} \aldpi \left(\frac{m_\mu}{M} \right)^2
L \:\:,\;\;
\frac{L}{M^2}=\frac{\alpha}{3\pi}\intzi \frac{\D s}{s^2}
R^\mathrm{NP}(s) \epo
\eea
An example is a heavy lepton given by Eq.~(\ref{heavyF}).
A heavy narrow vector meson resonance of mass $M_V$ and electronic width 
$\Gamma (V\to\epm)$ (which is $O(\alpha^2)$) contributes 
$R_V(s)=\frac{9 \pi}{\alpha^2}\,M_V \,\Gamma(V\to\epm)\,\delta(s-M_V^2)$
such that  $L=\frac{3\Gamma(V\to\epm)}{\alpha M_V}$ and hence
\be
\Delta \amu^{\mathrm{NP}}=\frac{m_\mu^2 \, \Gamma(V\to \epm)}{\pi M_V^3} 
=\frac{4\alpha^2\, \gamma_V^2\,m_\mu^2}{3M_V^2} \epo
\ee
For $\gamma_V=0.1$ and $M_V=200~\gv$ we get $\Delta a_\mu
\sim 2 \power{-13}$. The hadronic contribution of a 4th family quark doublet
assuming $m_{b'}=m_{t'}=200~\gv$ would yield $\Delta a_\mu \sim 5.6
\power{-14}$ only. Unless there exists a new type of strong
interactions like Technicolor\footnote{Searches for Technicolor states
like color--octet techni--$\rho$ were negative up to 260 to 480 GeV
depending on the decay mode.}~\cite{TC,EichtenLane07,Sannino07}, new
strong interaction resonances are not expected, because new heavy
sequential quarks would be too shortlived to be able to form
resonances. As we know, due to the large mass and the large mass
difference $m_t \gg m_b$, the top quark is the first quark which
decays, via $t \to W b$, as a bare quark before it has time to form
hadronic resonances. This is not so surprising as the top Yukawa
coupling responsible for the weak decay is stronger than the strong
interaction constant.

New physics effects here may be easily buried in the uncertainties of
the hadronic vacuum polarization. In any case, we expect
$O((\alpha/\pi)^2)$ terms from heavy states not yet seen to be too
small to play a role here.

In general the effects related to single diagrams, discussed in this
paragraph, are larger than what one expects in a viable extension of
the SM, usually required to be a renormalizable QFT and to exhibit
gauge interactions which typically cause large cancellations between
different contributions. But even if one ignores possible
cancellations, all the examples considered so far show how difficult
it actually is to reconcile the observed deviation with NP effects not
ruled out already by LEP or Tevatron new physics searches, and if we
adopt the phenomenologically preferred MFV restriction. The latter
means to try to avoid conflicts with other experimental
facts. Apparently a more sophisticated extension of the SM is needed
which is able to produce substantial radiative corrections in the low
energy observable $\amu$ while the new particles have escaped
detection at accelerator facilities so far and only produce small
higher order effects in other electroweak precision observables.  In
fact supersymmetric extensions of the SM precisely allow for such a
scenario, as we will discuss below.

   \subsubsection{Flavor Changing Processes}
We already have seen that flavor changing processes could give large
contributions to $\amu$. As pointed out in~\cite{CM01} taking into
account just the vertex diagrams could be very misleading. The argument is
that the same interactions and heavy states which could contribute to
$\amu^{\rm NP}$ according to Fig.~\ref{fig:nplo} would contribute to
the muon self energy, via the diagrams Fig.~\ref{fig:nplofse}. By imposing
\begin{figure}[t]
\centering
\IfFarbe{%
\includegraphics[height=1.5cm]{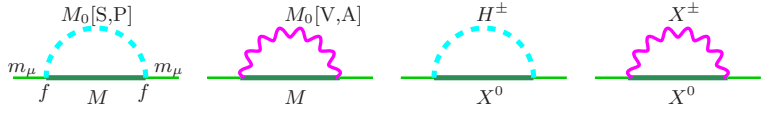}}{%
\includegraphics[height=1.5cm]{NPfermsel.eps}}
\caption{Lepton self--energy contributions induced by the new interactions
appearing in Fig.~\ref{fig:nplo} may generate $m_\mu$ as a radiative
correction effect.}
\label{fig:nplofse}
\vspace*{-1mm}
\end{figure}
chiral symmetry to the SM, i.e. setting the SM Yukawa couplings to
zero, lepton masses could be radiatively induced by flavor changing
$f\bar{\psi}_\mu \psi_FS+{\rm h.c.}$ and $f \bar{\psi}_\mu \,\I\,
\gafi\psi_ F P+{\rm h.c.}$ interactions ($F$ a heavy fermion, $S$ a
scalar and $P$ a pseudoscalar) in a hierarchy $m_\mu \ll M_F \ll
M_S,M_P$. Then with $m_\mu \propto f^2 M_F$ and $\amu
\propto f^2 m_\mu M_F/M_{S,P}^2$ one obtains $\amu
=\cC\, m^2_\mu/M_{S,P}^2$ with $\cC =O(1)$, and the interaction 
strength $f$ has dropped from the ratio. The problem is that a
convincing approach of generating the lepton/fermion spectrum by
radiative effects is not easy to accommodate. Of course it is a very
attractive idea to replace the Yukawa term, put in by hand in the SM,
by a mechanism which allows us to understand or even calculate the
known fermion mass-spectrum, exhibiting a tremendous hierarchy of about
13 orders of magnitude of vastly different couplings/masses [from
$m_{\nu_e}$ to $m_t$]. The radiatively induced values must reproduce
this pattern and one has to explain why the same effects which make up
the muon mass do not contribute to the electron mass.  Again the
needed hierarchy of fermion masses is only obtained by putting it in
by hand in some way. In the scenario of radiatively induced lepton
masses one has to require the family hierarchy like
$f_e^2M_{F_e}/f_\mu^2M_{F_\mu}\simeq m_e/m_\mu$, $f_P\equiv f_S$ in
order to get a finite cut--off independent answer, and $M_0 \to M_S\neq
M_P$, such that $m_\mu=\frac{f_\mu^2\,M_{F_\mu}}{16\pi^2} \,\ln
\frac{M_S^2}{M_P^2}$ which is positive provided $M_S>M_P\,.$

Another aspect of flavor changing transition in the lepton sector is
the following: after neutrino oscillations and herewith right--handed singlet
neutrinos and neutrino masses have been established, also lepton
number violating transitions like $\mu ^\pm \to e^\pm \gamma$, see
Fig.~\ref{fig:nplomeg}, are in the focus of further searches. The
corresponding contributions here read
\bea \bary{lcrclcl}
L^\mu_\mathrm{S} &\simeq& \frac{1}{6} ~~&,&~~~L^e_\mathrm{S} &\simeq&
~~\frac{m_\mu}{m_e}\left(\ln \frac{M_0}{m_\mu}-\frac{3}{4} \right),\\
L^\mu_\mathrm{P} &\simeq&\frac{1}{6}~~&,&~~~L^e_\mathrm{P} &\simeq&
-\frac{m_\mu}{m_e}\left(\ln \frac{M_0}{m_\mu}-\frac{3}{4}\right),\\
L^\mu_\mathrm{V} &\simeq& \frac{2}{3}~~&,&~~~L^e_\mathrm{V} &\simeq&
~~\frac{m_\mu}{m_e}, \\
L^\mu_\mathrm{A} &\simeq& -\frac{2}{3}~~&,&~~~L^e_\mathrm{A} &\simeq&
-\frac{m_\mu}{m_e} \epo
\eary
\eea
The latter flavor changing transitions are strongly constrained, first by
direct rare decay search experiments which were performed at the Paul
Scherrer Institute (PSI)
and second, with the advent of the much more precise measurement of $a_e$.
\begin{figure}[h]
\centering
\IfFarbe{%
\includegraphics[height=1.3cm]{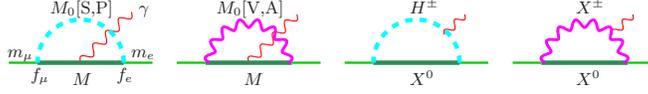}}{%
\includegraphics[height=1.3cm]{NPmeg.eps}}
\vspace*{0mm}
\caption{$\mu \to e \gamma$ transitions by new interactions (overall
flavor changing version of Fig.~\ref{fig:nplo}).}
\label{fig:nplomeg}
\end{figure}
For example, for a scalar exchange mediating $e \to \mu \to e$ with
$f^2/(4\pi^2)\simeq 0.01$ and $M_0\simeq 100~\gv$ we obtain $\Delta
a_e^{NP}\simeq 33 \power{-11}$ which is ruled out by
$a_e^\mathrm{exp}-a_e^\mathrm{the}\sim 1 \power{-11}$. Either $M_0$
must be heavier or the coupling smaller: $f^2/(4\pi^2) < 0.0003$.  The
present limit for the branching fraction $Br(\mu \to e \gamma)$ is
$1.2\power{-11}$, which will be improved to $10^{-13}$ at PSI by a new
experiment~\cite{RittPSI06} (see also~\cite{BarbieriHall94}) . 

At present the most important constraints from flavor changing
transitions are the ones obtained with quarks~\cite{HFAG06}. In
particular the $b \to s \gamma$ branching fraction given before
provides an interesting constraint for the SUSY parameter space as we
will discuss below. Note that models exhibiting tree level FCNCs are not 
in the class of MFV theories, nevertheless $a_e$ and $a_\mu$ 
can provide useful constraints on such processes.

   \subsubsection{Two-Higgs Doublet Models}
\label{sssec:THDM}
One possibility of extending the SM is to modify the Higgs sector where one
could add scalar singlets, an additional doublet, a Higgs triplet and
so on. From a theoretical point of view the case with two Higgs
doublets is very attractive. General two Higgs doublet models (2HDM)
are interesting as they predict 4 additional physical spin 0
bosons. In terms of the components of the two doublet fields $\Phi_i$
($i=1,2$) of fixed hypercharge $Y_i=(-1,+1)$, the new physical scalars
are the two scalars h and H, the pseudoscalar $A$ and the charged Higgses $H^{\pm}$:
\bea \bary{cclccl}
h &=& -\sial \;\eta_1 +\coal \;\eta_2 \;, &\hspace*{2.5cm}
A &=& -\sibet \;\chi_1 + \cobet \;\chi_2 \;, \\
H &=& \;\; \coal \;\eta_1 +\sial \;\eta_2\;, &\hspace*{2.5cm}
H^{\pm} &=& -\sibet \;\phi_1^{\pm} + \cobet \;\phi_2^{\pm} \epo
\eary
\eea
Two Higgs
doublets are needed in Minimal Supersymmetric extensions of the
SM (MSSM).  One reason is supersymmetry itself, the other is anomaly
cancellation of the SUSY partners of the Higgses. 
In the minimal SUSY models the masses of the extra Higgses at tree level are
severely constrained by the following mass- and coupling-relationships:
\bea
&&m^2_{\pm} = M_W^2+m_A^2\;,\;\;
m^2_{H,h} = \frac{1}{2}\left( M_Z^2+m_A^2 \pm 
\sqrt{(M_Z^2-m_A^2)^2+4M_Z^2 m_A^2 \sin^2 2\beta} \right), \nn \\
&&\tan(2\al) = \tan (2\beta)\;\frac{m_A^2+M_Z^2}{m_A^2-M_Z^2}\;,\;\; 
\sin^2 (\al - \beta ) = \frac{m_H^2}{m_A^2} 
\frac{\mz-m_H^2}{\mz+m_A^2-2m_H^2} \;.
\eea
Only two independent parameters are left, which we may choose to be $\tan
\beta$ and $m_A$. In the phenomenologically interesting region of
enhanced $\tan \beta$ together with a light Higgs for the CP-even part of
the Higgs sector  we have $\alpha \simeq \beta$, which we assume
in the following.

In 2HDMs many new real and virtual processes, like $W^\pm H^\mp
\gamma$ transitions, are the consequence. Present bounds on scalars 
are $m_{H^\pm}>80$~GeV and $m_{A}+m_{h}> 90$~GeV.  
In general, in type I models, fermions get
contributions to their masses from the vev's of both Higgs
scalars. Phenomenologically preferred and most interesting are the
type II models where a discrete symmetry guarantees that the upper and
the lower entries of the fermion doublets get their masses from different
vev's ($m_t\, {\propto}\, v_2$, $m_b\, {
\propto}\, v_1$) in order to prevent FCNC's~\cite{GW76}. Only the type
II models satisfy the MFV criterion.
Such models are also interesting because one easily may get $m_t \gg
m_b$ without having vastly different Yukawa couplings. Notice,
however, that the experimental bounds on $\der=1-\pi
\alpha/(\wz G_\mu M_Z^2 \cos^2 \Theta_W\,\sin^2\Theta_W)$, with $\cos^2
\Theta_W=M_W^2/M_Z^2$ and $\Delta \rho$, seem to require a top with a large Yukawa 
coupling, not just a large top mass. In addition if $\tan
\beta=v_2/v_1 \sim m_t/m_b$ the bottom Yukawa coupling is about equal
to the top Yukawa coupling and would practically cancel the top quark
contribution\footnote{The virtual top effect contributing to the radiative
corrections of $\rho=1+\Delta\rho$ allowed a determination of the top
mass prior to the discovery of the top by direct production at
Fermilab in 1995. The LEP precision determination of $\Delta
\rho=\frac{\sqrt{2}G_\mu}{16\pi^2}\,3\,|m_{t}^{2}-m_{b}^{2}|$ (up to
subleading terms) from precision measurements of $Z$ resonance
parameters yields $m_t=172.3^{+10.2}_{-7.6} ~\gv$ in excellent
agreement with the direct determination $m_t=172.6(1.4)~\gv$~\cite{TopMass08} at the
Tevatron.}. Anyway, the possibility of two
Higgs doublets is an interesting option and therefore has been studied
extensively~\cite{GW76}--\cite{Osland07} in the past.

The SM Higgs contribution Eq.~(\ref{SMhiggs}) is tiny, due to the fact
that the $H\bar{\mu}\mu$ Yukawa coupling $y_\mu\,=\,\sqrt{2}\,m_\mu/v$ is very small
because the SM Higgs VEV is large: $v=246.221(1)~\gv$. In 2HDMs of type II
the Yukawa couplings may be enhanced by large factors $\tan
\beta=v_2/v_1$. This is particularly important for the heavier fermions.
The relevant couplings read
\ba
\begin{array}{l l c}
Hf\bar{f}, & f=b,t & -\frac{g}{2} 
 \left( ~\frac{m_b}{M_W} \frac{\cos \al}{\cos \beta}, 
        ~\frac{m_t}{M_W} \frac{\sin \al}{\sin \beta} \right)\\
hf\bar{f}, & f=b,t & -\frac{g}{2} 
 \left( -\frac{m_b}{M_W} \frac{\sin \al}{\cos \beta}, 
        ~\frac{m_t}{M_W} \frac{\cos \al}{\sin \beta} \right)\\
Af\bar{f}, & f=b,t & -\gafi \frac{g}{2} 
 \left( \;\;\frac{m_b}{M_W} \tan \beta, 
  \;\;\frac{m_t}{M_W} \cot \beta \right)\\
H^+ b\bar{t} && \frac{g}{\wz} 
 \left( \;\;\frac{m_b}{M_W} \tan \beta \:\frac{1+\gafi}{2} 
  + \frac{m_t}{M_W} \cot \beta \:\frac{1-\gafi}{2} \right)\:V_{tb}\;\;.
\end{array}
\label{THDMyuk} 
\ea
The couplings for the other fermions are given by analogous expressions. For
example, the coupling for the $\tau$ may be obtained by substituting 
$m_t \ra 0,\;\; m_b \ra m_{\tau}$. 

For the contributions from the diagrams Fig.~\ref{fig:2HDM}a,
using Eqs.~(\ref{NPFFgen},\ref{NPGa}) for $M=m_\mu$ and $M_0=m_h,m_A$
and coupling $f^2=\sqrt{2}G_\mu m_\mu^2\,\tan^2 \beta$ we obtain, assuming
$m_h,m_A \gg m_\mu$
\ba
a^{(2)\:\mathrm{2HDM}}_\mu (h)
&\simeq& \frac{\wz G_\mu m_\mu^2}{4 \pi^2}\,\tan^2\beta\,
\frac{m_\mu^2}{m_h^2}\: \left(\ln \frac{m_h^2}{m_\mu^2 }-\frac{7}{6}
\right) >0\;, \crn
a^{(2)\:\mathrm{2HDM}}_\mu(A)
&\simeq& \frac{\wz G_\mu m_\mu^2}{4 \pi^2}\,\tan^2\beta\,
\frac{m_\mu^2}{m_A^2}\: \left(-\ln \frac{m_A^2}{m_\mu^2 }+\frac{11}{6}
\right)<0\epo   
\ea

\begin{figure}[t]
\centering
\includegraphics{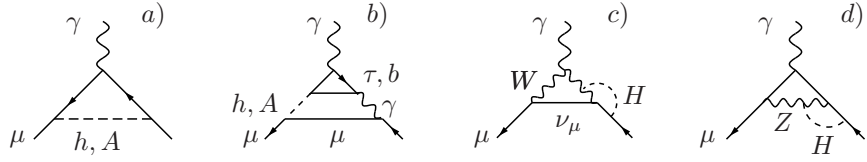}
\caption{Leading 2HDM graphs a) and b) contributing to
$\amu$. Diagrams c) and d), with $H \to h,H,A$, are examples of subleading bosonic contributions
which are modified with respect to the SM weak bosonic contributions
due to the extended Higgs structure.}
\label{fig:2HDM} 
\end{figure}

At 2--loops the Barr--Zee diagram Fig.~\ref{fig:2HDM}b yields an
enhanced contribution, which can exceed the 1--loop result. The
enhancement factor $m_b^2/m_\mu^2$ actually compensates the suppression
by $\alpha/\pi$ as $(\alpha/\pi) \times (m_b^2/m_\mu^2) \sim 4>1$
\be
a^{(4)\:\mathrm{2HDM}}_\mu(h,A) =\frac{\wz G_\mu\:m_\mu^2}{16
\pi^2}\:\frac{\alpha}{\pi} \:4\,\tan^2\beta\,\sum_{i=h,A;f}
N_{cf}Q_f^2\,F_{i}(z_{if}), 
\ee
with $z_{if}=m_f^2/m_i^2$ ($i=h,A$) and
\ba
\!\!\!F_h(z)&=&\,z\,\int_0^1\D x\, \frac{2x\,(1-x)-1}{x\,(1-x)-z}\,\ln \frac{x\,(1-x)}{z}\,
=\,-2z\,(\ln z + 2)+(2z-1)\,F_A(z), \nn \\
\!\!\!F_A(z)&=&\int_0^1\D x \,\frac{z}{x\,(1-x)-z}\,\ln \frac{x\,(1-x)}{z}\,=\,
\frac{2z}{y}\,\left\{\mathrm{Li}_2\left(1-\frac{1-y}{2z}\right)
-\mathrm{Li}_2\left(1-\frac{1+y}{2z}\right)\right\}, 
\label{BarrZeefun}
\ea 
with $y=\sqrt{1-4z}$.
The non--observation of processes like $\Upsilon \to H + \gamma$
sets stringent lower bounds on the scalar masses. Together with the LEP bounds
this prevents large 2HDM contribution to $\amu$.
As an illustration we present values for $\tan \beta=10[40]$,
$m_h=100~\gv$ and $m_A=100[300]~\gv$ in units of $10^{-11}$:\\ 

\begin{center}
\begin{tabular}{crrrrr}
\hline
$(m_h,m_A,\tan \beta)~~$ &$~a^{(2)}_\mu (h)$&$~a^{(2)}_\mu
(A)$&$~a^{(4)}_\mu (h)$&$~a^{(4)}_\mu (A)$ & sum\\
\hline
$(100,100,10)$ & $0.65$ & $-0.62$ & $-2.31$ &  $2.88$ & $0.61$\\
$(100,100,40)$ & $10.45$&$-9.89$& $-36.90$& $~46.09$&$9.74$\\
$(100,300,10)$ &$0.65$ &$-0.08$&$-2.31$&$0.55$&$-1.18$\\
$(100,300,40)$ &$~10.45$&$~~-1.30$&$~~-36.90$&$8.85$ &$~~-18.90$\\
\hline\\
\end{tabular}
\end{center}

If $m_A \sim m_h$ the contributions largely cancel. Assuming $m_h$ to be
in the 100 GeV region, to get a large $m_A-m_h$ mass splitting requires
a large $m_A$, which however yields a large contribution of the
disfavored negative sign. This means that the muon $g-2$ constraint gives a
bound on $m_A$ which, however, strongly depends on $\tan \beta$ 
(see e.g.~\cite{Krawczyk02}--\cite{Marchetti:2008hw} for a more detailed discussion).
Besides the dominant 2-loop contributions form Fig.~\ref{fig:2HDM}b
a 2--loop calculation of the 2HDM contributions, including diagrams
like Figs.~\ref{fig:2HDM}c,d, within the context of
the MSSM has been presented in~\cite{HSW04}. If one identifies $m_{h}$
with $m_H$ of the SM the correction is found to be small:
$a_\mu^{\mathrm{bos,2L}}(\mathrm{MSSM-SM})< 3\power{-11}$ in the
parameter range $m_A\gapprox 50~\gv$ and $\tan \beta \lapprox 50$. 
In fact, in the LL approximation, the 2HDM sector in the MSSM at 2--loops 
does not change the SM result. The reason is that at the 1--loop level the electroweak 
SM result numerically remains practically unchanged, because the
additional 2HDM diagrams all are suppressed by the small Yukawa coupling 
of the $\mu$ (like the SM Higgs contribution).

\subsubsection{Supersymmetry}
\label{sssec:supsym}
The most promising theoretical scenarios for new physics are
supersymmetric extensions of the SM, in particular the minimal
supersymmetric Standard Model. This ``minimal'' extension of
the SM, which doubles the particle spectrum of the SM equipped with an
additional Higgs doublet, is the natural possibility to solve the
Higgs hierarchy problem of the SM. It predicts a low lying Higgs close
to the current experimental bound and allows for a GUT extension where
$G_{\mathrm{GUT}}$ is broken to $G_{\mathrm{SM}}$ at a low scale, in the
phenomenologically interesting region around 1 TeV.

Supersymmetry implements a symmetry mapping $$\mathrm{boson}
\stackrel{\raisebox{.4ex}{Q}}{\leftrightarrow} \mathrm{fermion} $$
between bosons and fermions, by changing the spin by $\pm 1/2$
units~\cite{WessZumino74}. The SUSY algebra [graded Lie algebra]
$\left\{ Q_\alpha, \overline{Q}_\beta \right\}= -2\,\left( \gamma^\mu
\right)_{\alpha \beta}\,P_\mu\,;\:P_\mu=(H,\vec{P})$,
$P_\mu$ the generators of space--time translations, $Q_\alpha$ four
component Majorana (neutral) spinors and $\overline{Q}_\alpha=
\left(Q^+\gamma^0 \right)_\alpha$ the Pauli adjoint, is the only
possible non--trivial unification of internal and space--time symmetry
in a quantum field theory. The Dirac matrices in the Majorana
representation play the role of the structure constants.  The SUSY
extension of the SM associates with each SM state $X$ a supersymmetric
``sstate'' $\tilde{X}$ where sfermions are bosons and sbosons are
fermions. The superpartners for leptons, quarks, gauge and Higgs
bosons are called sleptons, squarks, gauginos and higgsinos,
respectively.  In addition there must be at least one extra Higgs
doublet which also has its SUSY partners. Thus it is the 2HDM (II)
extension of the SM which is subject to global
supersymmetrization. The minimal SUSY extension of the SM assumes that
the SM is ``completed'' by adding Majorana fermions and scalars, with
no new spin 1 bosons and no new Dirac fermions.

We restrict ourselves to a discussion of the MSSM which usually is
thought as a renormalizable low energy effective theory emerging from
a supergravity (SUGRA) model which is obtained upon gauging global
SUSY. SUGRA must include the spin 2 graviton and its superpartner, the
spin 3/2 gravitino. Such a QFT is necessarily non--renormalizable
\cite{SUGRA76}. Nevertheless, it is attractive to consider the MSSM as
a low energy effective theory of a non--renormalizable SUGRA scenario
with $M_{\mathrm{Planck}} \to \infty$~\cite{MSSM}.  SUSY is
spontaneously broken in the hidden sector by fields with no
$SU(3)_c\otimes SU(2)_L\otimes U(1)_Y$ quantum numbers and which
couple to the observable sector only
gravitationally. $M_\mathrm{SUSY}$ denotes the SUSY breaking scale and
the gravitino acquires a mass $$m_{3/2}\sim
M^2_\mathrm{SUSY}/M_\mathrm{Planck}$$ with $M_\mathrm{Planck}$ the
inherent scale of gravity.

SUSY is not realized as a perfect symmetry in nature. SUSY partners of
the known SM particles have not yet been observed because sparticles
in general are heavier than the known particles. Like in the SM, where
the local gauge symmetry is broken by the Higgs mechanism, SUGRA is
broken at some higher scale $M_\mathrm{SUSY}$ by a super--Higgs
mechanism. The Lagrangian exhibits global supersymmetry softly broken
at a scale $M_{\mathrm{SUSY}}$ commonly taken to coincide with the ``new
physics scale'' $\Lambda_{\rm NP}\simeq 1$ TeV, where the SM is expected
to loose its validity. If one assumes the sparticles all to have masses below
$M_{\rm SUSY}$, then relatively light sparticles around 100 GeV are
expected in the spectrum. The MSSM scenario is characterized by the following
features:
\bit
\item 
the gauge group is the SM gauge group with couplings $g_1=e/\cos
\Theta_W$, $g_2=e/\sin \Theta_W$ and $g_3=\sqrt{4 \pi \alpha_s}$ and 
no new heavy gauge bosons besides the W and Z exist;
\item 
there are no new matter fields besides the quarks and leptons and 
two Higgs doublets which are needed to provide supersymmetric masses to quarks
and leptons;
\item
it follows that gauge- and Yukawa-couplings of the
sparticles are all fixed by supersymmetry in terms of the SM couplings;
\item 
in spite of some constraints, masses and mixings of the sparticles remain
quite arbitrary.
\eit
In general, SUSY extensions of the SM lead to Flavor Changing Neutral
Currents (FCNC) and un-suppressed $CP$--violation, which are absent or
small, respectively, in the SM and known to be suppressed in nature.
Therefore one assumes that
\bit
\item
flavor- and CP-violation is as in the SM, namely
coming from the (now supersymmetrized) Yukawa couplings only.
\eit
This implies that at some grand unification scale $M_X$ there is a
universal mass term for all scalars as well as a universal gaugino mass term,
i. e. the SUSY-breaking Majorana masses of the gauginos are equal at $M_X$.
Note that an elegant way to get rid of the mentioned problems
is to impose that
\bit
\item
R-parity, even for particles, odd for sparticles, is conserved.
\eit
This is a strong assumption implying that sparticles must be produced
in pairs and that there exists an absolutely stable {\em lightest
supersymmetric particle} (LSP), the lightest neutralino. Thus all
sparticles at the end decay into the LSP plus normal matter. The LSP
is a Cold Dark Matter (CDM) candidate~\cite{EHNOS83} if it is neutral
and colorless. From the precision mapping of the anisotropies in the
cosmic microwave background, the Wilkinson Microwave Anisotropy Probe
(WMAP) collaboration has determined the relic density of cold dark
matter to~\cite{WMAP}
\be
\Omega_\mathrm{CDM} {\rm h}^2=0.1126\pm0.0081\epo
\label{DMconst}
\ee 
This sets severe constraints on the SUSY parameter
space~\cite{EOSS03,HBaeretal04,Ellis07} and defines the constrained
MSSM (CMSSM) scenario (see also~\cite{Cirelli:2008pk}).  Note that SUSY in general is
providing a new source for CP--violation which could help in
understanding the matter--antimatter asymmetry
$n_B=(n_b-n_{\bar{b}})/n_\gamma\simeq 6 \power{-10}$ present in our
cosmos. Low energy precision tests of supersymmetry and present
experimental constraints are reviewed and discussed
in~\cite{SUSYtests06}. For a topical review on supersymmetry, the
different symmetry breaking scenarios and the muon magnetic moment
see~\cite{Stock06}.

A question is: what should cause R--parity to be conserved? It just
means that certain couplings one usually would assume to be there are
excluded. If $R$ is not conserved, sparticles may be produced singly
and the LSP is not stable and would not provide a possible explanation
of CDM. Mechanisms which mimic approximate R--parity conservation
are known and usually based on a supersymmetric Froggatt-Nielsen model
which assumes a spontaneously broken horizontal local $U(1)_X$ 
symmetry~\cite{FroNi78,GM88,DPS95,BLR96,CCK96,MNR99,JVV00,DT03}.

The SUGRA scenario leads to universal masses for all SUSY partners:
\bit
\item s--matter: $m_{\tilde{q}} = m_{\tilde{\ell}} = m_{\tilde{H}} =m_{1/2}\sim m_{3/2}$
\item gauginos: $M_3 = M_2 = M_1=m_{0} \sim m_{3/2}$
\eit
where $M_3$, $M_2$ and $M_1$ are the mass scales of the spartners of the
gauge bosons in $SU(3)_c$, $SU(2)_L$ and $U(1)_Y$, respectively. 
The non--observation of any sparticles so far requires a mass bound of
about $m_{3/2} \sim 100 \div 1000~\gv\,,$ which is of the order of the
weak scale 246 GeV.\\ 
In general one expects different masses for the different types of gauginos:
\bit
\item $M'$ the $U(1)_Y$ gaugino mass\,,
\item $M$ the $SU(2)_L$ gaugino mass\,,
\item $m_{\tilde{g}}$ the $SU(3)_c$ gluino mass \,.
\eit
However, the grand unification assumption $$M'=\frac{5}{3}\tan^2
\Theta_W\,M\,=\frac{5}{3}\frac{\alpha}{\cos^2\Theta_W\,\alpha_s}\,
m_{\tilde{g}}$$ leads back to the minimal SUGRA (mSUGRA)
scenario\footnote{The difference between CMSSM and mSUGRA is that in the
latter one fixes the Higgsino mixing mass parameter $\mu$ by demanding a 
radiative breaking of the EW symmetry.}.  A very
attractive feature of this scenario is the fact that the known SM
Yukawa couplings now may be understood by evolving couplings from the
GUT scale down to low energy by the corresponding RG equations. This
also implies the form of the muon Yukawa coupling $y_\mu \propto \tan
\beta$, as
\ba
y_\mu=\frac{m_\mu}{v_1}=\frac{m_\mu\,g_2}{\sqrt{2}M_W\,\cos \beta}
\label{muYukawa}
\ea
where $1/\cos \beta \approx \tan \beta$. This enhanced coupling is
central for the discussion of the SUSY contributions to $\amu$. In
spite of the fact that SUSY and GUT extensions of the SM have
completely different motivations and in a way are complementary,
supersymmetrizing a GUT is very popular as it allows coupling constant
unification together with a low SUSY breaking scale which promises
nearby new physics.  Actually, supersymmetric $SU(5)$ circumvents the
problems of the normal $SU(5)$ GUT and provides a viable
phenomenological framework.  The extra GUT symmetry requirement is
attractive also because it reduces the number of independent
parameters.

While supersymmetrizing the SM fixes all gauge and Yukawa couplings of
the sparticles, there are a lot of free 
parameters to fix the SUSY
breaking and masses, such that mixings of the sparticles remain quite
arbitrary: the mass eigenstates of the gaugino--Higgsino sector are
obtained by unitary transformations which mix states with the same
conserved quantum numbers (in particular the charge):
\be 
\chi^+_i=V_{ij} \psi^+_j\,,\:\chi^-_i=U_{ij} \psi^-_j\,,\:\chi^0_i=N_{ij} \psi^0_j\;
\label{SUSYmixpar}
\ee
where $\psi^a_j$ denote the spin 1/2 sparticles of the SM gauge bosons
and the two Higgs doublets.  In fact, a SUSY extension of the SM in
general exhibits more than 100 parameters, while the SM has 28
(including neutrino masses and mixings). 

The main theoretical motivation for a supersymmetric extension of the
SM is the {\bf hierarchy} or {\bf naturalness}
problem. In the SM the Higgs mass is the only mass which is not
protected by a symmetry, which implies the existence of quadratic
divergences in the Higgs self--energy\footnote{In the SM the
quadratic divergence in the Higgs mass counterterm at 1--loop is $\delta m_H^2
\sim 6 (\Lambda/v)^2 (m_H^2+M_Z^2+2M_W^2-4m_f^2 )$ and is absent if the Higgs
mass is tuned to $m_H \simeq (4(m_t^2+m_b^2) -M_Z^2-2M_W^2)^{1/2}\sim 318~~
\mathrm{GeV}$~\cite{Veltman81}, which can be considered to be ruled out by
experiment.}. If we assume that, like in the SM, the Higgs boson is not a (quasi-) Goldstone
boson, then the only known symmetry which requires
this scalar particle to be massless in the symmetry limit is
supersymmetry\footnote{Other ``solutions'' of the hierarchy problem
are the little Higgs models, in which the Higgs is a
quasi-Goldstone boson which attains its mass through radiative
corrections, and the extra dimension scenarios where the effective cut--off can
be low (see below).}. Simply because a scalar is now always a supersymmetric
partner of a fermion which is required to be massless by chiral
symmetry. Thus only in a supersymmetric theory it is natural to have a
``light'' Higgs, in fact in a SUSY extension of the SM, the lightest
scalar $h^0$, which corresponds to the SM Higgs, is bounded to have
mass $m_{h^0} \leq M_Z$ at tree level. This bound receives large
radiative corrections from the $t/\tilde{t}$ sector, which changes the upper 
bound to~\cite{HaHe90} 
\be
m_{h^0} \leq
\left(1+ \frac{\sqrt{2}G_\mu}{2\pi^2\,\sin^2 \beta}\,3m_t^4\,\ln
\left(\frac{m_{\tilde{t}_1}\,m_{\tilde{t}_2}}{m_t^2} \right) + \cdots \right)\,M_Z
\label{SUSY_Hbound}
\ee
which in any case is well below 200 GeV. For an improved bound
obtained by including the 2--loop corrections we refer to~\cite{HiggsBound2}.

It is worthwhile to mention that in an exactly supersymmetric theory
the \amm must vanish, as observed by Ferrara and Remiddi in 1974~\cite{FeRe74}:
$$a_\mu^{\rm tot}=a_\mu^{\rm SM}+a_\mu^{\rm SUSY}=0\epo$$ Thus, since
$a_\mu^{\rm SM}>0$, in the SUSY limit, in the unbroken theory, we must
have $$a_\mu^{\rm SUSY}<0\epo$$ However, we know that SUSY must be
drastically broken. All super--partners of existing particles seem to be
too heavy to be produced up to now. If SUSY is broken $\amu$ may have
either sign. In fact, the 3 standard deviation
$(g_\mu-2)$--discrepancy requires $a_\mu^{\rm SUSY}>0$, of the same
sign as the SM contribution and of at least the size of the weak contribution
[$\sim 200\power{-11}$] (see Fig.~\ref{fig:amucontrib}).

\begin{figure}[t]
\vspace*{-2mm}
\centering
\IfFarbe{%
\includegraphics[height=3.2cm]{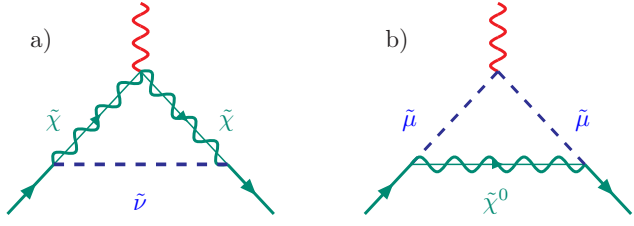}}{%
\includegraphics[height=3.2cm]{SUSY.eps}}
\caption{Leading SUSY contributions to
$g-2$ in supersymmetric extension of the SM. Diagrams a) and b)
correspond to diagrams a) and b) of Fig.~\ref{fig:nplo}, respectively.}
\label{fig:SUSYgraphs}
\end{figure}

The leading SUSY contributions, like the weak SM contributions, are
due to one--loop diagrams. Most interesting are the ones which get
enhanced for large $\tan \beta$.  Such supersymmetric contributions to
$a_\mu$ stem from sneutrino--chargino and smuon--neutralino loops, see
Fig.~\ref{fig:SUSYgraphs}, and yield\footnote{The precise result 
may be easily obtained from the generic 1-loop results of
Sect.~\ref{sssec:generic} with the appropriate choice of
couplings (see Eq.~(\ref{SUSYmixpar})). One obtains~\cite{Moroi95,MaWe01} 
\bea
a_\mu^{\chi^\pm}&=&\frac{m_\mu}{16\pi^2}\sum\limits_{k}
\left\{\frac{m_\mu}{12m^2_{\tilde{\nu}_\mu}}
(|c^L_{k}|^2+|c^R_{k}|^2)\,F_1^C(x_{k})+\frac{m_{\chi_k^\pm}}{3m^2_{\tilde{\nu}_\mu}}
\Repa [c^L_{k}c^R_{k}]\,F_2^C(x_k)\right\}\;,\\
a_\mu^{\chi^0} &=&\frac{m_\mu}{16\pi^2}\sum\limits_{i,m}
\left\{-\frac{m_\mu}{12m^2_{\tilde{\mu}_m}}
(|n^L_{im}|^2+|n^R_{im}|^2)\,F_1^N(x_{im}) 
+\frac{m_{\chi_i^0}}{3m^2_{\tilde{\mu}_m}}
\Repa [n^L_{im}n^R_{im}]\,F_2^N(x_{im})\right\}\;,
\eea
where $k=1,3$ and $i=1,...,4$ denote the chargino and neutralino 
indices, $m=1,2$ is the smuon index, and the couplings are given by
$c^L_k=-g_2\,V_{k1}$,
$c^R_k=y_\mu\,U_{k2}$,
$n^L_{im} =\frac{1}{\sqrt{2}}\,(g_1
N_{i1}+g_2N_{i2})\,U^{\tilde{\mu}~~*}_{m1}
-y_\mu N_{i3}\,U^{\tilde{\mu}~~*}_{m2}$ and
$n^R_{im} =\sqrt{2}\,g_1
N_{i1}\,U^{\tilde{\mu}}_{m2}
+y_\mu N_{i3}\,U^{\tilde{\mu}}_{m1}$.
The kinematical variables are the mass ratios 
$x_{k}=m^2_{\chi^\pm_k}/m^2_{\tilde{\nu}_\mu},~
x_{im}=m^2_{\chi^0_i}/m^2_{\tilde{\mu}_m}$,
and the one--loop vertex functions are given in
Eqs.~(\ref{LNPchirala}) and~(\ref{LNPchiralb}).} 
\ba
a_\mu^{\mathrm{SUSY}\,(1)}&=&a_\mu^{\chi^\pm}+a_\mu^{\chi^0}\cs
\label{amuMSSM1} \\
a_\mu^{\chi^\pm}&=&
~~\frac{g_2^2}{32\pi^2}~~\frac{m_\mu^2}{M^2_\mathrm{SUSY}}\,\mathrm{sign}(\mu
M_2) \:
\tan \beta\:\left[ 1+O(\frac{1}{\tan
\beta},\,\frac{M_W}{M_\mathrm{SUSY}})\right]\cs \\
a_\mu^{\chi^0} &=&
\frac{g_1^2-g_2^2}{192
\pi^2}\,\frac{m_\mu^2}{M^2_\mathrm{SUSY}}\,\mathrm{sign}(\mu M_2) \:
\tan \beta\:\left[ 1+O(\frac{1}{\tan
\beta},\,\frac{M_W}{M_\mathrm{SUSY}})\right]\cs 
\ea
where as usual we expanded in $1/\tan \beta$ and in
$M_W/M_\mathrm{SUSY}$ because we expect that SUSY partners of SM
particles are heavier. Parameters have been taken to be real and $M_1$
and $M_2$ of the same sign\footnote{In the MSSM the parameters
$\mu\,A_f$ and $\mu M_{1,2,3}$ in general are complex. However, not
all phases are observable. In particular, one may assume $M_2$ to be
real and positive without loss of generality.}. The couplings $g_1$
and $g_2$ denote the $U(1)_Y$ and $SU(2)_L$ gauge couplings,
respectively, and $y_\mu$ is the muon's Yukawa coupling
Eq.~(\ref{muYukawa}).  The interesting aspect of the SUSY contribution
to $\amu$ is that they are enhanced for large $\tan \beta$ in contrast
to SUSY contributions to electroweak precision observables, which
mainly affect $\Delta \rho$ which determines the $\rho$--parameter and
contributes to $M_W=M_W(\alpha,G_\mu,M_Z,\cdots)$. The
\amm thus may be used to constrain the SUSY parameter space in a
specific way.  Altogether one obtains
\begin{eqnarray}
\!\!\!\!\!\!\! a_\mu^{\rm SUSY}\!\!
 \simeq  \frac{\mathrm{sign}(\mu M_2)\,\alpha(M_Z)}{
8\pi\sin^2\Theta_W}\,}\,\frac{\left(5+\tan^2 \Theta_W \right)}{6}{
\frac{m_\mu^2}{M_{\rm SUSY}^2}\:}{ \tan\beta}{ \: \left( 1-\frac{4\alpha}{
\pi}\ln \frac{M_{\rm SUSY}}{ m_\mu}\right)
\label{improved SUSYcorr}
\end{eqnarray}
with $M_{\rm SUSY}$ a typical SUSY loop mass and the sign is
determined by the Higgsino mass term $\mu$. Here we also included the
leading 2--loop QED logarithm as an RG improvement factor~\cite{DG98},
which amounts to parametrize the one-loop result in terms of the
running $\alpha(M_{\rm SUSY})$.  Note that $M_{\rm SUSY}$ in the
logarithm is the mass of the lightest charged SUSY particle.  In
Fig.~\ref{fig:SUSYplot} contributions are shown for various values of
$\tan \beta$. Above $\tan \beta \sim 5$ and $\mu > 0$ the SUSY
contributions from the diagrams Fig.~\ref{fig:SUSYgraphs} easily could
explain the observed deviation Eq.~(\ref{amuNP}) with SUSY states of
masses in the interesting range 100 to 500 GeV.

In the large $\tan\beta$ regime we have
\begin{eqnarray}
\left|a_\mu^{\rm SUSY}\right|
 \simeq 123\times 10^{-11} \left( 100 {\rm\ GeV}\over
M_{\rm SUSY}\right)^2 \tan\beta \epo
\end{eqnarray}
$a_\mu^{\rm SUSY}$ generally has the same sign as the $\mu$--parameter.
The deviation Eq.~(\ref{amuNP}) requires positive $\mathrm{sign}(\mu)$
and if identified as a SUSY contribution
\begin{eqnarray}
M_{\rm SUSY} \simeq (65.5\mbox{ GeV})\sqrt{\tan\beta} \epo
\end{eqnarray}
Negative $\mu$ models give the opposite sign contribution to $a_\mu$ and
are strongly disfavored.  For $\tan\beta$ in the range $2\sim 40$ one obtains
\begin{eqnarray}
M_{\rm SUSY} \simeq 93-414 \mbox{ GeV} \cs
\end{eqnarray}
precisely the range where SUSY particles are often expected. 
For a more elaborate discussion and further
references we refer to~\cite{CM01,Stock06}.
\begin{figure}[t]
\centering
\IfFarbe{%
\includegraphics[height=5cm]{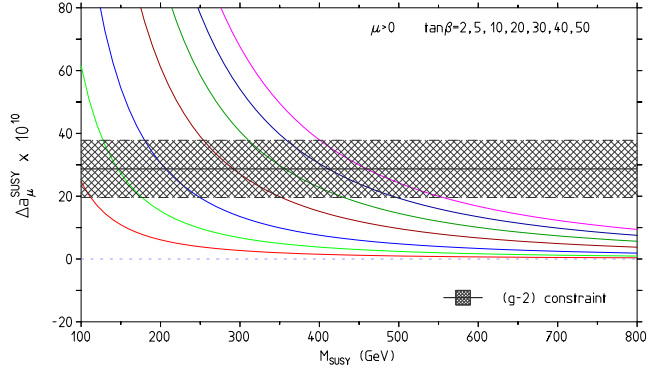}}{%
\includegraphics[height=5cm]{SUSYplot.eps}}
\caption{Constraint on large tan$\beta$ SUSY contributions as a
function of $M_\mathrm{SUSY}$.}
\label{fig:SUSYplot}
\end{figure}

A remarkable 2--loop calculation within the MSSM has been performed by
Heinemeyer, St\"ockinger and Weiglein~\cite{HSW_SUSY2,HSW04}. They
evaluated the exact 2--loop correction of the SM 1--loop contributions
Figs.~\ref{fig:oneloopdia} and \ref{fig:oneloopweak}. These are all
diagrams where the $\mu$--lepton number is carried only by $\mu$
and/or $\nu_\mu$. In other words SM diagrams with an additional
insertion of a closed sfermion-- or charginos/neutralino--loop. These
are diagrams like the electroweak ones Fig.~\ref{fig:twoloopweak} with
the SM fermion--loops replaced by closed $\tilde{\chi}$--loops plus
diagrams obtained be replacing $W^\pm$ with $H^\pm$. Thus
the full 2--loop result from the class of diagrams with closed
sparticle loops is known. This class of SUSY contributions is
interesting because it has a parameter dependence completely different
from the one of the leading SUSY contribution and can be large in
regions of parameter space where the 1--loop contribution is
small. The second class of corrections are the 2--loop corrections to
the SUSY 1--loop diagrams Fig.~\ref{fig:SUSYgraphs}, where the
$\mu$--lepton number is carried also by $\tilde{\mu}$ and/or
$\tilde{\nu}_\mu$. This class of corrections is expected to have the
same parameter dependence as the leading SUSY 1--loop ones and only
the leading 2--loop QED corrections are known~\cite{DG98} as already
included in Eq.~(\ref{improved SUSYcorr}). Results which illustrate
our brief discussion are shown in Fig.~\ref{fig:ammMSSM2}.  
The contributions of the 2HDM sector of the MSSM have been discussed
earlier in Sect.~\ref{sssec:THDM}.
\begin{figure}[t]
\centering
\includegraphics[height=5cm]{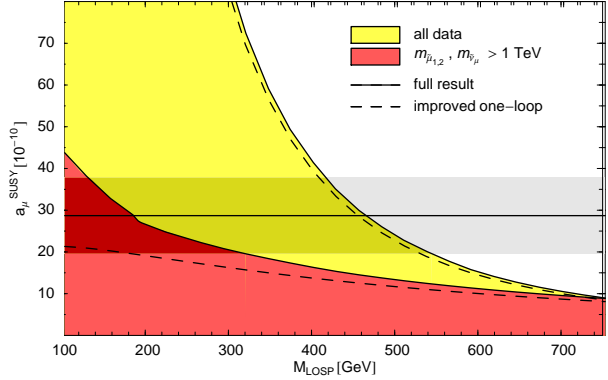}
\caption{\label{fig:ammMSSM2}
Allowed values of MSSM contributions to $a_\mu$ as a function of the
mass of the Lightest Observable SUSY Particle $M_{\rm LOSP}$, from an
MSSM parameter scan with $\tan\beta=50$. The $3 \sigma$ region
corresponding to the deviation (\ref{amuNP}) is indicated as a
horizontal band. The light gray region corresponds to all input
parameter points that satisfy the experimental constraints from
$b$--decays, $m_h$ and $\Delta\rho$. In the middle gray region, smuons
and sneutrinos are heavier than 1 TeV. The dashed lines correspond to
the contours that arise from ignoring the 2--loop corrections from
chargino/neutralino-- and sfermion--loop diagrams. Courtesy of
D. St\"ockinger~\cite{Stock06}.}
\end{figure}

The results for the SUSY contributions to $\amu$ up to the two-loops
is given by~\cite{Stock06}\footnote{All leading terms come from
Barr-Zee type diagrams. In terms of the functions
Eq.~(\ref{BarrZeefun}) the results read~\cite{Chang00,Cheung01,Arhrib01,Chen01,Feng06}
\bea
a_\mu^{(\chi \gamma H)}&=&\frac{\wz G_\mu m_\mu^2}{8\pi^2}\,\frac{\alpha}{\pi}
\sum_{k=1,2}\,\left[\, \Repa [ \lambda_\mu^{A}\lambda_{^{\chi_k^+}}^{A}]\,
F_A(m^2_{^{\chi_k^+}}/m_A^2)\,+ 
\sum_{S=h,H}\, \Repa [ \lambda_\mu^{S}\lambda_{^{\chi_k^+}}^{S}]\,
F_h(m^2_{^{\chi_k^+}}/m_S^2)\right]\;,
\crn
a_\mu^{(\tilde{f} \gamma H)}&=&\frac{\wz G_\mu m_\mu^2}{8\pi^2}\,\frac{\alpha}{\pi}
\sum_{\tilde{f}=\tilde{t},\tilde{b},\tilde{\tau}}\,\sum_{i=1,2}\,
\left[\sum_{S=h,H}\,  (N_cQ^2)_{\tilde{f}}\, \Repa [
\lambda_\mu^{S}\lambda_{\tilde{f}_i}^{S}]\,
F_{\tilde{f}}(m^2_{\tilde{f}_i}/m_S^2) \right]\;,
\eea
with $F_{\tilde{f}}(z)=z\,(2+\ln z- F_A(z))/2$ and couplings (see
Eqs.~(\ref{THDMyuk},~\ref{SUSYmixpar}))
\bea
\lambda_\mu^{h,H,A}&=&(-\sin \al/\cos \beta,\cos \al/\cos \beta,\tan \beta)\;,\crn
\lambda_{^{\chi_k^+}}^{h,H,A}&=&\wz M_W/m_{^{\chi_k^+}}\,
(U_{k1}V_{k2}\,(\cos \al,\sin \al,-\cos \beta)+U_{k2}V_{k1}(-\sin \al,\cos \al,-\sin \beta))\;,\crn
\lambda_{\tilde{\tau}_i}^{h,H}&=&2m_\tau/(m^2_{\tilde{\tau}_i}\cos \beta)\,
(-\mu^*\,(\cos \al,\sin \al)+A_\tau\,(-\sin \al,\cos \al))\,
(U^{\tilde{\tau}_{i1}})^*U^{\tilde{\tau}_{i2}}\epo
\eea
The last expression given for the $\tilde{\tau}$ applies to the $\tilde{b}$ with $\tau \to b$
everywhere, and for the $\tilde{t}$ with $\tau \to t$ together with 
$(\mu,\cos \beta,\cos \al,\sin \al) \to (-\mu, \sin \beta, \sin \al,
-\cos \al)$.}
\begin{eqnarray}
\amuSUSY &=&
\amuSUOL\left(1-\frac{4\alpha}{\pi}\log\frac{M_{\rm
    SUSY}}{m_\mu}\right)
+
\amu^{(\chi\gamma H)}+\amu^{(\tilde{f}\gamma H)}
\nonumber\\
&&+\amu^{(\chi\{W,Z\} H)}+\amu^{(\tilde{f}\{W,Z\}H)}
+\amuSUTLferm + \amuSUTLbos
+\ldots\,.
\label{amuSUSYknown}
\end{eqnarray}
The labels $(\chi \gamma H)$ etc identify contributions from
Fig.~\ref{fig:2HDM}b type diagrams which would be labeled by $(\tau h
\gamma)$, with possible replacements $\gamma \to V=\gamma,Z,W^\pm$, $h
\to H=h,H,A,H^\pm$ and $\tau^\mp \to X=\chi^\mp,\chi^0,\tilde{f}$. 
Contributions $(XVV)$ correspond to Fig.~\ref{fig:twoloopweak}a,d with
corresponding substitutions.  The remaining terms $\amuSUTLferm$ and
$\amuSUTLbos$ denote small terms like the fermionic contribution
Fig.~\ref{fig:2HDM}b and the bosonic contributions Figs.~\ref{fig:2HDM}c,d,
which differ from the SM result due to the modified Higgs structure.
The ellipsis denote the known but negligible 2--loop contributions as
well as the missing 2--loop and higher order contributions.

For the potentially enhanced Barr-Zee type contributions
the following simple approximations have been given~\cite{HSW_SUSY2,HSW04,Stock06}:
\begin{eqnarray}
\label{chaneuapprox}
\amu^{(\chi VH)} &\approx &
\ \ 11\tunit\ \left(\frac{\tan\beta}{50}\right)
\left(\frac{100\ {\rm GeV}}{\MSUSY}\right)^2\
{\rm sign}(\mu M_2),\\
\label{stopapprox}
\amu^{(\tilde{t}\gamma H)} &\approx &
-13\tunit\ \left(\frac{\tan\beta}{50}\right)
\left(\frac{m_t}{m_{\tilde{t}}}\right)
\left(\frac{\mu}{20 M_H}\right)
\ {\rm sign}(X_t)
,\\
\label{sbotapprox}
\amu^{(\tilde{b}\gamma H)} &\approx &
-3.2\tunit\ \left(\frac{\tan\beta}{50}\right)
\left(\frac{m_b\tan\beta}{m_{\tilde{b}}}\right)
\left(\frac{A_b}{20 M_H}\right)
\ {\rm sign}(\mu)\;.
\end{eqnarray}
The parameter $X_t$ is determined by the SUSY breaking parameter
$A_f$, $\mu$ and $\tan \beta$ by $X_t=A_t-\mu^*\cot \beta$.  Like for
the leading 1--loop case, the first approximation applies if all SUSY
masses are approximately equal (e.g. $\mu \sim M_2 \sim m_A$) (but the
relevant masses are different in the two cases), and the second and
third are valid if the stop/sbottom mixing is large and the relevant
stop/sbottom and Higgs masses are of similar size. We refer to the
review by St\"ockinger~\cite{Stock06} for a more detailed presentation
of the higher order SUSY effects.

In comparison to $g_\mu-2$, the SM prediction of $M_W$~\cite{MWSM04},
as well as of other electroweak observables, as a function of $m_t$
for given $\alpha$, $G_\mu$ and $M_Z$, is in much better agreement
with the experimental result (1 $\sigma$), although the MSSM
prediction for suitably chosen MSSM parameters is slightly favored by
the data~\cite{MWMSSM06}.  Thus large extra corrections to the ones of
the SM are not tolerated.  The leading radiative shift of $M_W$ by
SUSY contributions enters via $\Delta \rho$. As we know, $\Delta
\rho$ is most sensitive to weak isospin splitting and in the SM is
dominated by the contribution from the ($t,b$)--doublet. In the SUSY
extension of the SM these effects are enhanced by the contributions
from the four SUSY partners $\tilde{t}_{\rm L,R},\tilde{b}_{\rm L,R}$
of $t,b$, which can be as large as the SM contribution itself for
$m_{1/2}\ll m_t$ [light SUSY], and tends to zero for $m_{1/2}\gg m_t$
[heavy SUSY]. It is important to note that these contributions are not
enhanced by $\tan \beta$. Thus, provided  $\tan \beta$ enhancement is
at work, it is quite natural to get a larger
SUSY contribution to $g_\mu-2$ than to $M_W$, otherwise some 
tension between the two constraints would be there as $M_W$ prefers
the heavy SUSY domain.

\begin{figure}[t]
\centering
\parbox{1.2in}{%
\hspace*{-3.5cm}%
\includegraphics[height=6.4cm]{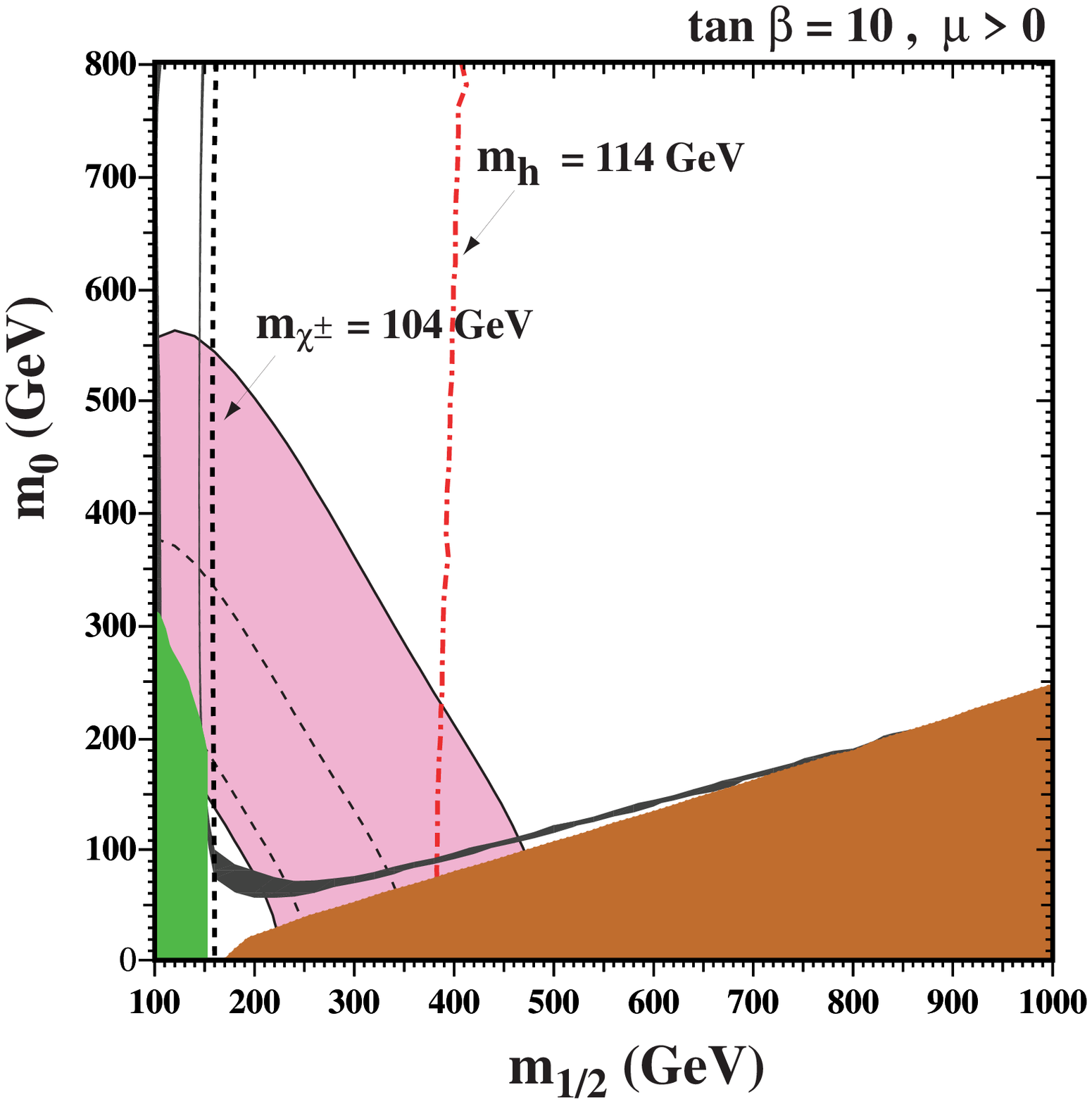}}
\qquad 
\begin{minipage}{1.2in}%
\vspace*{0mm}
\hspace*{-0.5cm}%
\includegraphics[height=6.4cm]{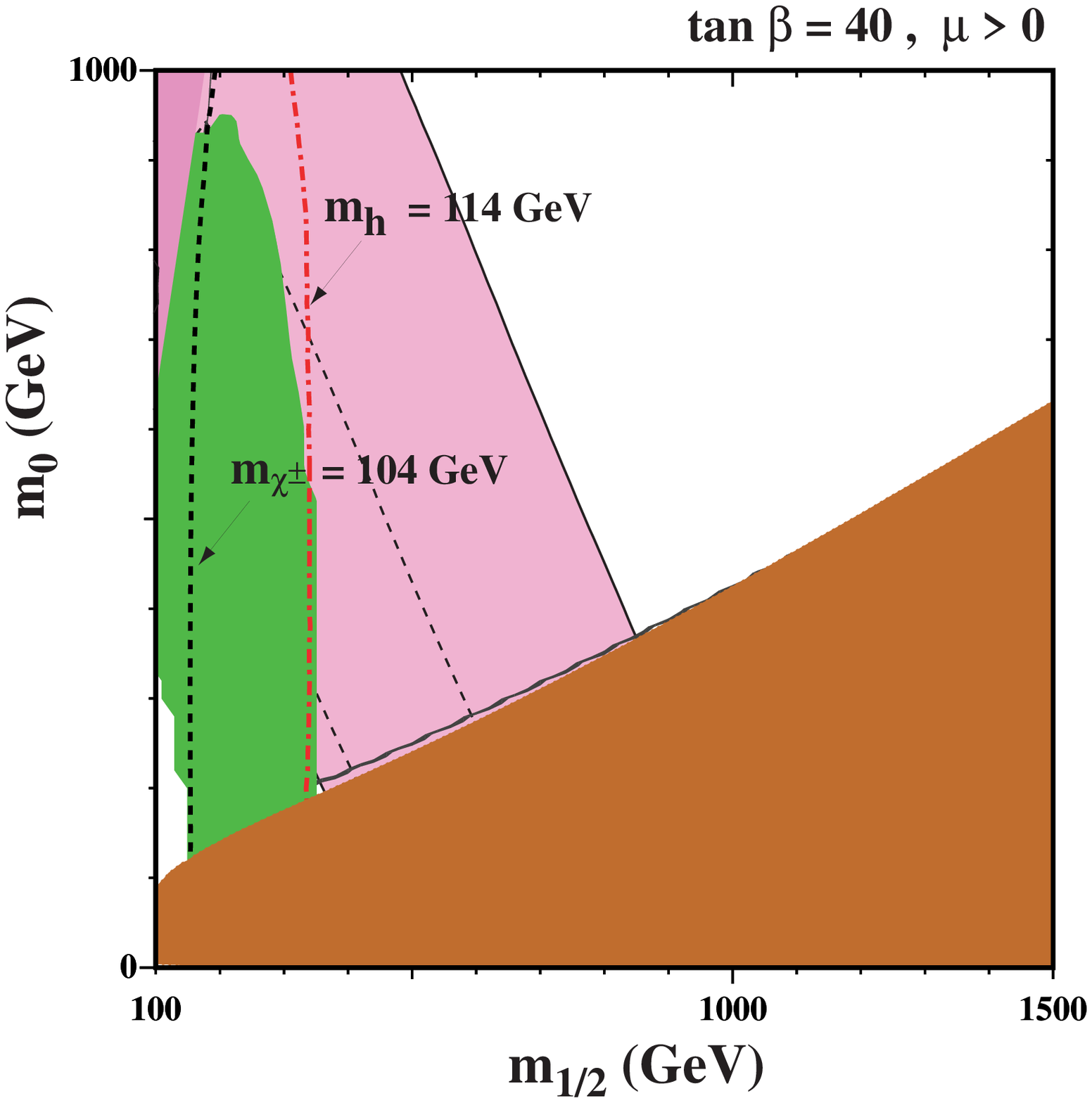}
\end{minipage}%
\vspace*{-2mm}
\unitlength1mm
\begin{picture}(116,62)(0,-58)
\SetOffset(-8,1)
\SetWidth{0.6}
\Text(17,27)[]{\bf g-2}
\Text(13,34)[]{\bf 2$\sigma$}
\Text(14,20)[]{1$\sigma$}
\Text(42,17)[]{\bf excluded for}
\Text(35,13)[]{\bf neutral dark matter}
\SetOffset(-8,1.6)
\Text(44,48)[]{\bf $\varOmega$h$^2$}
\Line(124,130)(70,48)
\ArrowLine(70.27,48.41)(69.73,47.59)
\SetOffset(0.6,0)
\SetWidth{0.3}
\LongArrow(41.55,48.3)(50.81,53.4)
\LongArrow(50.81,53.4)(41.55,48.3)
\SetWidth{0.6}
\SetOffset(61,2)
\Text(26,32)[]{\bf g-2}
\Text(22,43)[]{\bf 2$\sigma$}
\Text(19,27)[]{1$\sigma$}
\Text(42,17)[]{\bf excluded for}
\Text(36,13)[]{\bf neutral dark matter}
\SetOffset(58,0.7)
\Text(44,48)[]{\bf $\varOmega$h$^2$}
\Line(124,130)(81,72)
\ArrowLine(81.65,72.88)(81,72)
\SetOffset(60.6,1)
\SetWidth{0.3}
\LongArrow(50.35,68.35)(65.15,75.15)
\LongArrow(65.15,75.15)(50.35,68.35)
\SetWidth{0.6}
\unitlength1pt
\SetOffset(-28,0)
\rText(26,34)[b][l]{b{\scriptsize $\to$}s$\gamma$~{\bf ex}}
\SetOffset(165.71,0)
\rText(28.86,74.29)[l][l]{b{\scriptsize $\to$}s$\gamma$~{\bf ex}}
\end{picture}
\vspace*{-58mm}
\caption{The $(m_0,m_{1/2})$ plane for $\mu > 0$ for (a) $\tan \beta =
10$ and (b) $\tan \beta =40$ in the constrained MSSM scenario (which
includes mSUGRA). The allowed region by the cosmological neutral dark
matter constraint Eq.~(\ref{DMconst}) is shown by the black parabolic
shaped region. The disallowed region where $m_{\tilde{\tau}_1}<
m_\chi$ has dark shading. The regions excluded by $b\to s\gamma$ have
medium shading (left). The $g_\mu-2$ favored region at the 2 $\sigma$
[$(290\pm 180)\power{-11}$] (between dashed lines the 1 $\sigma$
[$(290\pm 90)\power{-11}$] band) level has medium shading. The LEP
constraint on $m_{\chi^\pm}=104~\gv$ and $m_h=114~\gv$ are shown as
near vertical lines. Plot courtesy of K. Olive updated
from~\cite{EOSS03}.}
\label{fig:CDMetc}
\end{figure}

Assuming the very restricted CMSSM scenario, besides the direct limits from LEP and
Tevatron, presently, the most important constraints come from
$(g-2)_\mu$, $b \to s \gamma$ and from the dark matter relic density
(cosmological bound on CDM) given in
Eq.~(\ref{DMconst})~\cite{EOSS03,HBaeretal04}.
Due to the precise value of $\Omega_\mathrm{CDM}$, the lightest SUSY
fermion (sboson) of mass $m_0$ is given as a function of the lightest
SUSY boson (sfermion) with mass $m_{1/2}$ within a narrow band. In
fact only a part of the relic cold dark matter need be neutralinos,
such that $\Omega_\mathrm{CDM}$ is an upper bound only. This
is illustrated in Fig.~\ref{fig:CDMetc} together with the constraints
from $g_\mu-2$ Eq.~(\ref{amuNP}) and $b \to s
\gamma$. Since $m_h$ for given $\tan \beta$ is fixed by $m_{1/2}$  
via Eq.~(\ref{SUSY_Hbound}) with $\mathrm{min}(m_{\tilde{t}_i};i=1,2) \sim
m_{1/2}$, the allowed region is to the right of the (almost vertical)
line $m_h=114~\gv$ which is the direct LEP bound. Again there is an
interesting tension between the SM like lightest SUSY Higgs mass
$m_h$ which in case the Higgs mass goes up from the present limit to
higher values requires heavier sfermion masses and/or lower $\tan
\beta$, while $\amu$ prefers light sfermions and large $\tan
\beta$. Another lower bound from LEP is the line characterizing
$m_{\chi^\pm}>104~\gv$.  The CDM bound gives a narrow hyperbola like
shaped band. The cosmology bound is harder to see in the $\tan \beta
=40$ plot, but it is the strip up the $\chi-{\tilde{\tau}}$ degeneracy
line, the border of the excluded region (dark) which would correspond
to a charged LSP which is not allowed. The small shaded region in the
upper left is excluded due to no electroweak symmetry breaking (EWSB)
there. The latter must be tuned to reproduce the correct value for
$M_Z$.  The $\tan \beta =40$ case is much more favorable, since
$g_\mu-2$ selects the part of the WMAP strip which has a Higgs above
the LEP bound. Within the CMSSM the discovery of the Higgs and the
determination of its mass would essentially fix $m_0$ and $m_{1/2}$.

Since the SM prediction~\cite{Misiaketal06}  for the $b \to s \gamma$ rate
$\mathrm{BR}(b\to s \gamma)= (3.15\pm0.23)\power{-4}$ is in
good agreement with the experimental value, only small
extra radiative corrections are allowed (1.5 $\sigma$).  In SUSY extensions of the
SM~\cite{BaGi93}, this excludes light $m_{1/2}$ and $m_0$ from light
to larger values depending on $\tan \beta$.  Ref.~\cite{Misiaketal06}
also illustrates the updated $b \to s \gamma$ bounds on $M_{H^+}$ ($>
295~\gv$ for $2 \leq \tan \beta $) in the 2DHM (Type II)~\cite{ASW79}.

It is truly remarkable that in spite of the different highly non--trivial
dependencies on the MSSM parameters, with $g-2$ favoring definitely
$\mu >0$, $\tan \beta$ large and/or light SUSY states, there is a
common allowed range, although a quite narrow one, depending strongly
on $\tan \beta$.
 
In general less constrained versions of the MSSM or other SUSY
extensions of the SM (see e.g.~\cite{Barger:2004mr}) are much harder
to pin down or even to disprove. Certainly, the muon $g-2$ plays an
important role for the upcoming LHC searches.  Note that the 3
$\sigma$ deviation (if real) requiring $\mathrm{sign}(\mu)$ positive
and $\tan \beta$ preferably large, poses constraints which hardly can
be obtained from a hadron collider.  In any case, the muon $g-2$ would
yield important hints for constraining the SUSY parameter space, if
SUSY would show up at the LHC.

\subsubsection{Little Higgs models} 

The Higgs mass in the SM is not protected by any symmetry and turns
out to be quadratically sensitive to the cutoff $\Lambda$. This leads
to the well-known hierarchy or fine-tuning problem, in particular, if
the SM is embedded in some grand unified theory or if gravity (and
thus the Planck scale) enters.

Less known is the the so-called ``little hierarchy
problem''~\cite{LEP_paradox}. We can view the SM as an effective field theory
(EFT) with a cutoff and parametrize new physics in terms of higher-dimensional
operators which are suppressed by inverse powers of the cutoff
$\Lambda$. Precision tests of the SM at low energies and at LEP/SLC point to a
small Higgs mass of order 100~GeV and have not shown any significant
deviations from the SM. This in turn translates into a cutoff of about
$\Lambda \sim 5-10~\mbox{TeV}$ which is slightly more than an order of
magnitude above the electroweak scale and thus requires fine-tuning at the
percent level only.

An attractive set of solutions to this little hierarchy problem are the little
Higgs models~\cite{LH_original,LH_reviews}. In these models, the Higgs boson
is a pseudo-Goldstone boson of a global symmetry which is spontaneously broken
at a scale $f$. This symmetry protects the Higgs mass from getting quadratic
divergences at one loop, even in the presence of gauge and Yukawa
interactions. The electroweak symmetry is broken via the Coleman-Weinberg
mechanism~\cite{Coleman_Weinberg} and the Higgs mass is generated radiatively,
which leads naturally to a light Higgs boson $m_H \sim (g^2 / 4\pi) f \approx
100~\mbox{GeV}$, if the scale $f \sim 1~\mbox{TeV}$.\footnote{If one
implements the Coleman-Weinberg mechanism directly in the SM, the Higgs mass
would be fixed at about 10 GeV~\cite{Ellis79} which as we know is ruled out by
experiment.} In contrast to supersymmetric theories, here the new states at
the TeV-scale which cancel the quadratic divergences arising from the top
quark, gauge boson and Higgs boson loops, respectively, have the same spin as
the corresponding SM particles. The little Higgs model can then be interpreted
as an EFT up to a new cutoff scale of $\Lambda \sim 4 \pi f \sim
10~\mbox{TeV}$.

Among the different versions of this approach, the littlest Higgs
model~\cite{Littlest_Higgs} achieves the cancellation of quadratic divergences
with a minimal number of new degrees of freedom.  The global symmetry breaking
pattern is $SU(5) \to SO(5)$ and an $[SU(2) \times U(1)]^2$ gauge symmetry is
imposed, that is broken down at the scale $f$ to the diagonal subgroup
$SU(2)_L \times U(1)_Y$, which is identified with the SM gauge group. This
leads to four heavy gauge bosons $A_H, W_H^\pm, Z_H$ with masses $\sim f$ in
addition to the SM gauge fields. The SM Higgs doublet is part of an assortment
of pseudo-Goldstone bosons which result from the spontaneous breaking of the
global symmetry. The multiplet of Goldstone bosons contains a heavy $SU(2)$
complex triplet scalar $\Phi$ as well.  Furthermore, a vector-like heavy quark
that can mix with the top quark is postulated.  It turns out, however, that
electroweak precision data put very strong constraints on the littlest Higgs
model.  Typically one obtains the bound $f \gapprox 3-5~\mbox{TeV}$ in most of
the natural parameter space, unless specific choices are made for fermion
representations or hypercharges~\cite{LH_EW_tests}. Since $f$ effectively acts
as a cutoff for loops with SM particles, this reintroduces a little hierarchy
between the Higgs boson mass and the scale $f$.

These constraints from electroweak precision data can be bypassed by imposing
a discrete symmetry in the model, called T-parity~\cite{T_parity}.  In the
littlest Higgs model with T-parity (LHT)~\cite{LHT_Low}, this discrete
symmetry maps the two pairs of gauge groups $SU(2)_i \times U(1)_i, i=1,2$
into each other, forcing the corresponding gauge couplings to be equal, with
$g_1 = g_2$ and $g_1^\prime = g_2^\prime$.  All SM particles, including the
Higgs doublet, are even under T-parity, whereas the four additional heavy
gauge bosons and the Higgs triplet are T-odd. The top quark has now two heavy
fermionic partners, $T_{+}$ (T-even) and $T_{-}$ (T-odd). For consistency of
the model, one has to introduce additional heavy, T-odd vector-like fermions
$u^i_H, d^i_H, e^i_H$ and $\nu^i_H$ ($i=1,2,3$) for each SM quark and lepton
field. For further details on the LHT, we refer the reader to
Refs.~\cite{LHT_Low,Hubisz_Meade,Hubisz_et_al,Chen_Tobe_Yuan}.

In the LHT there are no tree-level corrections to electroweak precision
observables and there is no dangerous Higgs triplet vev $v^\prime$ that
violates the custodial symmetry of the SM grossly. This relaxes the
constraints on the model from electroweak precision data and allows a
relatively small value of $f$ in certain regions of the parameter space.  As
shown in Refs.~\cite{Hubisz_et_al,Asano_et_al,Hundi_Mukhopadhyaya_Nyffeler}, a
scale $f$ as low as $500~\mbox{GeV}$ is compatible in the LHT with electroweak
precision data.  Furthermore, if T-parity is exact, the lightest T-odd
particle, typically the heavy, neutral partner of the photon, $A_H$, is stable
and can be a good dark matter
candidate~\cite{Hubisz_Meade,Asano_et_al,LH_Dark_Matter}. The LHT can thus
serve as an attractive alternative to supersymmetric or extra-dimensional
models. 

\textbf{$a_\mu$ in the Littlest Higgs model:}
The contribution to the muon $g-2$ in the Littlest Higgs model has
been calculated in
Refs.~\cite{Park_Song,Casalbuoni_etal,Choudhury_etal}. The relevant
diagrams are of the generic type shown in
Figs.~\ref{fig:nplo}c,~\ref{fig:nplo}d.  Ref.~\cite{Park_Song} only
considered the effect of loops containing the new heavy gauge bosons
$A_H, W_H$ and $Z_H$ and found that even the contribution from the
lightest of them, $A_H$, which could have a mass as low as $200$~GeV,
is very small. Typically, they obtain $a_\mu^{\mathrm{LH}} < 1
\times 10^{-11}$ in all of the allowed parameter space, assuming that
the symmetry breaking scale fulfills $f > 1$~TeV. Soon afterwards it
was, however, pointed out in Ref.~\cite{Casalbuoni_etal} that this
calculation was incomplete. In fact, the couplings of the ordinary SM
gauge fields $W$ and $Z$ to SM fermions are modified in the Littlest
Higgs model by factors $\left(1 + {\cal O}(v^2/f^2)\right)$ and thus
lead to a modification of the usual one-loop SM electroweak
contribution to the muon $g-2$. This leads in general to a bigger
effect than loops involving new heavy particles. For instance, for $f
= 1$~TeV, they obtain in some region of the allowed parameter space a
contribution of about
\begin{equation}
a_\mu^{\mathrm{LH}} \sim 10 \times 10^{-11}, 
\end{equation}
i.e. a factor 10 larger than the value quoted above. Nevertheless, even this
larger value cannot explain the observed discrepancy between the measured
value and the SM prediction of $a_\mu^{\mathrm{exp}} - a_\mu^{\mathrm{SM}} =
(290\pm 90)\: \power{-11}$ from Eq.~(\ref{amuNP}). Furthermore, as mentioned
above, electroweak precision data actually constrain the value of the symmetry
breaking scale $f$ to lie above $3-5$~TeV~\cite{LH_EW_tests}. For such high
values of $f$, the corrections to the $W$ and $Z$ couplings are much smaller
and the contribution to $a_\mu$ becomes again tiny. In addition, for higher
values of $f$, the new heavy particles also become heavier and thus their
contribution in loops is even more suppressed.

In Ref.~\cite{Choudhury_etal} a slightly modified version of the Littlest
Higgs models was studied, which included lepton number violating terms. These
terms which involve the complex Higgs triplet field $\Phi$ were first
introduced in Ref.~\cite{LH_LNV} to generate neutrino masses. The
corresponding new Yukawa couplings $Y_{ab}$, where $a,b$ denote different
generations, and the vev $v^\prime$ of the Higgs triplet have to satisfy the
relation $Y_{ab} v^\prime \sim 10^{-10}$~GeV in order to lead to neutrino
masses in the expected range. Assuming that $v^\prime$ is of the order of
$10^{-10}$~GeV, which requires, however, a substantial amount of (unnatural)
fine-tuning, one could in principle have $Y_{ab} \sim {\cal O}(1)$. The
contribution to the anomalous magnetic moment then reads, in the limit
$m_{\phi} \gg m_i, i=e,\mu,\tau$,
\begin{equation}
a_\mu^{\mathrm{LNV-LH}} = \sum_{i=e,\mu,\tau} {3 \over 16 \pi^2} {m_\mu^2
  \over m_\Phi^2} |Y_{\mu i}|^2.
\end{equation}
Thus for $m_\phi \sim 1$~TeV and $Y_{\mu i} \sim 1-3$ one can obtain
$a_\mu^{\mathrm{LNV-LH}} \sim (60 - 600) \times 10^{-11}$ and thus ``explain''
the observed discrepancy. On the other hand, in the same
paper~\cite{Choudhury_etal}, the authors study the lepton number violating
decay $\mu^- \to e^+ e^- e^-$ which is also induced by the new terms in the
Lagrangian. From the bound $\mbox{BR}(\mu^- \to e^+ e^- e^-) < 1 \times
10^{-12}$ on the corresponding branching ratio they then obtain bounds on the
Yukawa couplings $Y_{ab}$. For instance, for $m_\phi \sim 1$~TeV, the relation
$Y_{ee}^* Y_{\mu e} < 0.2 \times 10^{-5}$ has to hold in order not to violate
the bound on the branching ratio. It therefore seems questionable
whether one can really obtain the above quoted large value for $a_\mu$ where
$Y_{ab} \sim 1$ is needed. In any case a perturbative treatment would
be obsolete.

\textbf{$a_\mu$ in the Littlest Higgs model with T-parity:} 
The value of the muon $g-2$ in the Littlest Higgs model with T-parity
(LHT) was studied in
Refs.~\cite{Goyal,Choudhury_etal_LHT,Blanke_etal}. Leading
contributions are given by 1--loop diagrams of the type shown in
Figs.~\ref{fig:nplo}c,~\ref{fig:nplo}d. While the conclusions of the
papers concerning the anomalous magnetic moment agree qualitatively,
Ref.~\cite{Blanke_etal} disagrees with the analytic expressions for
the contributions given in the earlier two papers.

The main difference to the Little Higgs model without T-parity is,
that in the LHT the couplings of the SM gauge bosons $W,Z$ to SM
fermions are unchanged. Therefore only loops involving new heavy
particles contribute to $a_\mu$. In addition to diagrams involving the
(T-odd) heavy gauge bosons, the new T-odd heavy ``mirror leptons''
$\ell_H^i$ and neutrinos $\nu_H^i$ ($i=1,2,3$) can now appear in the
loops. Assuming that bounds on slepton searches at LEP apply roughly
also to these leptons in the LHT, they could be as light as
200-300~GeV.  Furthermore, since the electroweak precision tests allow
scales as low as $f \sim 500$~GeV, the ``heavy'' photon can be rather
light, $M_{A_H} \sim 65$~GeV (also $W_H$ and $Z_H$ will be around 300
GeV). Nevertheless, in both Ref.~\cite{Choudhury_etal_LHT} and
Ref.~\cite{Blanke_etal} it was found that\footnote{ In terms of the
generic functions given in Eqs.~(\ref{LNPchirala},\ref{LNPchiralb})
the results reads
\bea
a_\mu^{\mathrm{LHT}}=\frac{\wz G_\mu m_\mu^2}{32 \pi^2} \frac{v^2}{f^2}
\sum_{i=1}^3\left|V_{H\ell}^{i \mu} \right|^2\left[L_1(y_i)-L_2(y_i) +\frac{1}{5}L_1(y'_i)\right]
\eea
where $L_1(z)=-\frac{13}{24}F_3^C(z)$ and
$L_2(z)=-\frac{5}{3}F_3^N(z)$ with $y_i=m_{\ell_H^i}^{2}/M_{W_H}^2$
and $y_i'=a y_i$ where $a=5/\tan^2 \Theta_W$ and $V_{H\ell}^{i \mu}$
are the couplings (denoted by $f$ in Fig.~\ref{fig:nplo}).}
\begin{equation}
a_\mu^{\mathrm{LHT}} < 12 \times 10^{-11}, 
\end{equation} 
even for such choices of the parameters at the boundary of the region allowed
by EW precision test and direct searches. Therefore, the discrepancy between
the experimental value and the SM prediction for $a_\mu$ cannot be explained
within the LHT. In fact, it was pointed out in Ref.~\cite{Blanke_etal} that
the muon $g-2$ is actually one way to discriminate the LHT from the MSSM. The
main reason for the relatively small value of $a_\mu^{\mathrm{LHT}}$ compared
to the MSSM is the absence of any enhancement factor, like a large
$\tan\beta$.

On the other hand, the small value of $a_\mu^{\mathrm{LHT}}$ also implies that
one can, at present, not obtain any strong new bounds on the parameters of the
LHT. This is in contrast to many lepton number violating processes which were
also discussed in Refs.~\cite{Choudhury_etal_LHT,Blanke_etal}. Some of these
processes could be detectable with new experiments in the coming years. These
potentially large effects are due to the fact that the LHT, in contrast to the
Littlest Higgs model without T-parity (or the MSSM), does not belong to the
class of Minimal Flavor Violation (MFV)~\cite{MFV02} models.

   \subsubsection{Extra Dimensions}
String theory requires at least 6 extra spatial dimensions (beyond the
3 we know). The extra dimensions are assumed to be curled up
(compactified). The compactification radius $R$ is not (yet)
predictable from string dynamics and thus must be considered as a free
parameter which phenomenologically could be as low as $R^{-1} \sim
300~\gv$~\cite{Antoniadis90}--\cite{GiudiceWells06}.  Phenomenological
bounds on extra dimensions have been reviewed
in~\cite{Hewett02}. Originally proposed by Kaluza and
Klein~\cite{Kaluza21Klein26}, the rejuvenation of theories with extra
dimensions came with the observation~\cite{ArkaniHamed98} that such
scenarios could provide an alternative solution (besides supersymmetry
or little Higgs models) of the hierarchy problem, as for
phenomenological reasons $1/R$ can be low and pretty close to the
electroweak scale. A true solution of the hierarchy problem would
require a dynamical explanation for the low value of $1/R$, of
course. Extra dimension models projected down to the 4--dimensional
Minkowski space are non-renormalizable effective theories and in
general require a cut-off, usually identified with the effective
Planck constant, which will be defined below\footnote{In $D>4$
coupling constants like gauge couplings carry a dimension $m^{4-D}$
and therefore the $D$--dimensional field theories with extra
dimensions are non-renormalizable}. Naive momentum cut-off truncation
of the excitation spectrum usually spoils gauge
invariance~\cite{HPW00,AHCG01} and more sophisticated constructions
are necessary. However, as we will see, some quantities in theories
with one extra dimension, in particular the anomalous magnetic moment, are
actually finite.

If particles are created they can disappear into the extra dimension
and become invisible except, in some scenarios, from missing energy
and momentum, which would be observable in laboratory experiments. On
a phenomenological level there are many possibilities concerning the
number of dimensions, the size of the compactified directions and the
type of particles which may move into them. The prototype mechanism
has been proposed by Kaluza-Klein (KK), noticing that a 5-dimensional
theory is able to describe simultaneously gravitational and
electromagnetic interactions. In extra dimension scenarios one
considers a $D$ dimensional spacetime with $\Delta=D-4$ extra space
dimensions. The space is factorized into Minkowski space times a
$\Delta$--dimensional torus of radius $R$: $M^4 \times T^\Delta$. The
torus has finite volume $V_\Delta=(2\pi\,R)^\Delta$. The SM fields are
fields on $M^4$, which is called the brane or ``the wall''. Spacetime
vectors are of the form $z^A=(x^\mu,y^i)$ and a scalar field, for
example, takes the form
\be
\phi(x,y)=\sum_{\vec{n}}
\frac{\varphi^{(\vec{n})}\ofx}{\sqrt{V_\Delta}}\,\exp \left(\I \frac{\vec{n}\cdot\vec{y}}{R} \right)
\ee
and the sum is discrete because the extra dimensions are of finite
size $R$. When compactified on a circle of radius $R$, the $M^4$
fields $\varphi^{(\vec{n})}$ are the KK excitations and represent
particles propagating in $M^4$ with masses
$m^2_{(\vec{n})}=|\vec{n}|^2/R^2+m_0^2$.  $m_0$ is the mass of the
zero mode. The masses of the KK states depend on the geometry of the
extra dimensions. Gravity extends over $D$ dimensions and the Einstein
action reads
\be
S_G=\frac{\overline{M}_D^{2+\Delta}}{2}\,\int \,\D^4x\,\D^\Delta
y\sqrt{-\mathrm{Det}\,g}\,\cR (g)
\ee
where $g$ is the metric, $\cR$ the scalar curvature and
$\overline{M}_D$ the reduced Planck mass of the $D$--dimensional
theory. Integrating out the extra coordinates under the hypothesis of
factorization one obtains
$\overline{M}_{Pl}={M}_{Pl}/\sqrt{8\pi}=(\sqrt{8\pi\,G_N})^{-1}=2.4
\power{18}~\gv$ ($G_N$ Newton's gravitational constant) from the
relation
\be
\overline{M}_\mathrm{Pl}^{2}=\overline{M}_D^{2+\Delta}\,V_\Delta
=\overline{M}_D^{2+\Delta}\,(2\pi\,R)^\Delta =M_D^{2+\Delta}\,R^\Delta
\epo
\ee
Here, by a redefinition, factors $2\pi$ have been absorbed into $M_D$
which is to be considered as a fundamental $D$--dimensional Planck
mass.

The most direct consequence of extra dimensions is due to the graviton
(G) excitations, if we assume that SM fields are confined to the brane
and do not propagate in the additional dimensions~\cite{Graesser99}. The KK gravitons
then couple to ordinary matter like the muon via the energy--momentum tensor $T^{\mu
\nu}$ with coupling proportional to $\sqrt{G_N}$ such that
\bea
\Delta  a^{\mathrm{KK}}_\mu (\mathrm{G})&=&\frac{G_N
m_\mu^2}{2\pi} \,\sum_{(\vec{n})}\,w(m^2_\mu, m^2_{(\vec{n})}) \approx \frac{G_N
m_\mu^2\,S_\Delta\,R^\Delta}{2\pi}  \,\int \D s \,s^{\Delta/2-1}\,w(m^2_\mu,s)\cs 
\eea
where we have replaced the sum by an integral, because the mass
splittings are tiny especially for the heavier states which dominate
the integral. $S_\Delta=2 \pi^{\Delta/2}/\Gamma(\Delta/2)$ is the surface of the
$\Delta$--dimensional sphere. While the contribution from a KK state
is finite the integral is divergent due to the fact that the heavier
states appear with increasing multiplicity.  One can show that $w(m^2_\mu,s) \to c$ as $s \to
\infty$ with $c=5$~\cite{Graesser99}. With a cut-off
$\Lambda=\lambda\,M_\Delta$ one obtains
\be
\Delta  a^{\mathrm{KK}}_\mu (\mathrm{G})=\frac{c \lambda^{\Delta}}{4\pi^2}
\frac{\pi^{\Delta/2}}{\Delta \Gamma(\Delta/2)} \frac{m_\mu^2}{M^2_D}\epo
\label{EDgraviton}
\ee
In addition there exists a \textit{radion field} due to the
fluctuation of the radius, a spin 0 field which couples to the trace
$T^\mu_{~\mu}$. This radion yields a tiny contribution
only~\cite{Graesser99} and will not be considered further. Using
Eq.~(\ref{EDgraviton}), we in principle can get a constraint on $M_D$
which strongly depends on $\Delta$, however.  For $\Delta=2$ severe
constraints from cosmology and astrophysics exist, requiring $M_6 >$ a
few TeV~\cite{Hewett02}. If $\Delta =6$ (as required by string theory)
the bound is much weaker and actually comes from the colliders LEP and
Tevatron $M_{10}>$ 600 -700 GeV. The bound Eq.~(\ref{amuNP}) interpreted
as $\delta a_\mu < 380 \power{-11}$ requires $M_{10}>$ 980 GeV [for $M_5$
one gets 610 GeV as a limit].  

Another simple extra dimensions scenario is the one where only the gauge
bosons propagate into the extra dimensions, while fermions and the
Higgs remains in ordinary space. The Lagrangian in $M^4$ reads
\be
\cL_\mathrm{int}=\sum_{i=A,Z,W}\,g_i\/j_i^\mu\,\left( A_{\mu i}+ \wz
\,\sum_{n=1}^{\infty} A_{\mu i}^{(n)}\right) 
\ee 
where $g_i =\sqrt{\pi R}\,g_i^{(D)}$ are the 4--dimensional
effective couplings, rescaled from the $D$--dimensional
theory. The fields $A_{\mu i}^{(n)}$ denote the KK excitations, the
zero mode Lagrangian is the SM one.  The masses $M_n$ of the
gauge--boson KK excitations are related to the masses
$M_0$ of the zero-mode gauge bosons (i.e. the normal SM gauge bosons) by
\be
M_n^2=M_0^2+n^2/R^2 \mathrm{ \ with \ \ } n^2=\sum_{i=1}^{\Delta} n_i^2
\ee 
and their couplings $g_n$ are given by $g_n=\sqrt{2}\,g$, where $g$ is
the standard coupling of the zero mode. Relative to the SM, the extra
effects $\amu^{\mathrm{KK}}$ to $g-2$ from a KK-tower of states are
proportional to a factor $\sum_n D_n \frac{m_\mu^2}{M_n^2}$, where
$D_n$ is the degeneracy of the $n$th KK
mode~\cite{Graesser99}--\cite{CCC01}.  For $\Delta=1$ we have $D_n=1$
and thus
\ba
\!\!\!\!\!\!\sum_n D_n \frac{m_\mu^2}{M_n^2}/(m_\mu\,R)^2&=& \sum_n  \frac{D_n}{(M_0\,R)^2+n^2}
=\sum_n \frac{1}{(M_0\,R)^2+n^2} \approx 
\sum_n \frac{1}{n^2}=\frac{\pi^2}{6}  
\label{KKfactor}
\ea
In general all SM particles propagate into the extra dimensions and
affect physics in the $D=4$ subspace.  In the Appelquist, Chang and
Dobrescu model~\cite{ACD00} with one universal extra dimension (UED)
compactified as $S^1/Z_2$ (even number of additional dimensions are
compactified as $T^2/Z_2$) there is only one new parameter $1/R$ the
compactification scale. Universal refers to the property that all SM
fields propagate into the extra dimensions. The interaction of all KK
modes among themselves and with the SM fields are all expressed in
$1/R$ and the SM parameters. One thus obtains
the prediction~\cite{Rizzo01,Appelquist01}
\be
\Delta  a^{(2)\:\mathrm{KK}}_\mu =
-\frac{g^2}{8\pi^2}\,\frac{3-4s^2+\Delta/2\,(3+8s^2)}{12c^2}\,\sum_n
D_n \frac{m_\mu^2}{M_n^2}
\ee
using $s^2=\sin^2 \Theta_W=0.231$ one obtains
\be
\Delta  a^{(2)\:\mathrm{KK}}_\mu =-24.8 \power{-11}\,(1+1.2\,\Delta)\,S_{\mathrm{KK}}
\ee 
with $S_{\mathrm{KK}}=\sum_n \frac{6D_n}{\pi^2}\,\left[ \frac{300~\gv\,}{M_n}\right]^2.$
This sum is convergent only for a single extra dimension where $D_n=1$
and where the smallest value allowed for $1/R$ is about $ \sim
300~\gv\,$, giving $S_{\mathrm{KK}} \approx 1$ and one finds
\be
\Delta  a^{(2)\:\mathrm{KK}}_\mu =-0.276 \cdot a^{(2)\:\mathrm{EW}}_\mu
= -53.7[-24.8] \power{-11}\epo
\ee
which differs numerically from the value $-12.8 \power{-11}$ given
in~\cite{Appelquist01} and is closer to the $-40 \power{-11}$ estimated 
in~\cite{Rizzo01}. 
Given in brackets the value when scalar modes are
not included, which corresponds and essentially agrees with the result
in Eq.~(\ref{eq:CCC})  [at $1/R=300~\gv$ the result is
$-20.4 \power{-11}$] below. Note that the $\Delta$ scalar components of the
$D$ dimensional gauge fields for $D=5$ enhance the effect by a factor 2. 
Another interesting scenario proposed by Barbieri, Hall and Nomura~\cite{Barbieri00} 
assumes an extension of the SM to $D=5$ dimensions,
with N=1 supersymmetry, compactified on $M^4 \times S^1/(Z_2 \times
Z_2')$ with a compactification scale $1/R \sim 370 \pm 70~\gv$. Again, the model
allows for unambiguous predictions depending on $R$ and the SM
parameters only. The modification of $g-2$ has been worked out in~\cite{CCC01} 
and reads
\ba
\Delta  a^{(2)\:\mathrm{KK}}_\mu =
-\frac{g^2}{192}\,\frac{11-18s^2}{12c^2}\,(m_\mu\,R)^2
\ea
Numerically, for $1/R=370\pm70~\gv$, it is given by
\ba
\Delta  a^{(2)\:\mathrm{KK}}_\mu &=& - 0.07^{+0.04}_{-0.02} \cdot 
\amu^{(2)\:\mathrm{EW}}=-(13.6^{+7.1}_{-4.0})\power{-11}
\label{eq:CCC}
\ea
which again for any sensible value of $R$ yields a result well inside
the uncertainties of the SM prediction. 

A number of authors have obtained results of similar size but of opposite
sign~\cite{Graesser99,Nath99,Casadio00,Park01}.  The corrections of
large extra dimension scenarios are rather small, below $1\sigma$, for
sensible values of the compactification radius $R$, typically 10\% -
25\% of the weak SM contribution and inside the uncertainties of the
hadronic contribution. In more than one extra dimension a second
cut-off $M_S \gg 1/R$ must come into play in a calculation of $a_\mu$
and the predictive power is reduced in general. The positive graviton
KK contribution also for $D=5$ requires a cut--off.  Using $M_D \sim
1.2 \power{13}$ for $D=5$ and $1/R\sim300$ GeV a contribution is
practically absent; a phenomenological cut--off $M_5 \sim 3$ TeV,
however, would yield $\Delta a^{\mathrm{KK}}_\mu (\mathrm{G}) \sim
15\power{-11}$ which would compete with other type of extra dimensions
contributions.

   \subsubsection{Anomalous Gauge Couplings}
It is a common belief that the SM is a low energy effective theory
which results from an expansion in $E/\Lambda$, where $E$ is a
laboratory energy scale and $\Lambda$ is
the intrinsic cut--off of nature at the ``microscopic'' level. In this
case it is natural to ask for corrections to the SM as a renormalizable low
energy effective theory by operators of higher dimension. But also compositeness
scenarios would lead to the same structure of effective interactions.
Here we only consider P--conserving anomalous interactions of the 
$WW\gamma$--type as an illustration. It is a correction to
the weak 1--loop contribution from diagram Fig.~\ref{fig:oneloopweak}a.
The most natural non-standard couplings correspond to an anomalous magnetic dipole
moment (see~\cite{FJano94} and references therein)
\ba
\mu_W = \frac{e}{2m_W} (1+\kappa+\lambda)
\ea
and an anomalous electric quadrupole moment
\ba
Q_W = -\frac{e}{2m_W} (\kappa - \lambda) \epo
\ea
In the SM, local gauge symmetry, which is mandatory for
renormalizability of the SM, requires $\kappa = 1$ and $\lambda = 0$.
These anomalous couplings yield the following contribution to $a_\mu$~\cite{MMPR89}:
\ba
a_\mu(\kappa,\lambda) \simeq
\frac{G_\mu m_\mu^2}{4\sqrt{2} \pi^2}
 \left[ (\kappa-1) \ln\frac{\Lambda^2}{m_W^2} - \frac{1}{3}\lambda\right]\epo
\ea
Actually, the modification spoils renormalizability and one has to
work with a cut--off $\Lambda$ in order to get a finite answer as
usual in an effective theory. For $\Lambda
\simeq 1$ TeV  LEP data from $e^+e^-\to W^+W^-$ production find
bounds $\kappa-1 = -0.027\pm 0.045$, $\lambda = -0.028\pm
0.021$~\cite{PDG06,LEPTGC}.  Applying the LEP bounds we can get not
more than $a_\mu(\kappa,0) \simeq (-3.3\pm 5.3) \power{-10}$,
$a_\mu(1,\lambda) \simeq ( 0.2\pm1.6) \power{-10}$, and thus the
observed deviation cannot be due to anomalous $WW\gamma$
couplings. The constraint on those couplings from $g-2$ is at least an
order of magnitude weaker than the one from LEP. This also illustrates that 
$\amu$ is not per se the most suitable observable for testing physics
at the high end of scales.

\section{Outlook and Conclusions}
\label{sec:Conclusions}

The measurement of $\amu$ is one of the most stringent tests for ``new
physics'' scenarios, thanks to the current impressive precision.  The
Brookhaven muon $g-2$ experiment marks a new milestone in precision physics,
with a 14--fold improvement over the previous CERN experiment. The precision
achieved by the experiment E821 at BNL at the end was still statistics
dominated and just running the experiment a little longer could have reduced
the error down to the original goal of $\pm 40\times 10^{-11}$.  The recent
result of the Brookhaven experiment has highlighted a 3$\sigma$ discrepancy
with the present SM prediction and hence makes comparisons with reliable
predictions from extensions of the SM very interesting. Most interestingly,
there is now a high tension between the $g-2$ constraints and other precisely
measured observables, which provides very useful hints for where to look for
the ``new physics'' at the upcoming high energy frontier experiments at the
Large Hadron Collider (LHC) at CERN. This is a strong motivation for a
next-generation $g-2$ experiment~\cite{Hertzog_etal_07}.

\subsection{Future Experimental Possibilities} 

As a possible follow--up experiment of
E821 a proposal for an upgraded experiment E969 has been submitted some
time ago~\cite{LeeRob03,Hertzog08}.  Considerations aim at $\sim 15
\power{-11}$ experimental precision.  The BNL storage ring would
remain the key element. Different options are discussed for Brookhaven
and for Fermilab in the USA and for JPARC in Japan. A higher rate and
less background would be important and this requires several changes
in the experiment. At JPARC with a high-intensity 30 GeV proton beam
one could realize a backward decay beam which allows a better
separation of muons and pions. A similar effect could be achieved
using backward muons together with a muon accumulator ring at
Fermilab or with a longer beam line. At FNAL statistics could be
increased by a factor 25 due to the higher repetition rate of muon
fills in the ring.

In order to further reduce the systematic errors one would also require a more
uniform magnetic field and an improved centering of the muon beam and an
improvement of the calibration. For the detection and analysis of the
precession signal one could use a complementary method of determining
$\omega_a$ by measuring the energy versus time in place of events versus
time. The monitoring detectors would be improved by a better segmentation.  A
reduction of the pileup and muon losses (better kicker) would allow for
further improvements.

This all looks very promising and will likely happen in a not too far
future.  For the theory side this represents a tremendous challenge.
Some possible improvements and open problems will be discussed in the
following outlook.

\subsection{Theory: Critical Assessment of Theoretical Errors and Open
Problems} 

\subsubsection{QED}
A first type of problems concerns the extremely complex four-- and five--loop
calculations, so far almost exclusively done by Kinoshita and his
collaborators, which require cross checks by independent groups. Indeed new
more effective numerical methods are being developed at
present~\cite{QEDfuture}. The 5--loop calculations are just at the
beginning~\cite{KinoNio06,Aoyama:2008gy,Baikov:2008si}. Here the problem is to
master the exponentially growing complexity with increasing order of
perturbation theory. Only a very high level of automatization of such
calculations will allow us to get reliable results within a reasonable
time. Of course the increasing computational power available will also be a
key factor. As such kind of calculations are needed in many other branches of
particle physics, like for understanding and analyzing LHC data or for future
precision experiments with an $\epm$ linear collider like the ILC or at CLIC,
there is little doubt that big progress will be made in all kinds of
challenging perturbative calculations.

\subsubsection{Hadronic vacuum polarization}
The main conceptual problems remain with the hadronic contributions
which are limiting the theoretical precision.  The hadronic vacuum
polarization requires substantial improvement and will depend very
much on new improved $\epm$--annihilation experiments in particular in
the range up to 2.5 GeV. A serious effort in improving calculations of
radiative corrections to these hadron production processes will be
urgently needed. In the region including many--hadron final states
inclusive versus exclusive approaches can provide important cross
checks for estimating the true uncertainties. A major step in the
reduction of the uncertainties are expected from
VEPP-2000~\cite{VEPP2000} and from a possible ``high-energy'' option
DAFNE-2 at Frascati~\cite{DAFNE2}. Also important are forthcoming
hadronic cross section measurements at BES III (Beijing)~\cite{BES3},
CLEOc (Cornell/USA), and last but not least, additional very useful
results from radiative return measurements are expected from the
B--factories BaBar and Belle.

Since hadronic $\tau$--decay spectra and $\epm \to \mathrm{hadrons}$
data should be related by isospin symmetry one expects to be able to
further reduce the errors by including the $\tau$ data. However, one
observes substantial yet unexplained deviations between these data,
and therefore usually one does not include the $\tau$ data. A solution
of this $\epm$ vs. $\tau$ puzzle also would help to improve the determination 
of $\amuh$.

A better theory supported method in determining in particular the
dominating low energy $\pi\pi$ channel is possible. The theory of the
pion form--factor, based on exploiting analyticity, unitarity and the
low energy chiral expansion, looks very promising for future
improvements~\cite{LeCo02}.  This interesting framework systematically
takes into account the constraints from the experimentally well
determined $\pi\pi$--scattering phase shifts. However, in order to be
able to fully exploit this approach one has to reduce the presently
existing inconsistencies between the different measurements of the
modulus of the pion form factor. Furthermore, in such a theory-driven
approach to the lowest order hadronic vacuum polarization
contribution, higher order corrections shown in
Fig.~\ref{fig:ammhohad}d) would then have to be added ``by hand''
using some hadronic models for the corresponding 4-point function
$\langle VVVV \rangle$. In this context it is important to remind that
future experiments must attempt to actually measure these final state
radiation effects~\cite{Gluza:2002ui,Pancheri:2006cp}.

With more precise values for the QCD parameters, in particular the
quark masses $m_c$ and $m_b$~\cite{KSS07}--\cite{Heitger:2003nj}, one
will be able to make more precise pQCD calculations of the photon
self-energy in the Euclidean region, at $Q^2=-q^2\gapprox~(2.5~\gv)^2$,
say. There, the light quarks are essentially massless and also can be
treated perturbatively.  This will allow us to calculate $\amuh$ via
the Adler function~\cite{EJKV98,FJ03,FJ08}, by exploiting more
extensively and in a well controlled way pQCD. This would provide
determinations of $\amuh$ in a manner which is less dependent on
experimental data and their uncertainties.

An important long term project are non-perturbative calculations of
the vacuum polarization function in lattice
QCD~\cite{VP_lattice}. Since we know that the hadronic vacuum
polarization is dominated by the contribution from the
$\rho$--resonance, it is crucial to have the physical $\rho$ ``in the
box'', i.e. it requires to simulate at physical parameters (quark
masses) and in full QCD including the $u$, $d$, $s$ and $c$
quarks. The advantage is that for calculating $\amuh$ the photon
vacuum polarization is needed at Euclidean momenta only, which is
directly accessible to lattice simulations. The aim of a precision at
the 0.5\% level needed in future, is certainly very hard to
reach. But also less precise results would be very important in view
of the existing problems, which we have discussed in
Sect.~\ref{sec:hadvap}. Note, however, that again higher order radiative
effects are not included in a lattice study of the two-point function $\langle
VV \rangle$.

\subsubsection{Hadronic light-by-light scattering}

In contrast to the hadronic vacuum polarization contribution to $a_\mu$, no
direct connection with experimental data can be made for hadronic
light-by-light scattering and, very likely, we have to rely on models in the
foreseeable future.  Because of the increased accuracy of a potential future
$g-2$ experiment and the expected substantial reduction of the error of the
other hadronic contributions to $a_\mu$, in particular the hadronic vacuum
polarization, also a reconsideration of the hadronic light--by--light
contribution is certainly needed. Whereas some control over the pseudoscalar
and axial vector contributions has been achieved by implementing many
experimental and theoretical constraints, this is not the case for the other
contributions, like the dressed pion loop, the scalar exchanges and the
dressed quark loop. Unless one succeeds in getting a much better control on
the corresponding model uncertainties, it will be very difficult to reduce the
final overall error of the hadronic light-by-light scattering contribution
below the ``guesstimate'' of $\pm 40 \times 10^{-11}$ quoted in
Section~\ref{sec:LbLsummary}.

In view of recent and foreseeable progress in computer performance,
and using recently developed much more efficient computer simulation
algorithms, we expect that lattice QCD will be able to provide a 
useful estimate in coming years. Exploratory
studies~\cite{lbl_lattice,lbl_lattice2} actually look rather
promising. Recall that one has to integrate the relevant QCD
four-point function $\langle VVVV \rangle$ over the full phase space
of the three off-shell photons, which is much more complicated than
what has been done in the lattice studies for the hadronic vacuum
polarization contribution to $g-2$~\cite{VP_lattice}. However, even a
number for the light-by-light scattering contribution with an error of
50\% (but ``reliable'' !) would be very helpful to complement the
other approaches.

In the following, we suggest some potential ways of
improvement. Probably only a collaborative effort of all interested
people (maybe in the form of a hadronic light-by-light scattering
working group) can lead to real progress and to a final estimate for
the central value and its error, which is trustworthy.

The first, most important point is that one has to have one consistent
framework (model) for all contributions, since the splitting into different
contributions is model dependent as stressed in Sect.~\ref{sec:lbl}. A priori
this was already the case for the two existing full evaluations by Bijnens et
al.~\cite{BijnensLBL} using the ENJL model and Kinoshita et
al.~\cite{HKS95,HK98} employing the HLS model, but only to a certain
extent. In both cases the general framework had to be adjusted for some
contributions, in particular for the pseudoscalar exchange and the charged
pion loop. However, as soon as one tries to ``improve'' (``repair'') the model
for one contribution, one has to make sure that this is consistent for the
other contributions.

One purely phenomenological approach would be to use some resonance Lagrangian
which contains the lowest lying and possibly heavier states, where all the
couplings have been determined by some physical processes (e.g. from resonance
decays or from the contribution of these resonances to the production of a
certain number of pions in $e^+ e^-$
scattering~\cite{Eidelman:1994zc,Juran:2008kf,Ivashyn:2007yy}, etc.). It
should be kept in mind with such an approach that for instance different
formalisms can be used to describe vector mesons (vector field, antisymmetric
tensor field, massive Yang-Mills, etc.). Even imposing chiral and gauge
invariance does not uniquely fix the corresponding Lagrangian and certain
couplings might be absent in one particlar model, unless one adds terms with
more derivatives or more resonance fields. Furthermore, there is no consistent
chiral counting to discard certain terms as higher order in $p^2$ in such a
resonance Lagrangian. 

Note that even if all parameters in such a resonance Lagrangian can be fixed
by some decay or scattering processes, this does not guarantee that
integrating over the photon momenta in hadronic light-by-light scattering, or
by performing loops with resonances, will lead to ultraviolet finite
results. In general, these resonance Lagrangians are non-renormalizable and we
will need some momentum cutoff or some new counter-terms with a priori unknown
finite coefficients. The question is whether varying the cutoff in some
``reasonable'' range, like 1-2 GeV, will lead to a stable result, if the high
momentum region is modeled by a (dressed) quark loop. Note, however, that
Ref.~\cite{Prades:2009tw} does not include the dressed light quark loops as a
separate contribution. It is assumed to be already covered by using the
short-distance constraint from Ref.~\cite{MV03} on the pseudoscalar-pole
contribution. This issue certainly needs to be clarified.

Another approach is based on the large-$N_c$ framework advocated in
Ref.~\cite{LMD98}. One tries to write ans\"atze in terms of exchanged
resonances for all the relevant QCD Green's functions and matches them with QCD
short distance constraints and the chiral Lagrangian at low energies. In
principle, also in this case one can use some chiral and gauge-invariant
resonance Lagrangian and then fix (some of) the coefficients by matching with
QCD at high energies. The advantage of the Lagrangian approach, in contrast to
the LMD or LMD+V ans\"atze used in Refs.~\cite{KnechtNyffeler01,MV03}, is
that for instance crossing symmetry is automatically fulfilled and that the
connection between the parameters which enter in different Green's functions
is explicit. Also in this case one has to fix the other parameters in the
Lagrangian using some experimental input. The problem is that in such an
approach one has to start with the most general effective Lagrangian and this
will lead to many unknown parameters. This is in contrast with the above
phenomenological approach, where only those couplings are introduced, which
are ``needed'' to describe a particular process.

For both approaches it is crucial to have as much experimental information as
possible, e.g.\ data for various form factors (on-shell and off-shell), to
constrain these model Lagrangians or to check their consistency, if some
parameters are fixed by QCD short-distance constraints. A new
important constraint  for the internal $\pi^0\gamma\gamma$--vertex  
would be a measurement of $\FFc(m_\pi^2,-Q^2,-Q^2)$, which
should be possible at future high luminosity machines.

One problem with the resonance Lagrangian approach is how to go beyond the
leading order in $N_c$. In most cases, such resonance Lagrangians are only
used at tree-level, which corresponds to the leading order in $N_c$. How can
one consistently include loops with resonances (some progress has been made,
see Refs.~\cite{loops_with_resonances,Ivashyn:2007yy}) and how can we account for finite width
effects ? Maybe the latter point is not so relevant for hadronic
light-by-light scattering, since one can always evaluate the contribution to
$g-2$ in Euclidean space in order to avoid physical poles and thresholds. A
somehow related problem concerns the scalars. Can we really use a simple
resonance Lagrangian, which usually assumes narrow states ? Often the scalar
states are very broad or it is not clear, whether they are really some $q{\bar
q}$ bound states or rather some four quark states or meson
molecules. Furthermore, how can we constrain the scalar sector theoretically,
i.e. by matching with QCD short-distance constraints, since the scalars
directly couple to the non-perturbative vacuum of QCD.

One important step within such a resonance Lagrangian framework would then be
to check that the so far neglected contributions from heavier states, like
$\pi^\prime$ or some tensor mesons, are really small. Furthermore, one should
check how stable the final result is, if we include such heavier resonance
states and, at the same time, integrate in the (dressed) quark loop only the
high-momentum region above all those resonances, e.g.\ from 1.5 GeV upwards
(quark-hadron duality). One problem might be that the quark loop also serves
as some sort of counterterm to absorb the cutoff dependence of the integrals
including the resonances. If there is no matching between low and high
momenta, the final result will likely be very dependent on the exact choice of
the cutoff.

The same four-point function $\langle VVVV \rangle$ which is relevant
for hadronic light-by-light scattering also enters at higher order in
the hadronic vacuum polarization (hadronic blob with an additional
photon attached to itself, see Fig.~\ref{fig:ammhohad}d)), although of
course in a different kinematical region. It would be a good
cross-check on all models for hadronic light-by-light scattering, if
they could successfully predict this contribution. In the more recent
$\pi\pi$ cross-section measurements, this contribution is usually
included in the data for $e^+ e^- \to \mbox{hadrons}$. In fact
experiments always have to apply cuts to the photon spectrum and the
missing hard final state radiation is added ``by hand'', assuming sQED
to provide a reasonable approximation.

Finally, it might be useful to complement these approaches using resonance
models with a more theoretical framework, where one tries to identify
different contributions in the full light-by-light scattering amplitude in a
dispersive framework, looking for the one-pion cut, the two-pion cut, etc., in
the spirit of the theory of the pion form factor developed in
Refs.~\cite{LeCo02,LeCo08}. It might also be interesting whether one
could make some model-independent statements (exact low-energy theorems) in
some particular limits, e.g.\ to identify the dominant contribution, if $m_\mu
\to 0, m_\pi \to 0$ with the ratio $m_\mu / m_\pi$ fixed to its physical
value. For instance, can one prove without resorting to some particular
Feynman diagram with model-dependent form factors, that the leading term is
completely given by the (undressed) ``charged pion loop''. Even if one might
not be able, with such an approach, to obtain precise numerical predictions
for the light-by-light scattering contribution in the real world, the
knowledge of the result in some limits could help to check the consistency of
calculations within some models or on the lattice, e.g. the size and sign of
different contributions, like from scalar exchanges. Some progress in this
direction has been achieved in the effective field theory approach described
in Sect.~\ref{sec:LbLEFT}, where the leading logarithmically enhanced terms
for large $N_c$ have been identified.

It remains to be seen, whether such a concerted effort can reliably
predict the light-by-light scattering contribution to $a_\mu$ with an
error comparable with the final goal of a future experiment, i.e. about
$\pm 15 \times 10^{-11}$.  Of course, the corresponding contribution to
the electron $a_e^{\rm LbL; had}$ should also be evaluated at the same time.
We hope that some progress can be achieved by taking into account the
issues which have been brought up in this Section and in the recent
literature~\cite{KnechtNyffeler01,MV03,BP07}. If not, hadronic
light-by-light scattering might be the future roadblock to strongly
constrain new physics models, in case a new muon $g-2$ experiment will be
carried out in the future.

Another possibility, proposed by Remiddi some time ago, would be to increase
the precision of independent measurements of $\alpha$ by a factor 20 and then
use $a_e$ together with improved QED and SM calculations and determine $a_\mu
({\rm unpredicted})$ assuming that it is proportional to $m_\ell^2$:
thus from 
\ba
a_\mu & = & a_\mu({\rm predicted}) + a_\mu({\rm unpredicted}) \cs \nonumber \\ 
a_e   & = & a_e({\rm predicted}) + (m_e/m_\mu)^2\,\times\,a_\mu({\rm
  unpredicted}) \cs \nonumber
\ea
we could determine $a_\mu({\rm unpredicted})$.  This quantity would, however,
not only include the hard to pin down hadronic contributions, but also
unaccounted new physics effects which scale with $m_\ell^2$. Unfortunately,
the leading hadronic contributions are due to the $\rho^0$ resonance in the
hadronic vacuum polarization or to light states like $\pi^0$--exchange in
light-by-light scattering, which do not scale in this simple manner in
general. Nevertheless, such attempts to constrain whatever kinds of physics
would be certainly very interesting.

\subsection{Conclusions}

The BNL muon $g-2$ experiment has stimulated significant progress in
theory as well as in hadronic cross--section measurements. QED and
electroweak contributions are well established at the required level
of precision. Fortunately, substantial experimental improvements of
the $\epm$--annihilation data allowed to reduce the uncertainties of
the theoretical prediction to match the experimental precision,
although there remain some as yet not understood deviations in
particular between different $\pi\pi$ cross--section
measurements. Here also a clarification of the $\epm$ vs. $\tau$
puzzle seems to be crucial for a better understanding of possible
experimental and/or theoretical problems.  The main problem at present
is the limited precision of the $\epm$ data in the range between 1 and
2 GeV.  We expect further improvements by about a factor 2 in accuracy
of $\amuh$ after VEPP-2000, DA$\Phi$NE-II, CESRc and
$c-\tau$--factories [Beijing], as well as further results from
radiative return measurements at DA$\Phi$NE and the B--factories, will
be available. The hadronic light--by--light scattering contribution
will then be limiting further progress in the prediction of $g_\mu-2$.
The actual 3.2 $\sigma$ deviation between $a_\mu^\mathrm{exp}$ and
$a_\mu^\mathrm{the}$ opens plenty of room for speculations about its
origin and we hope it is a true effect, due to physics beyond the
Standard Model.

Even if there are some serious limitations in controlling the hadronic
effects (see also~\cite{Passera:2008jk}), the current very precise SM
prediction of the muon \amm is pretty safe and a real triumph for the
SM, which incorporates, except from gravity, all particle interactions
within the unifying framework of a spontaneously broken $SU(3)_c
\otimes SU(2)_L \otimes U(1)_Y$ gauge theory with 3 families of
leptons and quarks. The muon \amm is a pure quantum observable,
unambiguously predicted by the SM, and known at a precision which
unfolds the whole spectrum of physics incorporated in the SM. Only the
Higgs boson, due to its tiny coupling to light fermions, is far from
playing a relevant role.  The muon anomaly's sensitivity to new
physics is largest to nearby new states or new effective interactions,
while too heavy states yield too small effects to be seen. At the same
time light states are severely constrained in particular by the LEP and
Tevatron new physics searches. In any case the muon $g-2$ with its
3$\sigma$ disagreement / agreement between theory and experiment
provides important constraints to parameters of new physics models as
outlined in some detail above. It certainly also helps to disentangle
the plentitude of possibilities we have to go beyond the SM and can
complement upcoming new physics searches at the LHC.

An improved muon $g-2$ experiment would challenge new efforts to improve the
accuracy of the theoretical prediction and corroborate the existence
of physics beyond the SM which we belief must be there in any case.

\vspace*{0.5cm} 


\noindent 
{\bf Acknowledgments}

We would like to thank Oliver B\"ar, Gilberto Colangelo, J\"urg Gasser, Andrei
Kataev, Wolfgang Kluge, Heiri Leutwyler, Peter Minkowski, Stefan M\"uller,
Federico Nguyen, Giulia Pancheri, Ximo Prades, Rainer Sommer, Arkady
Vainshtein and Graziano Venanzoni for numerous stimulating discussions and for
correspondence.  A.N.\ is grateful to Marc Knecht, Michel Perrottet and
Eduardo de Rafael for many discussions and the pleasant collaboration on the
hadronic light-by-light scattering contribution.  F.J.\ gratefully
acknowledges the kind hospitality at Frascati National Laboratory and thanks
Staszek Jadach and his team from the Niewodniczanski Nuclear Physics Institute
at Krakow and Karol Ko\l odziej and his Colleagues from the Institute of
Physics of the University of Silesia at Katowice for the kind hospitality
extended to him. A.N.\ thanks the Institute for Theoretical Physics at the
University of Bern, the Centre de Physique Th\'eorique in Marseille, the
Theory Group at DESY Zeuthen and the Institute for Theoretical Physics at ETH
Z\"urich for the hospitality during some stages of writing this review. F.J.
gratefully acknowledges support by the Alexander von Humboldt Foundation
through the Foundation of Polish Science. This work was supported in part by
the EU grants MTKD-CT2004-510126, in partnership with the CERN Physics
Department and with the TARI Program under contract RII3-CT-2004-506078, and
by the Department of Atomic Energy, Government of India, under a 5-Years Plan
Project.

\appendix 
\section{Some Standard Model parameters, $\zeta$ values and polylogarithms}
\label{sec:appA}
\noindent
For calculating the weak contributions we need the Fermi constant
\be
G_\mu=1.16637(1) \power{-5}~ \gv^{-2}\cs
\label{gmu}
\ee
the weak mixing parameter (here defined by $\sin^2 \Theta_W=1-M_W^2/M_Z^2$)
\be
\sin^2 \Theta_W=s_W^2=0.22276(56)
\label{sin2W}
\ee
and the masses of the intermediate gauge bosons $Z$ and 
$W$ (see~\cite{PDG06,LEPEWWG06})
\be
M_Z=91.1876 \pm 0.0021 \ \gv \css M_W= 80.398 \pm 0.025 \ \gv \epo
\label{GBmasses}
\ee
In the SM the Higgs particle the mass is
constrained by LEP data to the range~\cite{LEPEWWG06}
\be
114 \ \gv < m_H <144 \ \gv (\mathrm{at \ } 95\% \mathrm{ \ C.L.}) \epo
\label{Hmassbound}
\ee

\noindent
In calculations of hadronic contributions we will use
the charged pion mass
\ba
m_{\pi^\pm}     &=&  139.570\,18\,(35) \mv \;,
\label{pionmassc}
\ea
and the neutral pion mass $m_{\pi^0}$ which has the value
\ba
m_{\pi^0}     &=&  134.976\,6\,(6) \mv \epo
\label{pionmassn}
\ea
In some cases quark masses are needed.
For the light quarks $q=u,d,s$ we give $m_q=\bar{m}_q(\mu=2~ \gv)$,
for the heavier $q=c,b$ the values at the mass as a scale
$m_q=\bar{m}_q(\mu=\bar{m}_q)$ and for $q=t$ the pole mass:
\ba \label{currentquarkmasses}
\begin{tabular}{rrclrrclrrcl}
 $m_u=$ & $3$     & $\pm$ & $1\,\mv$ & 
 $m_d=$ & $6$     & $\pm$ & $2\,\mv$ &
 $m_s=$ & $105$   & $\pm$ & $25\,\mv$ \\ 
 $m_c=$ & $1.25$  & $\pm$ & $0.10\,\gv$~~~~~&
 $m_b=$ & $4.25$  & $\pm$ & $0.15\,\gv$~~~~~& 
 $m_t=$ & $172.6$ & $\pm$ & $1.4\,\gv~.$\\ 
\end{tabular}
\ea
When using constituent quark masses we will adopt the values
\be
\!\!\!\!M_u=M_d=300\,\mv\cs M_s=500\,\mv\cs M_c=1.5\,\gv\cs M_b=4.5\,\gv \epo
\label{CQMasses}
\ee
It should be noted that for $u,d$ and $s$ quarks such large effective light
quark masses violate basic chiral \WTis of low energy QCD. The latter requires
values like those in Eq.~(\ref{currentquarkmasses}) for the so called
\textit{current quark masses} to properly account for the pattern of chiral
symmetry breaking. 

\noindent
Typically, analytic results for higher order terms
may be expressed in terms of the Riemann zeta function
\be
\zeta(n)= \sum\limits_{k=1}^{\infty} \frac{1}{k^n}
\label{Zetadef}
\ee
and of the polylogarithmic integrals
\be
\mathrm{Li}_n(x)=
\frac{(-1)^{n-1}}{(n-2)!}\int\limits_{0}^{1}\frac{\ln^{n-2}(t)\ln(1-tx)}{t}dt=
\sum\limits_{k=1}^{\infty}\frac{x^k}{k^n}\cs
\label{Lindef}
\ee
where the dilogarithm $\mathrm{Li}_2(x)$ is often referred to as the Spence
function. The series representation holds for $|x|\leq 1$. The dilogarithm is
an analytic function with the same cut as the logarithm. Useful relations are 
\ba
\Sp (x) &=& -\Sp (1-x) +\frac{\pi^2}{6}-\ln x \ln (1-x), \crn
\Sp (x) &=& -\Sp (\frac{1}{x}) -\frac{\pi^2}{6}- \ha \ln^2(-x), \crn
\Sp (x) &=& -\Sp (-x) + \ha \Sp(x^2)\epo
\label{spencerel}
\ea
Special values are:
\be
\Sp (0)=0\cs \Sp (1)=\frac{\pi^2}{6}\cs \Sp (-1)= -\frac{\pi^2}{12}\cs
\Sp ( \ha ) = \frac{\pi^2}{12}-\ha (\ln 2)^2 \epo
\ee
Special $\zeta(n)$ values we will need are 
\be
\!\!\!\!\!\!\zeta(2)=\frac{\pi^2}{6}\,,\; \zeta(3)=1.202~056~903\cdots \,,\; 
\zeta(4)=\frac{\pi^4}{90}\,,\; \zeta(5)=1.036~927~755\cdots \,.
\label{Zetanum}
\ee 
Also the constants
\ba
\mathrm{Li}_n(1)&=&\zeta(n)\cs \mathrm{Li}_n(-1)=-[1-2^{1-n}]\:\zeta(n), \crn
a_4 &\equiv&
\mathrm{Li}_4(\frac{1}{2})=\sum_{n=1}^{\infty}1/(2^nn^4)=0.517~479~061~674 
\cdots\;,
\label{Linum}
\ea
related to polylogarithms, will be needed for the evaluation of
analytical results.

\end{document}